\documentclass[twocolumn,secnumarabic,amssymb, nobibnotes, aps,prx]{revtex4}
\usepackage{CJKutf8}
\usepackage{comment}
\usepackage{amsmath,amsthm,amssymb}
\usepackage[pdftex]{graphicx}
\usepackage{color}
\usepackage[abs]{overpic}

\newcommand*{\Ja}[1]{\begin{CJK}{UTF8}{ipxm}#1\end{CJK}}
\begin{document}
\title{
% A minimum extension of the logistic equation to consistently describe the growth curves of appearances of new vocabularies on nation-wide online social media: An analysis of the diversity of growth in society
%A minor extension of the logistic equation for growth curves of word counts on online social media: A parametric description of the diversity of growth in society
%Minor extensions of the logistic equation for growth curves of word counts on online media: Parametric description of growth diversity
%A minor extension of the logistic equation for growth of word counts on online media: Parametric description of social growth diversity
%A minor extension of the logistic equation for growth of word counts on online media: Parametric description of diversity of growth phenomena in society
Minor extensions of the logistic equation for growth curves of word counts on online media: Parametric observation of diversity of growth in society
}
% Use letters for affiliations, numbers to show equal authorship (if applicable) and to indicate the corresponding author
\author{Hayafumi Watanabe$^{1,2,3}$}\email[E-mail: ]{hayafumi.watanabe@gmail.com}
%\author[b,1,2]{Author Two} 
%\author[a]{Author Three}
\affiliation{Department of Economics, Seijo University, Setagaya-ku, Tokyo 157-8519, Japan}
\affiliation{College of science and technology, Kanazawa University, Kanazawa-shi, Ishikawa 920-1192, Japan}
\affiliation{The Institute of Statistical Mathematics, Tachikawa-shi, Tokyo 106-8569,Japan}

\begin{abstract}
To understand the \textcolor{black}{growth phenomena in collective human systems}, we analyzed monthly word count time series of new vocabularies extracted from approximately 1 billion Japanese blog articles from 2007 to 2019.
In particular, we first introduced the extended logistic equation by adding one parameter to the original equation and showed that the model can consistently reproduce various patterns of actual growth curves, such as the logistic function, linear growth, and finite-time divergence.
Second, by analyzing the model parameters, we found that the typical growth pattern is not only a logistic function, which often appears in various complex systems, but also a non-trivial growth curve that starts with an exponential function and asymptotically approaches a power function without a steady state. 
We also observed a connection between the functional form of growth and the peak-out behavior.
Finally, we showed that the proposed model and statistical properties are also valid for Google Trends data (English, French, Spanish, and Japanese), which is a time series of the nationwide popularity of search queries.
\end{abstract}
\maketitle
%\dates{This manuscript was compiled on \today}
%\doi{\url{www.pnas.org/cgi/doi/10.1073/pnas.XXXXXXXXXX}}

%\begin{document}

%\maketitle
%\thispagestyle{firststyle}
%\ifthenelse{\boolean{shortarticle}}{\ifthenelse{\boolean{singlecolumn}}{\abscontentformatted}{\abscontent}}{}

% If your first paragraph (i.e. with the \dropcap) contains a list environment (quote, quotation, theorem, definition, enumerate, itemize...), the line after the list may have some extra indentation. If this is the case, add \parshape=0 to the end of the list.
%\dropcap{T}his PNAS journal template is provided to help you write your work in the correct journal format. The instructions for use are as follows: 
%Note: please start your introduction without including the word “Introduction” as a section heading (except for math articles in the Physical Sciences section); this heading is implied in the first paragraphs. 
\section{Introduction}
\textcolor{black}{For over 200 years, growth phenomena in complex systems such as human populations, biological populations, innovation, and language change have been studied quantitatively in social sciences, biology, and physics.}
%Growth phenomena in complex systems such as human populations, biological populations, innovation, and language change have been studied quantitatively for over 200 years \textcolor{black}{in social sciences, biology and physics}.
% and it has been often described by the logistic equation or its extensions. 
%Population growth is a typical complex system phenomenon that can be observed in both natural and social phenomena.
The most basic description of the growth \textcolor{black}{phenomena} is the logistic equation
proposed in 1838 by Belgian mathematician Pierre-François Verhulst to explain the population growth of some countries \cite{Verhulst1838, cramer2002origins},   
%生物や社会置ける普及現象や成長現象は複雑システムの典型的な例である。
%普及現象を記述する最も基本的かつ有名な応用が広く有名な微分方程式がロジスティック方程式がある。
%ロジスティック方程式は，
%数学者ピエール＝フランソワ・フェルフルストによって1823年に人口増加現象を説明されるために考案された方程式である。
%Pierre François Verhulst 
%後の1920年、アメリカのレイモンド・パールとローウェル・J・リードの研究によって、ロジスティック方程式が再発見されることになる[6]。
%In this last study, Verhulst put forward some criticisms of his own model of population growth, and this led to Verhulst's logistic equation being ignored for many years until it was rediscovered by Raymond Pearl and Lowell Reed in 1920.
%The logistic equation 
%proposed in 1838 by Belgian mathematician Pierre-François Verhulst to explain the population growth of some countries \cite{Verhulst1838, cramer2002origins}, 
%He used data from several countries, in particular Belgium, to estimate the unknown parameters.
%ベルギーの人口データ、どんな。
%ショウジョウバエの個体群サイズ成長
%The Refractory Model: The Logistic Curve and the History of Population Ecology
%https://www.jstor.org/stable/2825134
\begin{equation}
\frac{dy(t)}{dt}= r y(t)\left( 1-\frac{y(t)}{Y} \right). \label{eq_logi}
\end{equation}
Here, $Y>0$ is the carrying capacity (i.e., the maximum population that the environment can sustain indefinitely), and $r>0$ is the growth rate. 
 This equation is based on the exponential population growth effect and density effects (i.e., the effects of the population growth rate being suppressed by resource or environmental limitations). 
 The solution to the logistic equation is the well-known logistic curve, which is also called the S-shaped curve, S-shaped curve, or S-curve.
 \begin{equation}
y(t)=\frac{Y}{1+(\frac{Y}{y(0)}-1)\exp(-rt)}. \label{eq_logifn}
 \end{equation}
 %He used data from several countries to estimate the parameters.
 %This discovery had been forgotten for many decades until 1920, when Raymond Pearl and LowellJ rediscovered the logistic equation and showed experimentally that the equation could be adapted to the growth of the Drosophila population as well as the human population.
 This discovery was forgotten until 1920, when Raymond Pearl and Lowell J. Reed rediscovered the logistic equation of the human population and showed experimentally that the equation could be adapted to the growth of the Drosophila population \cite{pearl1920rate, cramer2002origins}.
For more than a century since this discovery, many researchers have applied the logistic equation or its extensions, such as bacterial populations and harbor seal populations (biological systems) \cite{skalski2010wildlife,Fujikawa2003}, language change and lexical diffusion (linguistic systems) \cite{blythe2012s,denison2003log}, diffusion of innovations, durable consumer goods, and growth of population and epidemic diseases (social and economic systems) \cite{rogers2010diffusion, bass1963dynamic, wu2020generalized,mendez2015stochastic}.  \textcolor{black}{Furthermore, extensions of the logistic equation have helped improve our understanding of the mathematics of nonlinear dynamics and complex systems, such as nonlinear oscillation and chaos \cite{strogatz2018nonlinear}.}
% to be adaptable to many growth and diffusion phenomena (including social phenomena) 
%such as 
% bacterial populations and Callorhinus ursinus populations in biological science,   
%language change and lexical diffusion in sociolinguistics, and the diffusion of innovations and durable consumer goods in sociology or business science. 
%The equation and its extensions have been found to have a vast array of applications (including social phenomena) as well as biological population growth phenomena.　For example, these equations have been applied to language change and lexical diffusion in sociolinguistics and to the diffusion of innovations and durable consumer goods in management science.
%The simple nonlinear equations and their extensions have also contributed to our understanding of the mathematics of complex systems, such as Lotka-Volterra equations (competitive phenomena and nonlinear oscillatory phenomena) and chaotic phenomena (discretized equations).　\par
%方程式は，生物学的な個体数成長現象のみなく，膨大な適用事例（社会現象も含む）があることが発見されてきた。　例えば，方程式は，社会言語学においては，方程式は，言語変化，語彙拡散，経営科学では，イノベーションの普及，耐久消費財の普及などに応用された。
%
%また、このシンプルな非線形方程式とその拡張は， Lotka-Volterra equations (競争現象や非線形振動現象)やカオス現象（離散化した方程式）など，複雑系の数理の理解に寄与してきた。　\par
%利用され最も有名かつ応用が広く基礎的な微分方程式である。さらに，その拡張まで広げれば膨大な応用がある。例えば，応用としては，世界的な感染症データのモデリングに。 また，複雑系科学としてさらに，離散系はカオスをもたらす力学系の最も知られたものの一つになる。２１世紀前半でも，世界中の感染症の拡大の研究に応用している例もある。
%Blumberg, A.A. Logistic growth rate functions. J. Theor. Biol. 1968, 21, 42–44. [Google Scholar] [CrossRef]
%Cohen, J.E. Population growth and earth’s human carrying capacity. Science 1995, 269, 341–346. [Google Scholar] [CrossRef]
\par
%様々な現象で数理的に観測解析されている。普及現象や増殖現象は典型的な複雑システムにおける動力学の問題であり，
%例えば，最も基本的なロジスティック方程式は多くの生物や耐久消費財の普及を記述する。さらにその拡張も含めればさらい応用が広がる。
%例えば，ロジスティック方程式に相互作用を含めた拡張であるロテカボルタラ方程式は複雑系の最も有名な方程式の一つである。
%本研究は，現実の社会における普及現象の一つであるオンラインの新語における普及現象をロジスティック方程式のできる限り小さな拡張で統一的に
%に説明することを目指す。複雑系科学の視点でのオンライン言語の普及現象の考察に対応する。
%具体的には，現象をロジスティック方程式をわずかに拡張することで良く説明できることを示し，また，その式を使った現象の解析を行う。
%社会系は複雑系の典型的なものであるので，その現実系の一つ現象の体系的な観測は科学として重用と考えられる。
%応用的に重要なテーマであると
%考えられる。
\textcolor{black}{
    In this study, to understand growth phenomena of large collective human systems, we systematically observe the growth phenomena in online languages and describe them with the least possible extension of the logistic equation, similar to previous studies. In addition, by using our introduced model with small parameters, we aim to clarify the typical macro growth dynamics and their diversity. 
}
%we aim to describe the growth phenomena in online languages with the least extension of the logistic equation, similar to previous studies. In addition, by using the introduced model with small parameters, we aim to clarify the typical macro growth dynamics and their diversity. \textcolore{red}{一般の複雑システム的なhな足}
% described as parameters.
% from the perspective of complex systems science.
%behaviors
%we also aim to find their macro characteristics from the perspective of complex systems science. 
%As a complex systems science, 
%The significance of this research as complex system science is to
This study provides systematically organized observations of collective human growth phenomena,  which is a typical complex system; and the extension of the logistic equation allows us to associate online language growth phenomena with growth phenomena in previously studied complex systems. \par
Traditionally, growth phenomena in languages have been studied for many years in \textcolor{black}{quantitative linguistics, historical linguistics and sociolinguistics}.
They have been studied in the context of ``language change'' and ``lexical diffusion''.
Many studies have shown that a typical growth pattern is an S-curve (slow start, accelerating period, and slow end), and the sociolinguistic mechanisms behind the emergence of the S-curve have also been discussed \cite{denison2003log, ghanbarnejad2014extracting}.
%S字カーブやS字カーブに言語学的な定性的な考察はPiotrovski law と知られ，例えば, \cite{xxx} , \cite{xxx}. 
%定量的にS字カーブをPiotrovski law と知られ　ロジスティク方程式(とその拡張)に当てはめたものは、
%iotrovskaja and Piotrovskij (1978) \cite{xxx} and later, Altmann et al (1983) \cite{xxx}.  
A quantitative fit of the S-curve to the logistic equation is known as Piotrovski's law, which was first reported by Piotrovskaja and Piotrovskij (1974) \cite{piotrovskaja1974mathematiceskie} and by Altmann et al. (1983) \cite{altmann1983law}.  
Subsequently, many studies have found various examples of the logistic-like S-curve, almost establishing that S-curves are the majority of language change in long-term time-scale (historical  time-scale) \cite{blythe2012s}. 
For example, the replacement of ``wszytec'' with ``wszystec (all)'' in Polish from the 15th to the 18th century has an S-curve that can be well approximated by a logistic function (when plotted with time on the x-axis and the ratio of the replacement from ``wszytec'' to ``wszytec'' on the y-axis) \cite{gorski2021modeling}. Recently, with the increase in data, more precise verifications of the logistic equation or its extension models have been examined \cite{burridge2021inferring,ghanbarnejad2014extracting,amato2018dynamics,gorski2021modeling,maybaum2013language}.  \par
%例えば、１５世紀から１８世紀までのポーランド語における、``wszytec'' から “wszystec” (all) の置き換えが，ロジスティック関数に良く近似できるS字の形になっている（x軸を時間、y軸を`wszytec''に対する``wszystec'' の使用の比率としてプロットしたとき）。 

%This study can be regarded as a study of diffusion phenomena in language. 
%Traditionally, the logistic-like equation has been studied in linguistic science in the context of “language change” and “lexical diffusion” for about 50 years.
%It is known as the Piotrovski law,  which was reported in  
%As to language change, which has been a much-studied issue in modern language
%Science (Labov 1994; Crystal 2000; Labov 2001),
%Piotrovskaja and Piotrovskij (1978) \cite{xxx} and later, Altmann et al (1983) \cite{xxx}.  
%より一般にS-shape cuvver or slow-quick slow-quick など定性的に色々な事例が議論されている。
%For example, S-shape curve or slow-quick-slow law \cite{xxx,xxx,xxx}. 
%本研究が対象する言語における成長現象は伝統的には，計量言語学，歴史言語学，社会言語学等において１００年程度，研究されてきている。
%``language change'' や``lexical diffusion''　という文脈で研究されている。
%それ以降も色々研究している。例えば、　最近では、情報科学の進歩で新しいデータベースでより
%データドリブンな研究が盛ん手である、xxxが議論し，
%xxxがポーランド語について、xxxはロジスティック方程式の拡張を議論している。
%本研究のこれらの伝統的な研究との相違は，インターネット上の言語の，そして，数１００年単位で伝統的に研究に比べて、より短い年単位の
%時間スケールの現象をあつかっていることである。　インターネットにおける先行研究は, \cite{xxx}があり，twiiterにおける単語の
%拡散をいくつかに形状により分類している。
%本研究はパメラトリックの分類になっている。
%本研究のこれらの伝統的な研究との相違は，(i)時間スケール，(ii)最近のオンライン上の言語を用いてることである(伝統できな歴史的研究は年単位の解像度であるが，本研究は月レベルの解像度である）。
The main features of our study compared to these traditional studies are (i) the time scale (the time scale of traditional studies is a historical scale or more than 100 years with a time resolution of years, whereas that of our study is 5 or 10  years and  its time resolution of months) and (ii) the diversity of vocabulary (e.g., the word set of our study includes the recent online language, and we aim to describe the diversity of growth curves as simply as possible).
%For a quantitative study of word growth in the online social media, Twitter, the author examines S-shaped curves for several words. Consequently, he reported that both S- and non-S-curve words exist \cite{xxx}. Our study also 
%can be regarded as a more precise version of this study.
An example of a previous study on word count growth in online social media, which satisfies the above two conditions, is Ref. \cite{maybaum2013language}. The author examined the S-curve for several words on Twitter, which is a well-known social media or microblogging service, and claimed that there are both S-curve and non-S-curve type growth curves depending on the words. 
% This study is a example of word growth in the online social media.
Our study can also be regarded as a more precise version of this previous study.
\par
%そこで実際にどのようなふるまいをするかをピークを1に規格化し，それ以降の減少の中央値で観測したものが図
% derived a law which is able to capture several
%forms of the development. The intuition of historical linguists, who presupposed that the
%curve should be S-shaped, was turned into the mathematical function of the form –
%(1) 𝑦 = 𝑐
%1 + 𝑎 ∗ 𝑒−𝑏𝑥 ,
%which is supposed to seize the evolution of a phenomenon in a language. The 
%言語変化は，数１００年スケールでの言語の変化を研究している。例えば，。これがS-shaped curve (slow first slow)とも呼ばれる。定量的な検証は,ｘｘｘである。 \par
%Traditionally, they are called language change or lexical diffusion in sociolinguistics, and have been studied for more than 100 years.
%本研究は単語の拡散現象に近い。これまで，言語における普及現象の研究，「言語変化」や「語彙拡散」の文脈で古く研究されてきている。特に、これまでは年のスケールで１００年単位の
%言語変化のとって多くの研究であり，S-curve である（遅い早い遅い）のカーブ。
%Sカーブのモデル化の最も基本的なものが，ロジスティック方程式の解であるロジスィックカーブ(S-curve)が使われている。
%\begin{equation}
%y(t)=\frac{Y}{1+exp(-rt)}
%\end{equation}
%ロジスティック方関数は，昔から言語変化の主要な説明モデルであり，最初は、xxxであり、次に、yyyであり、その後、多くの異なるデータでうまくふぃってキングできることが確認されている。また、いくつかの拡張をすることでより説明できる報告もある。 これらの研究に比べた本研究の特徴は，インターネット情報の量や時間解像度を利用したより短い時間スケールの解析であることであることである。また，ネットスラングなども含めて多くの新語を統一的かつ網羅的に扱ってることにある。 ネットを利用して言語拡散の研究は，xxxがある。 データの整備により複雑系科学者や物理学者も参入している。
%The target vocabulary for this study includes new products and new concepts. The phenomenon of the diffusion of ideas and products has traditionally been studied for more than a century in sociology and business science as “innovation diffusion” or “product diffusion”.
From a practical standpoint, the time series of word counts in nationwide social media is used to quantify temporal changes in social interests \cite{google_trends} and is also used as a marketing tool to observe 
diffusion of new things such as new ideas, technologies, and new products.
%a marketing tool or a social sensor
%. because the diffusion of ideas and products.
%The time series is expected to be reflected in the time series of word counts of a focused word
%of social interests
Diffusion (growth) phenomena such as ideas, products, and innovations have traditionally been studied for more than a century in sociology and business science as ``innovation diffusion'' or ``product diffusion''. 
%The phenomena of diffusion of ideas and products, which relate to our target vocabularies and may be reflected in the time series of word counts,  have traditionally been studied for more than a century in sociology and business science as “innovation diffusion” or “product diffusion”. 
%Ever since F. Stuart Chapin used the so-called logistic function to
%analyze the spread of certain new ideas of public administration among A
%can cities, mathematical growth functions have been widely utilized as
%gate-level models of innovation diffusion processes. In general, these diffuse
%functions state the instantaneous level of diffusion of an innovation and
%given set of prospective adopters in terms of a simple mathematical function
%of the time that has elapsed since the introduction of the innovation.
%formally a diffusion function is often defined as the solution y =y(t) to a
%ferential equation of the for 
%. For example, Griliches (1957) used logistic function
%to describe the diffusion of hybrid seed corn among farmers, Dodd (1
%generalize empirical findings about the processes through which certain le
%messages spread in a community and Mansfield (
%社会学やビジネス科学での普及とlogistic-like S字曲線との関連の研究はこれまで数多くおこなれてきた，例えば，1928年のSF. Stuart Chapin の研究，1943年のBryce Ryan and Neal C. Grossらの農業技術の普及の研究，そして，1962年のEverett M. Rogers のイノベーション研究（最も代表的で有名な研究）があげられる。　
%in which, there have been many studies of the relationship between diffusion and logistic-like S-curves.
Many studies have discussed the relationship between diffusion and logistic-like S-curves.
Pioneering work on logistic-like S-curve growth phenomena is the growth of new types of social institutions by SF. Stuart Chapin in 1928 \cite{chapin1928cultural}, agricultural technology by Bryce Ryan and Neal C. Gross et al. in 1943 \cite{ryan1943diffusion}, and one of the most famous and representative work is ``Diffusion of Innovations'' by Everett M. Rogers in 1962 \cite{rogers2010diffusion}.
%, which is the most representative and famous study.　
%社会学やマーケティングやビジネス科学においても商品の普及やイノベーションは重要なテーマの一つであり，定量的にはロジスティック関数によく近似できるS字曲線で表現されることしばしばある。 社会学では，１８２０年代のShopinらの研究よりS字のイノベーションがいわれ，1943年の
%農業研究からさらに、ロジャーのイノベーション理論が最も代表的なものである。マーケティング分野では，
%ロジスティック方程式の軽微に拡張したバスモデルが用いられ現在でも使用されいてる。
Influenced by Roger's work, an extension of the logistic equation, the Bass model, developed by Frank Bass in 1969, is still often used to describe the process of how new products are adopted by a population. 
\begin{equation}
		\frac{dy(t)}{dt}= (ry(t)+\alpha)\left(1-\frac{y(t)}{Y}\right), \label{bass}
\end{equation}
where $\alpha \geq 0$ and $Y>0$ are constants \cite{bass1963dynamic,wu2020generalized, brdulak2021bass}. 
%From a practical standpoint, online linguistic data is also used as a marketing tool to measure changes in the magnitude of social interest.
%It is called “social listening” and is applied to predict trends and product demand and to analyze changes in social interest. 
%Our study is expected to provide fundamental knowledge about dynamics for this application, such as the diffusion of ideas and the diffusion of products.
Our study also \textcolor{black}{corresponds} to quantitative measurements of the diffusion of new products and ideas using linguistic data. Hence, the proposed model and its findings may contribute to innovation diffusion research as a tool for quantitative observation.
%ほんけんきゅうは言葉のデータを通した，新商品や新しいコンセプトの拡散と間接的な計測に対応しているため，研究はこれらのイノベーション普及の研究の精密化や一つのつﾂｰﾙや知見に貢献できると考えられる。
%。商品の種類と売れ方のパターン、成長や衰退について半定量的な解析がおこなれている。
%例えば、商品ではxxxxのように言われる。商品ではｘｘｘでいわれる。その中で普及モデルの一番基本的なものがバズモデルであり、耐久消費財の普及を表す。
%ロジスティックカーブを導き，ロジステｋィック方程式の拡張である。今回の対処のWebデータのカウントは、ソーシャルリスリングといわれ商品や概念の
%普及等の把握の近似として利用されることも多い。Google Trends はgoogleの検索数のデータであり「関心」の増減を洗わずもっともよくツわれる有名なデータの一つである。 
%イノベーションディフュージョン、新技術ディージョン、newproductディヒュージョン。バズ diffusionモデルは最も引用されてるモデルの一つである。本研究はこのようなモデルへの寄与も期待される。
%時間スケール的についてはこちらに近い。ただし，言語データのため，「モノ」がうごかなくても増加するのがポイントである。
\par
%ネットを利用して言語拡散の研究は，xxxがある。
%社会における普及現象は，社会における商品の普及，感染症の普及，新しい概念やコンセプトの普及など，この深い理解は社会現象を理解
%・制御するための主要な現象の一つである。一般に社会における普及現象は観測は実験等ができず精密には困難な場合も多かった。しかし，
%今回扱うオンラインにおける普及現象は，また，ブログやtwitterなどSNSデータがでてきて１０年程度になるため、年スケールの普及現象も
%月や日の時間スケールでの観測が可能になってきた。　また，オンラインによる普及現象は社会の様々な側面を間接的に観測できる可能性もあるため，
%マーケティング等や社会調査等のソーシャルリスリングの技術の改善につながる可能性もある（流行予測や需要予測など）。　\par
%本研究では、大規模で一国規模のWeb上での言語カウントデータを使って，網羅的に月ベースの新登場語の関心の増加を体系的に観測し，それを記述できる
%モデルを提案する。とくに，ロジスティックモデルの簡単な拡張で多くの単語で統一的に記述できるモデルを示す。このモデルにより，マーケティングの普%及現象を
%状態の理解の記述や予測，また，社会言語学等における，より短い時間スケールにおける単語拡散や言語変化を解析するための一つのツールを提供することが本研究の目的である。  \par
%なお，、本研究のように動力学として言語をとらえる試みは複雑システム化学や統計物理学の文脈でもおこなわれてきた。xxxさんは[]。xxxさんは[]を示し%ている。
%より広くみれば本家級は物理学者による言語の研究の一つと位置付けられる。それは。等がある。 \par
Note that physicists have studied linguistic phenomena using the concepts of complex systems \cite{link1}, such as competitive dynamics \cite{abrams2003linguistics}, statistical laws \cite{altmann2015statistical,petersen2012languages,gerlach2013stochastic}, and complex networks \cite{cong2014approaching}. Our research can also be regarded as the study of language phenomena from the perspective of physics or complex systems.  
%From a more topic-specific perspective of a time series of word counts,  
In particular, restricting the field to the study of dynamics of words counts daily or monthly time series, 
%as previously shown
%we already studied 
the noise structure \cite{RD_base} and dynamics of a ``mature phase'' in the life trajectory of words  \cite{watanabe2018empirical,watanabe2021relations}, consisting of an ``infant phase'', an ``adolescent phase'' (i.e., the phase of growth in society) and a ``mature phase'' (i.e., the phase of well-established in society), are previously shown. In contrast, this study focuses on the infant and adolescent phases. It is notable that in Ref. \cite{gerlach2013stochastic}, authors discussed the relationship between vocabulary growth and linguistic statistics on historical time scales by using the word frequencies data in printed books from 1520 to 2000 from physics or complex systems perspective. \textcolor{black}{Also in Ref. \cite{perc2012evolution}, the author clearly observed the linear preferential attachment (i.e. the rich-get-richer phenomenon) in word-count time series on the same data, and showed its relation to Zipf's law. 
Furthermore, the same author showed that this relationship between preferential attachment and Zipf's law is also valid for words used in physics papers in the last 100 years \cite{perc2013self}.  The preferential attachment effect presented in Ref. \cite{perc2013self}, where the speed of growth is proportional to the word frequency, is one of essential elements of the logistic equation given by Eq. \ref{eq_logi}.}  \par
In this study, we first briefly describe the two types of word count time-series data used in our study: Japanese blog data and the Google Trends data.
Second, we introduce an extension of the logistic equation in Eq.\ref{base_eq} to describe the time series data. 
%language growth online.
%Third, by analyzing the data and comparing with other models, we show that the introduced model fits well on the word counts time series on Japanese blog data.
Third, by analyzing the data, we show that the model can consistently reproduce various patterns of actual growth curves, such as the logistic function, linear growth, and finite-time divergence.
%Fourth, we discuss the characteristics of the growth phenomenon in online words in terms of the introduced model. We investigated the functional form of typical growth and its predictive ability. 
\textcolor{black}{Fourth, by analyzing the model parameters of the actual data, we investigate the typical patterns and diversity of growth dynamics and the forecasting ability of the proposed model.}  
%we discuss the characteristics of the growth phenomenon in online words in terms of the introduced model. We investigated the functional form of typical growth and its predictive ability.
%Fifth, we confirm the adaptability of the proposed model to other than Japanese blog data by applying the results to Google search data in English, French, Spanish, and Japanese.
Fifth, we \textcolor{black}{confirm} that the proposed model and its properties are adaptable to Google search data in English, French, Spanish, and Japanese. 
Finally, we provide our conclusions and discussions.
\section{Data}
\label{sec_data}
%We employed two types of word-count time series for our study: (a) Japanese blog data and (b) Google Trends (English, French, Spanish, and Japanese).
We employed two types of online language data: (a) Japanese blog data, and (b) Google Trends (English, French, Spanish, and Japanese).  From this data, we extracted the word-count time series for analysis.
\subsection{Japanese blog data}
%\paragraph{{\bf Newspapers}}  \\ \par
%We obtained the time-series of word appearances per day in nation-wide Japanese newspapers by using the “Shinbun trend in NIKKEI Telecom” database, which was provided by Nikkei Inc. Using this database, we obtained the daily word appearances in 80 newspapers published in Japan  from January 2005 to September 2017 \cite{nikkei}. We used the top 10,000 ranked words in frequency order as keywords. To obtain word frequency, we referred to the pages entitled "Wiktionary: Frequency lists" in Wiktionary \cite{wikitionaly}.} 
%\paragraph{{\bf Blogs}}\par
%that is, 
%the time series of the number of word occurrences in Japanese blogs per day.
We obtained the time-series of word appearances per day in nationwide Japanese blogs using a large database of Japanese blogs ("Kuchikomi@kakaricho"), which was provided by Hottolink, Inc. This database contains nine billion articles on Japanese blogs, which covers 90 percent of Japanese blogs from November 1, 2006, to December 31, 2019 \cite{kakaricho, ishii2012hit, PhysRevE.87.012805}.  
%ウィキペディアのタイトル語達から観測期間における新語を20764語を抽出した。具体的には，私たちは以下のような手順を用いた:
%(i)ウィキペディアのすべてのタイトル語達からブログデータにおいて頻度が大きい上位１００万語を抽出。
%(ii) 2016年のブログ記事数が0であった単語を新語とする。
\paragraph{Word selection}
We extracted 20,764 new words from one million Wikipedia title words during the observation period. Specifically, we used the following procedure.
%First, we extracted the top 1 million words with the highest frequency in the blog data from all Wikipedia title words. 
First, we extracted the top one million high-frequency words from the list of titles of articles in the Japanese version of Wikipedia \cite{wikipedia_data}. 
%Here, the frequency of the title words is measured by the blog data set, namely, the total number of blog posts of a focused title word during the observation period.
Here, the frequency is measured by the blog dataset, namely, the total number of blog posts with a focused title word during the observation period.
Second, we extracted 20,764 newly appearing words from the 1 million candidate words, which were defined as words that did not appear in blogs before 2016 (i.e., words with a total frequency of 0 in 2016).
%Secondly, among the 1 million words, words with zero blog posts in 2016 were considered as new words.
%We used 1,771 basic adjectives and 60,476 nouns as keywords from Wikipedia \cite{idadic}. \par
The details of word selection are provided in Appendix \ref{app_sec_data}. 
%\paragraph{{\bf Wikipedia page views}}  \par
%We obtained daily Wikipedia pageviews by using PageviewAPI. This is a public API developed and maintained by the Wikimedia Foundation that provides analytical data about the pageviews of Wikipedia articles. We obtained the data for the English, French, Chinese, and Japanese pages from Jul. 2015 to Sep. 2017 \cite{wikipedia_pageview}. We used the top 10,000 ranked words in frequency order as keywords \cite{RD_base} with respect to each language. To obtain word frequency, we referred to the pages entitled "Wiktionary: Frequency lists" in Wiktionary, as is the case with newspaper data. \cite{wikitionaly}. 
\paragraph{ Normalized time series of word appearances}
%本研究では，以上のデータから単語の１日当たりの書き込み数を$g(t)$，$<m>=1$に規格化したブログ全数に対応する時系列$m(t)$としたとき，
%以下の全数規格化差分時系列$f(t)$を元データとして扱う．
%\begin{equation}
%\delta f(t)=f(t)-f(t-1)=\frac{g(t)}{m(t)}-\frac{g(t)}{m(t)}
%\end{equation}
% Fig. \ref{tseries} shows an example of the time series of the word appearances. 
%ワードカウント時系列について書く。月のことも書かなきゃ。
We define the notation of the time series of word appearances $x_j(t)$ and $y_j(t)$  as follows:
	\begin{itemize} 
        \item We set the time step at 30 days. When the time stamp advances by one, real time advances by 30 days (almost a monthly time series).  
		\item $x_j(t)$ $(t=1,2,\cdots T)$ $(j=1,2,3,W)$ is the raw count of the articles containing the j-th word for 30 days at the time $t$ within the dataset. 
		%scaled by the total numbers of blogs $m(t)$.
		\item $y_j(t)=x_j(t)/m(t)$ is time-series of the count of the articles containing the j-th word normalized by the (scaled) total number of articles $m(t)$  (see the black triangles in Fig. \ref{fig_examples} and Fig. \ref{fig_world}). 
    \end{itemize}
    where $T$ is the last observation time, $W$ is the number of words, $m(t)=M(t)/(\sum_{t=1}^{T}\frac{M(t)}{T})$ is the scaled total number of articles, and $M(t)$ is the time series of the total number of articles over 30 days.
	%Here, $m(t)$ is the normalised total number of blogs assuming that $\sum_{t=1}^{T}m(t)/T=1$ for normalisation (see Fig.\ref{tseries}(b)), 
	%where $m(t)$ is estimated by the ensemble median of the number of words at time t, as described in the Appendix \ref{app_m}.  
	%次に全数で規格化した時系列$\tilde{F}_j(t)=F_j(t)/m(t)$を扱う．
	%この量は，実用的には，全数の効果を除いた個別の効果
	Note that $y_j(t)$ corresponds to the original time deviation of the $j$-th word separated from the effects of deviations in the total number of articles $M(t)$ (see Fig. 1 in Ref. \cite{RD_base}).  
%\subsection{ブログ}
%b. Blogs We obtained the time-series of the daily
%number of articles containing a keyword in nationwide
%Japanese blogs using a large database of Japanese blogs
%(”Kuchikomi@kakaricho”) provided by Hottolink, Inc.
%This database contains 3 billion articles of Japanese
%blogs, which covers 90% of Japanese blogs from Nov.
%2006 to Dec. 2012 \cite{xxx}. Note that similar to the
%newspaper data, if one article contains more than two fo-
%cused keywords, the system counts it as one article. We
%used 1,771 basic adjectives and 60,476 nouns as keywords
%from ipa-dic \cite{xxx}.
\subsection{Google Trends}
\label{sec_google}
Google Trends is a monthly time series of the number of searches for a focused word using the Google search engine provided by Google Inc. \cite{google_trends}. 
%Google Trends is provided by Google Inc. \cite{xxx}. 
%Words with many searches tend to be words of high social interest. 
Similar to the number of blog posts, it is used to quantify social interests (see the red crosses in Fig. \ref{fig_world}). 
%Google Trends is provided by Google Inc \cite{xxx}. 
%Therefore, Google Trends, like the number of blog posts, is used to quantify social interest(See Fig. \ref{xxx} and Sec. \ref{xxx}).
Google Trends was normalized to 100 for the maximum value of the observation period. 
The data is available since May of 2015. \par 
%検索される数が大きい単語は，社会的な関心が高い単語である傾向がある。
%そのため，グーグルトレンドは，ブログ記事数同様に、社会の関心の定量化に使われる。
%グーグルトレンドは、グーグル社より提供され，そこから取得できる \cite{xxx}。
%Google Trends is data on the number of searches for a focused word in the search engine google.
%Google Trends はｘｘｘのデータである。 以下のサイトから取得できる \cite{xxx}。 
%Google Trends is search query data in Google Search, where the number of searches is standardized to the maximum number of 100 in the observation period.
%We targeted data that were expected to be new words.
%Specifically, we targeted words that had no page views in Wikipedia in year XX and had 100 or more page views in year XX. In other words, not all new words, but words that were Wikipedia entries in XX years were targeted.
%For details on data extraction and data, please refer to \ref{xxx}.戻る
%In the study, the focus words were taken from Wikipedia article titles in English, French, Spanish, and Japanese, respectively.
We selected new words from the lists of titles of articles in the English, French, Spanish or Japanese versions of Wikipedia \cite{wikipedia_data}. Specifically, we used the following procedure: we selected the titles of articles whose annual  Wikipedia page views were 0 on May 1, 2015, and more than 50 page views (for French, Spanish and Japanese) or 1000 page views (for English) on \textcolor{black}{January 1 of 2016, 2017,$\cdots$, 2021 or 2022} \cite{wikipedia_pageview}. 
%For Japanese, we used the same new word list as in the blog data. 
The details of word selection are provided in Appendix \ref{app_sec_data}.   
%研究では，着目単語はウィキペディアの英語、フランス語、スペイン語、日本語の記事タイトルからそれぞれ取得した。
%The new words were selected using this Wikipedia page view (number of views of the focused article) because it is difficult to get trends for a large number of words in a short period of time with Google Trends due to system limitations. More explicitly, the criteria used to select new words are as follows: the article for the focused word has 0 annual page views on May 1, 2015, and more than 50 page views (for French, Spanish, and Japanese) or 1000 page views (for English) on January 1 of every year after 2016.
%https://dumps.wikimedia.org/other/pageviews/
%より明解いうと， 新語の選出に用いた基準は，以下の条件を満たすことである： 焦点にされた単語の記事の20xx年での年次ページビューが０，かつ，20xx年以降のいずれかの年のページビューが１００以上ある。
%新単語たちは、ウィキペディアのページビュー（着目記事の閲覧数）を用いて決定した。
\section{Extension of the logistic equation}
\label{sec_model}
%ロジスティック方程式は象を説明する最も典型的な数式はである.
%The logistic equation is one of the most primary equations describing the population growth curve,
%\begin{equation}
%\frac{dy(t)}{dt} = r y(t) \left(1-\frac{y(t)}{Y} \right), 
%\end{equation}
%where $y(t)>0$ is the population,  $Y>0$ is the carrying capacity (i.e., the maximum sustainable population), and $r>0$ is the growth rate for $y(t)\ggY$ namely case of the population $y(t)$ is very small compared to the maximum capacity $Y$, the p, and the initial value $y(1)>0$.
% 今回の新語等の普及現象の説明でもよく使われてきた．
%本研究ではオンラインでのワードカウントの普及現象を説明するためにこの方程式を以下のような拡張を行った. 
To describe the growth of word counts in online media, we extend the logistic equation by adding a power-law exponent $\alpha$ as follows: 
\begin{equation}
\frac{dy(t)}{dt} = r y(t) \left(1+\frac{y(t)}{Y} \right)^\alpha \label{base_eq}
\end{equation}
where $y(t)>0$ is the word count at time $t$ and $Y \neq 0$ is the transition point. 
For the logistic equation ( $\alpha=1, Y<0$ ), $Y$ is the carrying capacity (i.e., the maximum sustainable population).
$r>0$ is the growth rate for $y(t) \ll Y$; that is, the case of the population $y(t)$ is very small compared to the transition point $Y$. The model for $Y<0$ corresponds to the special case of Blumberg's equation or generalized logistic equation \cite{wu2020generalized}. 
\textcolor{black}{In this paper, we call this model as the proposed model.}
\par
%where $Y \neq 0$． \par
From Appendix \ref{app_asy_eq}, the time evolution of the equation can be qualitatively classified into four categories based on the signs of parameters $Y$ and $\alpha$: 
%\begin{enumerate}
%\item 定数に収束（ロジスティック方程式型）　($Y<0$, $\alpha>0$)，
%\item 有限時間発散（締め切り効果）　（$Y>0$, $\alpha>0$）
%\item 無限時間後に発散（べき乗関数）　($Y>0$ ,$\alpha<0$) 
%\item 有限時間の1階微間発散　($Y<0$ ,$\alpha<0$) 
%\end{enumerate}
\begin{enumerate}
\item Convergence to a constant (S-curve) ($Y<0$, $\alpha>0$). 
\item Finite-time divergence (Deadline effects) ($Y>0$, $\alpha>0$)
\item Divergence after infinite time (Asymptotic power-law function) ($Y>0$, $\alpha<0$) 
\item Finite-time divergence of first-order derivatives ($Y<0$, $\alpha<0$).  
\end{enumerate}
\textcolor{black}{Examples of words in each category are given in Appendix \ref{app_example}. }
\par
\begin{figure*}
    \begin{minipage}{0.19\hsize}
    \begin{overpic}[width=3.3cm]{"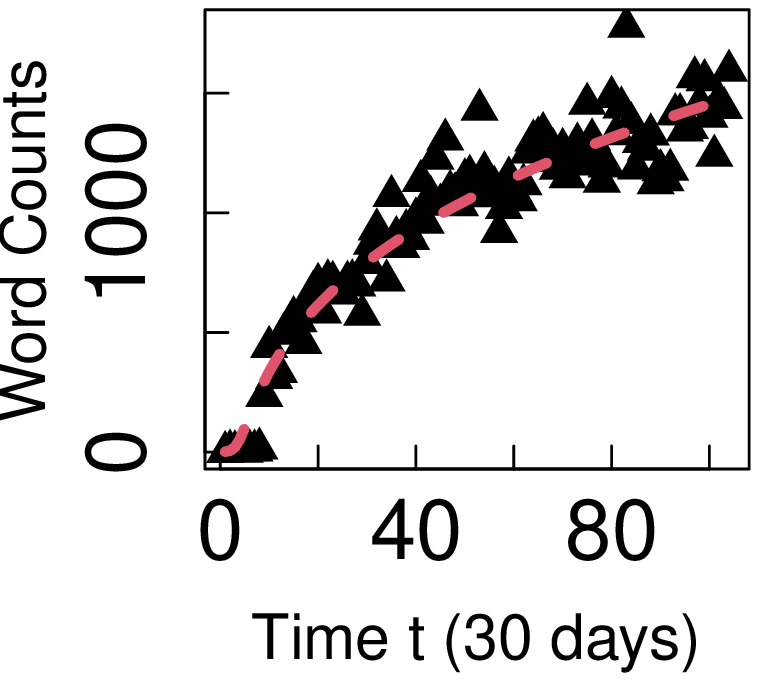"}
        %\put(22,55){(a)}
        \put(24,68){(a)}
        \put(44,33){Linear}
    \end{overpic}
    \end{minipage}
    \begin{minipage}{0.19\hsize}
            \begin{overpic}[width=3.3cm]{"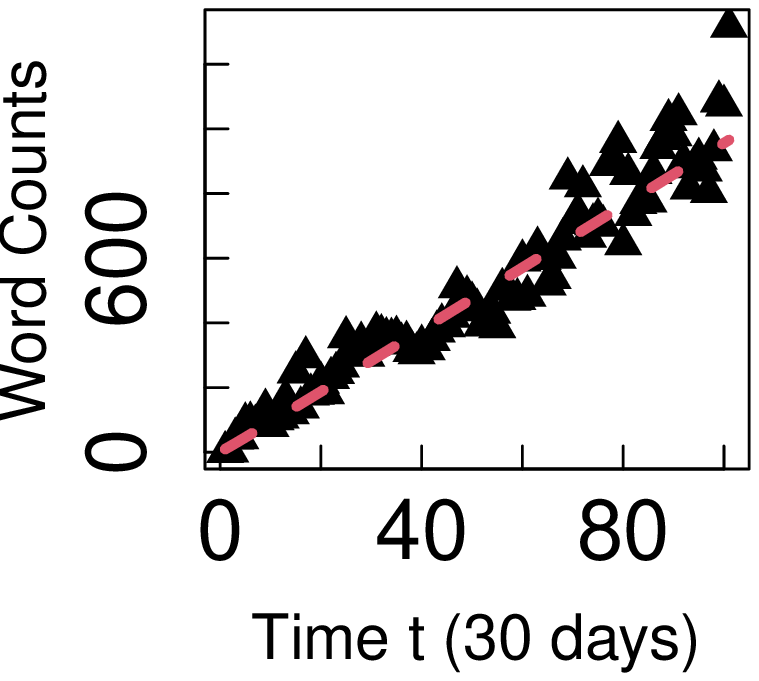"}
                %\put(22,55){(a)}
                \put(24,68){(b)}
            \end{overpic}
    \end{minipage}
    \begin{minipage}{0.19\hsize}
    \begin{overpic}[width=3.3cm]{"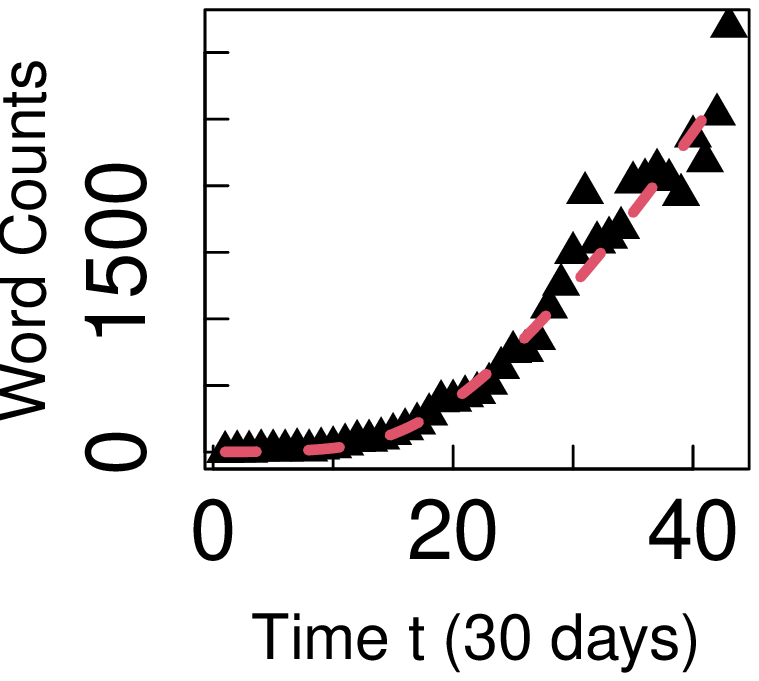"}
        %\put(22,55){(a)}
        \put(24,68){(c)}
    \end{overpic}
    \end{minipage}
    \begin{minipage}{0.19\hsize}
    \begin{overpic}[width=3.3cm]{"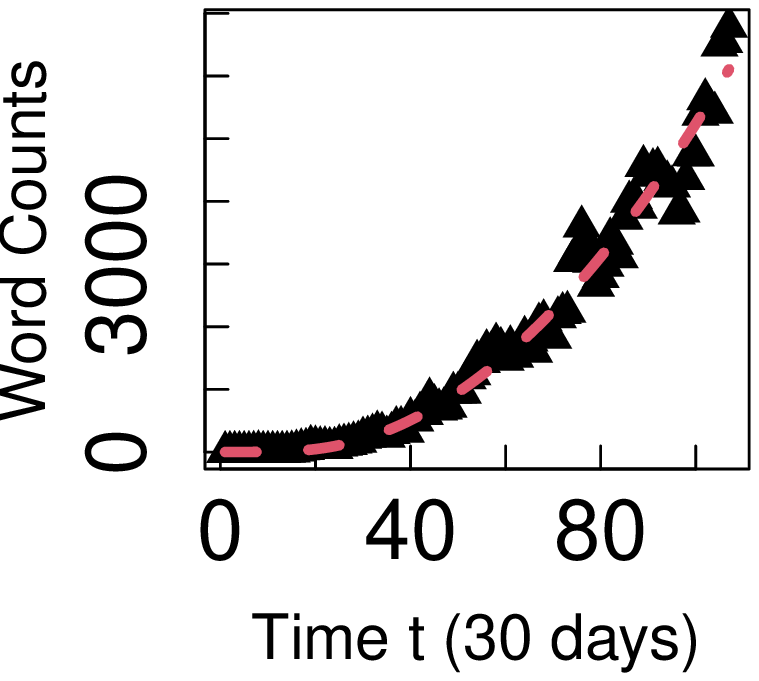"}
        %\put(22,55){(a)}
        \put(24,68){(d)}
    \end{overpic}
    \end{minipage}
    \begin{minipage}{0.19\hsize}
    \begin{overpic}[width=3.3cm]{"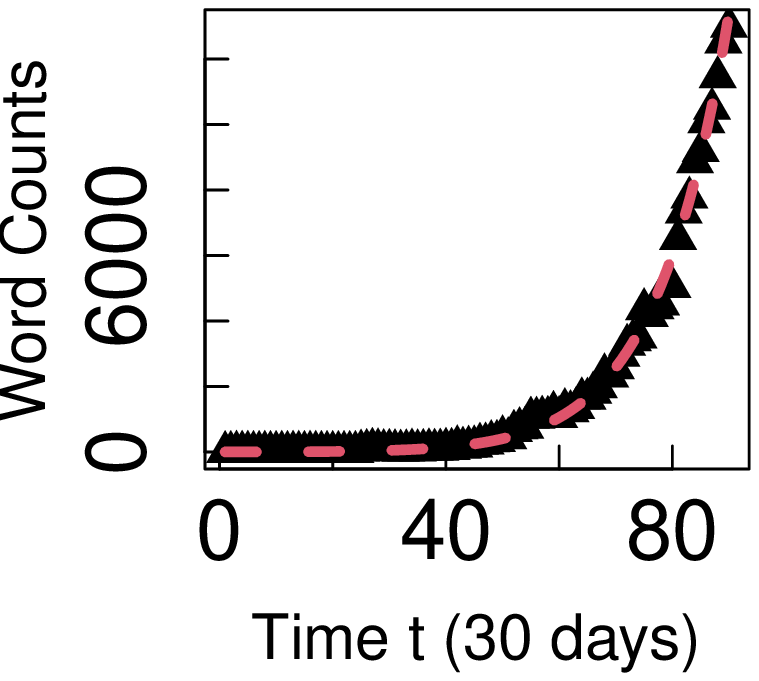"}
        %\put(22,55){(a)}
        \put(24,68){(e)}
    \end{overpic}
    \end{minipage}
    %\begin{minipage}{0.24\hsize}
    %%\includegraphics[width=4cm]{count_jyoseikatuyaku_1512.eps}
    %\includegraphics[width=4cm]{count_ogurayui__nolog_918.eps}
    %%\includegraphics[width=7cm,angle=270]{dif_a1_peri_ver2.eps}
    %\end{minipage}
    \begin{minipage}{0.19\hsize}
    \begin{overpic}[width=3.3cm]{"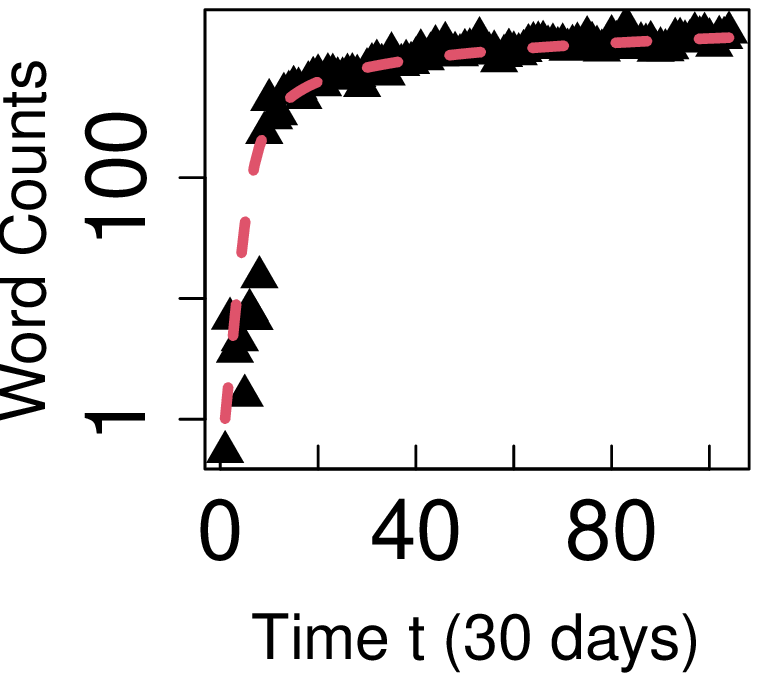"}
        %\put(22,55){(a)}
        \put(44,33){Semi-log}
    \end{overpic}
    \end{minipage}
    \begin{minipage}{0.19\hsize}
    \begin{overpic}[width=3.3cm]{"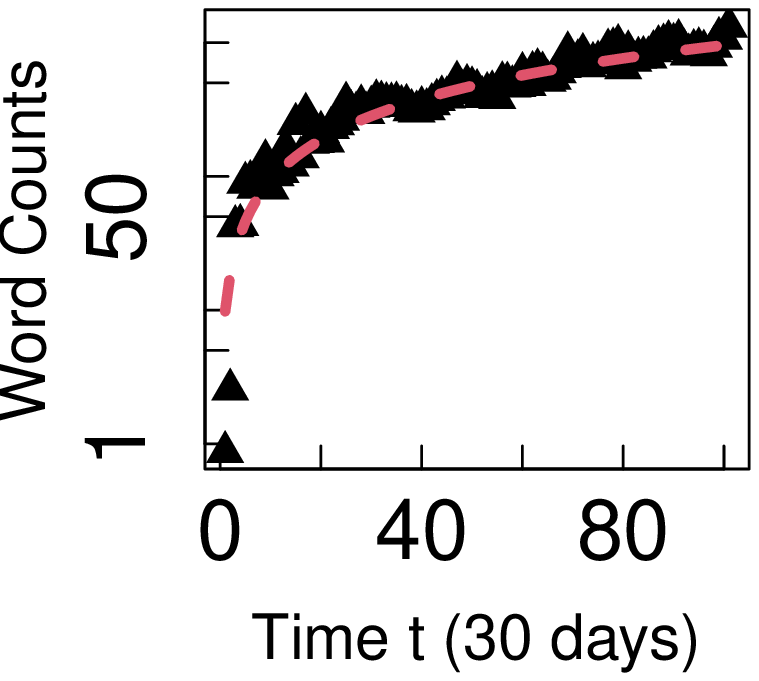"}
        %\put(22,55){(a)}
        \put(44,33){Semi-log}
    \end{overpic}
    \end{minipage}
    \begin{minipage}{0.19\hsize}
    \begin{overpic}[width=3.3cm]{"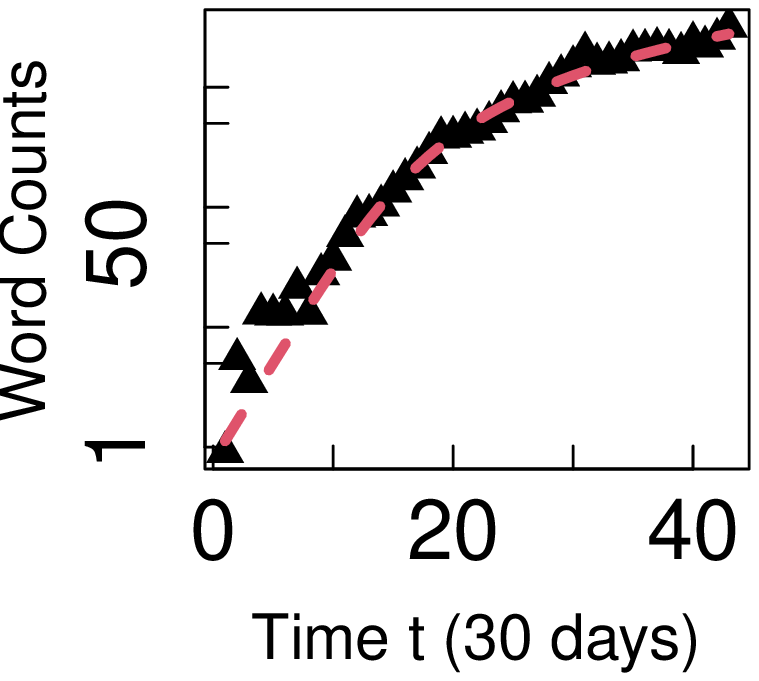"}
    %\begin{overpic}[width=3.3cm]{"count_rY03_tanakamimani_431_reg.eps"}
     
        %\put(22,55){(a)}
        \put(44,33){Semi-log}
    \end{overpic}
    \end{minipage}
    \begin{minipage}{0.19\hsize}
    \begin{overpic}[width=3.3cm]{"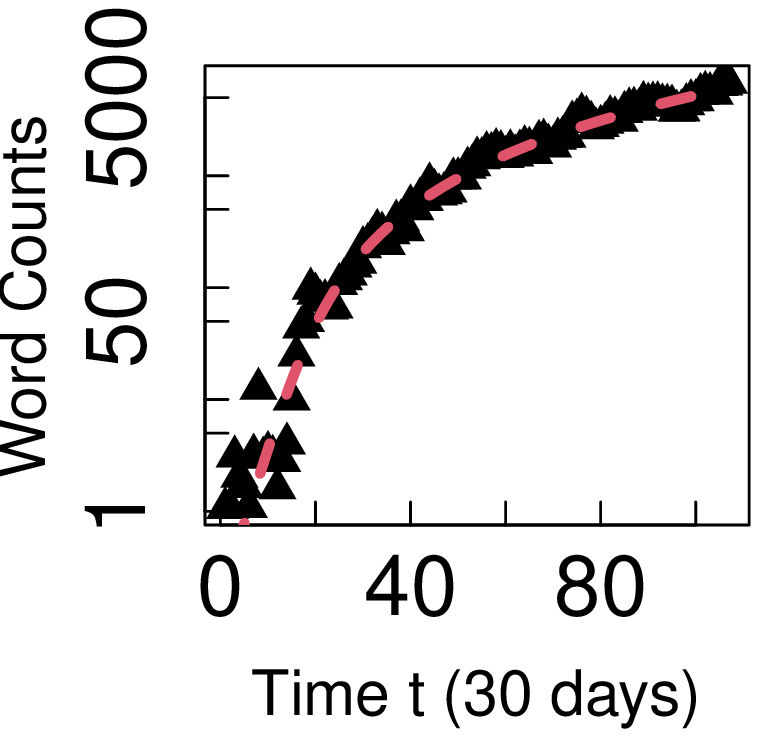"}
        %\put(22,55){(a)}
        \put(44,33){Semi-log}
    \end{overpic}
    \end{minipage}
    %\begin{minipage}{0.33333\hsize}
    %\includegraphics[width=8cm]{kizai_peri_test.eps}
    %\includegraphics[width=4cm]{count_windows_394.eps}
    %\end{minipage}
    \begin{minipage}{0.19\hsize}
    \begin{overpic}[width=3.3cm]{"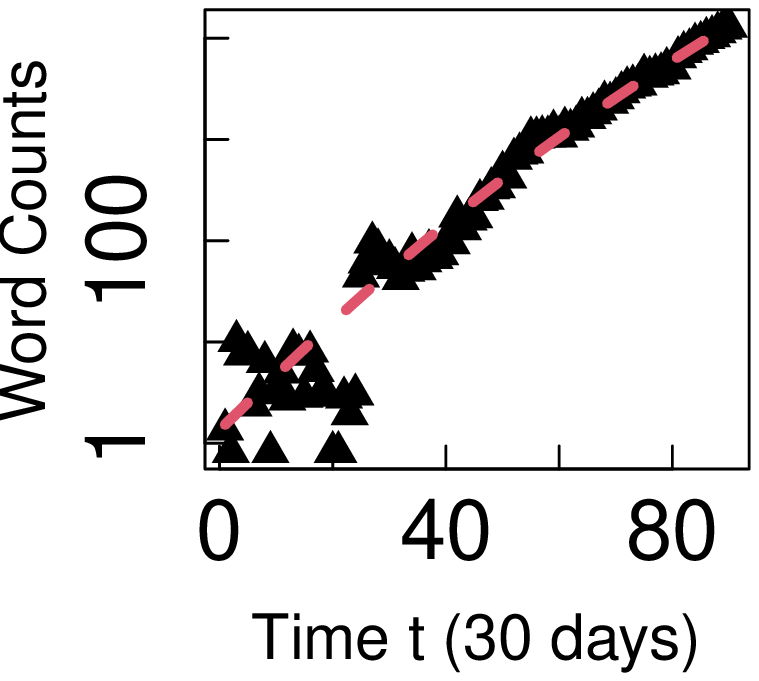"}
        %\put(22,55){(a)}
        \put(44,33){Semi-log}
    \end{overpic}
    \end{minipage}
    \begin{minipage}{0.19\hsize}
    \begin{overpic}[width=3.3cm]{"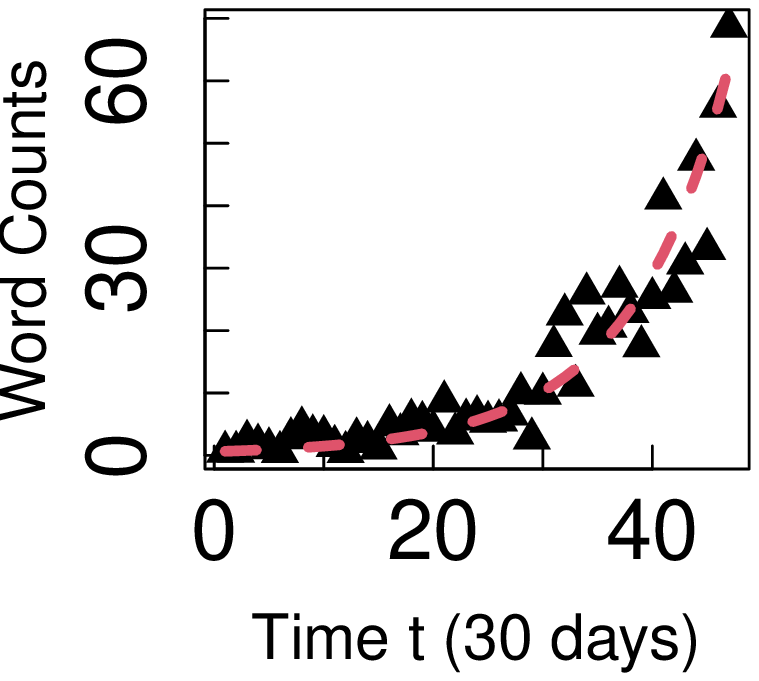"}
        %\put(22,55){(a)}
        \put(24,68){(f)}
        \put(24,52){Linear}
    \end{overpic}
    \end{minipage}
    \begin{minipage}{0.19\hsize}
    %\includegraphics[width=3.3cm]{"count_RometheaterKyoto_reg_nolog_3203.eps"}
    %\begin{overpic}[width=3.3cm]{"count_rY03_RometheaterKyoto_reg_nolog_3203.eps"}
    %\begin{overpic}[width=3.3cm]{"count_rY03_ionmoorutokoname__reg_nolog_6818.eps"} 
    \begin{overpic}[width=3.3cm]{"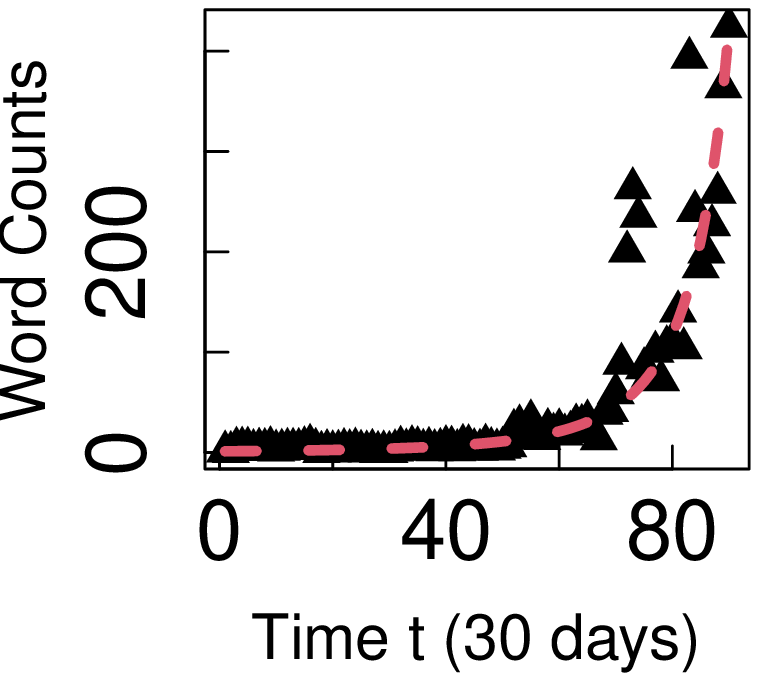"}    
    %\put(22,55){(a)}
        \put(24,68){(g)}
    \end{overpic}
    %\includegraphics[width=3.3cm]{count_DejitarruTruuka_nolog_3230.eps}
    %3230,header="count_DejitarruTruuka
    %\includegraphics[width=3.3cm]{count_sotiGorin_nolog_445.eps}
    %\includegraphics[width=3.3cm]{count_syeerugasu_nolog_367.eps}
    %\includegraphics[width=3.3cm]{count_sotiGorin_nolog_445.eps}
    %\includegraphics[width=3.3cm]{count_ted_nolog_2285.eps}
    \end{minipage}
    \begin{minipage}{0.19\hsize}
      \begin{overpic}[width=3.3cm]{"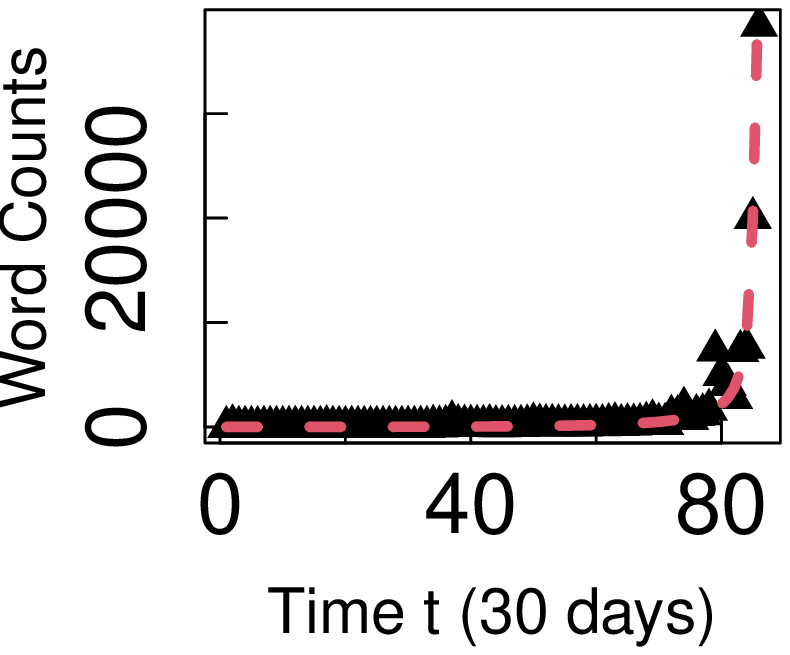"}
    %\begin{overpic}[width=3.3cm]{"count_rY03_tanakamimani__reg_nolog_431.eps"}
        %\put(22,55){(a)}
        %\put(24,68){(h)}
        \put(24,62){(h)}
    \end{overpic}
    \end{minipage}
    \begin{minipage}{0.19\hsize}
    \begin{overpic}[width=3.3cm]{"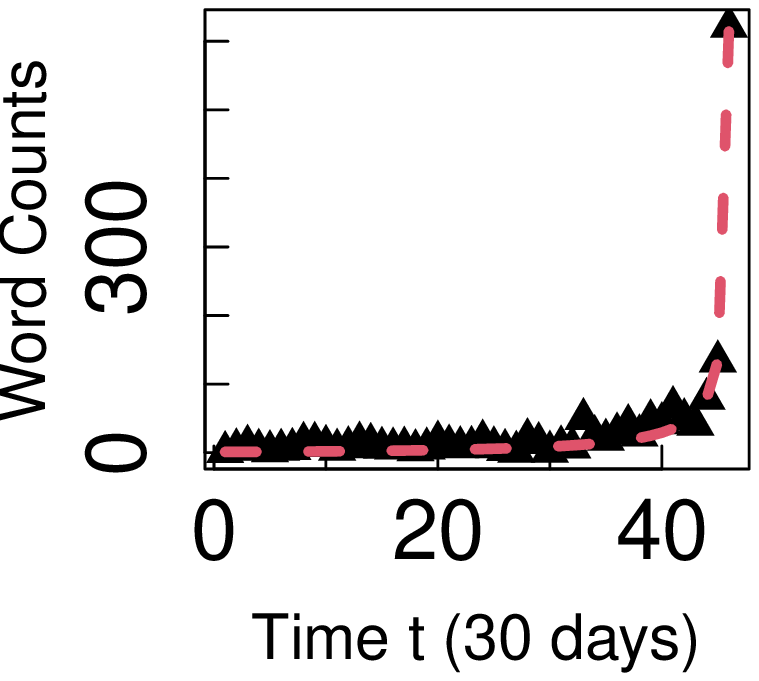"}
        %\put(22,55){(a)}
        \put(24,68){(i)}
    \end{overpic}
    \end{minipage}
    \begin{minipage}{0.19\hsize}
    %\includegraphics[width=3.3cm]{"count_kyawatan_reg_nolog_6949.eps"}
    %\begin{overpic}[width=3.3cm]{"count_rY03_kyawatan_reg_nolog_6949.eps"}
    %\begin{overpic}[width=3.3cm]{"count_rY03_NMD48__reg_nolog_25.eps"}
    %\begin{overpic}[width=3.3cm]{"count_rY03_SEALDs__reg_nolog_727.eps"}
    \begin{overpic}[width=3.3cm]{"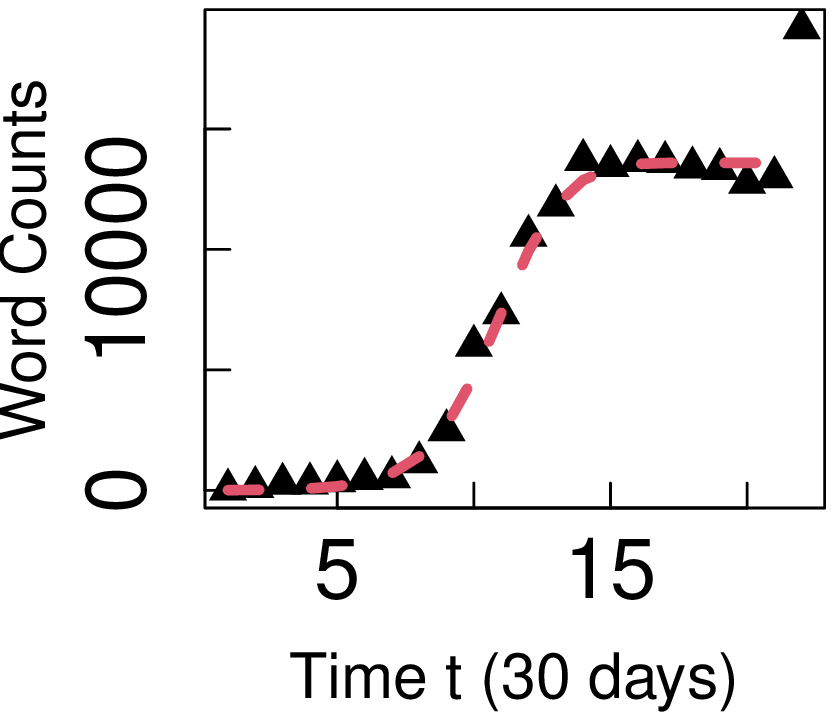"}
    %\begin{overpic}[width=3.3cm]{"count_rY03_SDXC__reg_nolog_398.eps"}
        %\put(22,55){(a)}
        \put(24,68){(j)}
    \end{overpic}
    \end{minipage}
    %\begin{minipage}{0.19\hsize}
    %\includegraphics[width=3.3cm]{count_jyoseiKatuyaukuSuisin528.eps}
    %\end{minipage}
    \begin{minipage}{0.19\hsize}
    \begin{overpic}[width=3.3cm]{"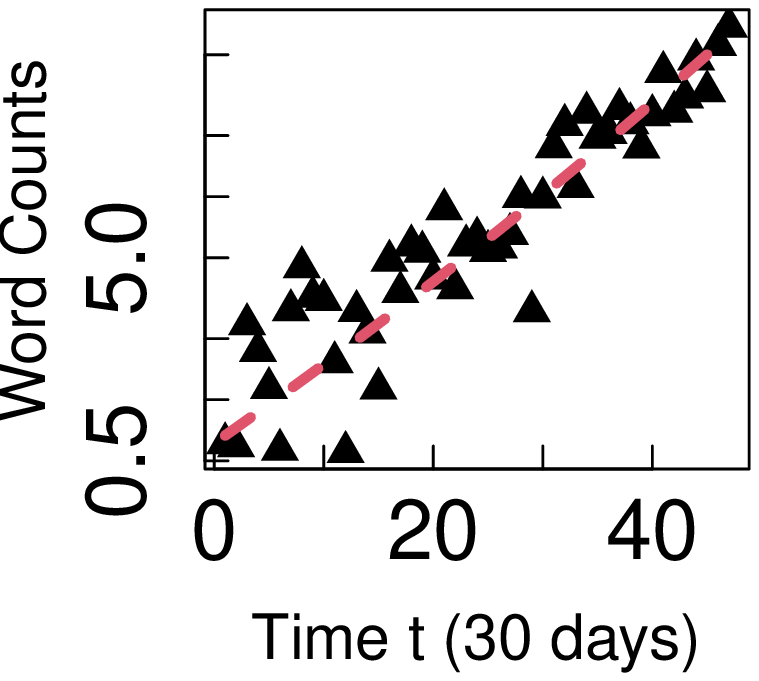"}
        %\put(22,55){(a)}
        \put(24,68){Semi-log}
    \end{overpic}
    \end{minipage}
    \begin{minipage}{0.19\hsize}
    %\includegraphics[width=3.3cm]{count_DejitarruTruuka3230.eps}
    %\includegraphics[width=3.3cm]{"count_RometheaterKyoto3203_reg.eps"}
    %\begin{overpic}[width=3.3cm]{"count_rY03_RometheaterKyoto3203_reg.eps"}
        %\begin{overpic}[width=3.3cm]{"count_rY03_ionmoorutokoname_6818_reg.eps"}
    \begin{overpic}[width=3.3cm]{"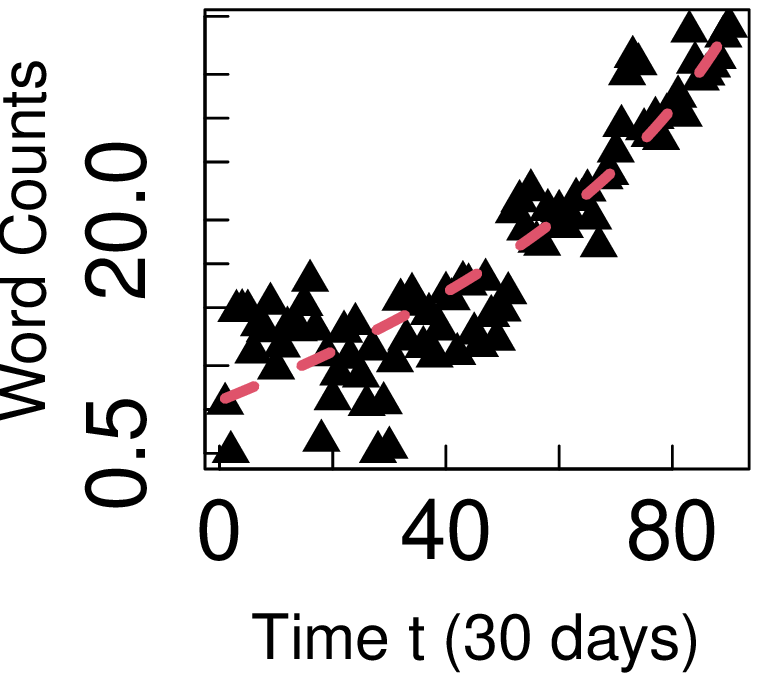"}       
        %\put(22,55){(a)}
        %\put(24,68){Semi-log}
        \put(24,62){Semi-log}
    \end{overpic}
    \end{minipage}
    \begin{minipage}{0.19\hsize}
    \begin{overpic}[width=3.3cm]{"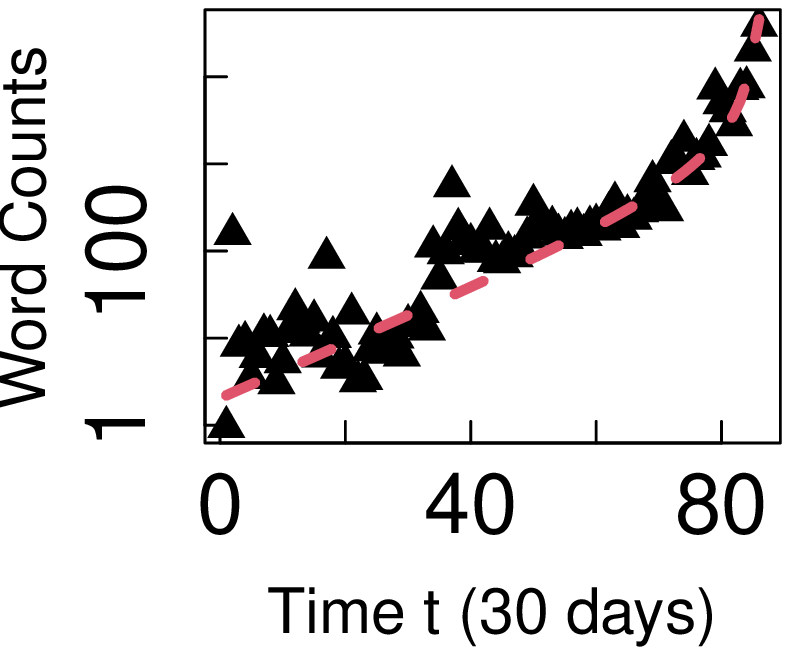"}
    %\begin{overpic}[width=3.3cm]{"count_rY03_tanakamimani_431_reg.eps"}
        %\put(22,55){(a)}
        \put(26,64){Semi-log}
    \end{overpic}
    \end{minipage}
    \begin{minipage}{0.19\hsize}
    \begin{overpic}[width=3.3cm]{"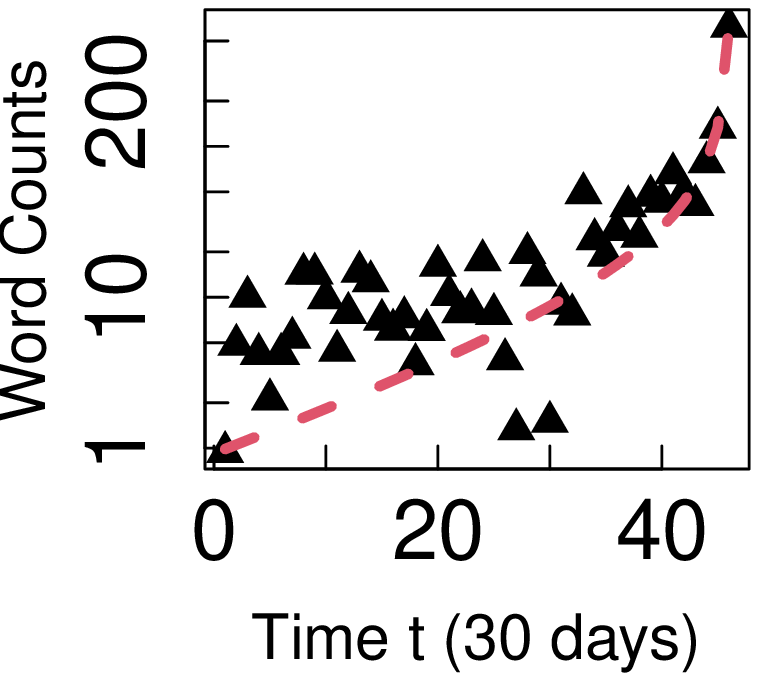"}
        %\put(22,55){(a)}
        \put(24,68){Semi-log}
    \end{overpic}
    \end{minipage}
    \begin{minipage}{0.19\hsize}
    %\includegraphics[width=3.3cm]{count_sabaani2927.eps}
    %\includegraphics[width=3.3cm]{count_pytyangorin8887.eps}
    %\includegraphics[width=3.3cm]{count_niborumabu9486.eps}
    %\includegraphics[width=3.3cm]{count_ted2285.eps}
    %\includegraphics[width=3.3cm]{"count_kyawatan6949_reg.eps"}
    %\begin{overpic}[width=3.3cm]{"count_rY03_kyawatan6949_reg.eps"}
    %\begin{overpic}[width=3.3cm]{"count_rY03_SDXC_398_reg.eps"}
    %\begin{overpic}[width=3.3cm]{"count_rY03_NMD48_25_reg.eps"}
    \begin{overpic}[width=3.3cm]{"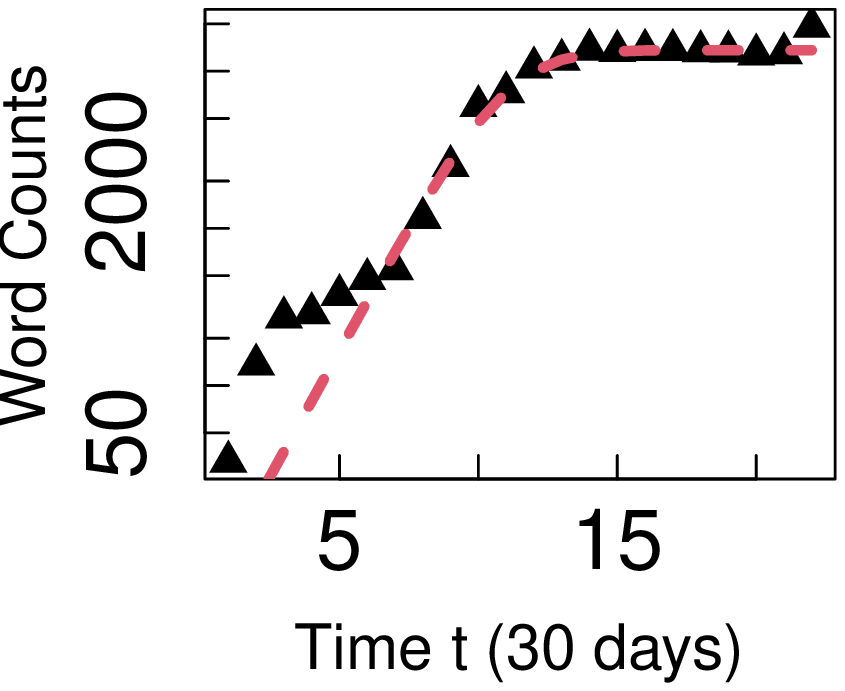"}
    %\begin{overpic}[width=3.3cm]{"count_rY03_SEALDs_727_reg.eps"}    
        %\put(22,55){(a)}
        %\put(24,68){Semi-log}
         \put(44,38){Semi-log}
    \end{overpic}
    \end{minipage}
    %\begin{minipage}{0.19\hsize}
    %\includegraphics[width=3.3cm]{count_shelu_367.eps}
    %\end{minipage}
    %\begin{minipage}{0.2\hsize}
    %\includegraphics[width=4cm]{count_rio_41.eps}
    %%\includegraphics[width=6cm,angle=270]{dif_a1_out_ver2.eps}
    %\end{minipage}
    %\begin{minipage}{0.2\hsize}
    %\includegraphics[width=4cm]{count_rio_41.eps}
    %\includegraphics[width=6cm,angle=270]{dif_a1_out_ver2.eps}
    %\end{minipage}
    \caption{Examples of time series of the growth process of word counts in Japanese blogs. 
    The black triangles are the real data $y_j(t)$ and the red dashed line is the corresponding theoretical line. Parameters were determined by Appendix \ref{app_sec_estimate}. 
    The upper figure is a linear plot, and the lower figure is the corresponding semilog plot. 
    (a) Pinterest (Web service) $\alpha=-3.07,Y=262.5,r=1.41,y(0)=0.374$ 
    (b) SagamiharaShiTyuuouKu (Sagamihara city central ward, new place name) $\alpha=-0.994,Y=1.14,r=8.13, y(0)=2.13$
    (c) Komyusyou (People who are not good at social interactions and get  nervous, Internet slang)  $\alpha=-0.799,Y=221,r=0.376, y(0)=0.775$
    (d) KuraudoFandingu (Crowd funding, Business or Internet Term)   $\alpha=-0.512,Y=31.0,r=0.317, y(0)=0.159$
    (e) OnlineSaron (Online salon, which is an online community service with a fixed fee, hosted by celebrities or well-known executives etc.) $\alpha=-0.0742,Y=31.4,r=0.121, y(0)=1.35$
    (f) Puurui (Pour Lui, Name of a new Japanese singer)   $\alpha=0.0884,Y=2.49,r=0.0872,y(0)=0.610$ 
   % (g) RomuSiataaKyoto (ROHM Theater Kyoto, Name of movie theater in Kyoto)  $\alpha=0.858,Y=34.2,r=0.152,y(0)=1.395$ 
    (g) JyoseiKatuyakuSuisin (Promotion of Women's Participation and Advancement in the Workplace, Name of Japanese Government Policy) $\alpha=0.262, Y=0.690, r=0.0282, y(0)=1.15$ 
    (h) RioGorin (Rio de Janeiro Olympics, Sports events) $\alpha=0.859,Y=806,r=0.0725,y(0)=2.05$ 
     %(h) TanakaMinami (Minami Tanaka, Name of a Japanese announcer who became a hot topic on TV and in magazines ) $\alpha=1.04,Y=93.4,r=0.0796,y(0)=2.13$ 
    (i) DesumaatikaraHajimaruIsekaiKyousoukyoku (Death March to the Parallel World Rhapsody, Titles of Japanese novels for young people) $\alpha=2.27, Y=90.4,r=0.0748,y(0)=0.922$
    %(j) Kyawatan (A word meaning "cute" used by teenage girls) $\alpha=0.614, Y=-4504,r=0.327,y(0)=0.206$
    (j) Pazudora (Puzzle and Dragons, Name of a puzzle video game) $\alpha=0.88, Y=-13593,r=0.741,y(0)=4.12$
    }
    \label{fig_examples}
\end{figure*}

For the parameters $\alpha>0$ and $Y<0$, which correspond to the first case, the equation corresponds to the S-curve.
For $\alpha=1$, the word count $y(t)$ increases the logistic function ( symmetric S-curve) given by Eq. \ref{eq_logifn} and for $\alpha \neq 1$,  $y(t)$ growth the asymmetric S-curve (Fig. \ref{fig_examples} (j)).   \par
%For $\alpha=-1$, the example of the first case, the extension function given by Eq. \ref{base_eq} is corresponds to the logistic equation given by Eq. \ref{xxx}, that is, the word counts $y(t)$ obeys the S-curve or the logistic function given by Eq. \ref{xxx}.  \par
%この拡張では，$\alpha=1$，$Y>0$のときロジスティック回帰になる。
For the parameter $\alpha>0$ and $Y>0$, which corresponds to the second case, $y(t)$ obeys the finite-time divergence known as the deadline effect. 
%有限時間発散（締め切り効果）も$\alpha>0$，$Y>0$のときに含む。
The deadline effect is a phenomenon in which interest increases with a power function toward a deadline $t^{*}$ \cite{alfi2007conference}, 
% 締め切り効果とは，オリピックや商品の発売など締め切り期日$t^{*}$のある現象に対して，べき関数
\begin{equation}
y(t) \propto \frac{1}{(t^*-t)^{\beta}},
\end{equation}
where $\beta>0$ is the power-law exponent. 
For instance, when $\alpha=1$, $y(t)\gg Y>0$, using the approximation of Eq. \ref{base_eq},  
\begin{equation}
\frac{dy(t)}{dt} \sim r y(t) \left(\frac{y(t)}{Y} \right)^1, 
\end{equation}
%と近似でき，
%，べき乗型の締め切りまでの時間に反比例して件数が増える有限時間発散の解となる.
we can easily obtain the deadline-effects solution 
\begin{equation}
y(t) \propto \frac{Y}{Y/(y(0) \cdot r)-t}.
\end{equation} 
 \par
\begin{figure}[t]
\begin{minipage}{0.48\hsize}
%\begin{minipage}{0.24\hsize}
%\includegraphics[width=4.0cm]{"all_1512_jyoseikatuyaku_all_rel.eps"}
\begin{overpic}[width=4.0cm]{"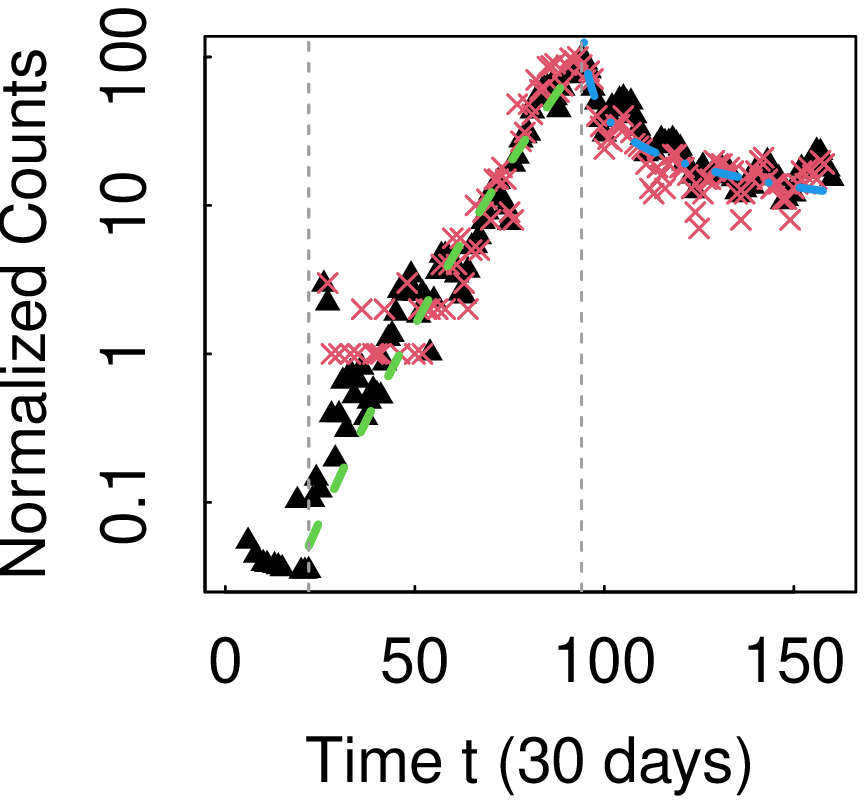"}
    %\put(22,55){(a)}
    \put(25,85){(a)}
\end{overpic}
\end{minipage}
\begin{minipage}{0.48\hsize}
%\includegraphics[width=4.0cm]{"all_42_rio_pow1_rel.eps"}
%\begin{overpic}[width=4.0cm]{"all_42_rio_pow1_rY03.eps"}
%\begin{overpic}[width=4.0cm]{"all_11302_aiura_rY03.eps"}
\begin{overpic}[width=4.0cm]{"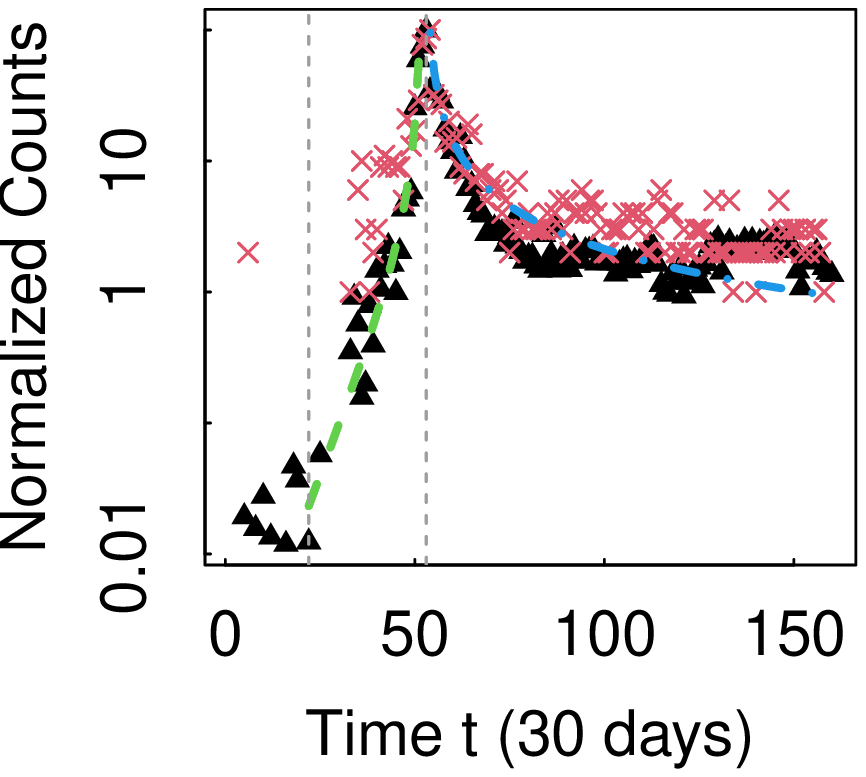"}
%\begin{overpic}[width=4.0cm]{"all_639_nihoerekiterurengo_rY03.eps"}

    %\put(22,55){(a)}
    \put(25,85){(b)}
\end{overpic}
\end{minipage}
\begin{minipage}{0.48\hsize}
\begin{overpic}[width=4.0cm]{"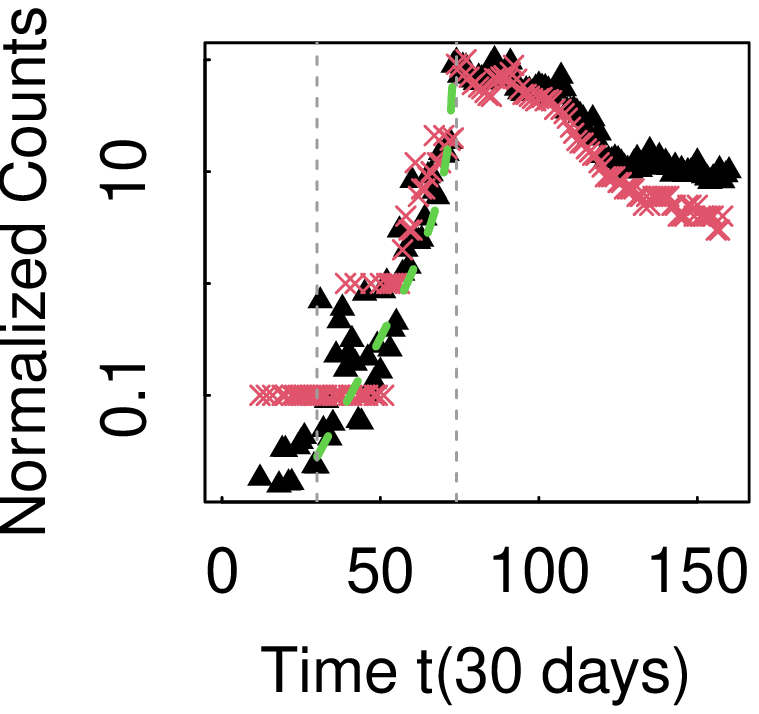"}
    %\put(22,55){(a)}
    \put(28,83){(c)}
\end{overpic}
\end{minipage}
\begin{minipage}{0.48\hsize}
\begin{overpic}[width=4.0cm]{"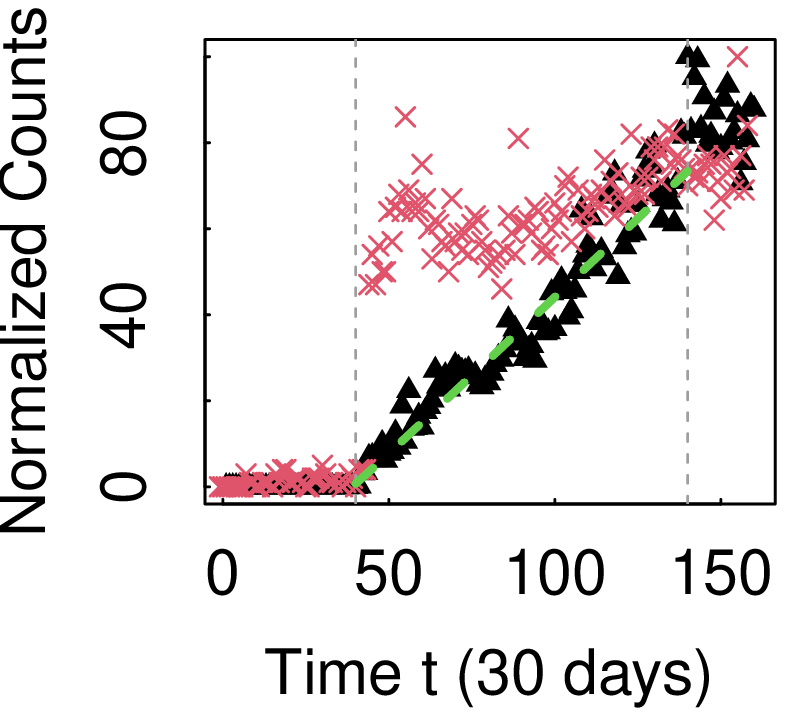"}
    %\put(22,55){(a)}
    \put(25,80){(d)}
\end{overpic}
\end{minipage}
\caption{
	Comparison of time series between blog data (black triangles) and Google Trends (red crosses).
	Blog data is scaled to a maximum value of 100. 
	The green dashed line indicates the theoretical curve given by the model Eq. \ref{base_eq}. 
	and blue dashed line is power-law function $\propto \tau^{-0.5}$ in the panne (a) and $\propto \tau^{-1.0}$ in the panel(b). The area between the grey dotted lines is the growth period detected by adapting the method described in Appendix \ref{app_sec_cut}.
	%線種を変える 
%(a) JyoseiKatuyakuSuisin (Promotion of Women's Participation and Advancement in the Workplace,Name of Japanese Government Policy ,$\alpha=0.35, Y=7.71, r=0.0403, x_0=1.15$)  
%(b) RiodejyaneiroGorin (Rio de Janeiro Olympics, Sports Events, $\alpha=1.59,Y=1304,r=0.0684,x_0=1.703$)
%(c) Windows8 (Windows 8, Famous software name, $\alpha=2.56,Y=1429,r=0.129,x_0=1.284$ )
%(d) SagamiharaShiTyuuooKu (Sagamihara city central ward, Place name, $\alpha=-1.00,Y=1.31,r=7.34, x_0=2.13$)
%(a) JyoseiKatuyakuSuisin (Promotion of Women's Participation and Advancement in the Workplace,Name of Japanese Government Policy ,$\alpha=0.262, Y=0.690, r=0.0282, y(0)=1.15$),
(a) AsaiiBouru (A\c{c}a\'{i} na tigela ,Brazilian dessert, $\alpha= -0.0722, Y=1.14, r=0.142, y(0)=1.09$),
%(b) RiodejyaneiroGorin (Rio de Janeiro Olympics, Sports Events, $\alpha=0.691,Y=263.5,r=0.0641,y(0)=1.70$),  
(b) RabuTyuunyuu (``Love Injection'', A silly joke phrase by an once-popular Japanese comedian, $\alpha=0.661, Y=158, r=0.179,y(0)=1.70$),
%(b) NipponErekiteruRengo (Name of a Japanese comedy duo, $\alpha=0.123, Y=0.184, r=0.0893,y(0)=0.367$),
(c) Windows8 (Windows 8, Famous software name, $\alpha=1.041,Y=276.2,r=0.120,y(0)=1.08$) and
(d) SagamiharaShiTyuuooKu (Sagamihara city central ward, Place name, $\alpha=-0.994,Y=1.15,r=8.13, y(0)=2.13$). 
}
\label{fig_world}
\end{figure}

This effect has been observed in \textcolor{black}{various phenomena} with scheduled deadlines, such as an application for an international conference  \cite{alfi2007conference} and numbers of blog posts for annual events like Christmas, Olympic Games and a launch of a new product that has been announced in advance (Figs. \ref{fig_examples} (g),(h),(i) and Figs. \ref{fig_world} (b), (c)). \par
%の式に従い関心や行動が増加がしていく現象である(図\ref{examples}(xxx)-(xxx)).
%\par
%実際に，日本語ブログにおいても，多くのロジスティック方程式が見られる（図 \ref{xxx}）.
%本研究で対象にするオンライン上での新規の単語の普及現象は，オンライン上に現れる１か月当たりでの単語カウント（一国規模でのブログでのその単語が使われる記事数，検索エンジンでの検索数）で観測される.
%このカウント（以下、ワードカウント）の増加の時間依存性を調べることで，普及の様子を知ることができる.
%ワードカウントにおける新語の普及には，ロジスティック回帰で説明できないものが知られている. 実際に，\ref{xxx}図に示すように，本研究のデータでも
%有限発散や線形増加などロジスティック回帰で説明できない多様な時系列のパターンが観測された.
% \par
%本研究では，ブログ等やネット上での単語の頻度や検索数$x(t)$に対応する. 
%これまで，このようなデータを用いた新語の普及ではロジスティック方程式式で記述できないものが知られている.
%また，図\ref{xxx}に示すように，本研究のブログデータ等での観測でもロジスティック方程式と異なる普及のパターンが観測できた.
%しかしながら，Web等における言語のカウントデータを利用した普及現象の観測では，ロジスティック方程式以外の普及のパターンも観測されている.
%（図 \ref{xxx}, \ref{xx}）.
%例えば，既存研究等でもたびたび指摘されるのは，有限時間発散のパターンである.　また，ブログデータから線形に近いパターンも見られた（図 \ref{xxx}）.\par 
%本研究は，以下のようなロジスティック方程式の拡張で，シンプル拡張にかかわらず，比較的よく多様なワードカウントの時間発展を説明できることを示す.%
%\begin{equation}
%\frac{dy(t)}{dt} = r y(t) \left(1+\frac{y(t)}{Y} \right)^\alpha \label{base_eq}
%\end{equation}
%ここで，$Y \neq 0$ を正負を含む実数とする． \par
%この方程式は，ロジスティック方程式を含む（$\alpha=1$, $Y<0$）．さらに，既存研究でよく議論されてきた締め切効果も説明できる.
%締め切り効果は，オリンピックような特定の日付に目的があるイベントのある言葉によくみられる典型的に見られる拡散パターンである（図\ref{xxx}）. \par
%実際，式 \ref{base_eq}は，$\alpha=1$，$y(t)\ggY$ の範囲では，

%一方，式 \ref{base_eq}は，発散ではなくロジスティック関数的な指数拡散より遅い拡散パターンも表現できる
The parameter $\alpha<0$, $Y>0$,  which corresponds to the third case,  can express growth asymptotically slower than exponential growth.
%The equation \ref{base_eq} can express not only the finite-time divergent solutions but also the growth slower than exponential growth. 
For example, in the case where $\alpha=-1, Y>0$, for $y(t) \gg Y$, the approximation of Eq. \ref{base_eq}, 
\begin{equation}
\frac{dy(t)}{dt} \sim r y(t) \left(\frac{y(t)}{Y} \right)^{-1} = rY  
\end{equation}
we can also obtain a linear solution 
\begin{equation}
y(t) \sim rYt +y(0) \propto t. 
\end{equation}
%線形に近似できる解も得られる.
% このような線形の解は，本データでも実際に，ローカル「行政区」の名前や「行政施設」名などの普及に見られた（図\ref{examples}(a)）. \par
This linear behavior was also confirmed in our data concerning the growth of word counts for the names of local boroughs or administrative facilities (see Fig. \ref{fig_examples}(b)).
\par 
%まとめると，Appendix \ref{xxx} より，パラメータの$Y$と$\alpha$の符号により，４つに式の時間発展は定性的に分類できる.
%In summary, from Appendix \ref{xxx}, the time evolution of the equations can be qualitatively classified into four categories on the basis of the signs of the parameters $Y$ and $\alpha$, 
%\begin{enumerate}
%\item 定数に収束（ロジスティック方程式型）　($Y<0$, $\alpha>0$)，
%\item 有限時間発散（締め切り効果）　（$Y>0$, $\alpha>0$）
%\item 無限時間後に発散（べき乗関数）　($Y>0$ ,$\alpha<0$) 
%\item 有限時間の1階微間発散　($Y<0$ ,$\alpha<0$) 
%\end{enumerate}
%\begin{enumerate}
%\item Convergence to a constant (S-curve) ($Y<0$, $\alpha>0$)，
%\item Finite-time divergence（Deadline effects）（$Y>0$, $\alpha>0$）
%\item Divergence after infinite time (Asymptotic power-low function）($Y>0$ ,$\alpha<0$) 
%\item Finite-time divergence of first-order derivatives ($Y<0$ ,$\alpha<0$) 
%\end{enumerate}
%\par
%なお，この方程式の解は一般に変数分離法を求めることで求めることができ，解$y(t)$は，
The exact solution of the equation given by Eq. \ref{base_eq} can be solved as a variable separation form, 
%\begin{equation}
%t= t_0 + \frac{1}{r}\left(-B(1-\frac{x(t)}{N};1-\alpha,0)+B(1-\frac{x(t_0)}{N};1-\alpha,0)\right)
%\end{equation}
\begin{eqnarray}
t&=&t_0+\frac{1}{r} \int^{1+y(t)/Y}_{1+y(t_0)/Y} \frac{1}{(1-x) \cdot x^{\alpha}}dx \\
&=& t_0+\frac{1}{r} (B_{\alpha}(1+y(t)/Y)-B_{\alpha}(1+y(t_0)/Y)). \label{th_t}
%&=&t_0+\frac{1}{r}B_{\alpha,N,y(t_0)}(y(t)) \label{th_t}.
\end{eqnarray}
By introducing the inverse function, we can formally write the solution as
%\begin{equation}
%y(t)=B^{-1}_{\alpha,N,y(0)}(r(t-t_0)) \label{th_y}.
%\end{equation}
\begin{equation}
	%x(t)=N \left( 1-B^{-1}_{\alpha}(-r(t-t_0)+b_0) \right)
	y(t)=Y \left(B^{-1}_{\alpha}(r(t-t_0)+B_{\alpha}(1+y(t_0)/Y)) -1 \right), 
\end{equation}
%と形式的にと書ける.
%\begin{equation}
%	t= t_0+\frac{1}{r} (B_{\alpha}(1+x(t)/N)-B_{\alpha}(1+x(t_0)/N)). 
%\end{equation}
	% \par
	% ここで，$B_{\alpha}(x)=B(x;a-\alpha)$.  
	 %なお，ベータ関数では虚部は$x$に依存せず差分を取ることでキャンルされるため，実部のみが現れる. \par
%これを$x(t)$に関して解けば，%
%Solving formally for $x(t)$, $x(t)$ can be written as% follows: 
%	\begin{equation}
%	%x(t)=N \left( 1-B^{-1}_{\alpha}(-r(t-t_0)+b_0) \right)
%	x(t)=N \left(B^{-1}_{\alpha}(r(t-t_0)+b_0) -1 \right).
%	\end{equation}
%where 
%\begin{eqnarray}
%&&B_{\alpha,N,y(0)}(y(t))=\int^{1+y(t)/Y}_{1+y(t_0)/Y} \frac{1}{(1-x) \cdot x^{\alpha}}dx \\
%&&\left\{  
%\begin{array}{cc}
%{}_1F_2(\alpha)-{}_1F_2(\beta) & (\alpha is noniteger) \\
%{}_1F_2(\alpha)-{}_1F_2(\beta) & (\alpha is integer)
%\end{array} \right.  
%\end{eqnarray}
where 
%\begin{eqnarray}
%	&&B_a(v) \equiv \nonumber  \\
%	&&\left\{ 
%	\begin{array}{ll}
%	\frac{v^{1-a}{}_2F_1(a,1-a;2-a;v)}{1-a} & (others)   \nonumber \\ 
%	 \frac{1-v} {|1-v|} \log(|1-v|)+\log(v)+\sum^{a}_{i=2}\frac{1}{-i+1}v^{-i+1} & (a=2,3,\cdots), \nonumber  \\
%	\end{array}
%	\right. , 
%\end{eqnarray}
\begin{eqnarray}
    &&B_a(v) \equiv \nonumber  \\
    &&\left\{ 
    \begin{array}{ll}
     (a=1)  & \nonumber \\
    \frac{1-v} {|1-v|} \log(|1-v|)+\log(v) & \nonumber \\
    (a=2,3,\cdots) & \nonumber \\
    \frac{1-v} {|1-v|} \log(|1-v|)+\log(v)+\sum^{a}_{i=2}\frac{1}{-i+1}v^{-i+1} &  \nonumber  \\
    (others)  & \nonumber \\
    \frac{v^{1-a}{}_2F_1(a,1-a;2-a;v)}{1-a} &  \nonumber \\ 
    \end{array}
    \right.
    \end{eqnarray}
%\begin{equation}
%	b_0=B_{\alpha}(1+x(t_0)/N).
%\end{equation}
${}_2F_1(a,b;c;x)$ is the Gauss hypergeometric function and the inverse function of $B_{\alpha}(x)$ is denoted by $B^{-1}_{\alpha}(t)$.
%\begin{equation}
%B^{-1}_{\alpha}(B_{\alpha}(x))=x. 
%\end{equation}
\textcolor{black}{Here, $B^{-1}_{\alpha}(B_{\alpha}(x))=x$, $0<B^{-1}_{\alpha}(t)<1$ for $Y<0$ and $B^{-1}_{\alpha}(t)>1$ for $Y>0$.}
Details of the derivation of the solution are given in Appendix \ref{app_seq_th}.
%\end{eqnarray}
%である。ここで，逆関数を導入すると，
%\begin{equation}
%y(t)=N \left( 1-B^{-1}_{\alpha}(-r(t-t_0)+b_0) \right)
%\end{equation}
%と書ける. 
%
%\begin{eqnarray}
%t&=&t_0+\frac{1}{r} B(1+y(t_0)/Y,1-y(t)/Y;1-\alpha,0) \\ \nonumber 
% &=&t_0+\frac{1}{r} \hat{B}_{\alpha,Y}(y(t))
%\end{eqnarray}
%従って，これを形式的に解くと
%\begin{eqnarray}
%y(t)= \hat{B}^{-1}_{\alpha,Y}(r(t-t_0))
%\end{eqnarray}
%となる.
%\begin{eqnarray}
%t&=&t_0+\frac{1}{r} B(1+y(t_0)/Y,1-y(t)/Y;1-\alpha,0) \\ \nonumber 
% &=&t_0+\frac{1}{r} \hat{B}_{\alpha,Y}(y(t))
%\end{eqnarray}
%
%
%ここで，$t_0=0$としても一般性を失わないので，
%\begin{equation}
%t=t_0+\frac{1}{r}B_{\alpha,N,y(0)}(y(t))
%\end{equation}
%とかける.
%\begin{equation}
%y(t)=N \left( 1-B^{-1}_{\alpha}(-r(t-t_0)+b_0) \right)
%\end{equation}
%と書ける. 
%ここで，
%\begin{equation}
%b_0=B_{\alpha}(1-x(t_0)/Y).
%\end{equation}
%であり，$B^{-1}_{\alpha}$は，関数$B_{\alpha}(x)$ の逆関数である.
%つまり，
%\begin{equation}
%B^{-1}_{\alpha}(B_{\alpha}(x))=x 
%\end{equation}
%を満たす関数である.
%ここで，$\Re(z)$は，複素数$z$の実部，$B(x;\beta_1,\beta_2)$は，複素数領域に拡張したベータ関数であった.
\par
%本研究は，最低限のロジスティック方程式の拡張にかかわらず，単語カウント時系列の時間発展を良く説明できることを示す. また，時系列の様々な多様性を性質をパラメータの変化で統一的に理解できることも示したい．
%ただし，この式はデータを良く説明できるが，多少冗長であり，完璧な式はない可能性もある（同じ時系列を説明する複数パラメータの組が存在する）.
%\par
%\section{データ}
%以下のデータを定義する. 
%$W$は文字列。$y_i(t)$は規格化した件数.
%
%In summary, the proposed model also improves on the logistic regression model in terms of prediction accuracy, and in addition, it is roughly on the same level as the more complex general-purpose time series model.

\begin{figure*}[t]
    \begin{minipage}{0.24\hsize}
        %\includegraphics[width=4cm]{count_sagamiharamidoriku__nolog_2533.eps}
        %%%%%%%\includegraphics[width=3.8cm]{transformed_count_ave.eps}
        %%%%%%%%%%\includegraphics[width=3.8cm]{"transformed_count_ave_t_th.eps"}
        %\begin{overpic}[width=3.8cm]{"transformed_count_ave_t_th.eps"}
            \begin{overpic}[width=3.8cm]{"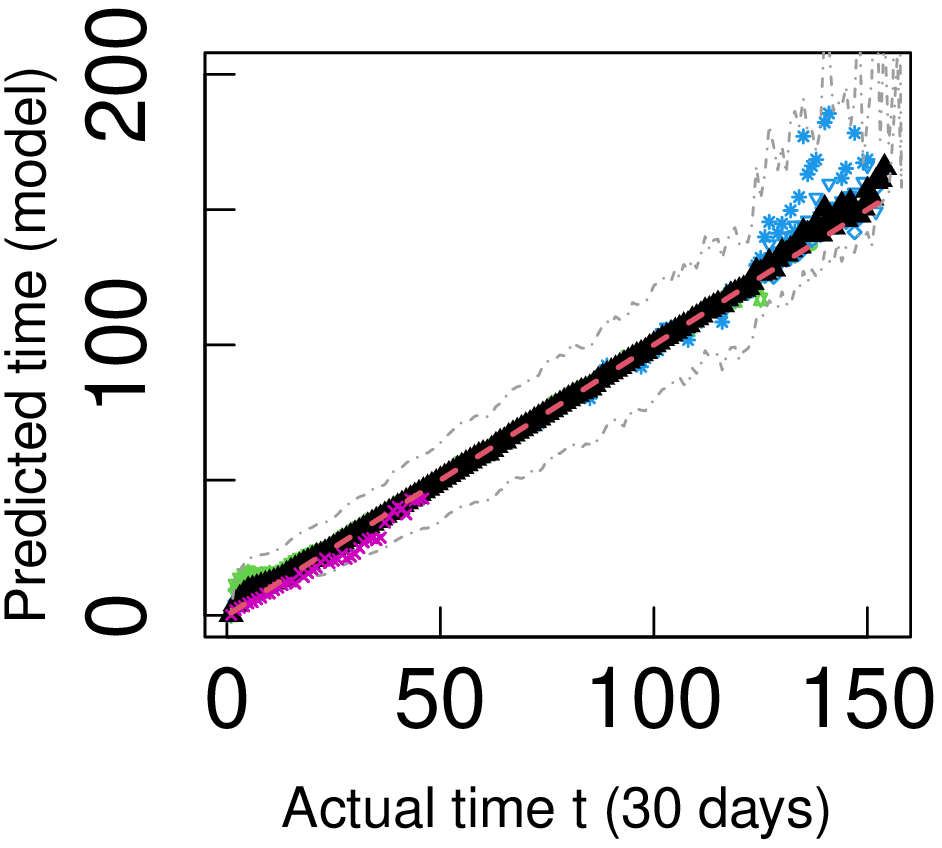"}
            %\put(22,55){(a)}
            \put(26,75){(a)}
            \put(26,65){Blog}
        \end{overpic}
        \end{minipage}
    \begin{minipage}{0.24\hsize}
        \begin{overpic}[width=3.8cm]{"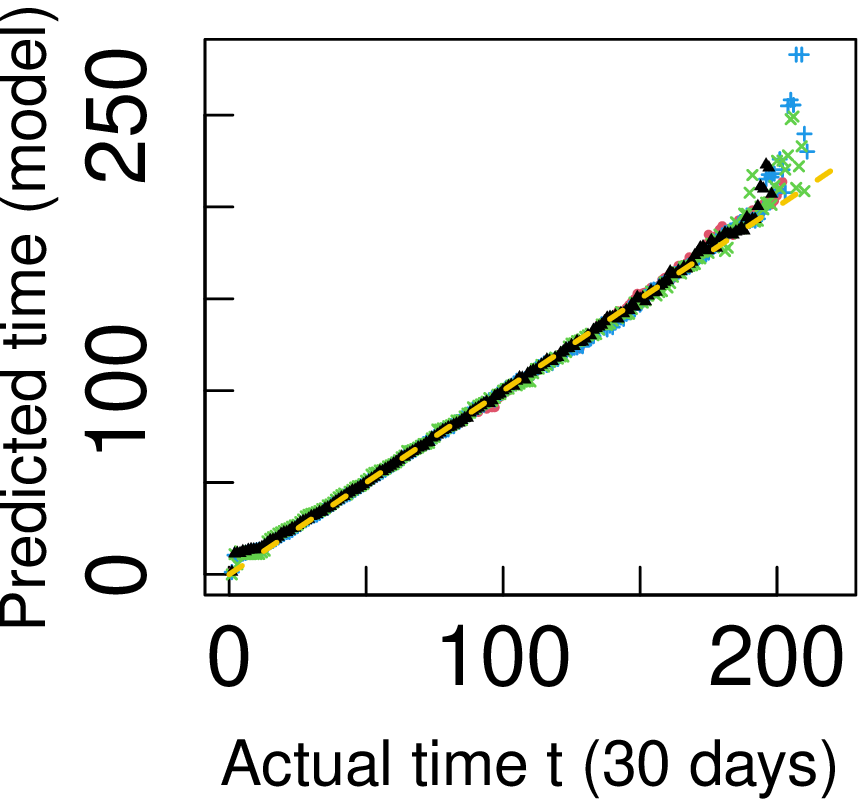"}
            %\put(22,55){(a)}
            %\put(24,69){(b)}
            %\put(24,60){Search query}
            \put(26,82){(b)}
            \put(26,72){Search query}
        \end{overpic}
    \end{minipage}
    %\includegraphics[width=8cm]{b_L.eps}
    %\begin{minipage}{0.48\hsize}
    \begin{minipage}{0.24\hsize}
            \begin{overpic}[width=3.8cm]{"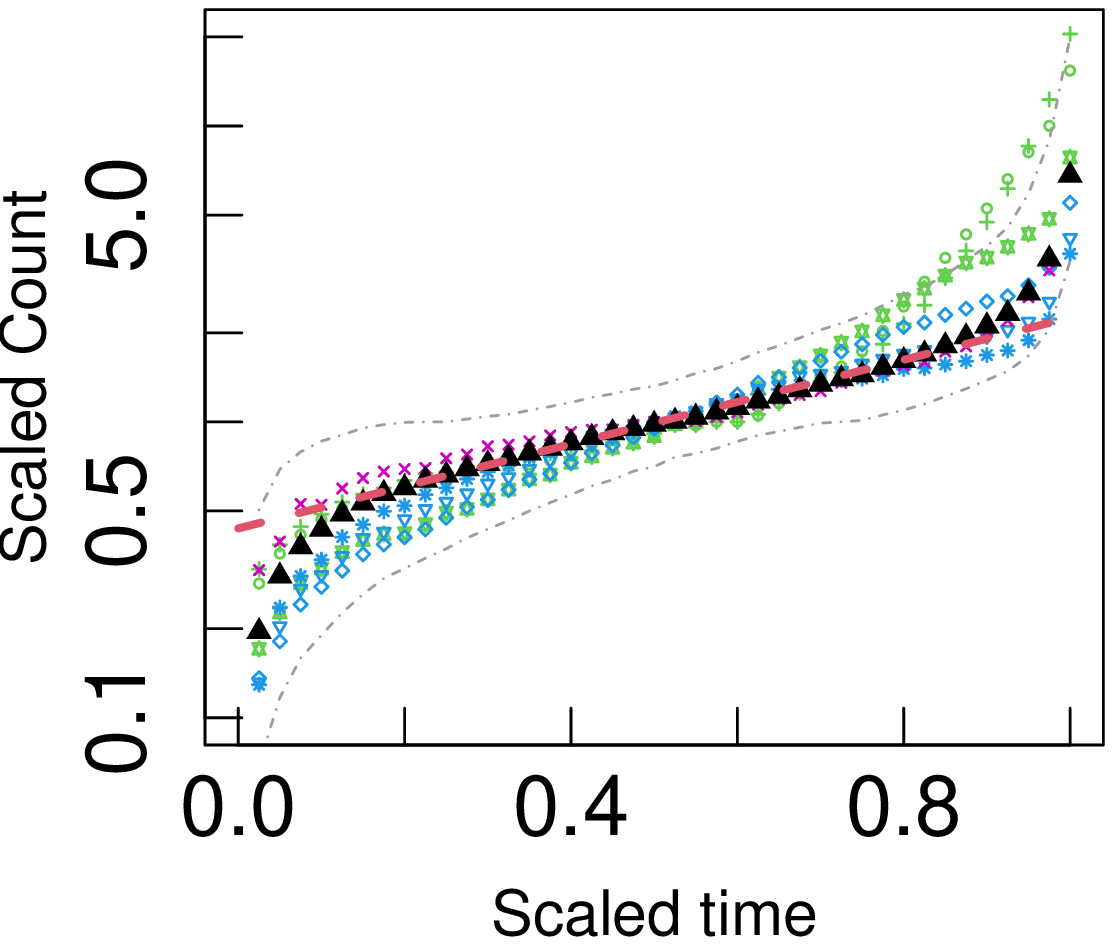"}
                %\put(22,55){(a)}
                \put(24,72){(c)}
                \put(32,30){Linear trans.}
            \end{overpic}
    \end{minipage}
    %\begin{minipage}{0.48\hsize}
    %\begin{minipage}{0.48\hsize}
    \begin{minipage}{0.24\hsize}
    %\begin{minipage}{0.48\hsize}
        %\includegraphics[width=4cm]{count_sagamiharamidoriku__nolog_2533.eps}
        %\includegraphics[width=3.8cm]{transformed_count_ave_bass_comp.eps}
         %%%%%%%%%\includegraphics[width=3.8cm]{"transformed_count_ave_t_bass_pure.eps"}
         \begin{overpic}[width=3.8cm]{"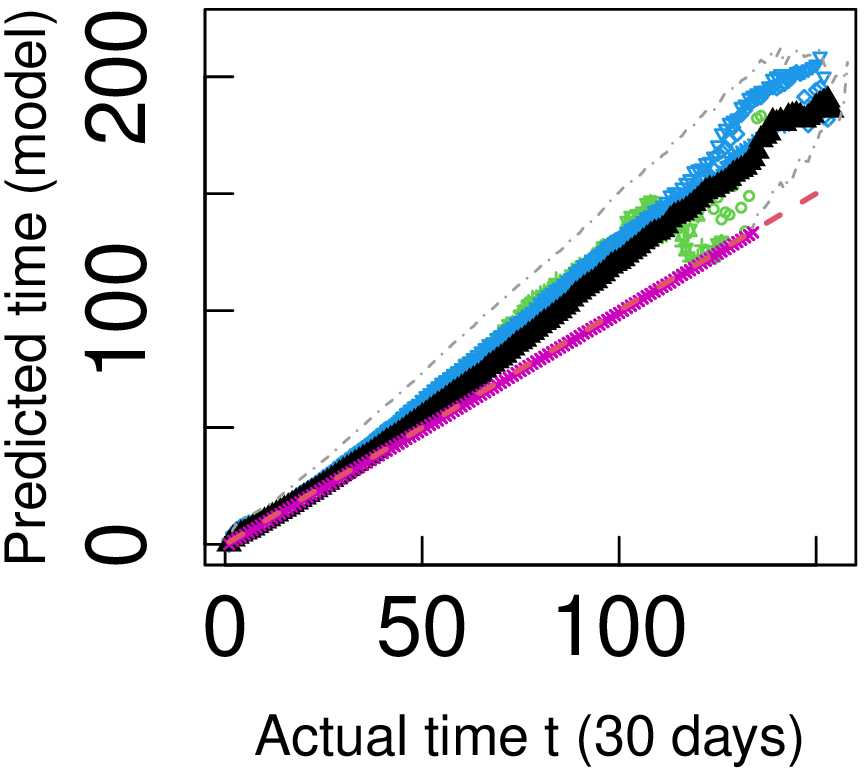"}
            %\put(22,55){(a)}
            %\put(24,69){(d)}
            %\put(38,28){Bass model}
             \put(26,75){(d)}
            \put(45,34){Bass model}
        \end{overpic}
    \end{minipage}
    \caption{
    Statistical verification of the proposed model given by Eq.\ref{base_eq}. \\
    % with the ensemble median time series.
    (a) Verification of blog data by using the statistical relation $\hat{z}(t|\alpha,s_Y)=t$ given by Eq. \ref{medi_th_alpha} (On calculating this relation, we predict $t$ from $y$ by using the proposed model given by Eq. \ref{th_t}). 
     x-axis corresponds to actual time, y-axis corresponds to predicted time by the model and
     the red dashed line is $y=x$. 
     In case of $Y>0$ ($s_Y=sign(Y)=1$), we plotted in green circles for $\alpha=0.5$, pluses for $\alpha=1.0$, for triangles up and down $\alpha=1.5$ 
     and blue diamonds for $\alpha=-0.5$, triangles point down for $\alpha=-1.0$ and  
     in the case of $Y<0$  ($s_Y=-1$), we plotted magenta crosses.
     %In the case of $Y<0$  ($s_Y=-1$), Blue diamonds for $\alpha=-0.5$, triangles point down for $\alpha=-1.0$ and, 
     %In the case of $Y<0$  ($s_Y=-1$), we plotted magenta crosses. 
    We also plot the ensemble median over all data $\hat{z}(t)$ given by Eq. \ref{eq_t_line}  in black triangles and the corresponding 10th and 90th percentiles are shown in gray dashed lines.
    The same straight line $y=x$, independent of $\alpha$ and $Y$ means that the word counts are statistically consistent with the dynamics described by the proposed model
    given by Eq. \ref{base_eq}. \\
    (b) The corresponding figure of the panel (a) for Google Trends data and proposed model given by Eq .\ref{base_eq}.   We plot the ensemble median over all the data $\hat{z}(t)$ given by Eq. \ref{eq_t_line}. The yellow dashed line is $y=x$.
     The data are shown in black triangles for Japanese, red circles for French, green crosses for Spanish, and blue pluses for English. \\
    (c) Ensemble median of time series scaled for time and word count by using the linear transformation $\hat{y'}(t'|\alpha,s_Y)$ given by Eq. \ref{simple_scale}. 
    The ensemble was taken conditionally by parameters $\alpha$ and sign of $Y$, $s_Y$ in the same way as the panel (a).  
     %In case of $Y>0$ ($s_Y=1$), we plotted in green circles for $\alpha=0.5$, pluses for $\alpha=1.0$ and triangles up and down $\alpha=1.5$  \\
     %Verification of actual data for blog data.
     % The time description (inverse functions of solution of differential
     %equation, F(−1)(t|Yj )).
    %We plot the ensemble median conditioned by parameters $\alpha$ and $Y$ in the same way as the panel (a). 
    From the panel, we can see that the x-axis and y-axis are both scaled, but the functional form is different for each $\alpha$, $s_Y$.  \\
    (d) The corresponding figure of the panel (a) for the Bass model given by Eq. \ref{bass}.  
    %The same straight line independent of $\alpha$ and $Y$ means that 
    From curves not obeying the straight line $y=x$, the word count data are conducted with the dynamics described by the bass model given by Eq. \ref{bass}. \textcolor{black}{Note that in panel (d), the parameters used for conditioning (color differences) are different from those used for prediction (i.e., Bass model). In the conditioning, we use the parameters of the proposed model and not the parameters of the Bass model.} \\
     \\
     %and blue diamonds for $\alpha=-0.5$; triangles point down for $\alpha=-1.0$
    %時間とワードカウントをスケールした時系列のアンサンブル中央値 given by Eq. \ref{xxx}。アンサンブルは，パラメータ$\alpha$と$Y$ごとにグループ分けしてとった（赤三角は、）。また、全データのアンサンブルはｘｘｘ色で与えた。パネルより、私たちはx軸とy軸はともにスケールされているが、関数形は$\alpha$ごとに異なってることがわかることができる。
    %(d) Corresponding time series transformed by the extended model given by Eq. \ref{xxx}. 
    %The model shows that the data can be transformed into the same straight line, independent of $\alpha$ and $Y$. 
    %This means that the word count data are statistically consistent with the dynamics described by the model.
    %The black triangles (which look like a black line) in Fig.\ref{fig_ave}(a) indicates $\hat{z}(t)$ given by Eq. \ref{medi} of the real data, where the median is taken 
    %for all words. Ensemble median of scaled time series by using the linear transformation $hat{y'}(t'|\alpha,s)$ given by Eq. \ref{simple_scale}. 
    %From the panel, we can see that the x and y axes are both scaled, but the functional form is different for each $\alpha$.
    %The straight line given by triangles implies the data reproduce Eq. \ref{th_t} well.
    %The straight line illustrated by triangles impl-ies the $\hat{z}(t)$ for actual data reproduce Eq. \ref{th_t}, namely, the time series of word counts are consistent with the dynamics given by Eq. \ref{base_eq} for not a few words.
    %Note that 
    }
    %拡張モデル given by Eq. \ref{xxx} による変換した対応する時系列 given by Eq. \ref{xxx}. 
    %モデルにより、$\alpha$や$Y$に依存せずデータが同一の直線に変換できていることがわかる。つまり、ワードカウントデータは，統計的には，モデルで記述されるダイナミクスと矛盾しないことを意味する。
    \label{fig_ave}
    \end{figure*}
    \begin{figure}
        \begin{minipage}{0.48\hsize}
        %\begin{minipage}{0.24\hsize}
        %\includegraphics[width=4cm]{"hist_alpha_reg.eps"}
        \begin{overpic}[width=4cm]{"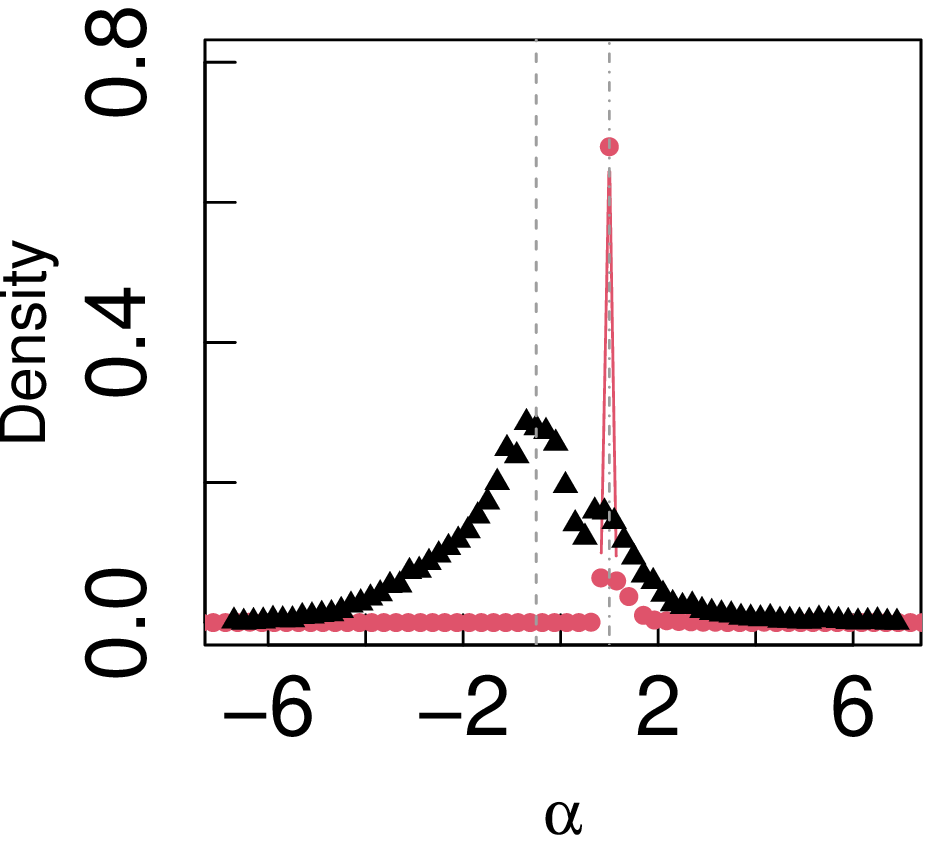"}
            %\put(22,55){(a)} 
        %\put(24,68){(a)}
        %\put(24,58){blog}
        \put(30,85){(a)}
        \put(30,75){blog}
        \end{overpic}
        \end{minipage}
        \begin{minipage}{0.48\hsize}
            \begin{overpic}[width=4cm]{"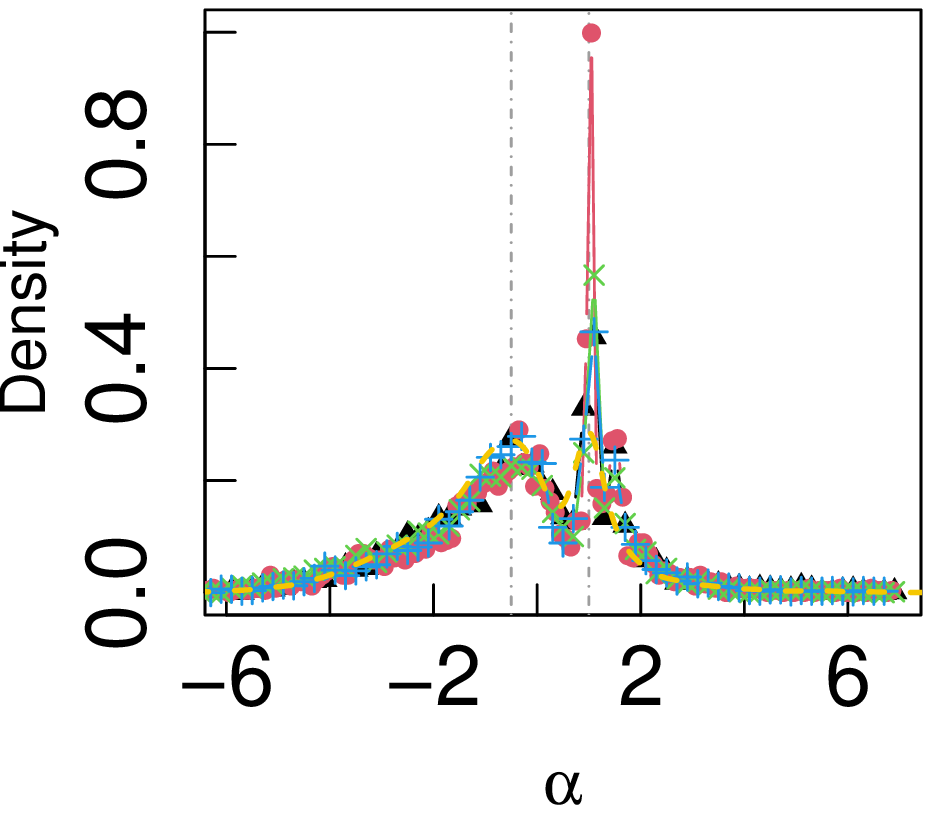"}
                %\put(22,55){(a)}
            %\put(24,69){(b)}
            %\put(24,58){Search}
            \put(30,85){(b)}
            \put(30,75){Search}
            \end{overpic}
        \end{minipage}  
        \begin{minipage}{0.48\hsize}
           \begin{overpic}[width=4cm]{"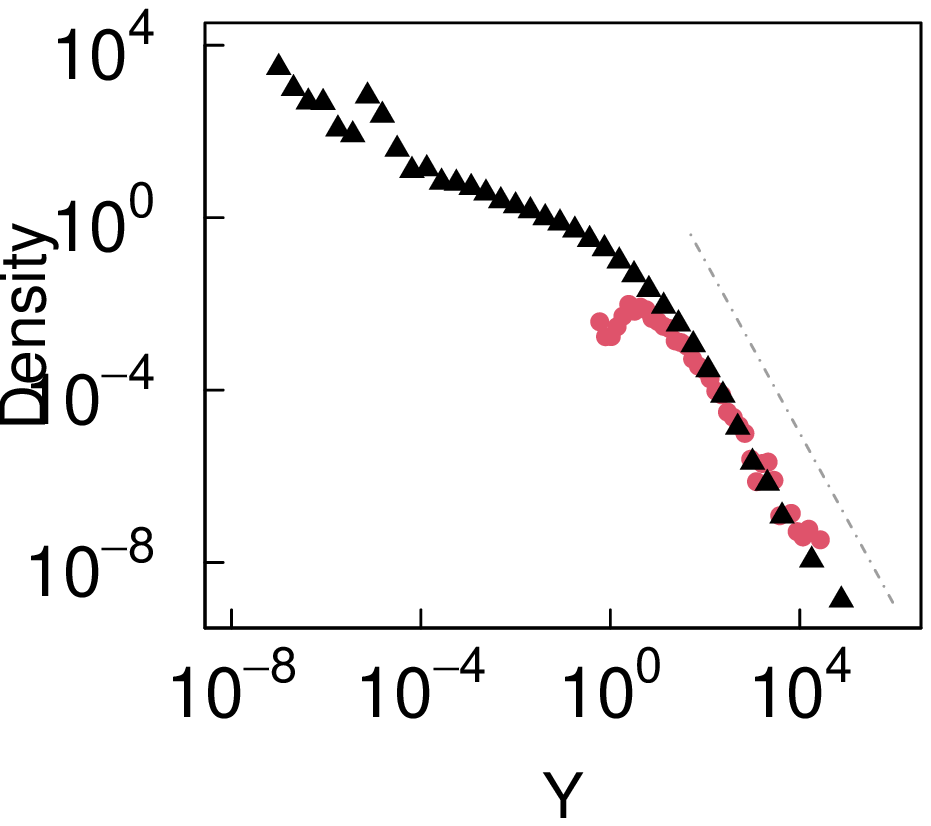"}
            %\put(22,55){(a)}
        %\put(24,31){(c)}
        %\put(36,31){blog}
        \put(24,35){(c)}
        \put(36,35){blog}
        \end{overpic}
        \end{minipage}
        \begin{minipage}{0.48\hsize}
            \begin{overpic}[width=4cm]{"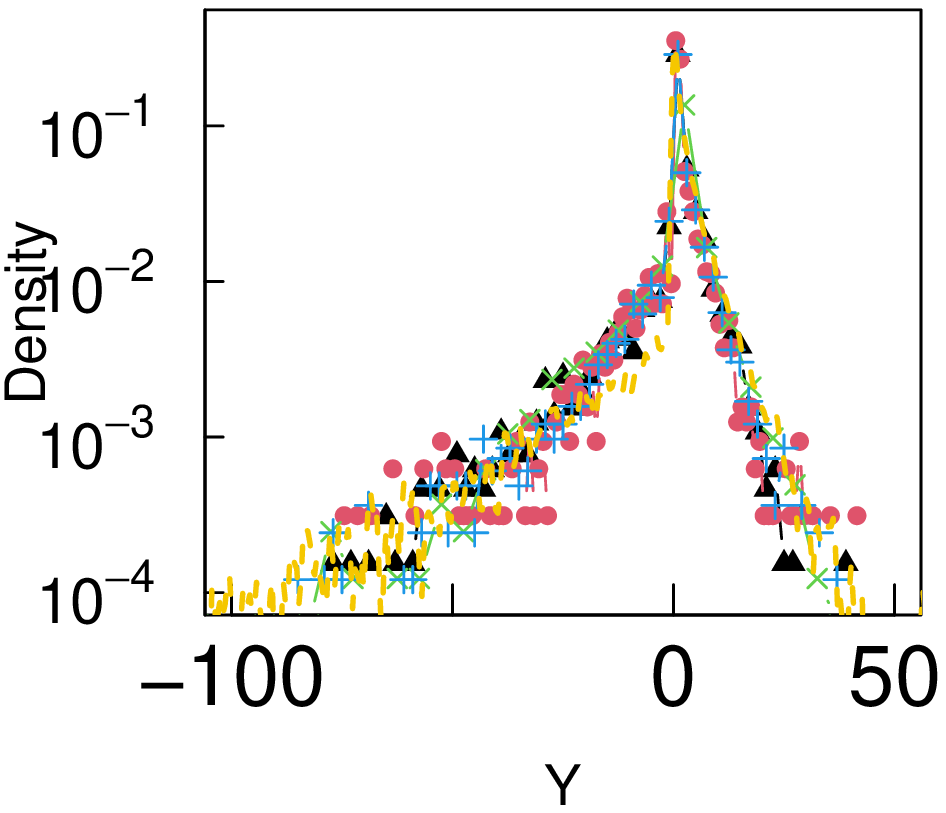"}
                %\put(22,55){(a)}
            %\put(24,69){(d)}
            %\put(24,58){Search}
             \put(30,85){(d)}
            \put(30,75){Search}
            \end{overpic}
        \end{minipage}
        \begin{minipage}{0.48\hsize}
        %\includegraphi
         %\includegraphics[width=4cm]{"hist_r_reg.eps"}
         \begin{overpic}[width=4cm]{"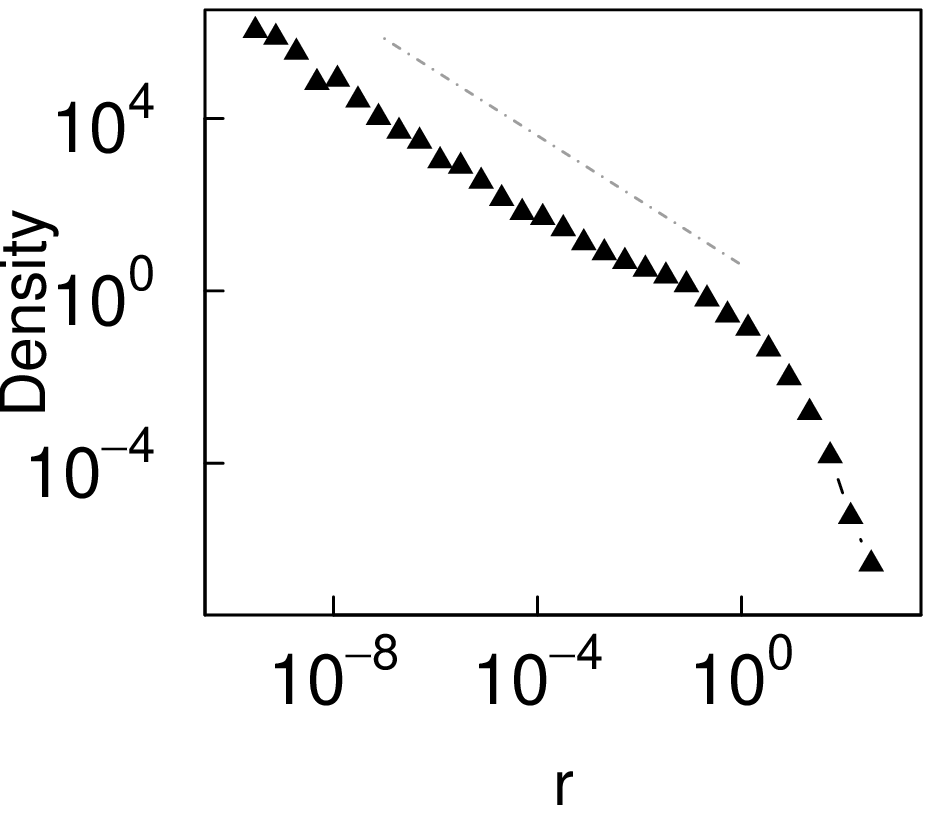"}
            %\put(22,55){(a)}
         %\put(24,31){(e)}
        % \put(36,31){blog}
         \put(24,35){(e)}
         \put(36,35){blog}
        \end{overpic}
        \end{minipage}
        \begin{minipage}{0.48\hsize}
            \begin{overpic}[width=4cm]{"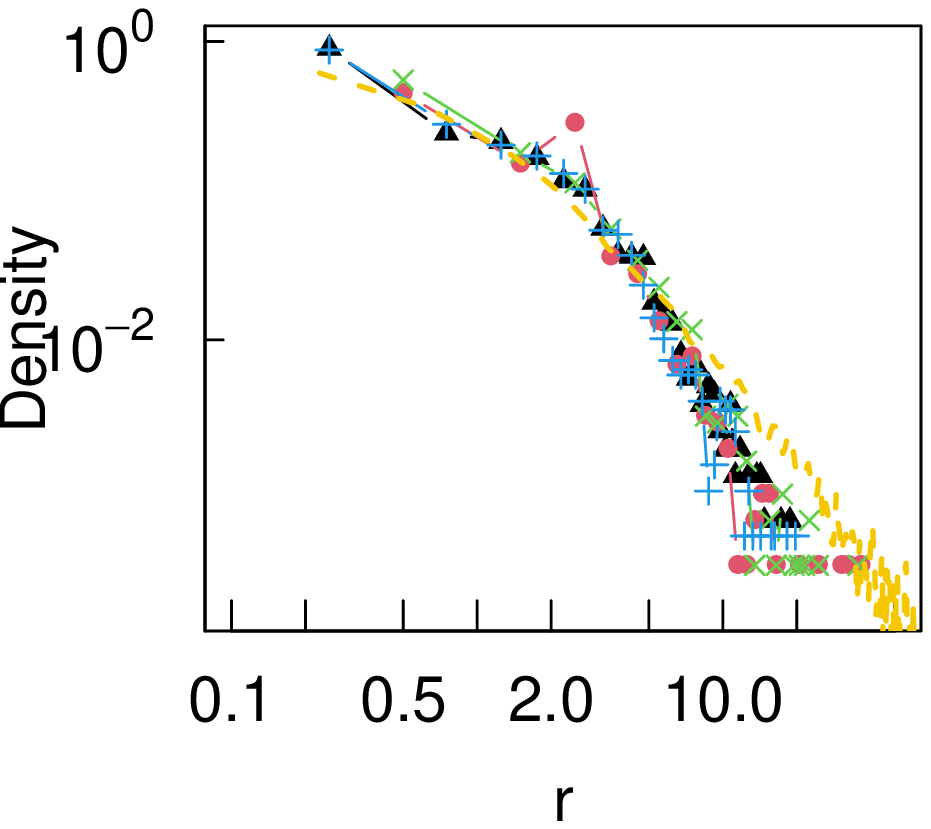"}
                %\put(22,55){(a)}
             %\put(24,31){(f)}
            % \put(58,68){Search}
             \put(24,43){(f)}
             \put(24,33){Search}
            \end{overpic}
        \end{minipage}
        \caption{Probability density function for $\alpha$, $Y$, and $r$. 
        (a) Distribution of $\alpha$ for blog data.  Black triangle indicate the data for $Y>0$, red circles for $Y<0$.
        The vertical grey dashed lines are $\alpha=-0.5$ and $\alpha=1$. 
        From the figure, we can see that the most typical values are  $\alpha \sim -0.5$ for $Y>0$, and  $\alpha \sim 1$ for $Y<0$ (i.e., the logistic function). 
        %The peak at $\alpha=0$ is an effect of regularization and is not essential (see Appendix \ref{app_sec_estimate}).
        (b) Distribution of $\alpha$ for Google Trends data and blog data. The data is shown in black triangles for Japanese, blue pluses for English, red circles for French, green crosses for Spanish, and the yellow dashed line for the blog data. The vertical grey dashed lines are $\alpha=-0.5$ and $\alpha=1$. The graph shows that all data have peaks at $\alpha \sim -0.5$ and $\alpha \sim 1$.
        (c) Distribution of $|Y|$ for blog data. We added the $\propto x^{-2}$ (the Zipf's law) shown in dotted lines. Black triangles indicate the data for $Y>0$, red circles for $Y<0$.
         %We can confirm that the distributions of $|Y|$ have the power law tail with exponent $-2$ corresponding to Zipf's law.
        (d) Distribution of $Y$ for Google Trends data and blog data. The colors and shapes correspond to the panel (b). For the blog data, we plot $Y'=100 \times Y/max(y(t))$ to match the Google Trends data maximum of 100 constraints.
        (e) Distribution of $r$ for blog data. The colors and shapes correspond to the panel (a). We added the $\propto x^{-0.75}$ shown in dotted lines.
        (f) Distribution of $r$ for Google Trends data and blog data. The colors and shapes correspond to the panel (b). 
        %The colors and shapes correspond to the panel (a). 
        }
        \label{fig_hist}
        \end{figure}
\section{Validation of the model with actual data}
\label{sec_vali}
We validated the proposed equation in three ways using real data.
Note that in this section, we analyze only the growth domain,  
which was extracted using the method described in Appendix \ref{app_sec_cut}. 
Examples of the growth domains are the domains between the gray vertical dashed lines in Fig. \ref{fig_world}.
Furthermore, words with a growth period of less than 12 points (approximately one year) were excluded from our analysis. \par
\textcolor{black}{In the Appendix \ref{app_sec_model}, we examine the proposed model in more detail.  For example, we compare the proposed model with models not discussed in this section.}  \par
%More detailed validation of the proposed model and comparison with other models is given in Appendix \ref{app_sec_model}.
%なお、ここでは時系列から成長部分のみに着目している。 
%Note that we have extracted only the growth part from the time series using the method described in Appendix \ref{app_seq_growth} and used only words with a growth period of at least about one year or 12 points (30 days $\times$ 12 points).
\subsection{Example of verification by individual time series}
Figs. \ref{fig_examples} show typical time series of word counts in the blog data. We can confirm that the curves given by Eq. \ref{base_eq} (the red dashed lines) are well in agreement with the most of the actual time series (black triangles), the proposed model Eq. \ref{base_eq} can comprehensively describes various patterns of growth of the word time series. 
For example, the linear growth for $\alpha \sim -1, Y>0$ is shown in panel (b), the finite-time divergence  $\alpha \sim 1, Y>0 $ in panel (h), and the S-curve $\alpha \sim 1, Y<0$ in panel (j).
The tuning parameters of the model, $\alpha$, $Y$, $r$, were estimated by minimizing the mean absolute error with regularization. 
The details of the parameter tuning are shown in Appendix \ref{app_sec_estimate}. \par
%図\ref{examples}はいくつかのブログデータにおける典型的な単語カウントの時系列について式\ref{xxx}を検証した例である.
% チューニングパラメータは，$\alpha$, $r$, $Y$ であり，$x_0$は実データから
%決めた.チューニング方法は，付録\ref{xxx}を参照されたい。
%図より，様々なパターンの単語時系列の発展の形状が式\ref{base_eq}で統一的に説明できているが確認できる.
% the equation \ref{base_eq} can comprehensively describe various patterns of growth of the word time series.
%Correspondingly, 
%In other words, the parameters $\alpha$ and $Y$ of Eq \ref{base_eq} also take various values. The statistics of the parameter 
%investigate more detail in Sec. \ref{xxx}. 
%\par 
%なお，パラメータはxxxの方法で推定した.
%まず 図\ref{xxx}により，様々なパターンの時系列が説明できることがわかる.
\subsection{Statistical verification of the shapes of a growth curve}
\label{sec_ave}
%次に，個々の時系列のみなく，データ全体の統計としてもオンラインでの単語の普及を \ref{base_eq}をよく表現できていることを示めす。
Second, we demonstrate that Eq. \ref{base_eq} holds not only for individual time series, but also for the statistics of the entire dataset.
%具体的には，もし，
%方程式が式\ref{base_eq}を満たしていれば，式\ref{th_t}が成立する.　そのため，式\ref{th_t}を用いて実データを変換すると，$t_0=0$としたとき，
By transforming the time series $y_j(t)$ and obeying the dynamics in Eq. \ref{base_eq}, we can derive the word-independent relationship, 
\begin{equation}
%z_i(t)=\frac{1}{r}B_{\alpha,N,x_0}(y_i(t))=t.
z_j(t)=t_0+\frac{1}{r} (B_{\alpha}(1+y_j(t)/Y)-B_{\alpha}(1+y_j(t_0)/Y))=t, 
%\label{eq_t_line}
\end{equation}
where we use Eq. \ref{th_t}.
%となり，単語に依存せず，時系列を単語に依存せず時間のみ依存する関数に標準化ができる。
%したがって，これが成立すれば，この量の時間ごとの単語に関する中央値をみても，
Taking the ensemble median for words in this transformed time series $z_j(t)$, we can also obtain  
the simple relation, 
\begin{equation}
\hat{z}(t)=Median_j[z_j(y_j(t)) ]=t. \label{eq_t_line}
\end{equation}
%これを検証に利用した. \par 
%We use this time series of the ensemble medians $\hat{z}(t)=t$ to validation for actual data.
This statistical relationship was used to validate the data.
\par
The black triangles (which appear as a black line) in Fig. \ref{fig_ave}(a) indicates $\hat{z}(t)$ given by Eq. \ref{eq_t_line} of the real data.
% where the median is taken for all words.
%まず，図\ref{fig_ave}(a)の黒三角はすべての対象単語について中央値をとったものである.
%The straight line given by triangles implies the data reproduce Eq. \ref{th_t} well.
The straight line illustrated  by triangles implies that $\hat{z}(t)$ for the actual data reproduces Eq. \ref{th_t}, mainly, the time series of word counts are consistent with the dynamics of the proposed model given by Eq. \ref{base_eq} statistically.
%for not a few words.
%結果，式\ref{th_t}を良く再現しほぼ直線になっていることが確認できる.　
%なお，中央値（５０％タイルだけなく）、
%90パーセンタイルと１０パーセンタイルもほぼ直線になっていることも確認できた（灰色線）。 \par
Note that the 10th and 90th percentiles shown in grey dashed lines are also almost straight lines, in the same way as the 50th percentile given by Eq. \ref{eq_t_line}.
 \par
%さらに，データ全体だけでなく，データごとに推定されたパラメータ$\alpha$と$N$ごとにわけても統計をとってみた。
%具体的には，
We also check that Eq. \ref{eq_t_line} (i.e., dynamics given by Eq. \ref{base_eq}) holds on actual data independently of the parameters $\alpha$ and $Y$.
We introduced an ensemble median conditioned on $\alpha$ and the sign of $Y$ denoted by $s_Y$ taking $1$ or $-1$ as follows:  
\begin{eqnarray}
&&\hat{z}(t|\alpha,s_Y)= \nonumber \\ 
&&Median_{\{j| \alpha-d \leq \alpha_j < \alpha+d, sign(Y_j)=s_Y \}}[\hat{t}_j(z_j(t))]=t, \nonumber \\ \label{medi_th_alpha}
\end{eqnarray}
where $d=0.5$ for $Y>0$ or $d=\infty$ for $Y<0$ is the box size for obtaining the statistics, and $Median_{\{S\}}[x]$ is the median $x$ over the set $S$. 
%理論式であることを示したい。
The fact shown in Fig. \ref{fig_ave}(a) that the lines with different colors and shapes are overlapping enough to hide each other means that the word count data consistent with the dynamics Eq. \ref{base_eq}, regardless of the main parameters $\alpha$ and $Y$, where green circles indicate [$\alpha=0.5, Y>0$]; green pluses [$\alpha=1.0, Y>0$],  triangles up and down [$\alpha=1.5, Y>0$]; blue diamonds [$\alpha=-0.5,Y>0$],blue triangles pointing down for [$\alpha=-1, Y>0$], blue stars for [$\alpha=-1.5, Y>0$]; purple crosses for [$Y<0$].
 \par 
 Finally, we confirmed that the simpler linear normalization and Bass model given by Eq. \ref{bass} cannot describe 
 the actual data. 
 %First, simpler linear normalization cannot convert data with different parameters into a common curve.
 %These quantities are plotted in Fig. \ref{fig_ave}(a).
 The simple linear normalizations are shown in Fig. \ref{fig_ave}(c).
 From this figure, we confirm that linear normalization converts data with different parameters into a common curve.
 %次に，比較として，異なるよりシンプルな標準化を用いたものが 
%図\ref{fig_ave}(b)である。具体的には，横軸を増加の上昇の全期間を１に，
%縦軸を時系列の中央値で下記のように標準化した.
%Specifically, the horizontal axis is standardized by the total growth period and the vertical axis by the median of the time series as follows:
In the figure, the horizontal axis is standardized by the total growth period, 
%the time series as follows:
\begin{eqnarray}
t_j'&=& t/T_j,  
\end{eqnarray}
where $T_j$ is the length of the growth period of $j$-th word. \par
The vertical axis is scaled by the median of the time series:
\begin{eqnarray}
y_j'(t) &=& y_j(t)/Median_{\{t=1,2,\cdots,T_j\}}[y_j(t)], 
\end{eqnarray}
%それを，式\ref{medi_th_alpha}と同様に単語の中央値をとりプロットした。
and in the same manner as in Eq. \ref{medi_th_alpha}, we plot the ensemble median conditioned on the parameters $\alpha$ and the sign of $s_Y$ (i.e, $s_Y=sign(Y)$):
\begin{eqnarray}
&&\hat{y'}(t'|\alpha,s_Y)= \nonumber \\
&&Median_{\{j| \alpha-\delta \leq \alpha_j < \alpha+\delta, sign(Y_j)=s_Y \}}[y_j'(z_j(t_j'))], \nonumber \\ 
\label{simple_scale}
\end{eqnarray}
where in the case that $y_j'(t')$ is not observed data, we estimate it by linear interpolation from the data before and after observing $y_j'(t')$.  This result implies the trivial fact that the shape of the growth curves depends on parameters $\alpha_j$ and $Y_j$. 
\par
\textcolor{black}{The black triangles in Fig. \ref{fig_ave}(d) are the statistics of the time series transformed by the Bass model given by Eq. \ref{bass}, corresponding to Fig. \ref{fig_ave}(a) for the proposed model given by Eq. \ref{base_eq}.}
% (also shown in green circles in Fig. \ref{fig_ave}(c)). 
The fact that this plot is not the straight line $y=x$ shown in the red dashed line indicates that the Bass model cannot accurately describe the real data.
%\begin{equation}
%    \hat{B}(t)=Median_{\{i=1,2,\cdots,W\}}[B_i(y_i(t)) ]=t, \label{eq_t_line}
%\end{equation}
%where 
%\begin{equation}
%    %z_i(t)=\frac{1}{r}B_{\alpha,N,x_0}(y_i(t))=t.
%    B_i(t)=t_0+\frac{1}{r} (B_{\alpha}(1+y(t)/Y)-B_{\alpha}(1+y(t_0)/Y)).
%    %\label{eq_t_line}
%\end{equation}
%where we use the Eq. \ref{th_t}.
    %となり，単語に依存せず，時系列を単語に依存せず時間のみ依存する関数に標準化ができる。
    %したがって，これが成立すれば，この量の時間ごとの単語に関する中央値をみても，
%Taking the ensemble median for words in this transformed time series $z_i(t)$, we can also get the simple relation, 
%\begin{equation}
%   \hat{B}(t)=Median_{\{i=1,2,\cdots,W\}}[B_i(y_i(t)) ]=t. \label{eq_t_line}
%\end{equation}
%Fig. \ref{fig_ave}(b) shows the corresponding figure of Fig. \ref{fig_ave2}(c), 
%ここで$t'$が観測データにない場合は，観測できるその前後のデータから線形補完して推定した. \par
%These quantities are plotted in Fig. \ref{fig_ave}(a).
%From this figure, it can be seen that the shapes of the curves differ depending on $\alpha$ and $Y$.
%This result means that the parameters $\alpha$ and $N$ estimated from the data have information on the shapes of the growth curves.
%プロットしたグラフでは，式\ref{th_t}を用いた標準化と異なり，$\alpha$や$N$ごとに異なる形状の曲線になっており，曲線の形状まで標準化
%できたないことがわかる.
% 一方，式\ref{xxx}で標準化した$\alpha$や$N$に依存せず同じ直線であり，個別にも式\ref{xxx}が成立していることも確認できた（図では重なっている）. 
\par
%\begin{equation}
%\hat{t}(t,\alpha)=Median_{i| i \in \alpha -\delta <\alpha_i' < \alpha+\delta ; Y>0 }\{\hat{t_i}(t)}
%\end{equation}
%現実に直線になっていることがある.
%図式のパラメータは，付録\ref{xxx}を用いて推定している.
%以下のように変換すると色々わかる.そこで，式を書いたら図で示す。\par
%where M edie[xj (t)] is the median over the words set
%{x1(t), x2(t), · · · xW (t)} and W is the size of the set. We
%take the median over the set of the mean frequency over
%30, ˆcj = ∑T
%t=1 fj (t)/T ≥ 30. We exclude words with a
%small mean because they have relatively large signal-to-
%noise ratios (see Eq. 30). Figs. 3 (a)–(f) show that 
%logarithmic 
\subsection{Verification by forecasting ability}
\label{sec_pred}
%拡張モデルはロジスティック回帰よりパラメータ数が多いモデルのためモデルのあてはまりは必ずよくなる。
%The fit of the extended model given by Eq. \ref{base_eq}  is always better than the logistic equation given by Eq. \ref{xxxx},  
%when estimated using all data 
The fit (i.e., training error) of the extended model given by Eq. \ref{base_eq} is always better than the logistic equation given by Eq. \ref{eq_logi} theoretically because the extended model  includes the logistic equation as the special parameter (i.e., $\alpha=1$, $Y<0$). 
%because the extended model given by Eq. \ref{base_eq} also includes the logistic equation as the special parameter (i.e., $\alpha=1$, $Y<0$). 
%Therefore, there is a possibility that the improved fit provided by the extended model could be an overfit. 
%Hence, we also confirm the improvement in predictability of the extended model compared to the logistic regression model.
Here, we compared the forecasting ability of the models to check for overfitting. If the forecasting ability of the extended model is lower than that of the logistic equation, the extension is meaningless, and the model can be considered overfitting.
%しかし，そのあてはまりの向上はモデルの冗長性により過学習の可能性も残される。
%そこで，モデルのあてはまりの向上が過剰学習でないことを確認するため，ロジスティック回帰モデルより予測能力の改善していることを確認する。
To check for overfitting, we estimated the model parameters using the first 70 percent of the time series from the beginning of the growth period and predicted the remaining 30 percent of the time series.
%具体的には，７５パーセントの時系列を使いパラメーターを推定し，残り２５パーセントの予測精度を提案モデルとそのほかの時系列モデルと比較した.
%表\ref{table_pred}は，単語のうち，提案モデルのほうが予測が中央値絶対誤差の意味で改善した割合を満たしたものである. 
%中央値誤差は、
The absolute mean error was used to measure the prediction accuracy:  
\begin{equation}
\delta_j^{(model)}=Mean_{\{t|0.70T_j \leq t \leq T_j \}}[|\hat{y}_j(t)-y_j(t)| ], \label{eq_delta}  
\end{equation}
%で定義した.
%where $Median_{\{t|t \in S\}}[x(t)]$は，集合$S$をみたす$t$に関するデータの中央値. 
where $Mean_{\{t|t \in S\}}[x(t)]$ is the mean of the data for $t$ that satisfies set $S$ and $\hat{y}_j(t)$ is the predicted value of the jth word at time $t$ from a model, such as the proposed model or the logistic equation. \par
% $\hat{y}^{(0)}_i(t)$ given by Eq. \ref{base_eq}, and we note $\delta_i^{(0)}$ as the absolute mean error of the extended logistic equation. \par
%$\hat{y}_i(t)$は時系列モデルからの予測値である. 
%提案モデルでは， $\hat{y}_i(t)=B^{-1}_{\alpha_i,Y_i,{y_0}_i}(r(t-t_0))$，つまり, Eq. \ref{th_y}から計算できる.
%表\ref{table_pred}は，単語のうち，提案モデルのほうが予測が中央値絶対誤差の意味で改善した割合を満たしたものである. 
Table \ref{table_pred} shows the winning ratio of the proposed model (i.e., the ratio of words  for which the proposed model has a higher prediction accuracy than the comparison model ). 
%This ratio corresponds to the winning ratio of the proposed model and was defined as 
The winning ratio of the proposed model to the other models is defined as 
\begin{eqnarray}
&&R^{(model)}= \\ \nonumber 
&&\frac{\sum_{\{j| \delta_j^{(model)} > \delta_j^{(0)}, j \in W_s \}}1}{\sum_{\{j| \delta_j^{(model)} > \delta_j^{(0)}, j \in W_s \}}1+\sum_{\{j| \delta_j^{(model)} < \delta_j^{(0)}, j \in W_s \}}1},
\label{eq_win}
\end{eqnarray}
where $W_s$ is the set of focused words with a sample size of 12 or more in the training data and $\delta_j^{(0)}=\delta_j^{(proposed)}$ is the absolute mean error of the proposed model given by Eq. \ref{base_eq}.  
For $R^{(model)}>0.5$, there are more words for which the proposed model has a smaller prediction error than the comparison model (roughly, the proposed model has a better prediction ability). Conversely, for $R^{(model)}<0.5$, there are more words for which the proposed model has a larger prediction error than the comparison model (roughly, the proposed model has a lower prediction ability).
\par 
%For this analysis, we used limited word count time series that have more than 12. \par
%For example, if the prediction errors of the extended logistic equation are lower than those of the comparison model for all words, the ratio takes the value 1; conversely, if the prediction errors are higher for all words, the ratio takes the value 0. \par
%提案モデルは，６割でロジスティック回帰に比べて予測精度が改善していることが確認できた. ただし，予測に用いるデータの最小の長さは１２ポイント（約１年）もしくは２４ポイント（約２年）とした.　表より，　
%From the table \ref{table_pred}, we can confirm that the winning ratio of the proposed model against the logistic equation $R^{(Logistic)}$ about 60\%, mainly, the predictive ability of the proposed model is improved over the logistic equation.
From Table \ref{table_pred}, we can confirm that the predictive ability of the proposed model is improved over that of the logistic equation in terms of the winning ratio ($R^{(logistic)}=0.64$).
We can also confirm that the Bass model, given by Eq. \ref{bass} also has less predictive ability than the proposed model ($R^{(Bass)}=0.67$).
% in terms of winning ratio ($R^{(Bass)}=0.70$). 
%propo against the logistic equation $R^{(Logistic)}$ about 60\%
%can be confirmed %
%that the about 60\% of word time series for which the prediction errors of the proposed model are improved compared to those of of the logistic equation. 
%For this analysis, we used limited word count time series that have more than 12 or 24 training points.
\par
In addition to the logistic equation and Bass model, 
we compared its predictive ability with commonly used general time-series models, 
 the SARIMA model (Seasonal Autoregressive Integrated Moving model) \cite{hyndman2008automatic}, which is one of the most well-known, traditional, and representative statistical time series models, and the Prophet model \cite{taylor2018forecasting}, which is a recently developed time-series model.
 These models, with a larger number of tuning parameters, can describe more complex structures than the proposed model.
 %In addition, 
The forecasting ability in terms of the winning ratio is \textcolor{black}{equivalent} or higher than that of both the SARIMA model ($R^{(SARIMA)}=0.52$) and Prophet model ($R^{(Prophet)}=0.59$).
Details and comparisons of predictive ability with other models are discussed in Appendix \ref{app_sec_model} and are shown in Table \ref{app_table_pred}. \par
\begin{figure}
\begin{minipage}{0.49\hsize}
%\begin{minipage}{0.24\hsize}
%\includegraphics[width=4.0cm]{"down_graph_rel.eps"}
\begin{overpic}[width=4cm]{"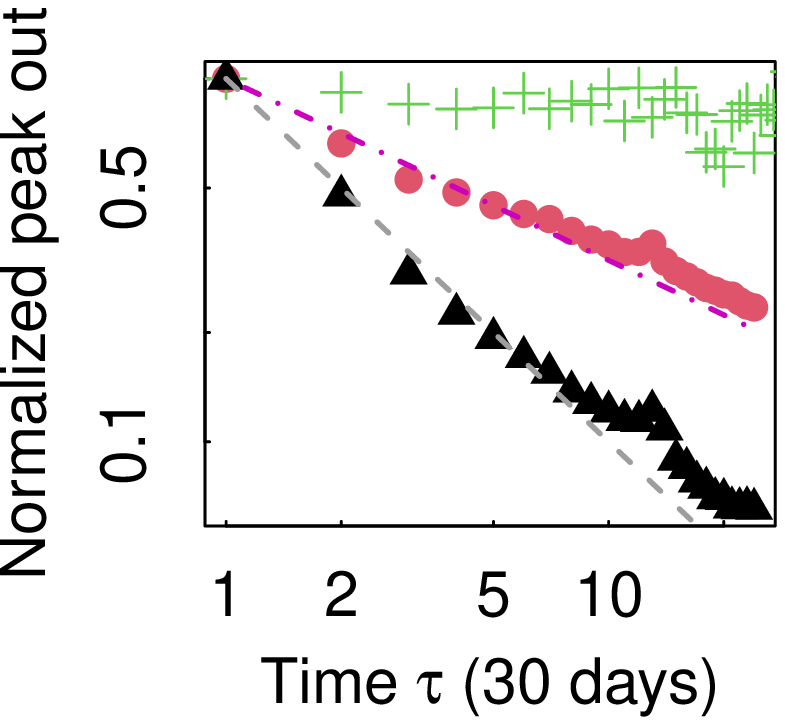"}
    %\put(22,55){(a)}
 %\put(24,31){(a)}
  \put(28,38){(a)}
\end{overpic}
\end{minipage}
\begin{minipage}{0.49\hsize}
\begin{overpic}[width=4cm]{"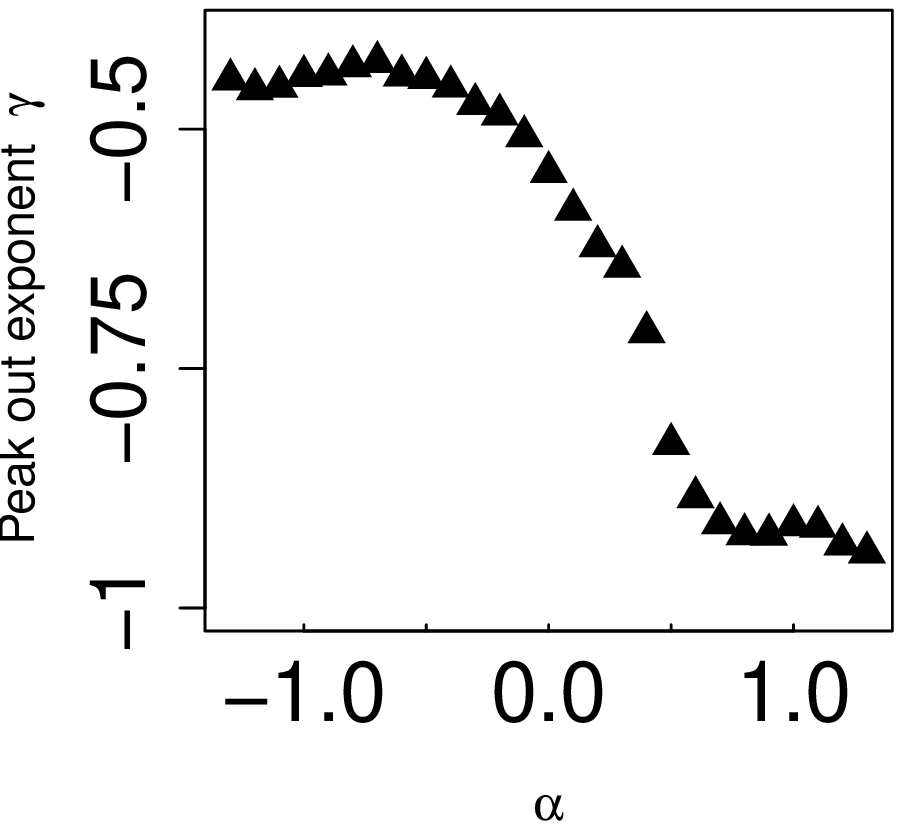"}
    %\put(22,55){(a)}
 %\put(24,31){(b)}
 \put(30,35){(b)}
\end{overpic}
\end{minipage}
\caption{
%Dependence of peak-out manner on growth model parameters.
(a) The behavor of peak-out in the log-log plot. 
Horizontal axis is time from peak.
The vertical axis is the ensemble median of peak-out standardized to 1 at the peak $\hat{v}(\tau|\alpha,s_Y)$
 given by Eq. \ref{eq_down}.  We plotted in black triangles conditioned by $\alpha>0$, $Y>0$, Red circles by $\alpha<0$, $Y>0$ and green pluses by $Y<0$. 
We can see that the power-law decay depends on the sign of $\alpha$ and $Y$.
(b) Relationship between $\alpha$ and the exponent $\gamma$ for $Y>0$, where the power-law decay is $\propto \tau^\gamma$. 
We can see the transition from $\gamma \sim -0.5$ to $\gamma \sim -1.0$.
%$\alpha$とピークアウトの関係。
%(a)ピークアウトの時系列。縦軸はピークを１に標準化したピークアウトのアンサンブル中央値の時系列データ。横軸がピークからの時間。私たちはべき乗減衰の仕方が$\alpha$の符号に依存することがわかることができる。
%(b)$\alpha$とべき乗則減衰の指数$\gamma$の関係。　私たちは$\gamma~0.5$から$\gamma~1.0$への遷移が見られることができる。
}
\label{fig_down}
\end{figure}
%
%\begin{figure}
%\begin{minipage}{0.49\hsize}
%\includegraphics[width=4.4cm]{alpha_google.eps}
%\end{minipage}
%\begin{minipage}{0.49\hsize}
%\includegraphics[width=4.4cm]{line_comp_world.eps}
%\end{minipage}
%\caption{
%Statistics on the growth time series of word counts in Google Trends.
%The data is shown in black triangles for Japanese, blue pluses for English, red circles for French, and green crosses %for Spanish. 
%(a)The probability density function of the $\alpha$ of the Google Trends data.  
%The corresponding data for Japanese blog shown in the yellow dashed line.
%pch 3 4 :Eng ,Es pch 4 col3  Fr col2 pch 16, Jap col1 pch17; col7 line Jap blog
%(b)Verification of the model's adaptability to Google Trends data \ref{xxx}.
%The red dashed line is $y=x$.
%From these graphs, we can confirm that the Google Trends data has properties that correspond to the Japanese blog data.
%モデルのグーグルトレンドデータへの適応可能性の検証 \ref{xxx} 。
%}
%\label{fig_google}
%\end{figure}
%\begin{table}
%    \begin{center}
%    \begin{tabular}{ccc}
%    $\alpha/K$ & -1 &  +1 \\
%    \hline
%    -1 &73 & 8140 \\
%    +1 & 3169 & 6985 \\
%    \hline
%    \end{tabular}
%    \end{center}
%    \caption{Number of samples for each parameter category}
%    \label{table_para}
%    \end{table}
\begin{table}
\begin{tabular}{cccc}
    Logistic & Bass & SARIMA &  Prophet \\
    \hline
    0.64 [0.63,0.65] &0.67 [0.66,0.68] & 0.52 [0.51,0.53]& 0.59 [0.58,0.60] \\
    %0.63 [0.62,0.64] &0.66 [0.65,0.67] & 0.48 [0.47,0.49] & 0.55 [0.54,0.56]  \\
    %Without regularization &0.60 [0.59,0.61] &0.47 [0.46,0.48] & 0.62 [0.61,0.63] \\
    %$\geq 24$ & 0.62 [0.61,0.63] &0.48 [0.46,0.49] & 0.62 [0.61,0.63] \\
    \hline
    \multicolumn{4}{l}{95\% confidence intervals in brackets.}  \\
    %\multicolumn{4}{l}{*The second column of the table shows the results without regularisation of $\alpha$.} 
    \end{tabular}
    \caption{Winning ratio of prediction errors of the proposed model to other models}
    \label{table_pred}
\end{table}

\begin{table}
    \begin{center}
    \begin{tabular}{ccc}
    $\alpha/K$ & -1 &  +1 \\
    \hline
    -1 &11 & 8868 \\
    +1 & 1323 & 4303 \\
    \hline
    \end{tabular}
    \end{center}
    \caption{Number of samples grouped by sign of $\alpha$ and $Y$}
    \label{table_para}
\end{table}

\begin{figure*}[t]
\begin{minipage}{0.19\hsize}
    \begin{overpic}[width=3.4cm]{"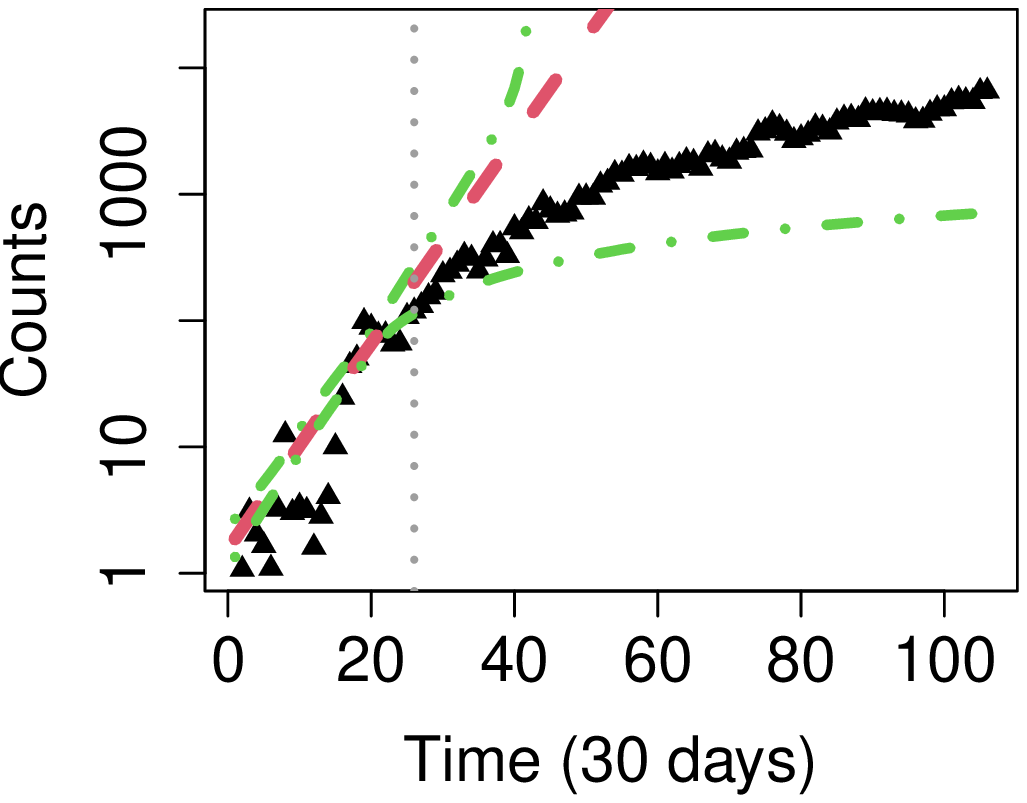"}
        %\put(22,55){(a)}
        \put(22,64){(a)}
    \end{overpic}
\end{minipage}
\begin{minipage}{0.19\hsize}
    \begin{overpic}[width=3.7cm]{"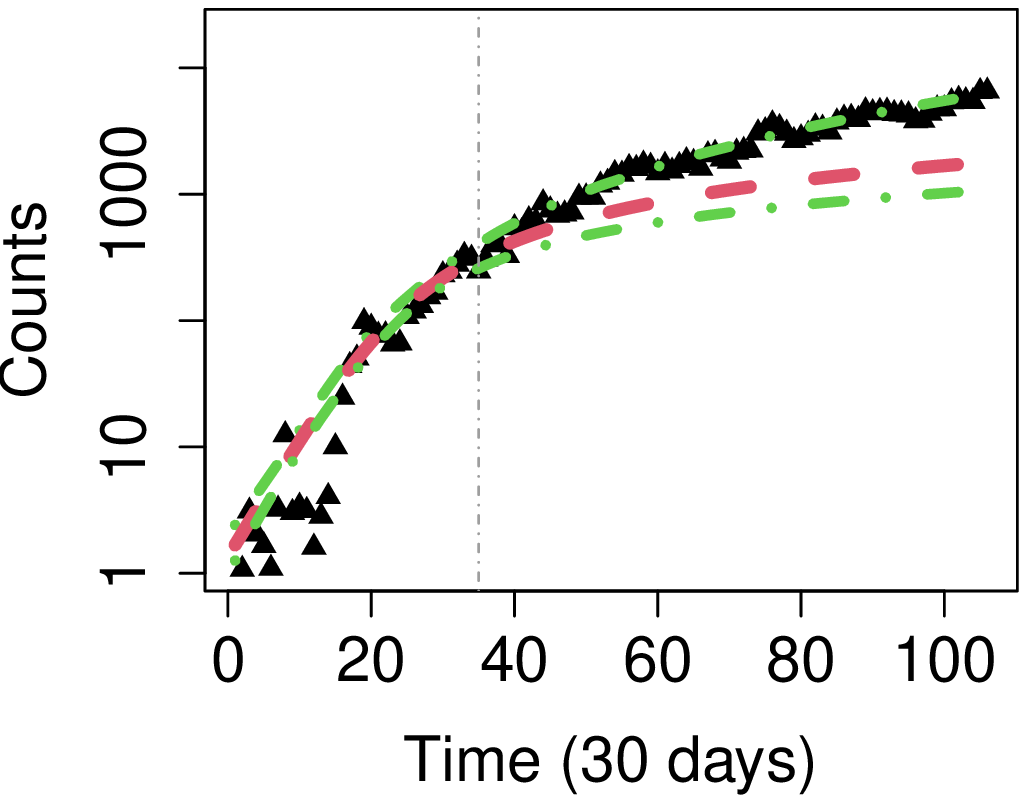"}
        %\put(22,55){(a)}
        \put(22,64){(b)}
    \end{overpic}
\end{minipage}
\begin{minipage}{0.19\hsize}
    \begin{overpic}[width=3.5cm]{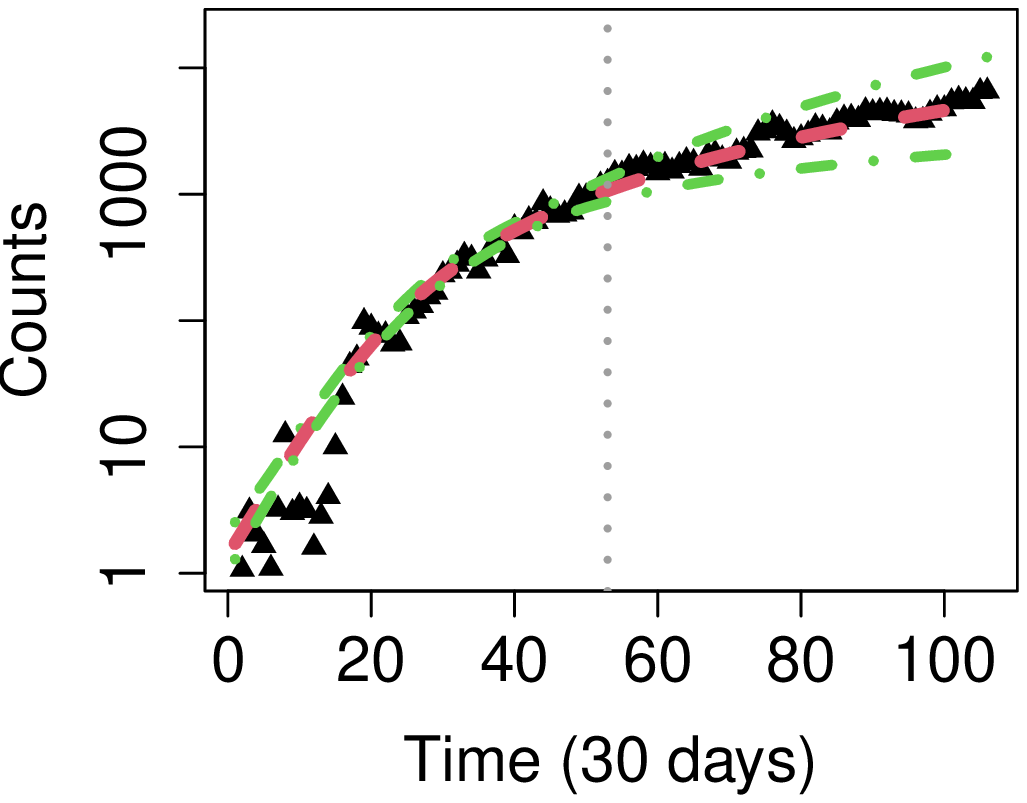}
        %\put(22,55){(a)}
        \put(22,64){(c)}
    \end{overpic}
\end{minipage}
\begin{minipage}{0.2\hsize}
%\includegraphics[width=3.4cm]{stan_75.eps}
%\begin{minipage}{0.48\hsize}
    %\includegraphics[width=4.0cm]{"all_4255_sagamiharasityuuouku_rel.eps"}
    \begin{overpic}[width=3.4cm]{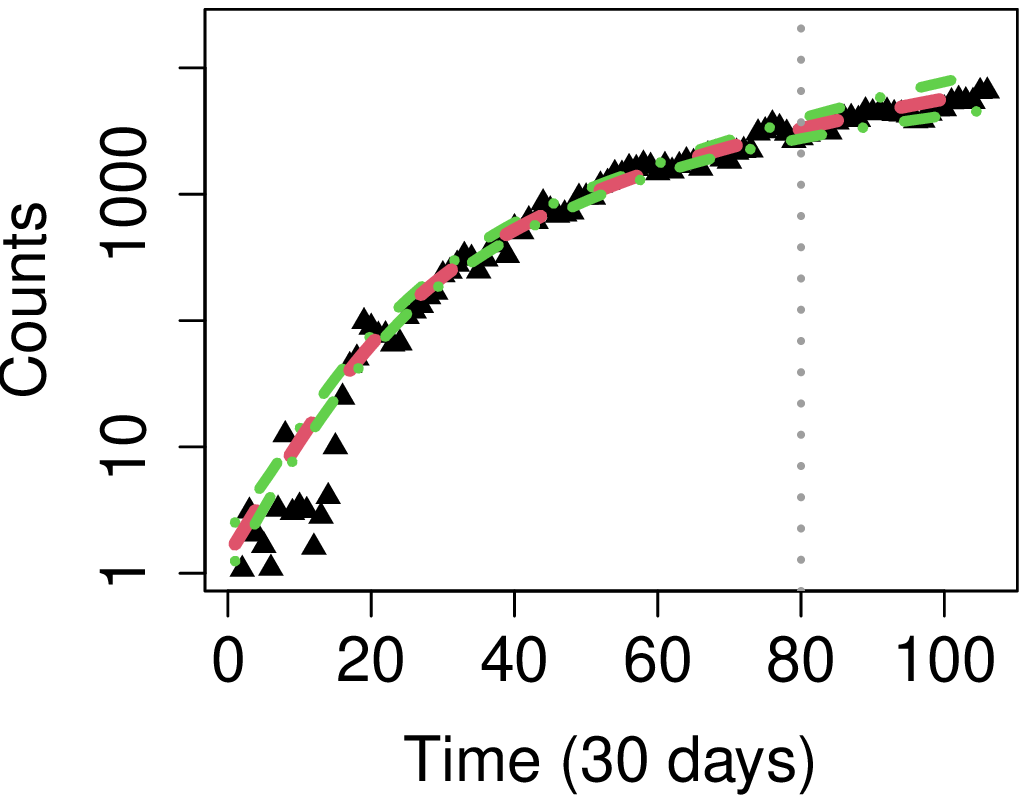}
        %\put(22,55){(a)}
        \put(22,64){(d)}
    \end{overpic}
\end{minipage}
\begin{minipage}{0.2\hsize}
%\includegraphics[width=4.3cm]{alpha_estimate_distb.eps}
%\begin{minipage}{0.48\hsize}
    %\includegraphics[width=4.0cm]{"all_4255_sagamiharasityuuouku_rel.eps"}
    \begin{overpic}[width=4.3cm]{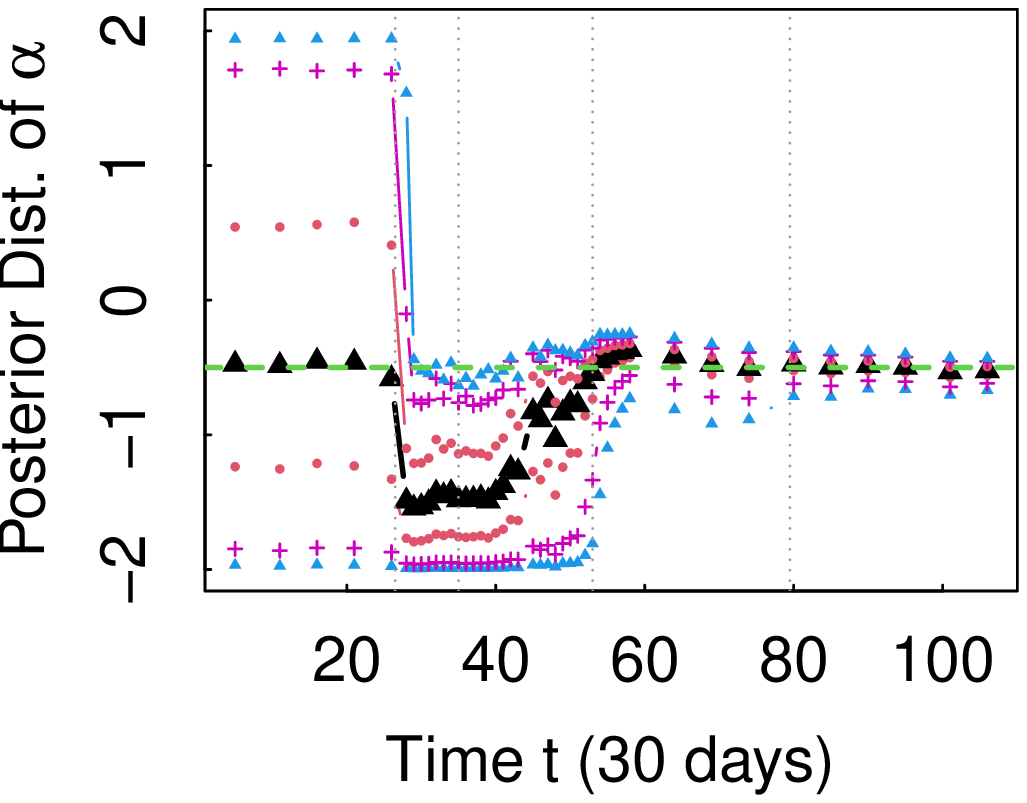}
        %\put(22,55){(a)}
        \put(95,77){(e)}
    \end{overpic}
\end{minipage}
\caption{Time series forecasting using the proposed model given by Eq. \ref{base_eq}.
(a)-(d) Comparison of time series of real data and time series predicted by the model. 
 The black triangles are real data for ``crowd funding'' (the same word as Fig. \ref{fig_examples} (d)).
 % the red dashed line for the median and the green dash-dotted lines for the average are predictions. 
 %The data are "crowd-funding" in the Japanese blog data (Fig. \ref{fig_examples} (d)). 
 The prediction lines are the red dashed line for the median, the green dashed line for the 1st and 99th percentile points, 
where the prediction uses Bayesian modelling, as shown in Appendix \ref{app_sec_param}, and the percentiles are calculated with their posterior distributions.
We used the data to the left of the grey dotted line to make our prediction.
The training data are $t \leq 26$ for the panel(a), $t\leq 37$ for the panel (b) $t \leq 53$ for the panel (c) and $t \leq 80$ for the panel (d).
We can confirm that the prediction improves with more training data. \\
(e)Dependence of the posterior distribution of $\alpha$ on the training data. 
The graph shows the posterior distribution of $\alpha$ when using the training data to the time indicated by the x-axis.
The data shown in black triangles is blue for 50 percentile, 
red circles for 25 percentile and 75 percentile, magenta pluses for 5 percentile and 95 percentile and 
blue triangles for 1 percentile and 99 percentile. The green horizontal dashed line is $y=0.5$ and the grey vertical dashed line corresponds to the training data in (a)-(d).
%1,99%変えたい
%q001,ylim=c(-2,2),pch=17,col=4)
%q005,ylim=c(0,1.2),pch=3,col=6)
%q025,ylim=c(0,1.2),pch=16,col=2)
%q050,ylim=c(0,1.2),pch=17,)
%q075,ylim=c(0,1.2),pch=16,,col=2)
%q095,ylim=c(0,1.2),pch=3,col=6)
%q099,ylim=c(0,1.2),pch=17,col=4)
The posterior distribution of $\alpha$ converges from a prior distribution (the uniform distribution of -2 to 2) to $\alpha \sim -0.5$ as the training data increases.
}
\label{fig_esitmation}
\end{figure*}
%h-wata@ismhpe runmed_nuxb_pow_new_start_end_th2_lasso0_0003_r2_para_c0_short_1]$ ./down_ana.R
\section{Applications of the model} 
%Statistical properties of the parameters of the model}
%ここでは，モデルの視点で，ウェブにおける普及現象の特性を解析する.
%In this section, we analyze the phenomenon of word count growth from the perspective of model parameters given by Eq. \ref{base_eq}.
In this section, we describe the application of the proposed model given by Eq. \ref{base_eq} to the analysis of growth phenomena in online social media. 
%In this section, we analyze the statistics of the parameters of the proposed model given by Eq. \ref{base_eq} in the actual data set. 
%Those parameters are expected to reflect the characteristics of the dynamics.
%We will attempt to quantify the properties of the diffusion phenomena by analyzing the parameters, which reflect the characteristics of the dynamics.
\subsection{Statistics of model parameters}
%次に，モデルのパラメータの統計を検討する.
%Next, we apply the proposed model to the analysis of growth phenomena in online social media. 
First, we examined the statistics of the parameters of the proposed model given by Eq. \ref{base_eq}.
These statistics of parameters are expected to reflect the degree of diversity in dynamics. \par
Fig. \ref{fig_hist} shows the probability density functions of $\alpha$, $Y$ and $r$.
From panel(a), in the case that $Y>0$, $\alpha \sim -0.5$ is the most frequent value (indicated by red circles), which  corresponds asymptotically to power-law growth ( $\propto t^{2}$). 
On the other hand, when $Y<0$, $\alpha=1$ is the most frequent  (shown in black triangles), which corresponds to the logistic equation. 
% 赤で示した($Y>0$)では，$\alpha \sim 0.5$が最頻値であることがわかる.また，$Y<0$では，$\alpha=1$が多いことがわかる（ロジスティック方程式に対応）. 
Moreover, from panels(c) and (e), the distributions of $Y$ and $r$ have power-law-like heavy tails, respectively. \par
%Note that the discontinuous peak at $\alpha=0$ in panel (a) is caused by the effect of regularization of $\alpha$ used when estimating the parameters. This peak disappears when the regularization of $\alpha$ is eliminated. The details are discussed in Appendix \ref{app_sec_estimate} and Fig. \ref{app_fig_hist}. \par 
%また，$K$と$r$はべき分布的な分布であることがわかる. 
%パラメータ間の相関としては，表より$\alpha<0$と$Y<0$と推定される例はほとんどなかった.
Second, we checked the correlations between the parameters.  From Table \ref{table_para}, we can confirm that there are relatively few words with both negative parameters $\alpha<0$ and $Y<0$.  
%Regarding correlations between parameters, 
However, no other \textcolor{black}{clear correlations} were found. \par
%それ以外は明確な関係性はなかった.
%Considering both Table \ref{xxx} and Fig. \ref{xxx}(a) together, the most typical parameters in the blog data used in this study are $\alpha \sim -0.5$ and $Y>0$. With this parameter, $y(t)$ asymptotes to $y(t) \propto t^2$.
%つまり，最も典型的な時系列は，$N>0$と$\alpha \sim 0.5$での時系列であった. 
%つまり、$y(t) \propto t^2$に漸近する。
\subsection{Behavior after peak}
%The proposed model only describes the time period between the grey vertical lines in the time series in Figs \ref{xxx} and 
The proposed model given by Eq. \ref{base_eq} does not tell us any information about the post peak-out behavior. 
In other words, the model describes only the growth period, such as the time period between the grey vertical lines in the time series in Fig \ref{fig_world}. 
%次に登り終わった直後の減少の仕方をパラメーターごとに調べた. 
%The extended model given by Eq. \ref{base_eq} describes only the growth of word counts and does not concern the post-growth period.
%今回の解析は主に件数の増加のみを扱い減少については対象にしていない（図\ref{xxx}の灰色の縦線にはさまれた部分のみを研究の主な対象にしている）
%For example, 
%The model describes the time period between the grey vertical lines in the time series in Figs \ref{xxx}.
%Theoretically, in the only case that $Y<0$, corresponding to the logistic equation, $y(t)$ asymptotically approaches to Y for $t \to \infty$.
%そこで登り方によって、ピークアウトしていく様子に違いがあるかがの問題がある。
%Theoretically, in the case that $Y<0$, corresponding to the logistic equation, $y(t)$ asymptotically approaches to Y for $t \to \infty$.
%理論的には，ロジスティック回帰モデルに対応する$Y<0$のケースでは，
%$t \to \infty$ で，$y \to Y$にゆるやかに漸近することがいえる。
%しかし，$Y>0$のケースでは，$y_i(t) \to \infty$となり、式\ref{base_eq}からはどのようにピークアウトしていくかは
%予測できない。
%However, in the $Y>0$ case, $y_i(t) \to \infty$ for $t \to \infty$, therefore, the extended logistic equation given by Eq. \ref{base_eq} does not tell us any information about the post peak-out behavior.
 Thus, we empirically investigated the relationship between peak behavior  and the behavior during growth (i.e., the  parameters of the proposed model). 
\par
%そこで実際にどのようなふるまいをするかをピークを1に規格化し，それ以降の減少の中央値で観測したものが図\ref{xxx}である.
%中央値による減少の典型的なふるまいは以下のように計算した。
Fig. \ref{fig_down}(a) shows the ensemble median of the peak-out behavior.
Specifically, we calculate that the median of the normalized time series, which is normalized to the peak at $t_j^{(max)}$ is 1,  
\begin{eqnarray}
&&\hat{v}(\tau|\alpha,s_Y) = \nonumber \\ 
&&Median_{\{j| \alpha-\delta_1 \leq \alpha_j < \alpha+\delta_2, sign(Y_j)=s_Y \}}\left[\frac{y_j(\tau+t_j^{(max)}-1)}{y_j(t_j^{(max)})}\right], \nonumber \\ \label{eq_down}
\end{eqnarray}
where $\tau$ is the time step from the peak, we take the median conditioned by the parameters $\alpha$, $s_Y$ taking $1$ or $-1$ (i.e., the sign of $Y$) and \textcolor{black}{$t_j^{(max)}=argmin_t\{y_j(t) \geq 0.999 \times y_j(T_j)\}$.} \par
From Fig. \ref{fig_down}(a), we can see that in the case of $Y<0$, the peak out is a gradual decrease, as expected because 
the logistic-type equation asymptotically approaches Y for $t \to \infty$.  
On the other hand, in the case of $Y>0$, the peak-out is a power-law decay whose power-law exponent depends on the parameter $\alpha$.
For $\alpha<0$ (i.e., the infinite-time divergence case), the decay can be approximated as $\hat{v}(\tau) \propto \tau^{-0.5}$, and for $\alpha>0$ (the finite-time divergence case), the decay can be approximated as $\hat{v}(\tau) \propto \tau^{-1}$. \par
%$Y<0$のケースでは予想通り比較的ピークからの減少が小さいことがわかる.一方、$Y>0$のケースでは，べき的に減衰している.　そして，$\alpha$に依存して，その減少のべき指数が異なった. 具体的には，$\alpha<0$では$\hat{v}(t) \propto t^-0.5$に近い緩やかに減衰，有限発散である$\alpha>0$のケースでは$\hat{v}(t) \propto t^-1$程度により早い減衰となっている.
%$Y<0$のケースでは予想通り比較的ピークからの減少が小さいことがわかる.一方、$Y>0$のケースでは，べき的に減衰している.　そして，$\alpha$に依存して，その減少のべき指数が異なった. 具体的には，$\alpha<0$では$\hat{v}(t) \propto t^-0.5$に近い緩やかに減衰，有限発散である$\alpha>0$のケースでは$\hat{v}(t) \propto t^-1$程度により早い減衰となっている. \par
%減衰のべき指数の
%$\alpha$依存性を図\ref{xx}(b)に示す.これは，データを$\alpha$ごとに分割し，両対数プロットに対して線形の最小二乗法でべき指数を推定したものである。
 %Fig. \ref{xxx}(b) shows the $\alpha$ dependence of the power exponent of attenuation, where we use the power-law approximation with the exponent $\gamma$, \hat{v}(t|\alpha,s) \propto to t^{-\gamma($\alpha)}.
 The Fig.  \ref{fig_down}(b) shows the dependence of the exponent $\gamma$ on $\alpha$ for $Y>0$ when the peak out is approximated by a power function $\hat{v}(\tau|\alpha,s=1) \propto \tau^{\gamma(\alpha)}$. In this figure, the data are grouped by $\alpha$, and the power exponents are estimated as coefficients of linear regression in the log-log plots.
  From this figure, the exponent of the peak-out  $\gamma$ transitions from approximately $-0.5$ to approximately $-1.0$ depending on $\alpha$. \par 
  %corresponds to the previous study in Ref. \cite{xxx} in which already well-established words were studied and their response to noise was also a power function with the exponent $-0.5$. \par
%Note that the power decay of exponent -0.5, $\propto t^{-0.5}$, corresponds to the power decay of the basic word \cite{xxx}, which is well-established in society with little change. 
% More precisely, the response corresponds to the response of a time series to normal random noise, not to special noise such as special news. This response is related to ultra-low diffusion (i.e., logarithmic diffusion)\cite{xxx}.  \par 
%この$\gamma \sim 0.5$ 程度は，十分定着した語に対するノイズへの応答も時間の$-0.5$に対応している。　一方，$\gamma \sim 1$の減衰はそれよりはやくなっている。\par
%図\ref{xxx}は実際の単語について$x^-1$の場合$\alpha \sim 0$と$x^-0.5$ $\alpha \sim 1$の場合でプロットしたものであり，おおまかには個別単語でも関数形がべき関数で一致してることがわかる。ただし，これはあくまで全体の統計的なふるまいであり，実際には図\ref{xxx}(c)(d)のように$Y>0$でもべき的に減衰しない単語も存在するため，個別の具体的なふるまいの解析は今後の課題である。
Fig. \ref{fig_world} displays example of the time series of the peak out.
Fig. \ref{fig_world}(a) and Fig. \ref{fig_world}(b) show the cases $\alpha=-0.0722$, $\gamma=-0.5$ and $\alpha=0.661$, $\gamma=-1.0$, respectively.
%, corresponding to the statistical results shown in Fig. \ref{fig_down}(a). 
These results \textcolor{black}{are not inconsistent} with the overall statistics shown in Fig. \ref{fig_down}.
% the peak out in panel (a) and panel (b) is a power function with the exponent $-0.5$ and the exponent $-1.0$ respectively, as shown in Fig. \ref{fig_down}.
On the other hand, panels (c) and (d) show examples where the peak-out differs from the power decay. 
%because some individual cases differ from the overall statistics.
This figure implies that individual time series sometimes differ from the overall statistics.
%Since the main theme of this paper is growth processes, a more detailed study of the patterns of peak-out behaviors remains for future work. 
\par
\textcolor{black}{Note that the power decay of exponent $\gamma = -0.5$ or $\propto \tau^{-0.5}$ was also observed in already well-established words (i.e., “mature phase” in the life trajectory of words) \cite{watanabe2018empirical}, in which it appears in responses to noise and is related to the ultraslow diffusion dynamics (i.e., logarithmic diffusion). } \par 
%次にパラメータと下り方の関係を調べた.図は、１としたときの中央値$Median(t/t(max))$である.  $Y<0$では，うまくいってる. $Y>0$$\alpha>0$ \par
\subsection{Sequential parameter estimation and prediction}
%実用上において予測にできる.また，科学的にもどの時点で予測が可能になるかがは興味深い問題である.
Forecasting is important for practical purposes.
The forecasting task requires sequential parameter estimation, that is, parameter estimation at the intermediate points leading to a peak. 
%from the start of growth to the peak. 
%To clarify the meaning of the model with respect to forecasting, we examined the predictability of forecasting at the midpoint of growth from the beginning to the peak.
We used the Bayesian estimation to make sequential parameter estimation. Details of the Bayesian estimation for the proposed model are provided in the Appendix \ref{app_sec_param}. 
%Using Bayesian estimation, we can evaluate the parameter uncertainty as a posterior distribution.
%The posterior distribution of $\alpha$, which describes the uncertainty of the parameter estimation, is shown in Figure \ref{xxx} (b) for its dependence on the timing of the estimation.
The time dependence of the posterior distribution of $\alpha$, which describes the uncertainty of parameter estimation, is shown in Fig. \ref{fig_esitmation} (e).  
Here, a prior distribution of $\alpha$ is assumed to have a uniform distribution from $-2$ to $2$. \par
 From Fig. \ref{fig_esitmation} (e), we can see that at the initial time point ($0 \leq t \leq 20$), the posterior distribution also maintains a uniform distribution.  
This is because exponential growth at the beginning  has no information about $\alpha$. 
The predictions are shown in Fig. \ref{fig_esitmation} (a), where the estimation is based on data before the grey vertical dotted line ($t=27$). 
From the figure, it can be seen that the range of the predicted distribution shown by the green dashed line (from 1 percentile to 99 percentile) is very broad. \par
%その後は，$t \sim 30$で，$-0.5$以上の可能性を排除し，その後、事後分布は$-0.5$に収束していく。
Returning to Fig. \ref{fig_esitmation} (e), for $t \sim 30$, the posterier distribution of $\alpha$ eliminates the possibility of $\alpha>0$, and thereafter gradually converges to $\alpha \sim -0.5$ for $t \leq 30$.
Based on the corresponding Figs. \ref{fig_esitmation}(b)-(d), we also confirm that the predicted posterior distribution becomes gradually narrower, i.e., more strongly predictable by the model. Note that for $Y>0$, \textcolor{black}{although the model can predict the growth process}, the model cannot predict the time at the end of growth, $T_j$ or $t_j^{(max)}$.
%それに対応する 図 \ref{xx}(b)-(d)でも、予測事後分布が徐々に狭くなっていくことが確認できる，つまり，モデルによりより強く予測可能になっていく。
%図 \ref{xxx} に戻ると，その後は，$t \sim 30$で，$\alpha>-0.5$の可能性を排除し，その後は、事後分布は$\alpha \sim -0.5$に収束していく。
% 本研究は事前分布として-2から2の一様分布を採用したため，ほとんど情報がない指数関数増加に近い，初期の時点での推定では一様分布のままで、確定できてないことがわかる。
%そこで、ベイズ推定を用いてパラメータを時系列を途中で推定を行い，どのようにパラメータが決まっていくかを
%調べた. 図\ref{xxx}はその結果である.　まず，図の領域では指数関数のため（２年目程度）、$\alpha$が特定できていなく、予測不可能にななっている
%しかし、後半のｘｘ（３年目程度）以降は、ほぼ$\beta=-0.5$が確定し，大まかには予測が可能になっていることがわかる。
%ただし，式$\ref{base_eq}$や今回の研究のパラメータの統計的性質のみでは，あくまで増加しつづけることの前提式であり，$Y<0$のロジスティック型の場合を除いては，いつピークアウトするかまでは予測できないことには注意されたい。
%\section{Comparison with the web search query}
\section{Web search query data}
%上記の統計は日本語ブログ時系列だけか，他のデータや言語でも成立するかを検討するため，
%検索クエリの人気度データも調査した。
\textcolor{black}{To confirm the generalizability of our study, we investigate web search query data (i.e., Google Trends).} 
%the properties that have investigated in the Japanese blog data, web search query data will also be investigated (Google Trends data). 
%The details of data
%Similar to the data on the number of blog posts studied previously, web search query data is one of the most commonly used data used to observe temporal changes in social interest. 
%An increase (or decrease) in the number of query searches is considered to correspond to an increase (or decrease) in social interest.
%ブログの時系列の同様に，検索数も社会の関心に対応し関心の時間的変化の観測に利用される.
%今回はGoogle Trendsデータを利用した。Google Trends は検索数を最大１００にするように
%企画化したデータを取得できる。\par 
%In this study, "Google Trends" was used as search query data.
%Google Trends is search query data in google search, where the number of searches is standardized to the maximum number of 100 in the observation period.
%We targeted data that was expected to be new words.
%Specifically, we targeted words that had no page views in Wikipedia in year XX and had 100 or more page views in year XX. In other words, not all new words, but words that were Wikipedia entries in XX years, were targeted.
The details of Google Trends data are provided in subsection \ref{sec_google} in section \ref{sec_data}.
\par
%対象の単語が同じならば，検索クエリデータは日本語ブログデータの時系列と類似することが多い。
%例えば，図 \ref{xxx}(a)と図 \ref{xxx}(b) 
%では，日本語ブログデータ（黒三角）と検索（赤丸）で増減の様子がほぼ共通してることが確認できる。
%ただし，もちろん，ブログと検索クエリで異なる場合もある。
%例えば、ある新しい地名を表す時系列のパネル(d)では、地名の誕生と同時に検索では検索数を突発的に増加するが，ブログデータでは線形に増加している。%これは新地名をみたときに調べる行動は先行するが，
%If the target words are the same, 
%The Google Trends data is often similar to the Japanese blog data.
Figs \ref{fig_world}(a)-(d) show examples of comparisons between Google Trends (red crosses) and the blog data (black triangles).
From Figs \ref{fig_world}(a)-(c), we can confirm that the increase and decrease in the number of search queries are almost the same for the Japanese blog data.
\textcolor{black}{Although the time-series variation is common for many words, there are cases in which search queries differ from the blog data.}
For example, for the case of ``SagamiharaShiTyuuouKu'' (Sagamihara city central ward, new place name)”, shown in the panel (d), the number of search queries increases abruptly when the place name is born (red crosses), while  the number of blog articles increases linearly (black triangles). 
%実際に定着するのは時間がかかることは意味しているかもしれない（調べて知った人が、旧地名から新地名へ認知が変化し，
%それが累積されることで実際にその単語を繰り返す使う人が増えていく）。
%旧地名から新しい地名へ認知が置き換きかえがコンスタントな検索クエリに対応し，実際に新地名を使う人，つまり，認知の置き換わった人達の累積が，線形的にふるまうブログデータに対応するかもしれない。
%The replacement of recognition from the old place name to the new place name may correspond to constant search queries, and the accumulation of people who actually use the new place name, i.e., those who have replaced the recognition, may correspond to blog data that behaves linearly.
The hypotheses that explain this difference are as follows: 
(i) A person searches only once, when the person has recognized the replacement of the new place name with the old place; (ii) On the other hand, the person has permanently used and written the new place name in the blog many times after the recognition. To summarize the two hypotheses, 
From the macro perspective of society as a whole, the word counts in the blog data are linear because of the accumulation of random constant recognition (observed in search queries). 
\textcolor{black}{Another example of a similar type of the difference is given in Fig. \ref{fig_toyota}. }
\par
\textcolor{black}{We compared the statistics of Google Trends data with blog data.} 
Fig. \ref{fig_ave}(b) shows the valification of the proposed model given by Eq. \ref{base_eq} to search query data in English, French, Spanish, and Japanese, corresonpding analysis of Fig. \ref{fig_ave}(a) for the blog data in subsection \ref{sec_ave} in section \ref{sec_vali}.
 %for search query data in English, French, Spanish, and Japanese. 
 The straight lines in this figure indicate that the model proposed by Eq. \ref{base_eq} is valid for all 4 languages statistically of Google Trends data. \par
%the expression \ref{xxx} corresponding to the validation of the extended logistic equation \ref{base_eq} is valid for all languages.
%For the search query data, we performed the analysis corresponding to the blog data Fig. \ref{fig_ave}(Eq. \ref{xxx}) and Fig. \ref{fig_hist}. 
Fig. \ref{fig_hist}(b) shows the distribution of the parameter $\alpha$ of the proposed model given by Eq. \ref{base_eq} for search query data. 
% in English, French, Spanish, and Japanese. 
%First, from this figure, we can see that the expression \ref{xxx} corresponding to the validation of the extended logistic equation \ref{base_eq} is valid for all languages. 
From Fig. \ref{fig_hist}(b), we can also confirm that the distributions of $\alpha$ for search query data (black triangles for Japanese, blue pluses for English, red circles for French and green crosses for Spanish) have peaks at $\alpha \sim -0.5$ and $\alpha \sim 1$ in the same as the Japanese blog data indicated by the yellow dashed line.
Moreover, from panels (d) and (f), the distributions of $Y$ and $r$ for Google Trends data and the blog data  also have almost the same shape density distributions. \par
%The distribution also confirms that the statistical distribution corresponds to that of Japanese blogs, indicated by the yellow line. 
%Note that from Eq. \ref{xxx}, $\alpha$ is a constant independent of the scale transformation.
%このような検索データに対して，図\ref{fig_ave}（式 \ref{xxx}）と図\ref{fig_hist}に対応する解析を行った。
%図\ref{fig_world}は英語、フランス語、スペイン語、日本語データついて解析を行ったものである。.
%まず、図\ref{fig_world}(a)より，式\ref{base_eq}の検証に対応する式\ref{xxx}がすべての言語で成立していることがわかる。次に，図\ref{fig_world}(b
%より，$\alpha$の統計は日本語ブログ同様に$alpha=0.5$と$\alpha=-1$にピークがあることも確認できる. \par
%また、分布は黄色で示した日本語ブログとも統計分布が対応していることも確認できる.　これらの結果より，今回のモデルや統計解析の結果が
%日本語のブログデータだけでなく，多言語や多媒体でも成立している可能性があることが示唆される. 
%ただし，今研究の対象としている語はウィキペディアの記事項目になった語である。
%つまり，私たちの結果の範囲はオンライン辞書に書かれるレベルに社会に普及した語に限定される。
%The distribution also confirms that the statistical distribution corresponds to that of Japanese blogs, as indicated by the yellow line.
 These results imply that the model and statistical analysis in \textcolor{black}{this study are probably} valid not only for Japanese blog data but also for other languages and media.  
 %In other words, the scope of our results is limited to words that have become popular in society to the level of online dictionaries.
%ただし，この統計はすべての新語ではなくwikipediaの記事タイトルになることを基準している。つまり，辞書に書かれるレベルに
%一時的にも一般定着した語ということに限定されてることに注意されたい。
Note that the words studied in this research are not all words but rather article entries in Wikipedia, namely, words that have successfully spread in society to the level that they are published in online dictionaries.
%In other words, it is limited to words that have spread in society to the level that they are written in online dictionaries.
\par
\section{Conclusion and Discussion}
\textcolor{black}{In this study, through systematic data analysis, we showed that the growth process of word counts of online can be well described by the slight extension of the logistic equation (Eq. \ref{base_eq}) and is useful in analyzing online growth phenomena.}
% (Figs. \ref{xxx},\ref{xxx}). 
First, the proposed model given by Eq. \ref{base_eq} can consistently describe the functional form of the growth curves observed by the actual time series data, such as the logistic function, the finite-time divergence, the linear function and the power-law function, etc., with two parameters $\alpha$, $Y$ essentially (Figs. \ref{fig_examples}, \ref{fig_ave}). \par
%モデルは，実際の成長時系列のデータにより観測される成長の関数形，ロジスティック関数，有限時間発散，線形，べき関数形などのような、を本質的に２つのパラメータで一貫して記述できる。
%本研究では，系統的な網羅データ解析によって，私たちはワードカウント of オンラインの成長過程をロジスティック方程式の わずかな拡張により、よく記述できることを示した.
%本研究では，網羅的な日本語ブログデータ解析により，
%ebのおける新語の単語カウントの誕生から定着するまでの成長過程をロジスティック方程式の簡単な拡張によりよく
%記述できることを示した（図\ref{xxx},\ref{xxx}）。統一的的な方程式の２つのパラメータの相違により指数的成長，S字カーブ(ロジスティック関数)，有限時間発散，線形的な増加など多くの時間発展のパターンやその中間のようなパターンを記述できる。 \par
Second, we examined the statistics of the model parameters that reflect information on the dynamics of word-count time series. As a result, in the Japanese blog data, we found that the most typical values for the combination of parameters $\alpha$ and $Y$ are $\alpha=-0.5$, $Y>0$ (\textcolor{black}{the word counts $y(t)$ asymptotical to the power-law function with exponent 2}), or $\alpha=1, Y<0$ (corresponding to a logistic function or S-curve), as shown in Fig. \ref{fig_hist}.
%次に，モデルのパラメータのヴューポイントから，ワードカウント時系列を調査した。
%次に，モデルのパラメータの視点で時系列を把握した，その結果， $\alpha$（分布の時間質的の相違様子を特徴づけるパラメータ）の分布の特徴（$\alpha=-0.5& N>0$および$\alpha=1, N<0$）が支配的であることがわかった。また，成長時のパラメータとピークアウトの仕方が無関係でないことも
%示した\ref{xxx},\ref{xxx}。
%また，私たちは、成長モデルのパラメータとその後のピークアウトの仕方が無関係でないことを暗示した。　\ref{xxx} \ref{xxx}。
In addition, we also implied that parameter $\alpha$ is related to the subsequent peak after growth (Fig. \ref{fig_down}).
%More precisely, we observed that when the decay is approximated by a power function, its power exponent transitions from about $-0.5$ to about $-1.0$ depending on the growth model's parameter $\alpha$ \ref{xxx}.
%より詳細には、私たちは，減衰をべき関数で近似したとき，そのべき指数が約$-0.5$から約$-1.0$に遷移することを成長モデルのパラメータ$\alpha$に依存してることを観測した。
 \par
%特に、全体の統計でみると、上昇時のパラメータに応じて、減衰をべき関数で近似したべき数が$0.5$から$1.0$程度に遷移することを確認した。 \par
Third, we analyzed search query data (i.e., Google Trends data) in English, French, Spanish and Japanese and we confirmed that the data had properties similar to the Japanese blog data (Fig. \ref{fig_ave}(b) and Fig. \ref{fig_hist}(b)). \par
%in regarding the fit of the model and the statistics of the $\alpha$ parameter Fig. \ref{xxx}.  \par
%第３に，フランス語，英語、日本語のgoogle トレンドデータの解析を行った。その結果，モデルの当てはまりや$\alpha$のパラメータの統計に関して，日本語ブログデータと共通する性質を持つことが確認できた \cite{xxx}。
%第３に，モデルの結果が日本語ブログ以外の通用するかを確認した。その結果，
%フランス語，英語、日本語のgoogle 検索のデータで$\alpha$の特徴やモデルの当てはまりなどブログデータと
%共通する性質を持つことが確認できた。\par 
%ただし，本研究の成果はすべての新語でなく，特定の単語に偏る限界がある。具体的には，単語収集をWikipediaの項目を参考にしているため，新語がWikipediaの辞書に掲載された単語ということで，つまり、すべての単語でなくてある程度
%Note that the statistics in this study are not for all new words but for the set of words that have been included in the Wikipedia dictionary. In other words, these are limited to words that have been established to some extent. 
%本研究の統計はすべての新語でなく，新語がウィキペディアの辞書に掲載された単語の集合でとっていることに注意されたい。別の言葉でいうと，ある程度には定着した単語に限定されることである。
%定着した単語に限定されることである。また，短期的な成長する単語は研究の対象外であり，１年以上の成長期間がある単語に限定している。 つまり，
%ある程度定着し,ゆっくり増加する単語に限定した結果である。 \par
%In addition, short-term growing words are out of the scope of the study and are limited to words with a growth period of at least one year. In summary, in this study, the statistics are limited to words that are, to some extent, established and not of short-term growth of less than one year. 
%また，短期的な成長する単語は研究の対象外であり，１年以上の成長期間がある単語に限定している。 まとめると，本研究において，統計は，ある程度は定着し、かつ、１年以下の短期成長でない単語に限定されている。
%\par
%第４に，応用にむけて成長途中でのデータの予測を試みた。結果，成長の途中までは予測は難しく（モデルパラメータは確定でない），徐々にモデルパラメータが確定しいく様子を確認した。
Fourth, for forecasting applications, we attempted to predict parameters during growth using Bayesian statistics. As a result, we observed that the model parameters could not be limited owing to the wide range of possibilities in the early stages of growth, and after that the range of possibilities of parameters gradually narrowed down (Fig. \ref{fig_esitmation}). This implies that when forecasting using only the proposed model and the time series data, there are words that are difficult to forecast the time series in the early stages of growth, but then it becomes gradually predictable.
%第４に，予測への応用に向けて，ベイズ統計により，成長途中でのパラメータの予測を試みた。結果，モデルパラメータは成長初期では広い可能性があり限定できず，徐々にモデルパラ%メータが絞り込まれていく様子を観測した。つまり，モデルと時系列データのみを用いて予測する場合，成長の初期では時系列の予測は難しく，その後，徐々に予測可能になってくるこ%とが暗示された。
 \par
 %Our study suggests the possibility that a variety of growth curves in word-count time series can be described and categorized by fewer parameters' model. 
 %Though model 
 %
 %and this findings could contribute to understand word-counting time series or growth phenomena in society (e.g., diffusion of new products or concepts). 
 %Although the model does not tell us anything directly about the micro growth mechanisms, the systematic description of the various macro growth curves in this study is expected to provide one clue to their understanding, and these findings could contribute to understand word-counting time series, or growth phenomena in society (e.g., diffusion of new products or concepts). 
 %\par
%モデルはミクロについては何も語らないが成長についてかなり限定する。これがダイナミクスの理解のヒントになると考えらえる。
 % このモデルや知見は，ワードカウント時系列，もしくは，社会における成長現象（例えば、新商品や新概念の拡散）を理解や予測するために貢献できる可能性がある。
%\paragraph{Limitation}
The limitations of the data are as follows: 
\begin{itemize}
\item The focused words that have been included in the Wikipedia dictionary are not for all new words namely words that have successfully spread in society to the level that they are published in online dictionaries. 
\item Short-term growing words are out of the scope of the study (limited to words with a growth period of at least one year).
\item We used only online data and we don't use non-online words such as newspapers due to data acquisition limitations. However, the characteristics of the word-count time series may be common between online and non-online words, such as newspapers, because there are a lot of similarities in previous studies \cite{watanabe2018empirical}. 
%Therefore, confirmation of this hypothesis is also a future task. 
%Therefore, it is expected that the model can be adapted to non-online words as well, and confirmation of this is also a future task. 
\end{itemize}
%Although the model does not tell us anything directly about the micro growth mechanisms, the systematic description of the various macro growth curves in this study is expected to provide one clue to their understanding and these findings could contribute to understand word-counting time series or growth phenomena in society (e.g., diffusion of new products or concepts). 
%Due to data acquisition limitations, the study was limited to online words only. However, 
%The characteristics of the word-count time series may be common between online and non-online words, such as newspapers. Therefore, it is expected that the model can also be adapted to non-online words, and confirmation of this is a future task.
%Note that the statistics in this study are not for all new words but for the set of words that have been included in the Wikipedia dictionary. In other words, these are limited to words that have been established to some extent. 
%本研究の統計はすべての新語でなく，新語がウィキペディアの辞書に掲載された単語の集合でとっていることに注意されたい。別の言葉でいうと，ある程度には定着した単語に限定されることである。
%定着した単語に限定されることである。また，短期的な成長する単語は研究の対象外であり，１年以上の成長期間がある単語に限定している。 つまり，
%ある程度定着し,ゆっくり増加する単語に限定した結果である。 \par
%In addition, short-term growing words are out of the scope of the study and are limited to words with a growth period of at least one year. In summary, in this study, the statistics are limited to words that are, to some extent, established and not of short-term growth of less than one year. 
The limitations of this model are as follows: 
\begin{itemize}
\item The proposed model cannot tell us anything about the micro-mechanisms although it restricts the macroscopic properties of the time series. 
\item The proposed model may still be redundant because parameter estimation is sometimes unstable without the regularization (see Appendix \ref{app_sec_estimate}). Therefore, even as a macro model, there may be more essential time series models with fewer parameters than the proposed model.
\end{itemize}
%Although the model does not tell us anything directly about the micro growth mechanisms, the systematic description of the various macro growth curves in this study is expected to provide one clue to their understanding, and these findings could contribute to understand word-counting time series, or growth phenomena in society (e.g., diffusion of new products or concepts). 
%提案モデルはオンライン言語における成長動力学の第一原理モデルでなく，あくまで近似モデルにすぎない。　
%しかし，近似モデルであったとしても，非常に多様性に見える成長現象をシンプルな提案モデルによって体系的に記述可能なことまでは示せた。
%それゆえ，本研究は，多様性も重要な要因である社会の成長現象の理解へ貢献できると信じている。
%モデルとしては完全でないかもしれないが、かなり体系的に成長現象をマクロモデルは社会における中期的な成長現象の記述や理解に貢献できると信じている。

The proposed model is not a first-principles model of growth dynamics in online languages but only an approximate model.
%However, even if it is an approximate model, we have shown that a simple proposed model can systematically describe growth phenomena that superficially appear to be very diverse.
However, even with an approximate model, we were able to show that a simple proposed model can systematically describe growth phenomena that appear superficially to be very diverse. 
%In addition, 
In addition, by applying the proposed equation, we could analyze non-trivial mechanical properties.
For example, we found that one of typical peak-out forgetting property $\propto \tau^{-0.5}$ is consistent with noise responses for well-established words observed in previous studies \cite{watanabe2018empirical}.  
%For example, typical peak-out forgetting property $\propto t^{-0.5}$ is consistent with noise responses for well-established words observed in previous studies \cite{watanabe2018empirical}. 
The coincidence in the aspects of two different forgetting phenomena, peak-out of new words and response to noise, may suggest the existence of a more universal dynamic property of social forgetting, whose entire picture has not yet been revealed.  
\textcolor{black}{We hope that this research can contribute to the understanding of growth phenomena in society or in general complex systems, where diversity is one of the important factors.}
%However, the model cannot tell us anything about the micro-mechanisms of the time series. 
%Hence, a future task is to relate the macro parameters of the model to the micro mechanisms.
% 本研究により，ワードカウント時系列の成長の多様な関数形をモデルにより少ないパラメータを用いて分類・記述できる可能性を示唆した。
%しかし，モデルは，時系列のミクロメカニズムについては何も語ることができない。　それゆえ、今後の課題は，モデルのマクロパラメータとミクロなメカニズムの関係づけることがある。\par
%The model may still be redundant because parameter estimation is sometimes unstable without Lasso regularization. Therefore, even as macro models, there may be more essential time-series models with fewer parameters than the proposed model.
%Also, due to data acquisition limitations, the study was limited to online words only. However, the characteristics of word-count time series may be common between online and non-online words, such as newspapers.
%Therefore, it is expected that the model can be adapted to non-online words as well, and confirmation of this is also a future task.
%本研究により，モデル少ないパラメータで多様なオンライン上での記述、整理できる可能性を示唆した。
%研究より，式\ref{xxx}はフィッティングモデルとして悪くはない。しかし，メカニズムについては何も語っていないので、
%それが今後の課題である。例えば，線形モデルと。また，モデルとしてはパラメータの冗長性があるモデルありようにも割れるため，現象の自由度は
%もう少し小さく，このモデルの先に現象の本質的をより記述できるモデルが存在するかもしれない。また，研究はオンラインの単語のみに限定したが
%例えば，オンラインと同様な現象が新聞等の非オンライン単語でも観測される報告もある。そのため，本モデルのオンラインメディア以外での
%検討も今後のk代である。
%\acknow{
\begin{acknowledgments}
The authors would like to thank Hottolink, Inc. for providing the data. This work was supported by JSPS KAKENHI, Grant Numbers 21K04529, 17K13815, and the Leading Initiative for Excellent Young Researchers MEXT Japan.  We would like to
thank Editage (www.editage.com) for English language editing. Computations were partially performed on the supercomputer system at ROIS Institute of statistical mathematics.
%}
%}
\end{acknowledgments}
%\showacknow{} % Display the acknowledgments section
\bibliography{newword}
\clearpage
\appendix
\section*{Appendix}
\def\thesection{A\arabic{section}}
\def\thesubsection{A\arabic{section}.\arabic{subsection}}
\renewcommand{\theequation}{A\arabic{equation}}
\renewcommand{\thefigure}{A\arabic{figure}}
\renewcommand{\thetable}{A\arabic{table}}
\setcounter{equation}{0}
\setcounter{figure}{0}
\setcounter{table}{0}

\section{Solutions of the extended logistic equation}
\label{app_seq_th}
%拡張されたロジスティック方程式の解を導出する. 
We derived the solution to the extended logistic equation  given in Eq. \ref{base_eq}. 
\begin{equation}
\frac{dy(t)}{dt} = r y(t) \left(1+\frac{y(t)}{Y} \right)^\alpha \label{app_base_eq}
\end{equation}
%まず，変数を分離すると
Using the separation of variables, we obtain
\begin{equation}
\frac{\frac{dy(t)}{dt}}{y(t) \left(1+\frac{y(t)}{N} \right)^\alpha} = r. 
\end{equation}
%両辺を積分すると，
By integrating both sides, we obtain 
\begin{equation}
\int^{y(t)}_{y(t_0)} \frac{dy}{x \left(1+\frac{y}{Y} \right)^\alpha } = \int^{t}_{t_0} r  dt. 
\end{equation}
% これを$t$について解くと，
We solve with respect to t, 
\begin{equation}
t= t_0 + \frac{1}{r} \int^{y(t)}_{y(t_0)} \frac{dy}{x \left(1+\frac{y}{Y} \right)^\alpha } .
\end{equation}
With the new variable, $v=1+y/N$, 
\begin{eqnarray}
\int^{y(t)}_{y(t_0)} \frac{dy}{y \left(1+\frac{y}{Y} \right)^\alpha} 
% \nonumber \\
=\int^{1+y(t)/Y}_{1+y(t_0)/Y} \frac{dv}{\left( 1-v \right) v^\alpha}.  \nonumber \\
%&&=B_{\alpha,N,y(0)}(y(t))
\end{eqnarray}
%\subsection{$2-\beta$が正整数でないとき}
%不定積分
The following indefinite integral formula can be used  
\begin{eqnarray}
&&\int{\frac{v^{-a}}{1-v}dv}= \nonumber \\
&&\left\{  
\begin{array}{ll}
\frac{v^{1-a}{}_2F_1(a,1-a;2-a;v)}{1-a} & (others)   \nonumber \\ 
\frac{1-v} {|1-v|} \log(|1-v|)+\log(v) & (a=1)   \nonumber \\ 
\frac{1-v} {|1-v|} \log(|1-v|)+\log(v)+& \nonumber \\
\sum^{a}_{i=2}\frac{1}{-i+1}v^{-i+1} & (a=2,3,\cdots) \nonumber  \\
\end{array}
\right. \nonumber \\
&&+C ,
\end{eqnarray}
%を用いると，
%ここで、${}_2F_1(q_1,q_2;q_3;x)$はガウスの超幾何関数である。$C$は定数.$sign(x)$は符号。また，
%$B(q_1,q_2)$は複素数領域に解析接続したベータ関数である。 また、下の式は
where ${}_2F_1(q_1,q_2;q_3;x)$ is a Gauss hypergeometric function, $C$ is an arbitrary constant, and the second and third cases are derived from the partial fraction decomposition, 
\begin{equation}
\frac{1}{(1-v)v^a}=\frac{1}{1-v}+\sum^{a}_{i=1}\frac{1}{v^i}. 
\end{equation}
%これを以下のように定義すると
Using the function, 
\begin{eqnarray}
&&B_a(v) \equiv \nonumber  \\
&&\left\{ 
\begin{array}{ll}
\frac{v^{1-a}{}_2F_1(a,1-a;2-a;v)}{1-a} & (others)   \nonumber \\ 
\frac{1-v} {|1-v|} \log(|1-v|)+\log(v) & (a=1) \nonumber \\
\frac{1-v} {|1-v|} \log(|1-v|)+\log(v) &  \nonumber \\
+\sum^{a}_{i=2}\frac{1}{-i+1}v^{-i+1} & (a=2,3,\cdots), \nonumber  \\
\end{array}
\right.
\end{eqnarray}
%とすると，
we can obtain 
\begin{eqnarray}
%&&B_{\alpha,N,y(0)}(x(t))= \nonumber  \\ 
%&&-B(x(t); 1-\alpha,1+x(t)/N)+B(x(t_0); 1-\alpha,1-x(t)/N) \nonumber \\
%B_{\alpha,N,y(0)}(x(t))=-B_\alpha(1+x(t)/N)+B_\alpha(1-x(t)/N) \nonumber \\
%B_{\alpha,N,y(0)}(x(t))=B_\alpha(1+x(t)/N)-B_\alpha(1+x(0)/N). \nonumber \\
&&\int^{1+y(t)/N}_{1+y(t_0)/N} \frac{dv}{\left( 1-v \right) v^\alpha} \nonumber \\
&&=B_\alpha(1+y(t)/Y)-B_\alpha(1+y(t_0)/Y). \nonumber \\
\end{eqnarray}
%と書ける。 \par
%ここで，$B(x;\beta_1,\beta_2)$は複素数領域に解析接続した不完全ベータ関数である. 
%\begin{equation}
%B(z;\beta_1,\beta_2)= \frac{x^{\beta_1}{}_2F_1(\beta_1,1-\beta_2;\beta_1+1;x)}{a-1}
%\end{equation}
%ここで、${}_2F_1(a,b;c;z)$はガウスの超幾何関数で定義される.
Therefore, we obtain the following solution,  
\begin{equation}
t= t_0+\frac{1}{r} (B_{\alpha}(1+y(t)/Y)-B_{\alpha}(1+y(t_0)/Y)). 
\end{equation}
% ここで，$B_{\alpha}(x)=B(x;a-\alpha)$.  
 %なお，ベータ関数では虚部は$x$に依存せず差分を取ることでキャンセルされるため，実部のみが現れる. \par
%これを$x(t)$に関して解けば，
Formally solving for $y(t)$, $y(t)$ can be written as follows: 
\begin{equation}
%x(t)=N \left( 1-B^{-1}_{\alpha}(-r(t-t_0)+b_0) \right)
y(t)=Y \left(B^{-1}_{\alpha}(r(t-t_0)+b_0) -1 \right). \label{eq_app_solution}
\end{equation}
%となる.
%なお，式$B^{-1}_{\alpha}(x)$は，$B_{\alpha}(x)$の逆関数であり,$B^{-1}_{\alpha}(B_{\alpha}(x))=x$.
% 式\ref{xxx}に定義されている.
%また，
The expression $B^{-1}_{\alpha}(x)$ is the inverse function of $B_{\alpha}(x)$, $B^{-1}_{\alpha}(B_{\alpha}(x))=x$, $0<B^{-1}_{\alpha}(t)<1$ for $Y<0$, $B^{-1}_{\alpha}(t)>1$ for $Y>0$ and $b_0$ is defined by 
\begin{equation}
b_0=B_{\alpha}(1+y(t_0)/N).
\end{equation}
%であった.
%\subsection{$2-\beta$が正の整数のとき}
%$2-\beta$が整数の場合は、超幾何関数が発散してしまうので、部分分数分解で解く。
%
%これを積分すると
%\begin{eqnarray}
%&&\int \frac{1}{(1-v)v^a} dv= \int \frac{1}{1-v}+\sum^{a}_{i=1}\frac{1}{v^i} dx  \\
%&&= sign(1-v) \log(|1-v|)+ \log(v)+\sum^{a}_{i=2}\frac{1}{-i+1}v^{-i+1}+C \nonumber \\
%\end{eqnarray}

\section{Asymptotic properties of the model for $t \to \infty$}
\label{app_asy_eq}
We compute four main categories of asymptotic properties of the extended logistic equation given by Eq. \ref{app_base_eq}.
%式の漸近特性を４つの分類を求める.
%\begin{equation}
%\frac{dy(t)}{dt} = r y(t) \left(1+\frac{y(t)}{N} \right)^\alpha \label{app2_base_eq}.
%\end{equation}
%\subsection{$\alpha=0$}
%First, we can caluculate trivial case of $\alpha=0$  $x(t)=y(t_0) \cdot \exp(r(t-t_0))$. 
Note that for $\alpha=0$, which is not included in the four cases, we obtain the trivial solution $x(t)=y(t_0) \cdot \exp(r(t-t_0))$. 
%Before dealing with the main 4-category, as a special trivial case, 
%\begin{equation}
%\frac{dy(t)}{dt} = r y(t) \left(1+\frac{y(t)}{N} \right)^0 = r y(t).
%\end{equation}
%we can get
%\begin{equation}
%y(t)=y(t_0) \cdot \exp(r(t-t_0)).
%\end{equation}
\subsection{$Y>0$, $\alpha>0$} 
Because $\frac{dy(t)}{dt}>0$, $y(t)$ infinitely increases with no upper bound.
Thus, for $y(t) \gg Y$, that is, $1 \ll\frac{y(t)}{Y}$, 
we can approximate Eq. \ref{app_base_eq} as follows:
\begin{equation}
\frac{dy(t)}{dt} = r y(t) \left(1+\frac{y(t)}{Y} \right)^\alpha \sim \frac{r}{Y^\alpha} y(t)^{\alpha+1}
\end{equation}
By solving this equation, we obtain 
\begin{equation}
y(t) \approx \left(\frac{\alpha r (t_0-t)}{Y^\alpha}+y(t_0)^{-\alpha}  \right)^{-1/\alpha} \label{eq_na},
\end{equation}
where $y(t_0) \gg Y$.
%以下の時刻$t^{*}$ で，
This equation diverges $y(t) \to \infty$ in finite time $t^{*}$, 
\begin{equation}
t^{*} \approx t_0 + \frac{Y^\alpha}{\alpha r} y(t_0)^{-\alpha}. 
\end{equation}
%つまり，$t\to t^{*}$で， $x(t) \to \infty $となる. \par
\subsection{$Y>0$, $\alpha<0$} 
%式\ref{eq_na} より, $\alpha<0$ のとき，
From Eq. \ref{eq_na}, for $\alpha<0$, we can approximate Eq. \ref{app_base_eq}  for $t \to \infty$ as follows:
\begin{eqnarray}
y(t) \approx \left(\frac{|\alpha| r (t-t_0)}{Y^\alpha}+y(t_0)^{-\alpha}  \right)^{1/|\alpha|}.
\end{eqnarray}
%したがって，$t \gg 1$ で，
By approximating this equation, we obtain the growth in the power law function
\begin{eqnarray}
y(t) \propto t^{1/|\alpha|}. 
\label{eq_na2}
\end{eqnarray}
%でべき乗の増加に漸近する.
\subsection{$Y<0$, $\alpha>0$} 
%$dx(t)/dt \geq 0$より，$x(t)$は無限に増加する.
%したがって，$x(t) \gg N$のとき，$1 \gg \frac{x(t)}{N}$より，
For $Y<0$, Eq. \ref{app_base_eq} is written by 
\begin{equation}
\frac{dy(t)}{dt} = r y(t) \left(1-\frac{y(t)}{|Y|} \right)^\alpha. 
\end{equation}
As $\frac{dx(t)}{dt} = 0$ for $y(t)=|Y|$, $y(t)$ asymptotically approaches $y(t)=|Y|$ for $t \to \infty$.
%\begin{equation}
%\frac{dx(t)}{dt} = 0.
%\end{equation}
%となるため，$x(t)=|N|$は不動点である.
%そこで，$x(t)=|N|-v(t)$ とおくと，
By rewriting the equation using $v(t)$, $v(t)=|Y|-y(t)$, we obtain 
\begin{equation}
\frac{dv}{dt} = - r (|Y|-v)(v/|Y|)^\alpha. \label{eq_v}
\end{equation}
For $v(t) \ll1$, Eq. \ref{eq_v} as follow:
\begin{equation}
\frac{dv}{dt} = -r|Y|^{1-\alpha}v^\alpha.  
\end{equation}
Solving this equation for $\alpha \neq 1$ we obtain
\begin{equation}
v(t) = \left( (\alpha-1)r(t-t_0)|Y|^{1-\alpha}+ v(t_0)^{1-\alpha}  \right)^{1/(1-\alpha)}, 
\end{equation}
and for $\alpha=1$,  
\begin{equation}
v(t)=v(t_0) \exp(r(t_0-t)/|Y|).
\end{equation}
Thus, by turning back $y(t)$ from $v(t)$,
for $\alpha \neq 1$, 
\begin{eqnarray}
%x(t) \approx |Y|- \left( (\alpha-1)r(t-t_0)}|Y|^{1-\alpha}+ (|Y|-y(t_0))^{1-\alpha}  \right)^{1/(1-\alpha)}.
&&y(t) \approx  \nonumber \\
&& |Y| \left\{1- \left( (\alpha-1)r(t-t_0)+ (1-y(t_0)/|Y|)^{1-\alpha}  \right)^{1/(1-\alpha)}  \right\}, \nonumber \\
 \label{app_yc_eq}
\end{eqnarray}
and for $\alpha= 1$, 
\begin{equation}
y(t) \approx |Y| - (N-y(t_0)) \exp(r(t_0-t)/|Y|). \label{app_yc_eq_case_1}
\end{equation} 
\par
%漸近特性は，$\alpha  > 1$では，$t \gg 1$ で, 
For $t \gg 1$ and $\alpha>1$, Eq. \ref{app_yc_eq} approaches $Y$ asymptotically in the power law function of $t$, 
\begin{equation}
y(t) \approx |Y| \left\{1- \left( (\alpha-1)rt \right)^{-1/(\alpha-1)}  \right \},  
\end{equation}
%となり，
%$N$に$t$のべき乗で漸近していく.
and for $\alpha=1$, Eq. \ref{app_yc_eq_case_1} approaches $Y$ asymptotically in the exponential law function of $t$, 
\begin{equation}
y(t) \approx |Y| - (|Y|-y(t_0)) \exp(r(t_0-t)/|Y|), 
\end{equation}
and for $0<\alpha<1$ Eq. \ref{app_yc_eq}  approaches $Y$ in infinite time $t^{*}$, 
\begin{eqnarray}
&&y(t) \approx |Y| \cdot \nonumber \\
&&\left\{1- \left(  (1-y(t_0)/|Y|)^{1-\alpha}-(1-\alpha)r(t-t_0)  \right)^{1/(1-\alpha)} \right\}, \nonumber \\ 
%y(t) \approx |Y| - \left( (|Y|-y(t_0))^{1-\alpha} -(1-\alpha)r(t-t_0)|Y|^{1-\alpha} \right)^{1/(1-\alpha)}, 
\end{eqnarray}
%有限時間$t^*$で$N$に漸近する
where 
\begin{equation}
t^*=t_0+\frac{(|Y|-y(t_0))^{1-\alpha}}{r(1-\alpha)|Y|^{1-\alpha}}. \label{app_n_eq_t}
\end{equation}
\subsection{$Y<0$, $\alpha<0$} 
%$x(t)=N$で，式\ref{xxx} より，$dx(t)/dt$は発散するので，
%From Eq. \ref{xxx}, when $x(t)=N$ $dx(t)/dt$ is diverse. 
From Eq \ref{base_eq}, $\frac{dy(t)}{dt}$ diverges when $y(t^{*})=Y$. $t^{*}$ was determined using Eq. \ref{app_n_eq_t}. 
 
%$x(t)<N$ の範囲で考える. そのとき，式\ref{xxx}で$N$に近づき，式 \ref{xxx}より有限時間で$t^{*}$で$N$に達する.
%その時間後には発散する.
\section{Methodology for Sampling Words}
\label{app_sec_data}
The sampling of words for this study was based on the titles of Wikipedia articles. 
Note that because of this sampling, our study is limited to established enough to be the title of a Wikipedia article rather than temporary words that are quickly forgotten. 
%Note that, in addition to the word choices described in this section, only words with a continuous growth period of at least 12 months (using the method in Appendix \ref{xxx}).were selected in the study, not short-term growth words that grew and decayed within a few days.
\subsection{Blog data}
%まず，ウィキペディア日本語版の項目から日本語ブログデータでの記事件数が多い上位１００万語を抽出した（カテゴリ等の単語ではない単語は除く）。
%次に，１００万語のうち、2006年１１月～１２月でブログ記事件数が0の単語の２万xxx語を新語として絞り込んだ。
We extracted new words using the following steps.
In step 1, we extracted the top one million words that had the highest frequency of articles in Japanese blog data from the list of titles in the Japanese version of Wikipedia \cite{wikipedia_data}. 
% (excluding words that are not categories or other words).
In step 2, we filter out 20,764 words that had 0 blog entries in both November and December 2006 among the 1 million words. 
%These words have 0 blog %entries from November-December 2006. 
\subsection{Google Trends}
%Wikipedia ページビューの２０１５年５月から２０２２年１月までの各年月の１日のデータを収集。最初の２０１５年５月で０の
%それ以外の月で１０以上になるものを抽出（１０ベトナム、５０スペイン、５０日本、５０フランス、１０００英語）。その単語のGoogle Trends
%を調べる。
%英語版、フランス語版、スペイン語版、日本語版の全タイトル項目の単語の２０１５年５月から2０２２年１月までの期間の各月の初日のページビューデータ%を収集。
%次に、観測初月の２０１５年５月でページビューが０ビューで，それ以降のいずれかの月で50view 以上（フランス語、スペイン語、日本語）、１０００view 以上（英語） となった新単語を抽出。
%Google Trends API　を用いて，選ばれた単語のGoogle Trendsを取得した。
%最後にgoogle  トレンドAPI　を用いて，選ばれた単語のgoogle　トレンドを取得した
For Google Trends data, we employed Wikipedia page views to extract new words from all titles in the English, French, and Spanish editions of Wikipedia \cite{wikipedia_data}. 
First, we collected page view data for the first day of each month from May 2015 through January 2022. 
%In step 1, data on first-day Wikipedia page views for each month from May 2015 to January 2022 for all words in all title entries in the English, French, Spanish, and Japanese editions. 
%Page view data for the first day of each month for the period from May 2015 through January 2022 was collected. 
Next, we extracted new words such that the word had 0 page views on May 1, 2015 (i.e., the first month of observation) and 50 or more views (French, Spanish and Japanese) or 1,000 or more views (English) on January 1 of 2016, 2017, $\cdots$, 2021 or 2022.
% For Japanese, we used the same new word list as in the blog data. 
%(French, Spanish, and Japanese) and 1000 or more views (English) in any month since then.
Finally, we used the Google Trends API to obtain Google Trends for selected words \cite{google_trends}.
\section{Methodology for Extracting an uptrend}
\label{app_sec_cut}
%一次的な最大値を上昇トレンドの最終値としないことを目的に以下の方法で
%トレンド最終点を抽出した.
Here, we show how we extracted a global growth period that is not a temporary local trend.
%global growth periods that are not temporary local trends. 
\textcolor{black}{The example of the global growth period is the period enclosed by the grey vertical lines in Fig. \ref{fig_world}.}  First, we describe the detection of the starting point of growth and next the detection of the end point of growth.
%ここでは，私たちは、一時的なローカルなトレンドでないグローバル成長時期を機械的に抽出するための方法を示す。つまり、図 \ref{xxx}で灰色の縦線に囲まれた期間を抽出する方法について示す。
\subsection{Extracting the beginning of growth}
\label{app_seq_growth}
%基本的には，１３か月移動中央値が２５％になる一番最初の時間までの間で、0を除いた最小値を
%トレンドの開始点とする（２５パーセント点は上昇後の下降トレンドを選ぶことを避けるため）.
%だだし，最小点は，１３か月移動中央値と生データで候補をだし，候補のうち，基本的には後の時間のほうを採用する
%（ただし、２つの時刻の間に明確な上昇トレンドがある場合，例外的に先に来るほうを選ぶ）.
%最小値が１０以下の場合は、時間規格化しないデータを使う。
%基本的には，トレンドの開始点は，１３か月移動中央値の時系列の最小点とする。
%より詳細には，最小点を計算するときに以下の条件をつける。
%（１）１３か月の移動中央値時系列が０の点は除く。
%（２）最小点を選ぶのは，１３か月移動中央値が２５％タイルになる一番最初の時間までの範囲とする。
%２つ目の条件は、上昇後の下降トレンド時の最小点を避けるため導入した。
\textcolor{black}{Basically, the starting point of the growth trend $t^{s}_0$ is the minimum point of the 13-points moving median time series or the minimum point  of corresponding row time series, $t^{s}_0=\max(t^{s}_1, t^{s}_2)$, where the minimum of the 13-points moving median of word counts $y_j(t)$ is denoted as $t^{s}_1=arg min_t\{MovingMedian_{13}({y_j}(t))\}$ and the minimum of corresponding row time series is denoted as $t^{s}_2=arg min_t\{MovingMedian_{13}({y_j}(t))\}$ and $y_j(t)=0$ is excluded. }
%Basically, we regard the starting point of growth, $t^{s}_0$, as the point (time) that the 13-points moving median of word counts $y_j(t)$ is the minimum, denoted as $t^{s}_1=arg min_t\{MovingMedian_{13}({y_j}(t))\}$  or  
%corresponding row time series $t^{s}_2=arg min_t\{{y_j}(t)\}$, where $y_j(t)=0$ is excluded.
% and we choose $t^{s}_0=\max(t^{s}_1, t^{s}_2)$ conservatively. 
%or the raw time series $y_j(t)$, where, on taking the minimum, $y_j(t)=0$ is excluded. 
%(i.e., it is regarded as approximately 13-month moving median)
%Basically, we determined the starting point of growth to be the point of the minimum of the 13-point moving median of $y_j(t)$.
%(i) The point at which the 13-month moving median time series is 0 is excluded.
%(ii) The minimum point is selected in the range up to the first time when the 13-month moving median reaches the 25th percentile tile.
%The second condition was introduced to avoid minimum points during a downtrend after an uptrend.
More precisely, the following procedure was used to determine the starting point of the growth:
%Translated with www.DeepL.com/Translator (free version)
\begin{enumerate}
\item %トレンドスタート点の上限時刻を決定.
%１３か月移動中央値が２５％に達する時間を計算し，それを上限時刻とする.
%開始点の上限時刻を計算する。具体的には，１３か月移動中央値が２５％点に最初に達する時刻を計算する。
We calculate the upper limit of candidates, $T^{s}$. 
The upper limit is set so that 13-points moving median first reaches the 25th percentile point, 
$T^{s}=min_t\{t|y(t) \geq Quantile25\{y(t)\} \}$.
%Calculate the upper time limit for the starting point. Specifically, the time when the 13-points moving median first reaches the 25th percentile point is calculated. 
This procedure is introduced to avoid selecting minimum points during a downtrend after an uptrend.
\item We calculate the first candidate of the starting point $t^{s}_1$. 
The first candidate is set as the time such that the 13-points moving median time series is minimum $t^{s}_1=arg min_{\{t \leq T_s\}}\{MovingMedian_{13}({y_j}(t))\}$. Here, the minimum value is less than 10, we recalculate using the 13-points moving median time series that is not normalized by the total number of articles $MovingMedian_{13}(x_j(t))$, where $x_j(t)$ is raw time series defined in the section \ref{sec_data}.
%上限時刻までのうち値が１３か月移動中央値で最小になるものを計算.候補１.
%（着目時系列の最小値が１０以下の場合は，時間規格化しないデータ，そうでない場合は規格化したデータを使う）.
%一つ目の成長開始点候補を計算する。 具体的には，１３か月移動中央値時系列が最小になる時刻を計算,where 時系列の最小値が１０以下の場合は，全数で規格化しない時系列を用いて再計算する.
\item We calculate the second candidate of starting point $t^{s}_2$. 
This second candidate is set as the time such that the raw time series (i.e., not taking the moving median) is the smallest and is calculated $t^{s}_2=arg min_{\{t \leq T_s\}}\{y_j(t)\}$.  Here, if the smallest value of the time series is less than 10, so we recalculate using the time series not normalized by the total number of articles $x_j(t)$.
% 上限時刻までのうち値が生データ最小になるものを計算.候補２.
%（着目時系列の最小値が１０以下の場合は，時間規格化しないデータ，そうでない場合は規格化したデータを使う）.
%二つ目の成長開始点候補を計算する。 具体的には，生の時系列（すなわち，移動中央値をとらない）が最小になる時刻を計算,where 時系列の最小値が１０以下の場合は，全数で規格化しない時系列を用いて再計算する.
\item We determine the starting point of growth $t^{s}_0$.
%トレンド開始点の決定.
Basically, we chose the starting point of the trend as the larger time of the two candidate points, conservatively, $t^{s}_0=\max(t^{s}_1,t^{s}_2)$. 
%Basically, we chose the starting point of the trend as the time of the larger of the two candidate points conservatively, $t^{s}_0=\max(t^{s}_1,t^{s}_2)$. 
%$t^{s}_0=argmin_{t \in \{t^{s}_0, t^{s}_1}\}{\y(t)\}$.
However, if there is a clear upward trend between the two candidates $\{t^{s}_1,t^{s}_2\}$, we set the starting point $t^{s}_0$ as the smaller time $t^{s}_0=\min(t^{s}_1,t^{s}_2)$ is selected as the growth starting point  
(The trend is identified based on the positive rank correlation between times $\{\min(t^{s}_1,t^{s}_2),\min(t^{s}_1,t^{s}_2)+1,\cdots,\max(t^{s}_1,t^{s}_2)\}$ and the corresponding counts $\{y_j(min(t^{s}_1,t^{s}_2)) ,y_j(min(t^{s}_1,t^{s}_2)+1) \cdots y_j(max(t^{s}_1,t^{s}_2))\}$.  Here, we recognize the trend when a p-value is less than 0.01 for the test of correlation.).
%２つの候補のうち，基本的には遅い方の時刻をトレンド開始点として採用する.　
%２つの候補のうち，基本的には後の時間のほうをトレンド開始点として採用する.　
%ただし、２つの時刻の間に明確な上昇トレンドがある場合，例外的に先に来るほうを選ぶ（時刻と時系列の正の順位相関をチェックしp値 一定以下.本研究で%は0.01採用）.
%ただし、２つの時刻の間に明確な上昇トレンドがある場合は，例外的に時刻が小さいほうを成長開始点として選ぶ。
%（本研究では，トレンドの認定は，時刻と時系列の正の順位相関をチェックしp値 が0.01採用）.
%ただし、２つの候補の間に明確な上昇トレンドがある場合は，時刻が小さいほうを成長開始点として選ぶ。
%（トレンドの認定は，時刻と対応するカウントの正の順位相関を基準とした。本研究は，基準は，相関の検定のp値 が0.01とした）.
\end{enumerate}
\subsection{Extraction of end of growth}
%明確な上昇トレンドの終わり（一二か月連続下降）を検出し，スタート点以降，下降トレンド開始より前の
%うちとと突発的な一時点な最大点と予想される点を除いたうちの一番大きい値をトレンドの最終点とする.
%検出法は，まず一二か月移動中央値で大まかなトレンドを検出し，そのトレンドでの上昇の最終点を選択，
%次に，徐々に短い時間スケールの近傍での最大点を探していく.
Basically, the endpoint of growth, $t^{e}_0$, is detected as the point at which a clear downward trend begins.
A clear downward trend is defined as a continuous decrease in word count for at least 12 months after the starting point ($t^{s}_0 \leq t^{e}_0$). 
%そこでは１３日移動平均でダウン路トレンドの開始点を探索し、移動中央値の時間スケールを小さくしながら、その近傍でより良いダウントレンドの開始点を探索している。
Specifically, we first roughly search for the starting point of the downtrend at the 13-points moving median to avoid being fooled by local trends. Next, we improve candidate points using progressively smaller time scale information (5-points moving median, 3-points median and original data). Finally, we examine whether the starting point of the downtrends found by the above method or the global maximum point is more suitable as the endpoint of the global uptrend.
%When detecting the end point of growth, the maximum value is searched for by gradually changing to a shorter time scale, from a 12-month moving median to a raw time series (no moving median).
%基本的には，成長の最終点は、明確な下降トレンドが始まる点として検出する。
%明確な下降トレンドは，開始点以降で一二か月間以上のワードカウントの連続的な減少とする。
%成長の終わりの点の検出の際は、１２か月移動中央値から生の時系列（移動中央値なし）まで，徐々に短い時間スケールに変えながら最大値を探索していく。
The details of the procedure are as follows:  
\begin{enumerate}
\item Detecting the starting points of a clear downtrend for the 13-points moving median $t^{e}_1$.
%Detecting the starting period of a clear downtrend.
Specifically, we select the point such that the starting point is where the 13-points moving median of the time series falls continuously for at least 12 time-points consecutively (i.e., approximately 12 months or a year).
% after .
Here, to avoid the error of detecting local downtrends, the start of a downtrend is detected at a point after the point at which the time series reaches the 90th percentile $T^{e}= min_t\{t|y(t) \geq Quantile90\{y(t)\} \}$, namely, we choose $t^{e}_1>T^{e}$.
If there is no point at which there are 12 consecutive downtrends, the last point of observation is taken as the end point of growth,  
$t^{e}_1=T$.
%具体的には，時系列の１３か月移動中央値が値が全期間の９０％タイル点到達した以降，１３か月移動中央値で連続１２回以上下がる点の間。
%トレンドが連続１２回下がる期間がなければ、最終値とする.
%明確な下降トレンドの開始期間の検出.
%具体的には，時系列の１３か月移動中央値が１２か月以上の間連続的に下がる開始点。
%成長途中でのローカルな下降を検出する誤りを避けるため、ダウントレンド開始時点は、時系列が９０パーセンタイルに達する点以降の時点で検出する。
%１２回連続下降する点がなければ、観測の最終時点を仮の成長終了点とする.
\item Exploring around the $t^{e}_1$. 
To determine the end of the growth trend more precisely, we explore in more detail the area around the $t^{e}_1$ which calculated the previous step.
Specifically, in this step, we make several other candidates set for the end time of the growth trend $\{t^{e}_2\}$.
The candidates in the set $\{t^{e}_2\}$ satisfy the following condition: the 13-points moving median is greater than 90\% of the maximum value between $t^{s}_0$ and $t^{f}_1$.  Particularly we calculate the set of candidates, 
\begin{eqnarray}
&&\{t^{f}_2\}= \{t|t^{s}_0 \leq t \leq t^{e}_1,  MovingMedian_{13}(y(t)) \geq \nonumber  \\
&& 0.9 \times \max_{\{t^{s}_0 \leq t \leq t^{e}_1\}}(MovingMdedian_{13}(y(t)))  \}. \nonumber
\end{eqnarray} 
%Here, this process is performed between the start point of the growth trend (calculated in the previous section) and the start point of the downtrend (calculated in the previous step).
%For example, if the maximum is 100 articles, the candidates are all times that take 90 articles or more.
%トレンド最終点候補の抽出
%具体的には，トレンド開始点と下降トレンド開始点の間で１３か月移動中央値が，最大値の９割以内より大きい値を候補とする.
%成長トレンド最終点候補の抽出。
%成長トレンドの終わりをより正確に決めるため，私たちは下降トレンドの初めの時刻（前のステップで計算した）の周辺をより詳細に探索する。
%具体的には，このステップでは，成長トレンドの終わり時刻のいくつかの候補を挙げる。　　
%成長トレンド終わりの点の候補達は、以下の条件を満たす：１３か月移動中央値が，最大値の９割より大きい。ここで、この処理は，成長トレンドの開始点（前節で計算）と下降トレンド開始点（前のステップで計算）の間で行う。例えば，最大値が１００記事の場合は，候補は，９０記事以上をとるすべての時刻達となる。
\item Adding information about shorter time scales (5-points moving median). 
%候補点近傍でより短い時間スケールで最大値を抽出
%すべての候補点について，候補の５か月移動平均で連続的に件数を増加できる点まで増加する。次に、その上昇した点からさらに３か月移動平均で件数を増加できる点まで増加する。
%We search for around candidates who set $\{t^{e}_2\}$ on shorter time scales.
%We extract the maximum on shorter time scales in the neighbourhood of the candidate points (computed in the previous step).
%Specifically, 
For all candidate points $\{t^{e}_2\}$, we search the neighbourhoods and replace the point such that point is reached by continuously increasing in 5-points moving median. 
%for the maximum value using a 5-month moving average, moving in the direction of a continuously increasing number of cases from the candidate point.  
%We search around the candidates on shorter time scales.
%$t_0=arg max_{\{t \in t-1,t,t+1\}}(MovingMedian_{5}(y_j(t)))$ 
Particularly, for all candidate points $t \in \{t^{e}_2\}$, we replace $t \to t^*$ using the following formula, 
\begin{eqnarray}
t^{*}_0&=&t_0+ arg max_{\{t \in \{t_0+m^{-},t_0+m^{+}\}\}}(q_j(t)) \label{app_move_end} \\
m^{+}&=&\max(\{m|m,s \in \mathbb{N}, m \geq 0, 0 \leq s \leq m \nonumber \\ 
&& ,\forall{s}[q_j(t_0+s)\leq q_j(t_0+s+1)]\})  \nonumber \\
m^{-}&=&\min(\{m|m,s \in \mathbb{N}, m \leq 0, m \leq s \leq 0 , \nonumber \\ 
&&\forall{s}[q_j(t_0+s) \leq q_j(t_0+s-1)] \}).  \nonumber
\end{eqnarray}
Here, we use the $q_j(t)=MovingMedian_5(f_j(t))$, $\{t_0\} \to \{t^{e}_2\}$ and $\{t^{*}_0\} \to \{t^{e}_3\}$. 
This transformation corresponds to a correction for the candidate points in the 13-points moving median with the addition of information on a shorter time scale (i.e., the 5-points moving median).
%This transformation means that the candidate points move from a 13 points moving median to 5 points moving median.
%We extract the maximum on shorter time scales in the neighbourhood of the candidate points (computed in the previous step).
%Specifically, for all candidate points, we first search for the maximum value using a 5-month moving average, moving in the direction of a continuously increasing number of cases from the candidate point.
\item Adding information of 3-points moving median.
Next, a similar to previous step's transform process given by Eq. \ref{app_move_end} is performed for the 3-points moving median, $q_j(t)=MovingMedian_3(f_j(t))$, with the starting point of this process being the candidate point in the 5-points moving median, $\{t^{e}_3\}$ calculated in the previous step,  $\{t_0\} \to \{t^{e}_3\}$.
The candidate point using the 3-points moving median is calculated as  $\{t^{*}_0\} \to \{t^{e}_4\}$.
%より短い時間スケールでの探索。私たちは，候補点たち（前のステップで計算した）の近傍でより短い時間スケールで最大値を抽出する。
%具体的には，すべての候補点について，まず，５か月移動平均を用いて，候補点より連続的に件数が多きなっていく方向に進み，最大値を探索する。
%次に、同様な上昇処理を３か月移動平均でも行う，この処理の、開始点は、５か月移動平均における候補点とする。
\item 
%成長トレンドの終わりの点の決定。
%成長トレンドの終わりの時刻は，成長トレンド候補点の中から最大の件数の点をとして決める。
We determine the tentative end point of the growth trend, $t^{e}_5$.
The end time of the growth trend is determined by the point with the largest number of growth trend candidate points $t^{e}_5=arg max_{t \in \{t^{e}_4\}}(MovingMedian_{3}(y_j(t)))$.
\item %微調整 
Fine-tuning with raw data (adding information of raw data).
%借決定点から生データでその点から上昇できる点まで上昇する.さらに，その上昇点の前後４か月最大点があれば，それを候補点とする.
%Fine-tuning with raw data.
We use raw data (data without a moving median) $y_j(t)$ to fine-tune the end points of the growth trend.
Specifically, we change the candidate point to a larger point near the candidate point determined by Eq. \ref{app_move_end} for $q_j(t)=f_j(t)$ and $t_0 \to t^{f}_5$. We denote the transformed time as $t^{e}_6=t^{e}_5+ arg max_{\{t \in \{t^{e}_5+m^{-},t^{e}_5+m^{+}\}\}}(y_j(t))$. 
%In this case, the neighbourhood is the point in the area that can rise continuously from the candidate point.
%生データでの微調整。
%私たちは生データ（移動中央値をとらないデータ）を用いて，成長トレンドの終わりの点の微調整を行う。
%具体的には，前ステップで決めた候補点近傍に大きい点に候補点に変更する。　この際、近傍は候補点から連続的に上昇できる領域にある点である。
\item %トレンド最終候補点とデータの最大値比較
Comparison between the candidate $t^{e}_6$ and the global maximum point $t^{max}=arg max_{t}[y_j(t)]$. 
%候補点の前後６点（すなわち６か月以内に）に時系列全体における最大値があるケースにおいて、私たちはデータの最大値のほうが現候補点より，成長トレンドの終わりの点として適するかをチェックする.
In cases where there is a maximum in the whole time series at 6 points before or after the candidate point (i.e., the case $t^{max}$ is satisfied with $t^{e}_6-6 \leq t^{max} \leq t^{e}_6+6$), we check whether the maximum of the data $t^{max}$ is more suitable as the end point of the growth trend than the current candidate point $t^{e}_6$. 
%具体的には，最大値はニュースや外力などによる１自的でローカルトレンド最大点でないかを調べる（最大点と最終候補の間に明確なトレンドがある）.
\textcolor{black}{Specifically, we make sure that the maximum is not a temporary maximum due to news or external forces or there is a clear upward trend from the candidate point to the maximum point.}
%具体的には，最大値はニュースや外力などによる一時的なで最大点でないことを確認する。つまり，候補点から最大点に向かって明確な上昇トレンドがあるかを調べる.　
%判断基準は，データを対数変換したときに，最終候補点と最大値の間が、１次関数、もしくは、２次関数で近似できる明確なトレンドがある場合は、最大値を最終のトレンド候補点とする（採用した基準：正規分布ノイズの線形トレンド決定係数０．４以上，係数のp値が１％以下（もしくは，二項検定のp値が0.05パーセント以下.二次関数の場合は決定係数0.85 以上）.
%We used linear and quadratic function approximations in log-transformed time series to determine if there was a clear upward trend.
%明確な上昇トレンドは，候補点から最大点まで抜き出した時系列について，以下の２つのどちらかを満たす場合:
%(i)線形近似のあてはまりよく回帰係数も０といえない(決定係数0.4以上かつ係数のp値が１％以下)。
%(ii)２次関数への近似のあてはまりよい（決定係数0.85以上）。
%(iii)二項検定(p=0.05)？
A clear uptrend is defined as a time series extracted between the candidate points $t^{e}_6$ and the maximum point $t^{max}$,  $\{y(\min(t^{e}_6,t^{max})),y(\min(t^{e}_6,t^{max})+1),\cdots,y(\max(t^{e}_6,t^{max}))\}$ that satisfies either of the following three conditions:
(i) A linear approximation is well fitted, and the regression coefficient is not zero (coefficient of determination is greater than 0.4, the sign of the regression coefficient is $sign(t^{(max)}-t^{e}_6)$ and the p-value of the coefficient is less than 1\%).
(ii) A good fit of the approximation to a quadratic function (coefficient of determination greater than or equal to 0.85) and the function's differential is always positive in the focus period for $t^{e}_6 < t^{max}$ or always negative for $t^{max} < t^{e}_6$.
(iii)The binomial test of differential for the positive ratio for $t^{e}_6 < t^{max}$  or negative ratio for $t^{max} < t^e_{6}$  taking 0.6 (the one-sided test p-value of the coefficient is less than 5\%). 
If the conditions are satisfied, the final candidate point is $t^{e}_0=t^{max}$. If the condition is not satisfied, then $t^{e}_0=t^{e}_6$. 
%私たちは，明確な上昇トレンドかどうかの判断は，対数変換した時系列において，１次関数近似と２次関数近似を用いた。
%\item スタート点の再調整.
%もし，トレンド終了点がスタート点より前に来た場合は，トレンド点より前のデータに関してトレンド開始点を探索しなおす.
\end{enumerate}

\begin{figure}[t]
    \begin{minipage}{0.98\hsize}
    %\begin{minipage}{0.24\hsize}
    %\includegraphics[width=4.0cm]{"all_1512_jyoseikatuyaku_all_rel.eps"}
    \begin{overpic}[width=6.0cm]{"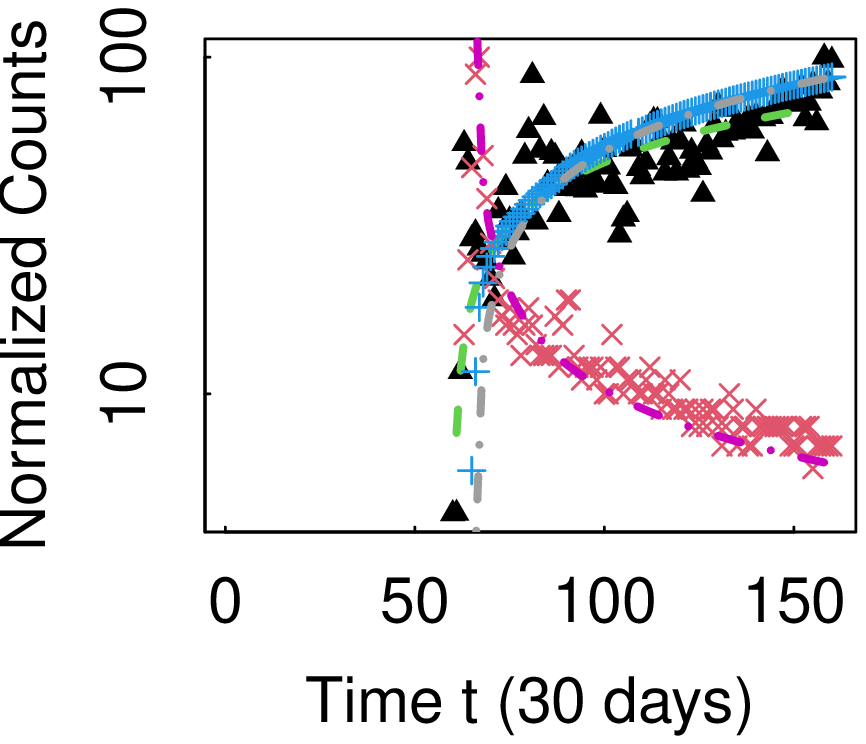"}
        %\put(22,55){(a)}
        %\put(22,68){(a)}
    \end{overpic}
    \end{minipage}
    \caption{
        Comparison of time series between blog data (black triangles) and Google Trends (red crosses) 
        for ``Toyota$\cdot$Akua'' (Toyota Aqua, Name of a car, $\alpha=-2.56,Y=15.7, r=16.8, y(0)=4.47$).
        Blog data is scaled to a maximum value of 100.   The green dashed line indicates the theoretical curve given by the model Eq. \ref{base_eq}.
        The gray dash-dotted line is power-law function $\propto t^{0.5}$, purple red dash-dotted line is power-law function $\propto t^{-0.5}$ and green pluses represent scaled cumulative values of Google Trends. From the figure, we can confirm that the blog data corresponds well to the cumulative values of Google Trends. 
        % in the panne (a) and $\propto \tau^{-1.0}$ in the panel(b).
        % The area between the grey dotted lines is the growth period detected by adapting the method described in Appendix \ref{app_sec_cut}.
       %(a) Toyota$\cdot$Akua (Toyota Aqua, $\alpha=-2.56,Y=15.7, r=16.8, y(0)=4.47$). 
    }
    \label{fig_toyota}
\end{figure}
\section{Estimation of Parameters}
\label{app_sec_estimate}
%以下の中央値を最適化した。ただし、推定を安定させるために正則化の項を加える.以下を最小にするようにパラメータをチューニングする
%$\alpha$と$Y$と$r$を３つをチューニングした。
We determined the parameters $r$, $\alpha$, and $Y$ to minimize the following median absolute error:
A regularization term (lasso or least absolute shrinkage and selection operator) is added to stabilize the estimation \cite{bishop2006pattern}.
\begin{equation}
Median_{\{t=1,2,\cdots,T_j\}}[y_j(t)-F(t|\alpha,r,Y))]+\lambda (|r|+|Y|),   
\label{app_eq_esitmate} 
\end{equation}
where the solution to $\frac{dy(t)}{dt}=f(t|\alpha,r,Y)$ is denoted by $y(t)=F(t|\alpha,r,Y)$ (in the case of the proposed model, $f(t|\alpha,r,Y)$ is given by Eq.\ref{app_base_eq} and 
$F(t|\alpha,r,Y)$ is given by \ref{eq_app_solution}).
 In this study, we employ a differential evolution algorithm for global optimization and $\lambda=0.03$ \cite{mullen2011deoptim}. 
The initial state $y(0)$ is determined by the initial value $\bar{y}(t_0)$ of the time series of smoothed splines $y(t)$ (in the case of $\bar{y}(t_0)<0$, we use $y(t_0)=0.8$) \cite{perperoglou2019review}. \par
Note that for $\alpha>0$ and $y(t) \gg Y>0$, the proposed model can be approximated by 
\begin{equation}
    \frac{dy(t)}{dt}=ry(t)\left(1-\frac{y(t)}{Y} \right) \approx \frac{r}{Y^\alpha} y(t)^{\alpha+1}=Cy(t)^{\alpha+1},
\end{equation}
where $C=\frac{r}{Y^\alpha}$ is the constant. 
\textcolor{black}{The equation shows that $r$ and $Y$ are not uniquely determined.} 
This indeterminacy may be one reason why estimation is unstable without regularization.
\par
%From this equation, $r$ $Y$がいちいに決まらず不定性が生じる問題がある。 これが色々ありそう。　 \par
%In addition, for validation, we examined the following quantities to estimate the parameters in which we remove $\alpha$ for regularization parameters: 
%\begin{equation}
%    Median_{\{t=1,2,\cdots,T_j\}}[y_j(t)-F(t|\alpha,r,Y))]+\lambda (|r|+|Y|).  
%    \label{app_eq_esitmate_no_reg} 
%\end{equation}
%The essential influence of the regularization of $\alpha$ on the results discussed in this paper was hardly observed, except for the forecasting problem (see Figs. \ref{app_fig_examples},\ref{app_fig_hist}).  Regarding the prediction, the performance will be worse if $\alpha$ is not regularized (see Table \ref{app_table_pred}).  One of the reasons for the inferiority of the proposed model without the regularization of $\alpha$ is overfitting; the proposed model is sometimes fooled (overfitted) by breaking news at the end of the training data, and incorrectly fits a finite-time divergence solution to this abrupt increase.
%$x_0$はデータから以下のように決定した.
%The parameter $x_0$ is determined by the actual data.

\section{Parameter Estimation with Bayesian Statistics}
\label{app_sec_param}
%推定するためにモデル\ref{xxx}を確率モデルに拡張する.
To perform the estimation using Bayesian statistics, we extended the model given by Eq. \ref{app_base_eq} to a stochastic model.
Specifically, we add independent noise $y(t;\alpha,Y,r) \epsilon(t|\sigma)$ with a normal distribution whose standard deviation is proportional to $y(t)$, 
\begin{equation}
\tilde{y}(t|\alpha,Y,r,\sigma) = y(t;\alpha,Y,r)+y(t;\alpha,Y,r) \cdot \epsilon(t|\sigma) 
\end{equation}
Here, $y(t;\alpha,Y,r)$ is calculated by Eq. \ref{eq_app_solution} or Eq .\ref{app_base_eq} and the noise $\epsilon(t|\sigma)$ is a normal distribution with a mean of 0 and standard deviation $\sigma$, 
\begin{eqnarray}
\epsilon(t|\sigma) \sim Norm(0,\sigma). \\
\end{eqnarray}
%$0 \leq r \leq 1.5$とし，$-2 \leq \alpha \leq 2$、また，$0 \leq \sigma \leq 0.2$に限定している.
%また，弱事前分布として，
For the convergence of the Bayesian estimation, we use the following prior distributions:
The prior distributions of parameters $r$ and $\sigma$ are uniform distributions 
\begin{eqnarray}
r \sim Unif(0,1.5) \\
\sigma \sim Unif(0,0.2), \\
\end{eqnarray}
where $Unif(a,b)$ denotes a uniform distribution that takes values from $a$ to $b$.
 The prior distribution of $\alpha$ is asymmetric and uniform. 
\begin{eqnarray}
\alpha|q \sim Acymetric Unif(-2,0;q), \\
\end{eqnarray}
%ここで，$Unif(a,b)$はaからbの値をとる一様分布である。
Here, the probability density distribution of the asymmetric uniform distribution is defined as
\begin{eqnarray}
p(\alpha|q)= \left\{
\begin{array}{cc}
q/2  & (0 \leq \alpha \leq 2) \\
(1-q)/2 & (-2 \leq \alpha \leq 0) \\
0  & (otherwise), \\ 
\end{array}
\right.
\end{eqnarray}
where the prior distribution of $q$ is given by a uniform distribution,  
\begin{equation}
q \sim Unif(0,1).
\end{equation}
This distribution is a mixture of $Unif(0,2)$ and $Unif(-2,0)$ with a mixing ratio parameter of $0 \leq q \leq 1$. 
%$\alpha>0$と$\alpha<0$の分布の混合パラメータであり, MCMCで$\alpha$が正負にうごきやすくし，収束しやすいように導入した
This mixed distribution was introduced to prevent the Markov chain Monte Carlo  method (a method for estimating parameters) from falling into a local solution. Without this distribution, $\alpha$ is likely to fall into either a positive or negative local solution because the likelihood around $\alpha \approx 0$ is often very low. \par
%この混合分布は，マルコフ連鎖モンテカルロ法（パラメータを推定する手法) による解が，局所解に陥ることを防ぐために導入した。 $\alpha=0$が尤度が多%くの場合とても低いため，この分布を導入なしには，$\alpha$が 正負のどちらかの局所解に陥りやすくなる。
%事前分布は，一様分布
Finally, we employ the following mixture two-sided exponential distribution as a prior distribution for $Y$, 
%最後に，$Y$は以下の混合の両側指数分布による弱事前分布を用いた. 
%$Y$は必要なければできる限り$0$に近くする，
%また，$\alpha<0$と$Y<0$が少ないという事前情報(see 表 \ref{xxx})を推定に考慮する.
\begin{eqnarray}
&&Y|\alpha \sim \nonumber \\
&&\left\{  
\begin{array}{ll}
 DoubleExponential(0,10^6) & (\alpha \geq 0)  \\
 AcymetricDoubleExpnential(0,10^6;0.95) &  (\alpha<0) 
\end{array}
\right. \nonumber  \\
\end{eqnarray}
where the probability density distribution of this prior distribution is given by
\begin{eqnarray}
P(Y|\alpha)=   
\left\{
\begin{array}{ll}
\frac{1}{2\mu} \cdot \exp(-|Y|/\mu) & (\alpha \geq 0) \\
\frac{0.95}{2\mu} \cdot \exp(-|Y|/\mu) & (\alpha < 0, Y>0) \\
\frac{0.05}{2\mu} \cdot \exp(-|Y|/\mu) & (\alpha < 0, Y<0), \\
\end{array}
\right.
\end{eqnarray}
and $\mu \sim Unif(0,10^6)$.
This prior distribution is introduced for two reasons: (i) $Y$ is as close to $0$ as possible if not necessary (Bayesian lasso) and (ii) prior information that few words have $\alpha<0$ and $Y<0$ (see Table \ref{table_para}).
\par
%この事前分布の導入理由は以下の２点である: (i) $Y$は必要なければできる限り$0$に近くする(ベイジアンラッソ), (ii) $\alpha<0$と$Y<0$となる単語は少ないという事前情報(see 表 \ref{xxx}).
%今回は、全データの件数の最大値をもとに$\mu=10^6$とした.
%このモデルにハミルトニアンもてかるロ法stanを用いて，$y(t)$の事後分布を計算し，その分布の中央値と99パーセントと１パーセントを
%図ではプロットしている。\par
\textcolor{black}{We compute the posterior distribution of $y(t)$ using the Hamiltonian Monte Carlo method with the computer library ``Stan'' \cite{carpenter2017stan}. }
\section{Comparison of models}
\label{app_sec_model}
We compared the proposed model given by Eq. \ref{base_eq} with other models related to the logistic equation.
% regarding the descriptive power of the dynamics of the data.
%ここではEq. \ref{base_eq}, 提案モデルが他のシンプルなモデルと比べて，よくデータを説明することを示す。
%まず，シンプルな２パラメータモデルを確認する。次に，２種類（線形型とべき乗型）の３パラメータのモデルを比較する。最後に，
%提案モデルの符号変化を限定した場合についても確認する。説明力の図的な比較は，図\ref{xxx}, 図\ref{xxx}, 図\ref{xxx}, 図\ref{xxx} 
%に，予測精度の数値的な比較は表\ref{xxx}に示す。
\subsection{Two parmeter model}
\label{app_sec_two}
\begin{figure*}
\begin{minipage}{0.19\hsize}
    \begin{overpic}[width=3.8cm]{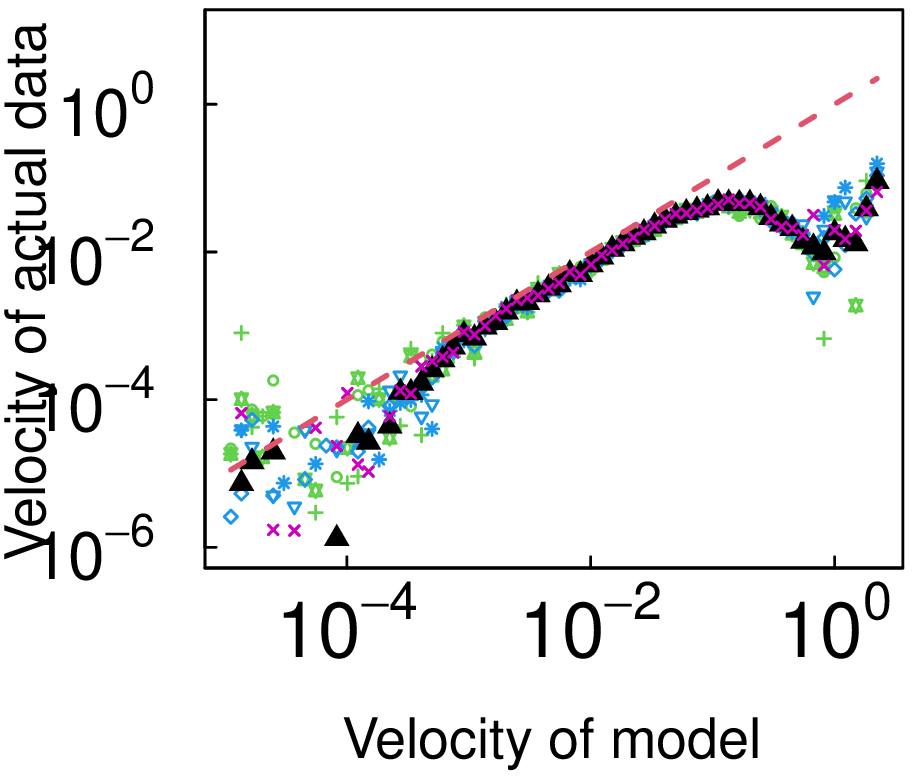}
        %\put(22,55){(a)}
        \put(23,72){(a)}
    \end{overpic}
\end{minipage}
\begin{minipage}{0.19\hsize}
    \begin{overpic}[width=3.8cm]{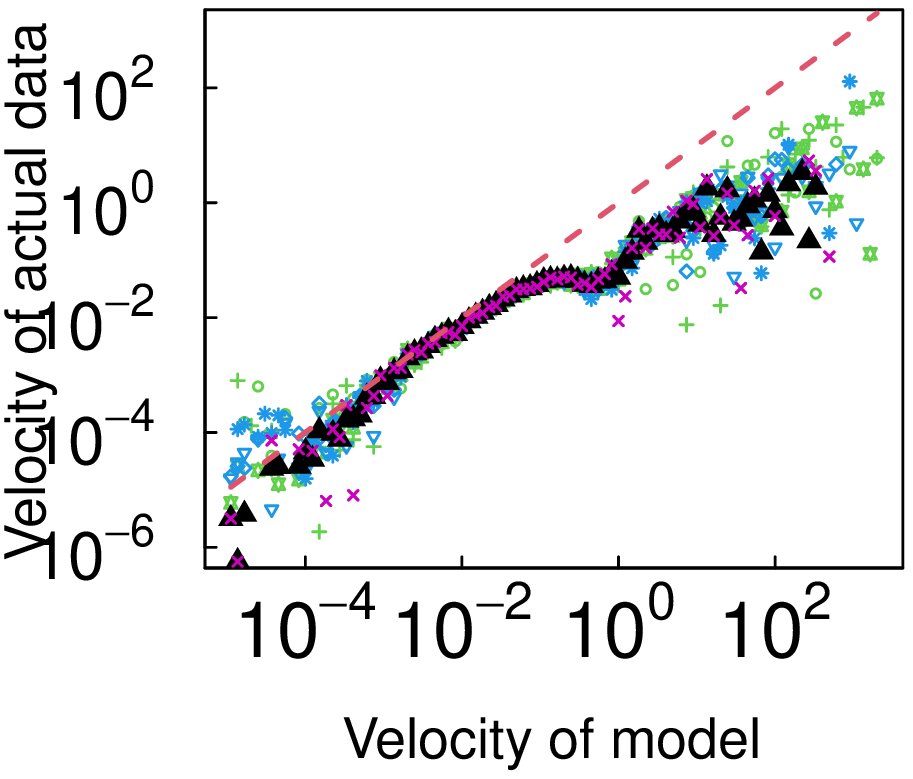}
        %\put(22,55){(a)}
        \put(23,72){(b)}
    \end{overpic}
\end{minipage}
\begin{minipage}{0.19\hsize}
\begin{overpic}[width=3.8cm]{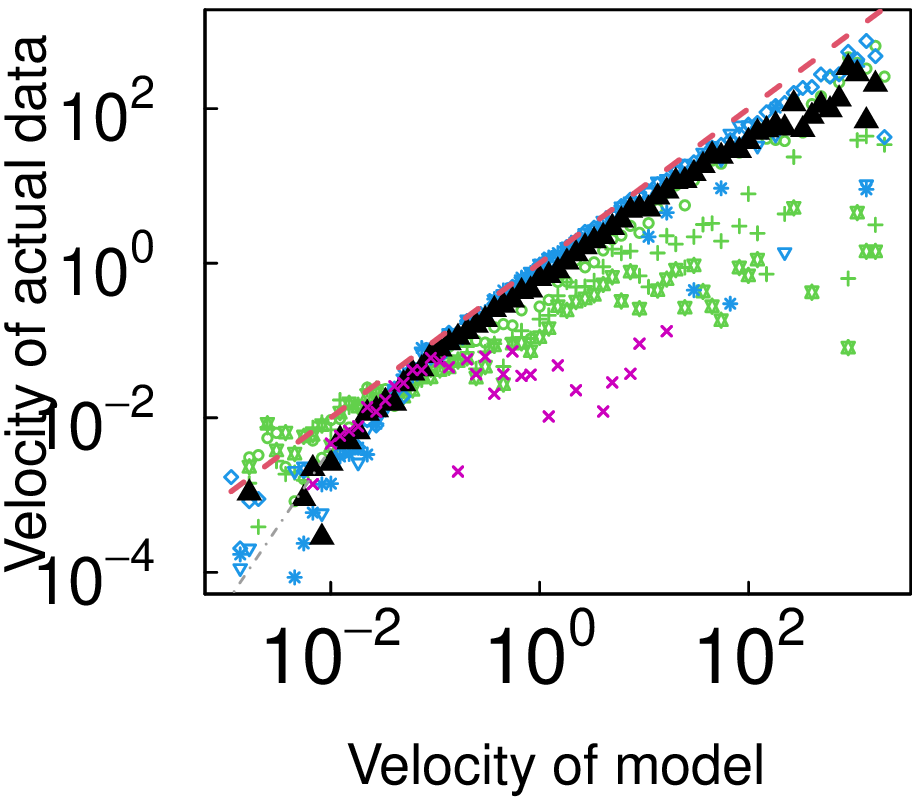}
    %\put(22,55){(a)}
    \put(23,72){(c)}
\end{overpic}
\end{minipage}
\begin{minipage}{0.19\hsize}
\begin{overpic}[width=3.8cm]{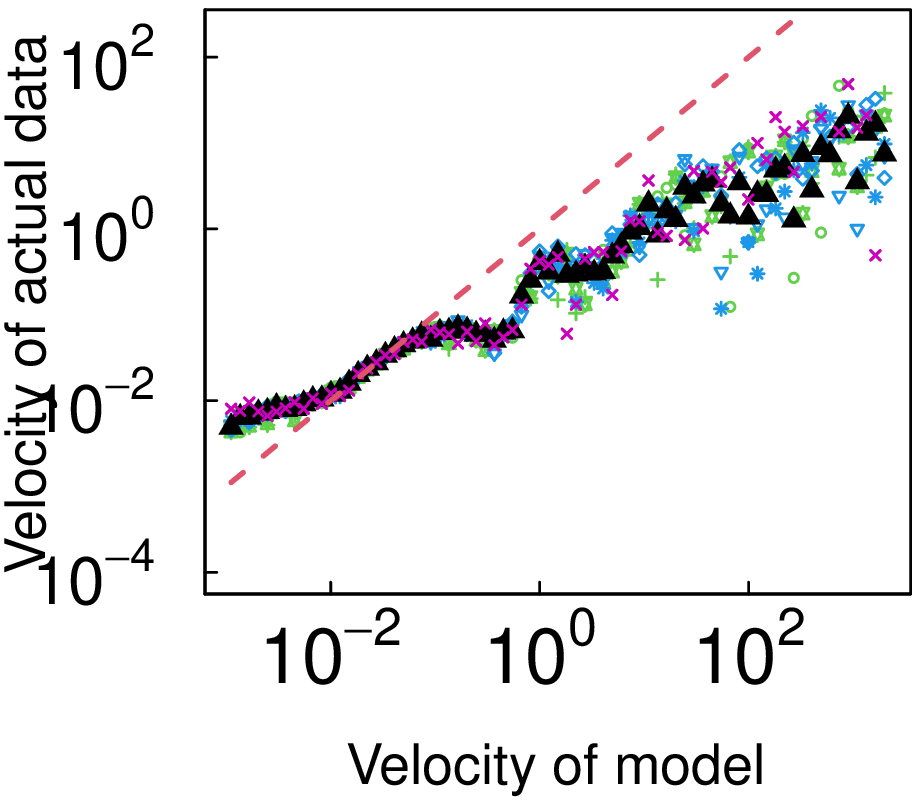}
    %\put(22,55){(a)}
    \put(23,72){(d)}
\end{overpic}
\end{minipage}
\begin{minipage}{0.19\hsize}
\begin{overpic}[width=3.8cm]{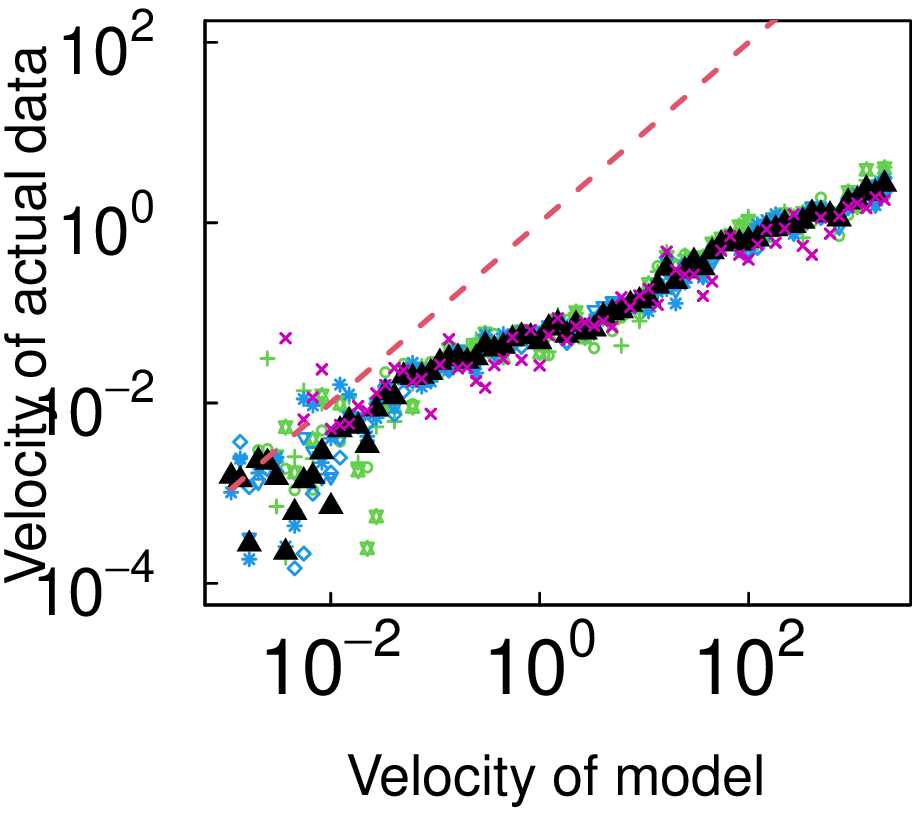}
    %\put(22,55){(a)}
    \put(23,72){(e)}
\end{overpic}
\end{minipage}
\begin{minipage}{0.19\hsize}
    %\includegraphics[width=3.8cm]{transformed_count_ave_v_genb.eps}
    %\begin{overpic}[width=3.8cm]{transformed_count_ave_v_genb.eps}
        \begin{overpic}[width=3.8cm]{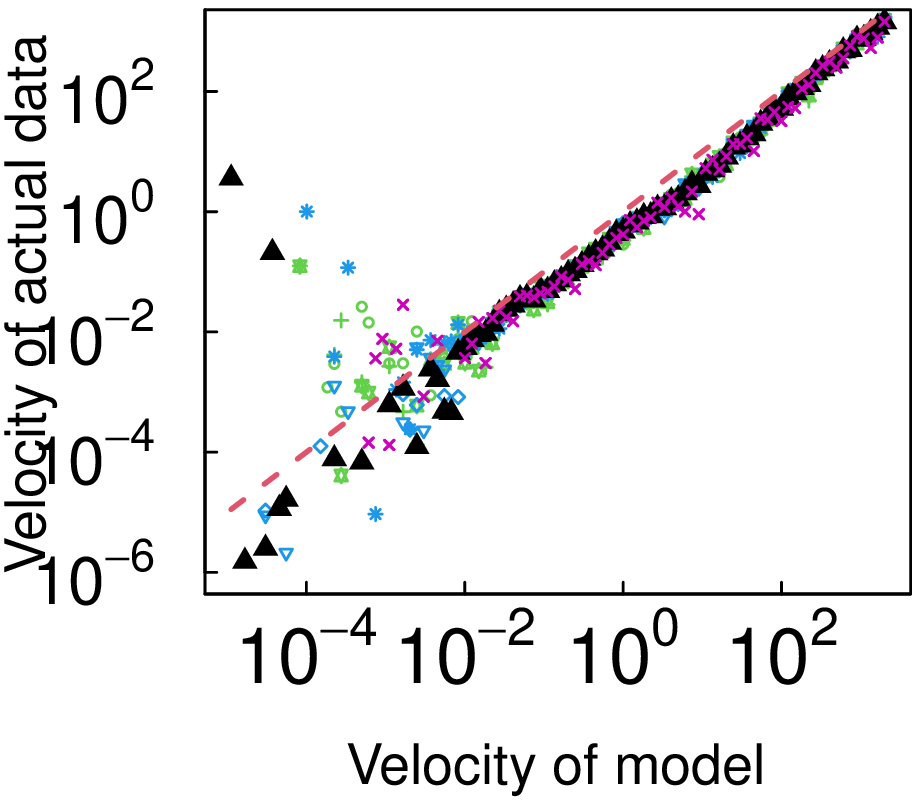}
        %\put(22,55){(a)}
        \put(23,72){(f)}
    \end{overpic}
\end{minipage}
\begin{minipage}{0.19\hsize}
    %\includegraphics[width=3.8cm]{transformed_count_ave_v_thb.eps}
    %\begin{overpic}[width=3.8cm]{transformed_count_ave_v_thb.eps}
        \begin{overpic}[width=3.8cm]{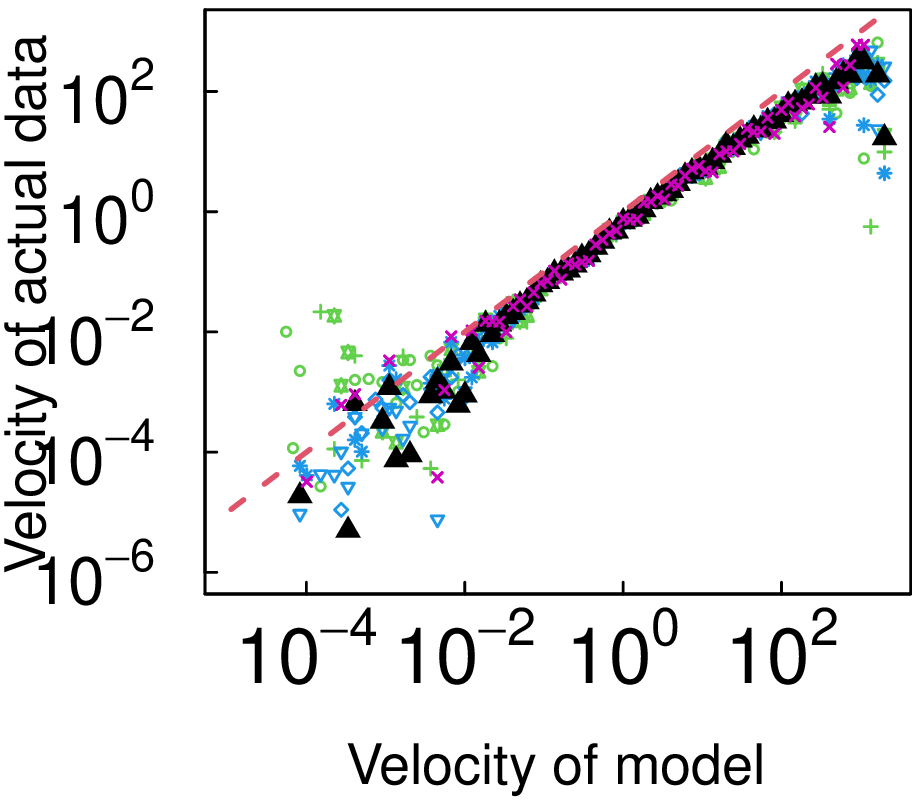}
        %\put(22,55){(a)}
        \put(23,72){(g)}
    \end{overpic}
\end{minipage}
\begin{minipage}{0.19\hsize}
    \begin{overpic}[width=3.8cm]{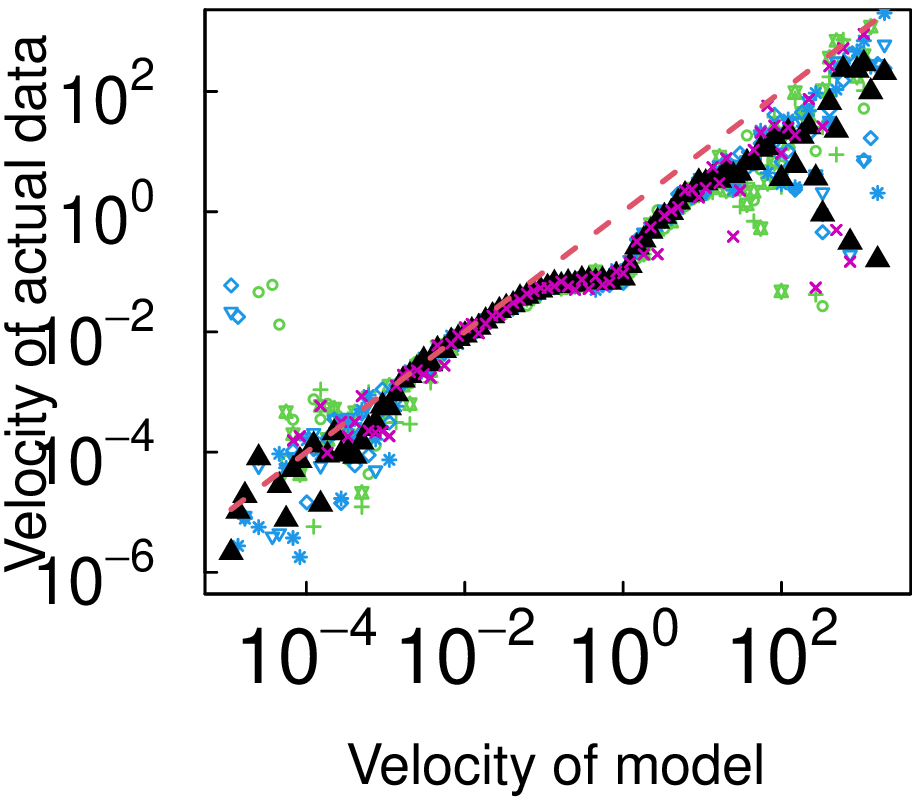}
        %\put(22,55){(a)}
        \put(23,72){(h)}
    \end{overpic}
\end{minipage}
\begin{minipage}{0.19\hsize}
    \begin{overpic}[width=3.8cm]{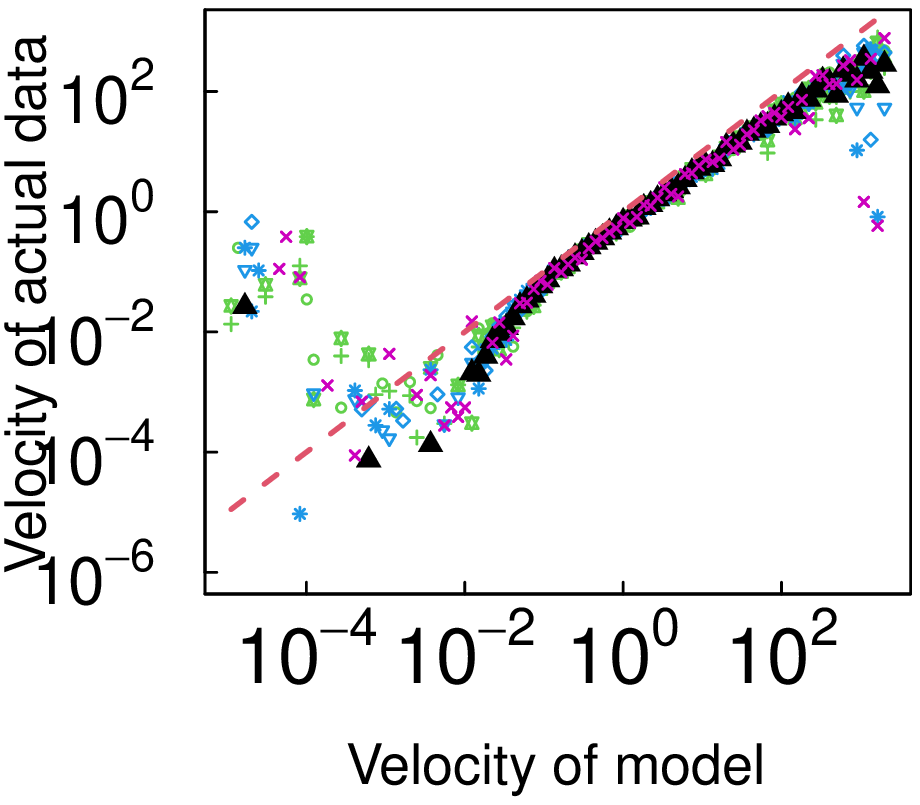}
        %\put(22,55){(a)}
        \put(23,72){(i)}
    \end{overpic}
    \end{minipage}
\begin{minipage}{0.19\hsize}
\begin{overpic}[width=3.8cm]{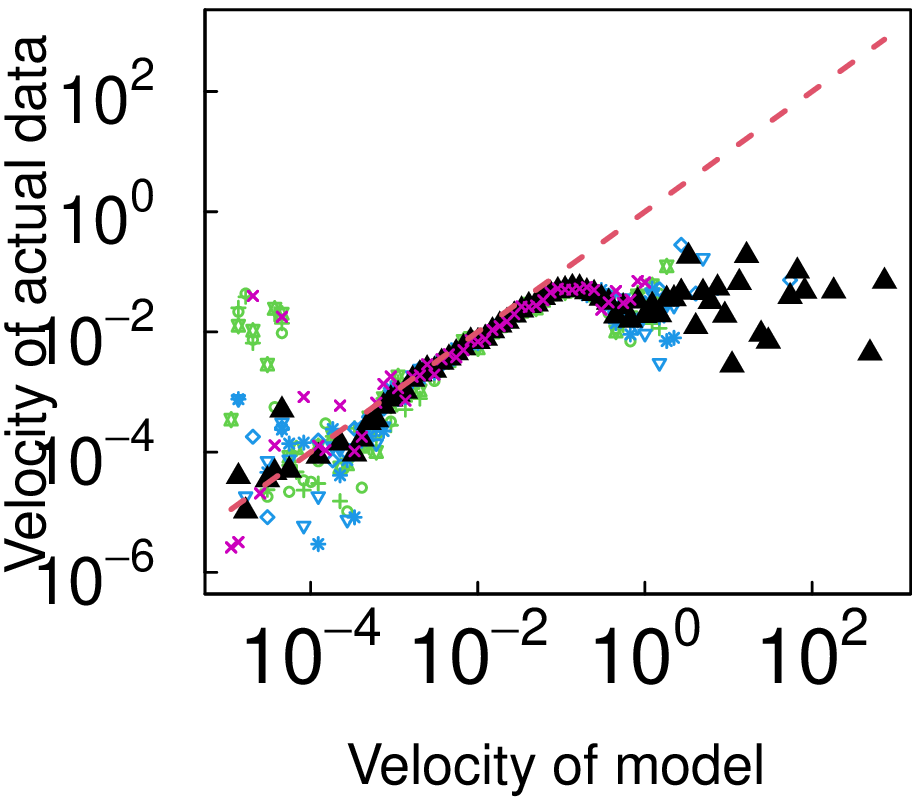}
    %\put(22,55){(a)}
    \put(23,72){(j)}
\end{overpic}
\end{minipage}
\caption{
    %Comparison of the descriptiveness of the dynamics of word count time series in various models by the differential equation. 
    Comparison of models given in Appendix \ref{app_sec_model} by using the differential equations. 
    If the model explains the data well, the graph will be close to the $y=x$ line shown by the red dashed line. 
    The y-axis corresponds to the left-hand side, $dy(t)/(dt) \cdot 1/|Y|$ of a model and the x-axis to the right-hand side $f(y_j(t)|\alpha,Y)/|Y|$ of a model, where a model is written by a differential equation, $\frac{dy(t)}{dt}=f(y(t)|\alpha,Y)$.
    Specifically, we plot the median quantity with respect to words given by Eq. \ref{app_eq_medi}, which is scaled by the model's scale parameter $Y$.
   The ensemble median over all data is plotted in black triangles and also the ensemble median grouped by proposed model's  (Eq. \ref{app_base_eq}) parameters $\alpha_j^{(0)}$ and $Y_j^{(0)}$ is plotted.
In case of $Y_j^{(0)}>0$, we plott in green circles for $\alpha_j^{(0)}=0.5$, green pluses for $\alpha_j^{(0)}=1.0$, for green triangles up and down $\alpha_j^{(0)}=1.5$, blue diamonds for $\alpha_j^{(0)}=-0.5$, blue triangles point down for $\alpha_j^{(0)}=-1.0$ and in the case of $Y_j^{(0)}<0$ magenta corsses. 
(a) Logistic equation given by Eq. \ref{app_base_logi},
(b) $Y$-sign extended logistic equation given by Eq. \ref{app_base_logi2},
(c) Single factor power-law model given by Eq. \ref{app_base_pow},
(d)($Y$-sign extended) Bass model given by Eq. \ref{app_base_bass},
(e) Constant model given by Eq. \ref{app_base_const},
(f) First factor power-law model given by Eq. \ref{app_base_first},
(g) Second factor power-law model (Proposed model) given by Eq. \ref{app_base_second},
(h) Inside power-law model given by Eq. \ref{app_base_inside},
(i) $Y$-sign positive-restricted second factor power-law model given by Eq. \ref{app_base_posi} and
(j)$Y$-sign negative-restricted second factor power-law model given by Eq. \ref{app_base_nega}.
  %様々なモデルのデータ動力学の記述可能性の比較。直線に近いほどデータがモデルに適合している。
   % 微分方程式を$\frac{dy(t)}{dt}=f(y(t|Y))$としたとき，y軸が左辺$dy(t)/dt$, x軸が右辺$f(x)$に対応する。ただし，左辺と右辺はモデルのスケールパラメータ$Y$でスケールしている。具体的には式\ref{xxx}をプロットしている。
%The ensemble median of all data was also given by Eq. \ref{xxx} and 10th and 90th percentiles shown in grey dashed lines are also almost straight lines given by Eq. \ref{xxx}.
}
\label{app_fig_comp_v}
\end{figure*}

First, we discuss the two-parameter models, which are special cases of the proposed model given by Eq. \ref{app_base_eq}.
% with three parameters. 
We check the following models.
%まず、提案モデルよりシンプルな２パラメータのモデルの説明力を確認する。　
%以下の３つのモデルを確認した。ロジスティック方程式
\begin{itemize}
    \item (a) The logistic equation: 
    \begin{equation}
        \frac{dy_j(t)}{dt}= r_j y_j(t)\left(1-y_j(t)/Y_j \right) \label{app_base_logi}, 
    \end{equation}
where $Y_j>0$. This is the basic logistic equation and the special case of the proposed model given by Eq. \ref{app_base_eq} for $\alpha=1$ and $Y<0$.  
\par
\item  (b) $Y$-sign extended logistic equation: 
\begin{equation}
    \frac{dy_j(t)}{dt}= r_j y_j(t)\left(1+y_j(t)/Y_j \right),  \label{app_base_logi2}
\end{equation}
where $Y_j \neq 0$. This equation extends the logistic equation to the carrying capacity $Y_j$ taking the positive and negative sign.
In the case that $Y_j<0$, the equation corresponds to the basic logistic equation given by Eq. \ref{app_base_logi}; and the special case of the proposed model given by Eq. \ref{app_base_eq} for $\alpha=1$. 
\par
\item (c) Single factor power-law model: 
\begin{equation}
    \frac{dy_j(t)}{dt}= r_j \cdot y_j(t)^{\alpha_j+1}. \label{app_base_pow}
\end{equation}
This equation is a simple power-law type differential equation, corresponding to the proposed model given by Eq. \ref{app_base_eq} for $Y \gg 1$.
\end{itemize}
Figs. \ref{app_fig_comp_v}(a)-(c) show a direct validation of the differential equations of above-mentioned models given by Eqs. \ref{app_base_logi}, \ref{app_base_logi2} and \ref{app_base_pow}. 
The x-axis corresponds to the right-hand side of the model, and the y-axis corresponds to the left-hand side. Thus, the closer the plotted curve is to the line $y=x$, as shown by the red dashed line, the better the correspondence with the models.
Here, to take the statistics regarding the words $j$, we use quantities normalized by the scale parameter $Y_j$, $p_j(t)= \frac{dy_j(t)}{dt}/|Y_j|$ in the y-axis and $q_j(t)=f(y_j(t)|\alpha_j,r_j,Y_j)/|Y_j|$ in the x-axis, where the differential equation is denoted as $\frac{dy}{dt}=f(y)$. If a model can explain the data, the graph remains linear after scale transformation. \par
% If $y=x$ holds, then $q=p$ also holds. \par 
%赤点線で示した$y=x$直線に近ければ微分方程式が現実のデータが近く，外れれていればあっていないといえる。　
%ただし，両辺が$y=x$の関係ならばx軸とy軸は同じ定数で割っても関係は維持するので，
%$Y$でスケールしている。$Y$はそれぞれのモデルでのパラメータを使う。べき乗モデルはパラメータ$Y$がないため，提案も出るの$Y$でスケールしている\。
\textcolor{black}{More specifically}, to make Fig. \ref{app_fig_comp_v}, we calculate the following values.
First, for the x-axis, the scaled right-hand side of the model is calculated as
\begin{equation}
p_j(t)= f(y_j(t)|\alpha_j,r_j,Y_j)/|Y_j|,  \label{app_eq_p}
\end{equation}
where the differential equation is denoted by $\frac{dy}{dt}=f(y|\alpha,r,Y)$ and the model parameters are $Y$, $r$, $\alpha$ are estimated by minimizing Eq.\ref{app_eq_esitmate}   in Appendix \ref{app_sec_estimate} . \par
% Here, we replace $G^{(-1)}(t|\alpha,r,Y,y_0)$ in Eq. \ref{app_eq_esitmate} with $F(t|\alpha,r,Y)$, where $F(t|\alpha,r,Y)$ is the solution of the differential equation. . \par
Second, for the y-axis, we approximate the derivative on the left-hand side of the models using the difference, 
%$f^{(model)}(y_i(t)|...)$ はモデルの右辺に対応する。例えば，$f^{logisti}(y_i(t)|...)$ は，式\ref{app_base_logi}の右辺に対応する。 
%パラメータ$N$, $r$, $\alpha$ は，セクション \ref{xxx}の方法でデータ$y_i(t)$の誤差最小化から推定したものを利用する。
\begin{eqnarray}
&&q_j(t)=  \frac{1}{|Y_j|} \cdot \nonumber \\
&&(MovingMedian_3(y_j(t+1))- MovingMedian_3(y_j(t))), \nonumber \\ \label{app_eq_q}
\end{eqnarray}
where the moving median of the three points is used to remove noise from the time series. \par
%この量について，xごとに条件付きの中央値をとり$p_i$をx軸に$q_i$をy軸にプロットしている。
    %それを，式\ref{medi_th_alpha}と同様に単語の中央値をとりプロットした。
%    and in the same way as Eq. \ref{medi_th_alpha}, we plot the ensemble median conditioned on the parameters $\alpha$ and the sign of $Y$:
%これらの量をもとに時間と単語についての中央値をとった量をプロットしている。なお，プロットの色は，パラメータによる条件づけである。パラメータは，テイアンモデルのパラメータを利用している。
Finally, to create a graph, we introduce an ensemble median for $t$ and $j$ conditioned on $\alpha^{0}_j$ and $s^{(0)}_j$,  
\begin{eqnarray}
  &&  \hat{q}(p|\alpha^{(0)}_j,s^{(0)})= \nonumber \\
   && Median_{\{j,t|, d^{-}_p < p_j(t) < d^{+}_p,  |\alpha^{(0)}_j-\alpha| < d_\alpha, sign(Y_j)=s^{(0)}_j \}}[q_j'(t))],  \nonumber \\
   \label{app_eq_medi}
\end{eqnarray}
where $d_\alpha=0.5$ for $Y>0$ or $d_\alpha=\infty$ for $Y<0$ is the box size of $\alpha$ to obtain the statistics, $d^{+}_p=p(\exp(0.1))$, $d^{-}_p=p(\exp(-0.1))$ is the box size of $p$, $\alpha^{0}_j$, $s^{(0)}_j=sign(Y^{(0)}_j)$, $Y^{(0)}_j$ are the parameters of $j$-th word of the proposed model given by Eq. \ref{app_base_eq}. For this calculation, in the case of the single factor power-law model given by Eq. \ref{app_base_pow}, we use $Y_j=Y^{(0)}_j$ for scaling in Eq. \ref{app_eq_p} and \ref{app_eq_q}  because the model does not have the scale parameter $Y_j$. \par
In Fig. \ref{app_fig_comp_v}, we plot $\hat{q}(p|\alpha^{(0)}_j,s^{(0)})$ as a function of $p$, with $p$ on the x-axis and $\hat{q}$ on the y-axis. 
The black triangle is the statistic for all data, whereas the other colors and shapes are medians conditioned on $\alpha^{(0)}$ or $Y^{(0)}$. \par
%$d^{+}_p=p(\exp(1/100)-1)$, $d^{-}_p=p(1-\exp(-1/100))$ 
%この量について，$p_i$をx軸に$q_i$をy軸にプロットしたの図\ref{xxx}である。\par%図\ref{xxx}よりロジスティックモデルの\ref{xxx}(b)に示した符号拡張ロジスティックモデルでは、yが小さいときはあっているが、
%後半ではあっていことがわかる。これは，これらのモデルが線形などのべき乗的な漸近を表せないことによる。
For the logistic equations given in Eq. \ref{app_base_logi} shown in Fig. \ref{app_fig_comp_v}(a) and the sign-extended logistic equation given by Eq. \ref{app_base_logi2}  shown in Fig. \ref{app_fig_comp_v}(b), for small $p$, the data are almost consistent with the theoretical line $\hat{q}=p$, but for large $p$ they are not. 
The reason for the disagreement with theory is that these models cannot represent the growth functions of the powers of $t$ such as $y_j(t) \propto t$ and $y_j(t) \propto t^2$. \par
Conversely, in the single factor power-law model given by Eq. \ref{app_base_pow}, for the large $p$, the statistics for the whole data, indicated by the black triangle, correspond to the theoretical line $\hat{q}=p$ (except for the largest outliers), but not for the small $p$. 
Furthermore, the figure also confirms that the model cannot be explained by the growth curve for $Y^{(0)}<0$ (i.e., S-shaped or logistic-like curve) indicated by the peach crosses. 
The reason for the disagreement with theory is that the single factor power-law model cannot represent the growth curve, which asymptotically converges to a constant value. $y_j(t) =const$ for $t  \gg 1$. 
\par 
%さらに，桃色で示した$Y^{(0)}<0$のロジスティックカーブ型のS字型がで説明でてないこともわかる。 \par
The proposed model given by Eq. \ref{base_eq} or \ref{app_base_eq} is a combination of the (sign-extended) logistic equation (Eq. \ref{app_base_logi2}) and the single factor power-law model (Eq. \ref{app_base_pow}).  
For $Y>0$ and $y_j(t) \gg |Y|$, the model can be approximated as the single factor power-law model, and where $y(t) \ll |Y|$ is small or $Y<0$, it behaves like the logistic equation.  \par 
\subsection{Three parmeter model I (Linear extension)}
Next, we examine the three-parameter models. First, we check the Bass model and a model with a constant term, which are typical extensions of the logistic equation:
\begin{itemize}
    \item (d) Bass model ($Y$-sign extended): 
    \begin{equation}
        \frac{dy_j(t)}{dt}= (r_jy_j(t)+\alpha_j)\left(1+y_j(t)/Y_j \right), \label{app_base_bass}
    \end{equation}
    where $\alpha_j>0$ and $Y_j \neq 0$.
    \item (e) Constant term model: 
    \begin{equation}
    \frac{dy_j(t)}{dt}= r_j y_j(t)\left(1+y_j(t)/Y_j \right)+\alpha_j, \label{app_base_const}
    \end{equation}
    where $\alpha_j>0$  and $Y_j \neq 0$.
%ロジスティックモデルを線形に拡張した以下の３パラータのバズモデルと定数項モデルもよく使用されるが，
\end{itemize}
%The black triangles in Fig. \ref{app_fig_comp_v} are not straight lines, and the reduced prediction accuracy shown in table \ref{app_table_pred} confirm that these extensions are not effective in explaining and predicting the data.
\textcolor{black}{The points in Figs. \ref{app_fig_comp_v}(d) and (e) are not straight lines.
Thus, we can confirm that these extensions are not effective in explaining the data.}
%, and the reduced prediction accuracy shown in table \ref{app_table_pred} confirm that these extensions are not effective in explaining and predicting the data.
%両方とも図が\ref{xxx} より，直線とはことなりあっていないことがわかる。また，予測精度も提案モデルより下がっている。
%さらにロジスティック方程式より下がっているため，オーバーフィッティングしている可能性もあり，有効な拡張でないことがわかる。
\subsection{Three parmeter model II (Power-law extension)}
We discuss the three-parameter model with the power-law factor,  which is a special case of the ($Y$-sign extended) Blumberg equation or the ($Y$-sign extended) generalized logistic equation \cite{wu2020generalized}.     
Note that, while the original models are for $Y_j<0$, in this study, it is extended to $Y_j \neq 0$.
\begin{itemize}
\item (f) First factor power-law model:
\begin{equation}
    \frac{dy_j(t)}{dt}= r_j y_j(t)^{\alpha_j}\left(1+y_j(t)/Y_j \right), \label{app_base_first}
\end{equation}
where $Y_j \neq 0$ and $\alpha_j$ take both negative and positive real numbers. We call this model a ``First factor power-law model'' because the model has the power term in the first factor.
\item (g) Second factor power-law model (Proposed model):
\begin{equation}
    \frac{dy_j(t)}{dt}= r_j y_j(t) \left( 1+y_j(t)/Y_j \right)^{\alpha_j}, \label{app_base_second}
\end{equation}
where $Y_j \neq 0$ and $\alpha_j$ take both negative and positive real numbers. We call this model a ``Second factor power-law model'' or ``Proposed model'' because  the model has the power term in the second factor and corresponds proposed model given by Eq. \ref{app_base_eq}.
\item (h) Inside power-law model ($Y$-sign extended Richards' Equation):
\begin{equation}
    \frac{dy_j(t)}{dt}= r_j y_j(t)\left( 1+sign(Y_j)\cdot (y_j(t)/|Y_j|)^{\alpha_j} \right) \label{app_base_inside}
\end{equation}
where $Y_j \neq 0$ and $\alpha_j$ takes both negative and positive real number. We call this model as ``Inside power-law model'' because the model has the power term inside the second factor.
\end{itemize}
From Figs. \ref{app_fig_comp_v} (f) and (g) for the first factor power-law model given by Eq. \ref{app_base_first} and the second factor power-law model given by Eq. \ref{app_base_second},  the data agree well with the theoretical straight line $p=q$. The results imply that these models are not inconsistent with real growth dynamics.  \par
We were unable to provide conclusive evidence of the difference between the first factor power-law model (f) and the second factor power-law model (g) in our data analysis. 
%そこで，以下の理由でほんけんきゅうではSecond factor power-model given by Eq. \ref{xxx} を採用した。
Therefore, we adopted the second factor power-law model, given by Eq. \ref{app_base_second} as the proposed model for the following reasons:
\begin{itemize}
\item The second factor power-law model (the proposed model) had better predictive ability than the first factor power-law model given in table \ref{app_table_pred}.
\item  The proposed model is consistent with the data analysis results of the two-parameter models, and the relationship with those two-parameter models is easy to interpret (see Appendix in section \ref{app_sec_two}.
%in sec. \ref{app_sec_model}).
\end{itemize}
%Note that Fig. \ref{app_fig_comp_v}, the first factor power-law model given by Eq. \ref{app_base_first} does not match the theoretical line $q=p$ for the small $p$, which may inherit a weakness of the single factor power-law model given by Eq. \ref{app_base_pow} discussed in Appendix in the subsection \ref{app_sec_two}.
% in sec. \ref{app_sec_model}.
%データ解析では，この２つモデルの良さの差異を明確に評価する方法はみつけられなかった。 
% \par 
%これは，べき乗モデルと
%振る舞いが似ており，最初の比例的な速度を説明できてないことが原因と予想される。ただし，
%そこまで大きくないため、この図やその他の図や説明力でもほぼ提案モデルと同等であるとも
%思われる。　今研究では，予測能力が提案モデルのほうが高いこと，２パラメータモデルとの
%整合性により，後ろべきモデルを採用した。　ただし、この２つのモデルの良し悪さの明確な
%違いは示せなかった。 \par
% \par
From Fig. \ref{app_fig_comp_v}(f) we can see that the inside power-law model given by Eq. \ref{app_base_inside} deviates from the theoretical line. One reason may be that this model cannot explain the growth curve of the power asymptote for $\alpha^{(0)}<0$, such as $y(t) \propto t$.
\subsection{Confirmation of the sign effect of the proposed model}
%最後に提案モデルの符号を限定した拡張を確認した。$Y>0$に限定した場合は，速度の比較では
%そこまで違いはでなかった。ただし，予測能力は，テイアンモデルより下がった。 $Y<0$では，
%べき乗的な漸近を説明できなために大きいほうでずれている。 \par
Finally, we verified the sign-restricted version of the proposed model, 
\begin{itemize}
\item (i) $Y$-sign positive-restricted second factor power-law model (Positive model):
\begin{equation}
    \frac{dy_j(t)}{dt}= r_j y_j(t) \left( 1+y_j(t)/Y_j \right)^{\alpha_j}, \label{app_base_posi}
\end{equation}
where $Y_j>0$.  This model corresponds to the proposed model (Eq. \ref{app_base_second}) restricted to $Y_j>0$.
\item (j) $Y$-sign negative-restricted second factor power-law model (Negative model):
\begin{equation}
    \frac{dy_j(t)}{dt}= r_j y_j(t) \left( 1+y_j(t)/Y_j \right)^{\alpha_j} , \label{app_base_nega}
\end{equation}
where $Y_j<0$. This model corresponds to the proposed model (Eq. \ref{app_base_second}) restricted to $Y_j<0$.
\end{itemize}
In Fig. \ref{app_fig_comp_v} (i),  for the model restricted to $Y_j>0$ given by Eq. \ref{app_base_posi}, the curve is linear and does not clearly differ from the proposed model. 
%However, its predictive ability is lower than that of the proposed model given in table \ref{app_table_pred}.
However, Fig. \ref{app_fig_comp_ts}(i) also indicates that this model does not explain the actual data better than the proposed model (The detail is mentioned in the section \ref{app_sec_medi}).  
When restricted to $Y<0$, as expressed by Eq. \ref{app_base_nega}, the model shows that it cannot explain the data (see Fig. \ref{app_fig_comp_v}(j)). This is because the model can only represent the growth curves of the S-shape type. \par 
These results imply that the proposed model can explain the growth curve relatively well, even when restricted to $Y>0$ (but does not explain the S-shape type curves). 
\subsection{Comparisons using other statistics}
\label{app_sec_medi}
\begin{figure*}
    \begin{minipage}{0.19\hsize}
        \begin{overpic}[width=3.8cm]{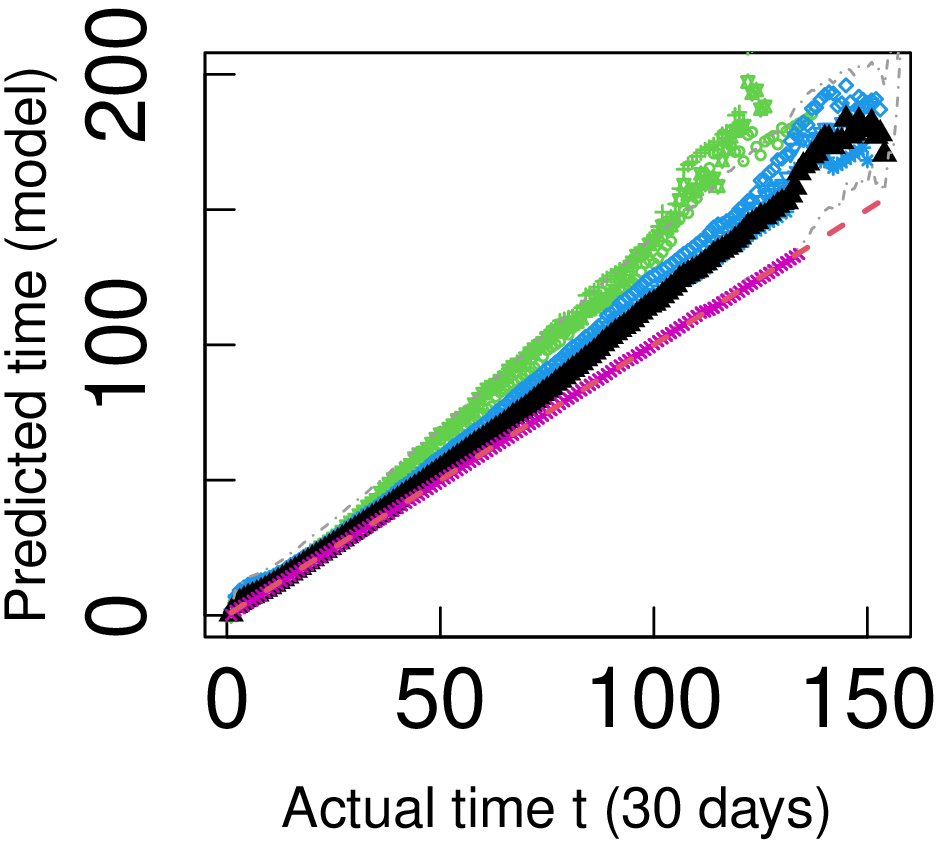}
            %\put(22,55){(a)}
            \put(23,72){(a)}
        \end{overpic}
    \end{minipage}
    \begin{minipage}{0.19\hsize}
        \begin{overpic}[width=3.8cm]{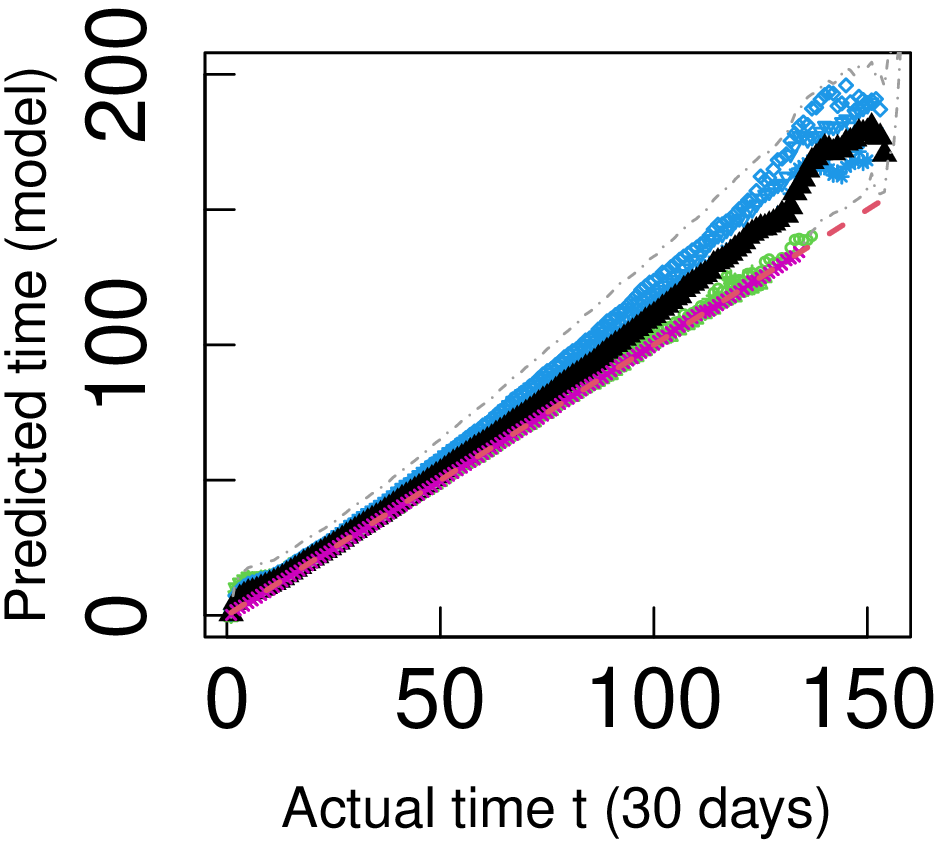}
            %\put(22,55){(a)}
            \put(23,72){(b)}
        \end{overpic}
    \end{minipage}
    \begin{minipage}{0.19\hsize}
        \begin{overpic}[width=3.8cm]{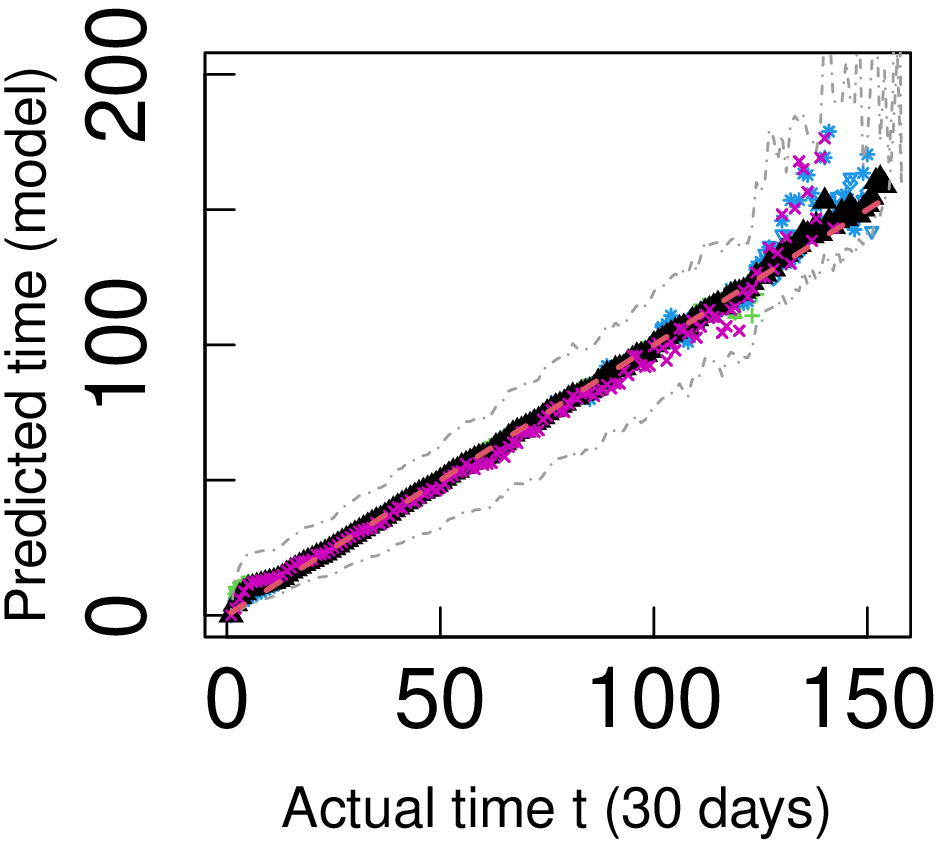}
            %\put(22,55){(a)}
            \put(23,72){(c)}
        \end{overpic}
    \end{minipage}
    \begin{minipage}{0.19\hsize}
        \begin{overpic}[width=3.8cm]{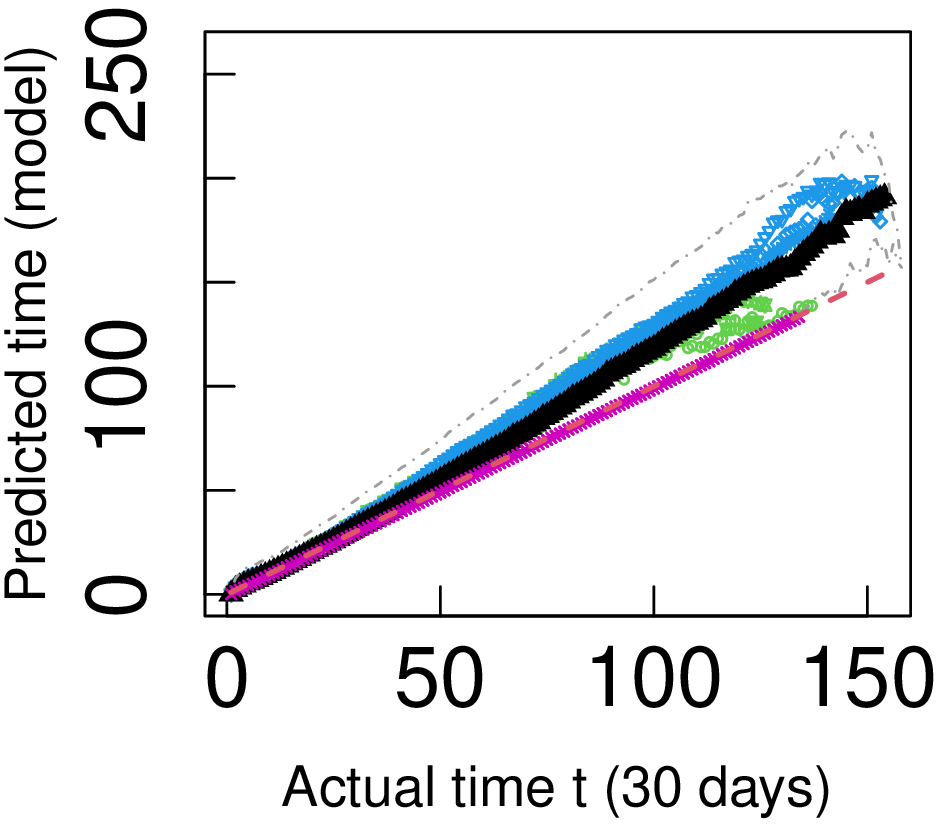}
            %\put(22,55){(a)}
            \put(23,72){(d)}
        \end{overpic}
    \end{minipage}
    \begin{minipage}{0.19\hsize}
        \begin{overpic}[width=3.8cm]{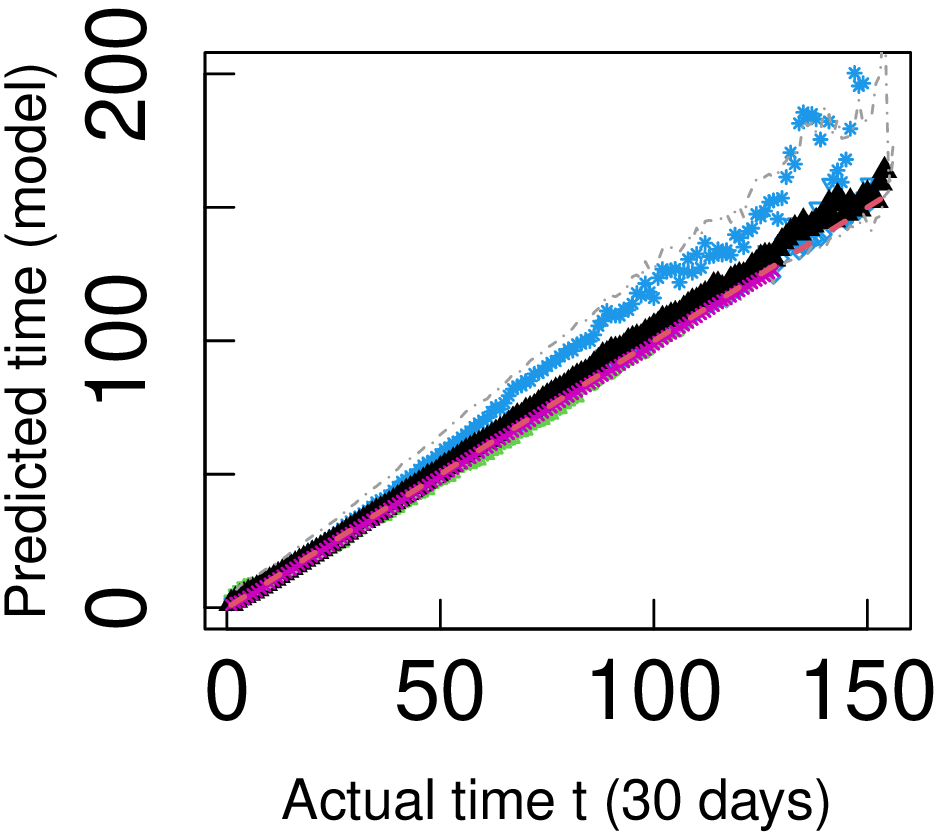}
            %\put(22,55){(a)}
            \put(23,72){(e)}
        \end{overpic}
    \end{minipage} 
    \begin{minipage}{0.19\hsize}
        \begin{overpic}[width=3.8cm]{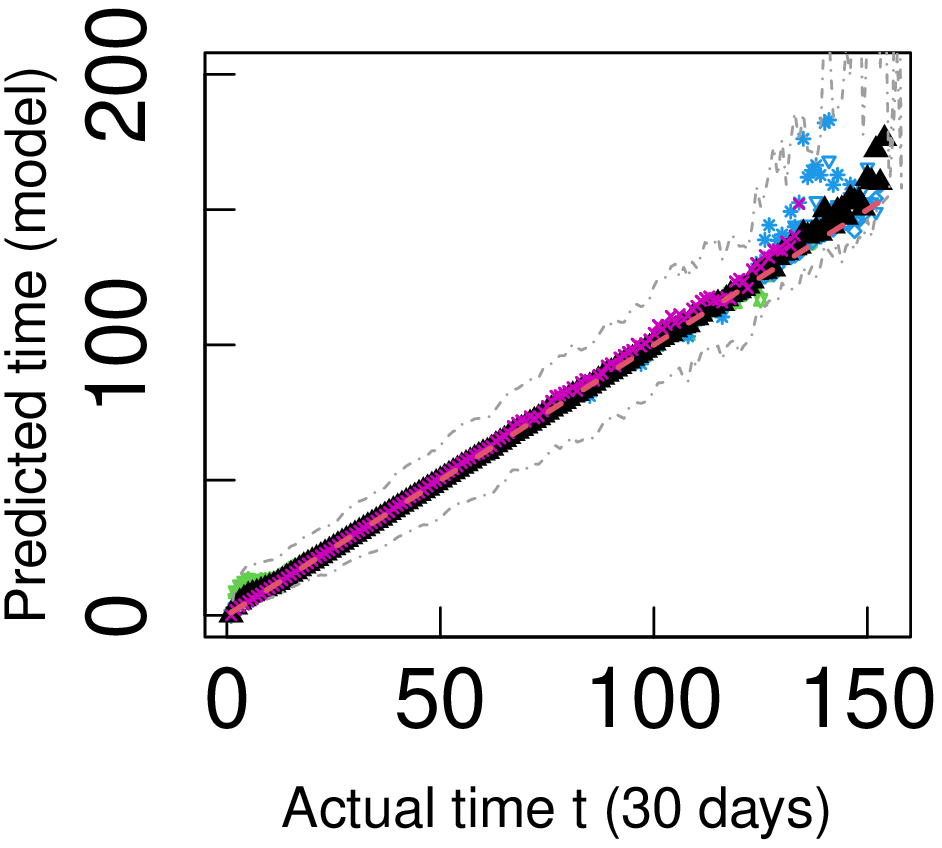}
            %\put(22,55){(a)}
            \put(23,72){(f)}
        \end{overpic}
    \end{minipage}
    \begin{minipage}{0.19\hsize}
        %\includegraphics[width=3.8cm]{transformed_count_ave_t_th.eps}
        %\begin{overpic}[width=3.8cm]{transformed_count_ave_t_th.eps}
            \begin{overpic}[width=3.8cm]{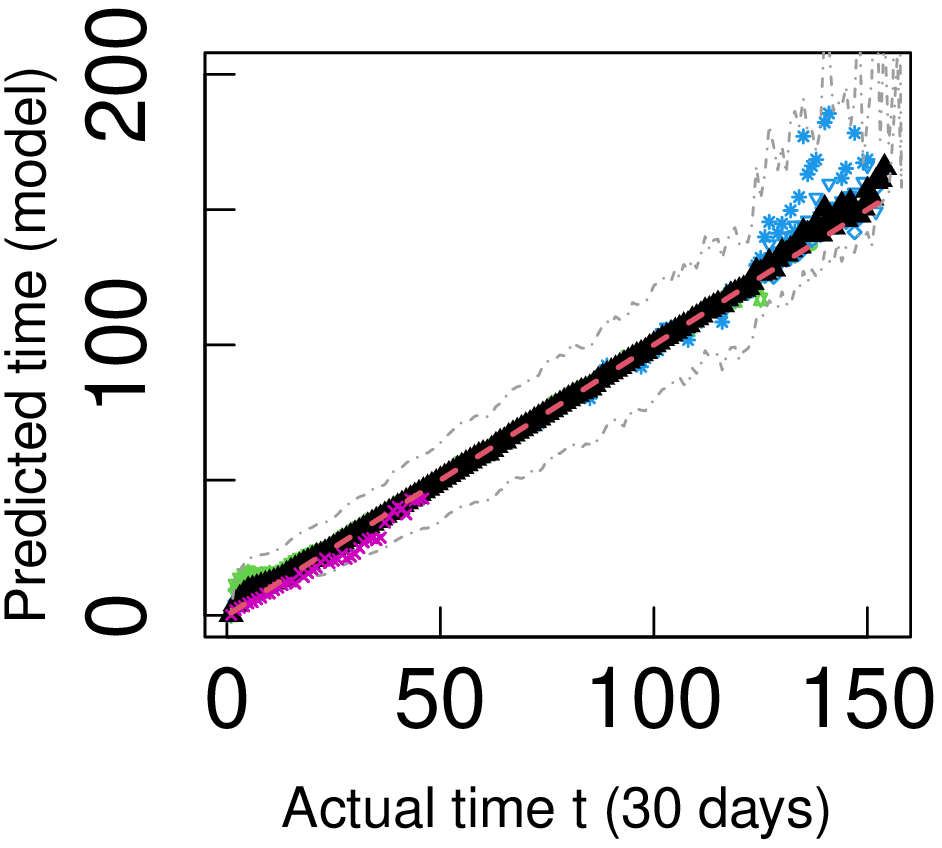}
            %\put(22,55){(a)}
            \put(23,72){(g)}
        \end{overpic}
    \end{minipage}
    \begin{minipage}{0.19\hsize}
        \begin{overpic}[width=3.8cm]{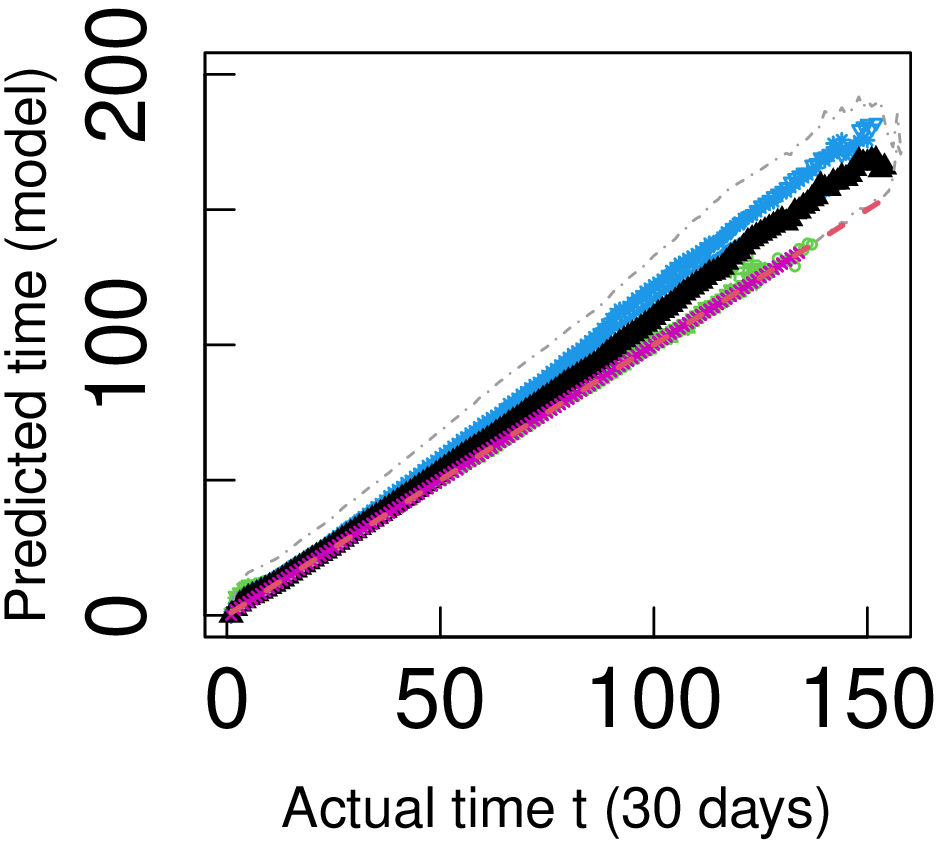}
            %\put(22,55){(a)}
            \put(23,72){(h)}
        \end{overpic}
    \end{minipage}
    \begin{minipage}{0.19\hsize}
        \begin{overpic}[width=3.8cm]{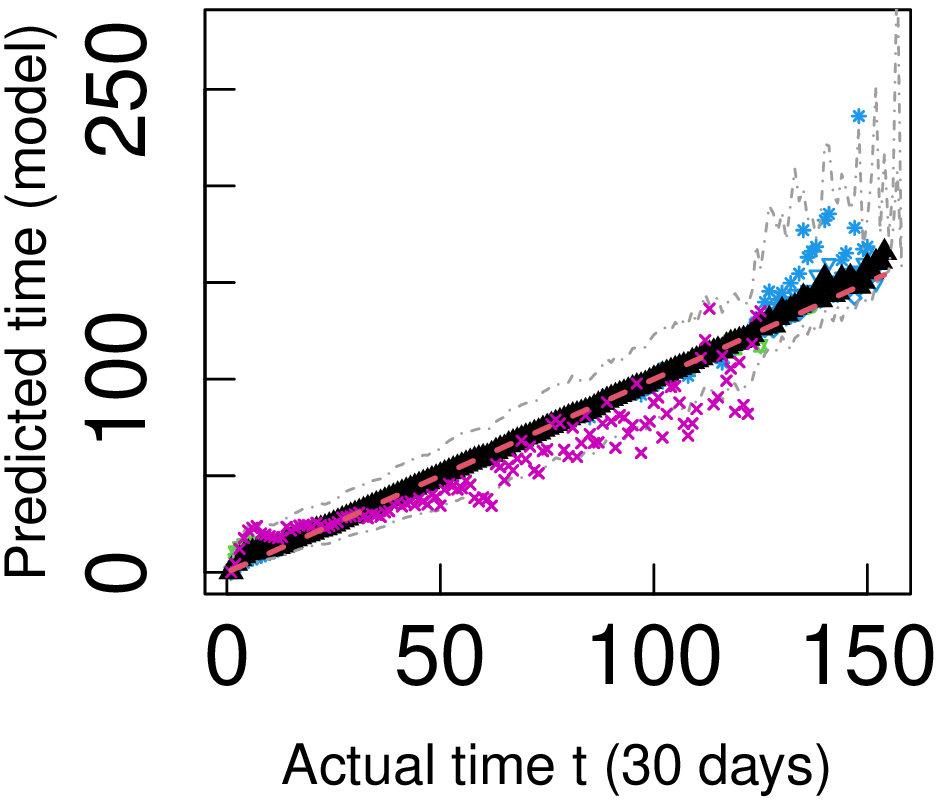}
            %\put(22,55){(a)}
            \put(23,72){(i)}
        \end{overpic}
    \end{minipage}
    \begin{minipage}{0.19\hsize}
        \begin{overpic}[width=3.8cm]{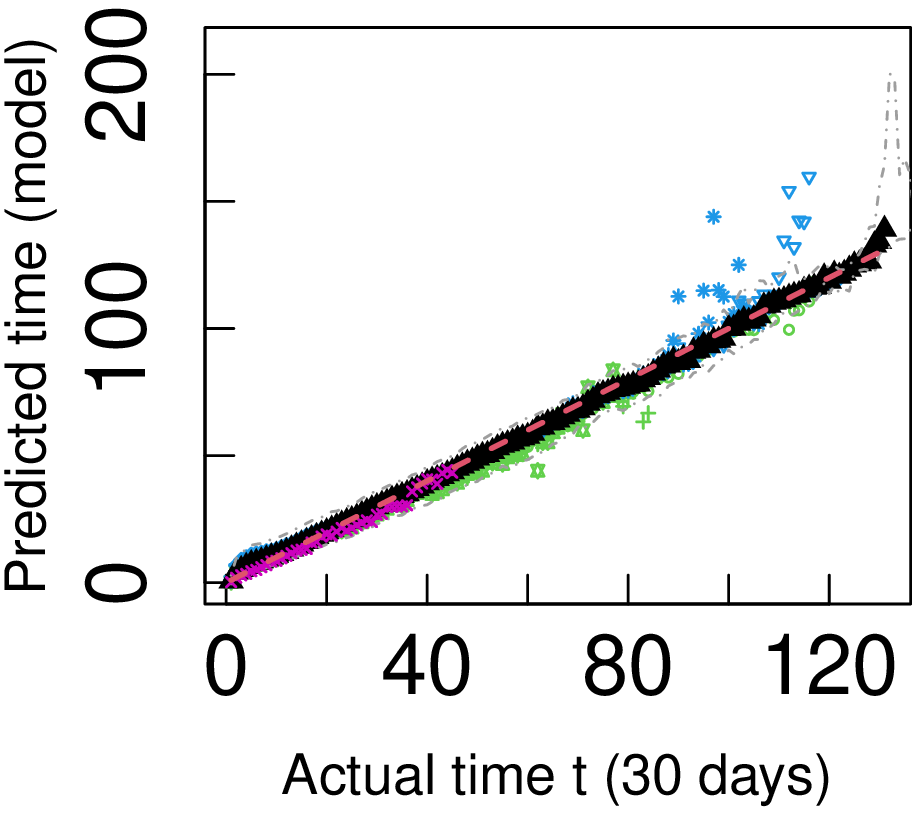}
            %\put(22,55){(a)}
            \put(23,72){(j)}
        \end{overpic}
    \end{minipage}    
    \caption{
        Comparison of the descriptiveness of the dynamics of word count time series in various models using the time description (Inverse functions of solution of differential equation, $t=F^{(-1)}(y|\alpha,Y)$). If the model explains the data well, the graph will be close to the $y=x$ line shown by the red dashed line. 
        The y-axis corresponds to the left-hand side $F^{(-1)}(y_j(t)|\alpha_j,Y_j)$ and the x-axis to the right-hand side $t$, where the differential equation is written as $\frac{dy(t)}{dt}=f(y|\alpha,Y)$ and the its solution as $y(t)=F(t|\alpha,Y)$ and its inverse function as $t=F^{(-1)}(y|\alpha,Y)$.
        Specifically, we plot the median quantity with respect to words given by Eq. \ref{app_eq_medi} using Eq. \ref{app_eq_p_t} (See also the subsection \ref{app_sec_medi}). 
        %scaled by the model's scale parameter $Y_j$ with respect to words given by Eq. \ref{xxx}.
        The ensemble median over all data is plotted in black triangles and also the ensemble median grouped by proposed model's  (Eq. \ref{app_base_eq}) parameters $\alpha_j^{(0)}$ and $Y_j^{(0)}$ is plotted.
        In case of $Y_j^{(0)}>0$, we plott in green circles for $\alpha_j^{(0)}=0.5$, green pluses for $\alpha_j^{(0)}=1.0$, for green triangles up and down $\alpha_j^{(0)}=1.5$, blue diamonds for $\alpha_j^{(0)}=-0.5$, blue triangles point down for $\alpha_j^{(0)}=-1.0$ and in the case of $Y_j^{(0)}<0$ magenta crosses. 
    (a) Logistic equation given by Eq. \ref{app_base_logi},
    (b) $Y$-sign extended logistic equation given by Eq. \ref{app_base_logi2},
    (c) Single factor power-law model given by Eq. \ref{app_base_pow},
    (d) ($Y$-sign extended) Bass model given by Eq. \ref{app_base_bass},
    (e) Constant model given by Eq. \ref{app_base_const},
    (f) First factor power-law model given by Eq. \ref{app_base_first},
    (g) Second factor power-law model (Proposed model) given by Eq. \ref{app_base_second},
    (h) Inside power-law model given by Eq. \ref{app_base_inside}, 
    (i) $Y$-sign positive-restricted second factor power-law model given by Eq. \ref{app_base_posi} and
    (j) $Y$-sign negative-restricted second factor power-law model given by Eq. \ref{app_base_nega}.
    }
    \label{app_fig_comp_t}
\end{figure*}

\begin{figure*}
    \begin{minipage}{0.19\hsize}
        \begin{overpic}[width=3.8cm]{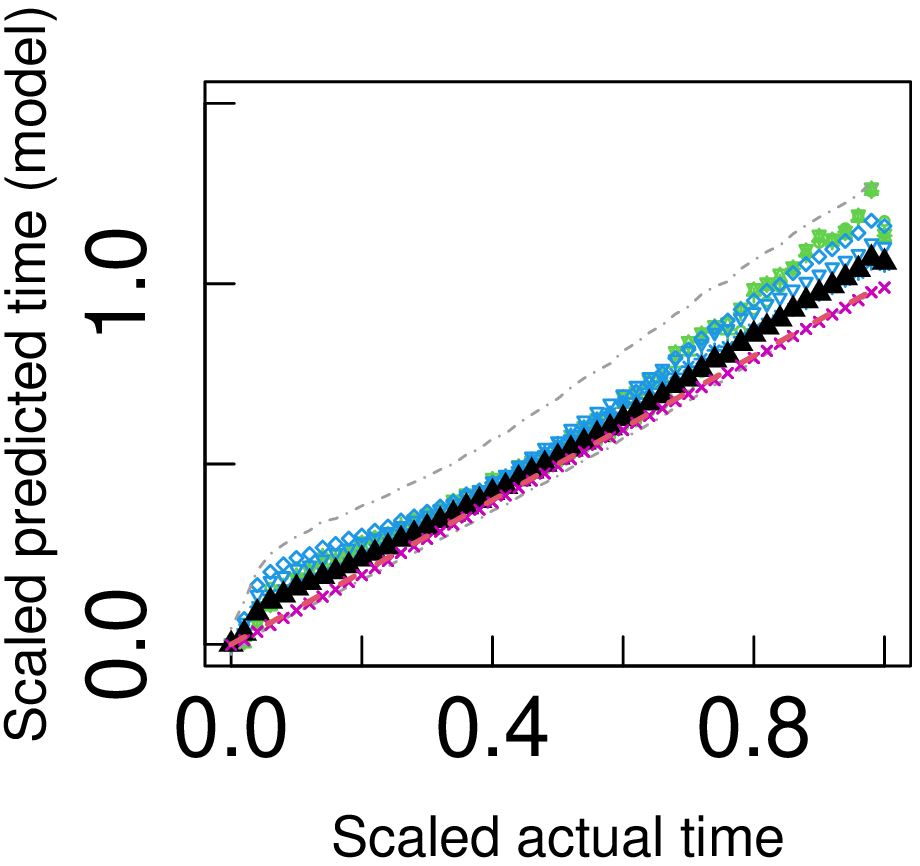}
            %\put(22,55){(a)}
            \put(23,72){(a)}
        \end{overpic}
    \end{minipage}
    \begin{minipage}{0.19\hsize}
        \begin{overpic}[width=3.8cm]{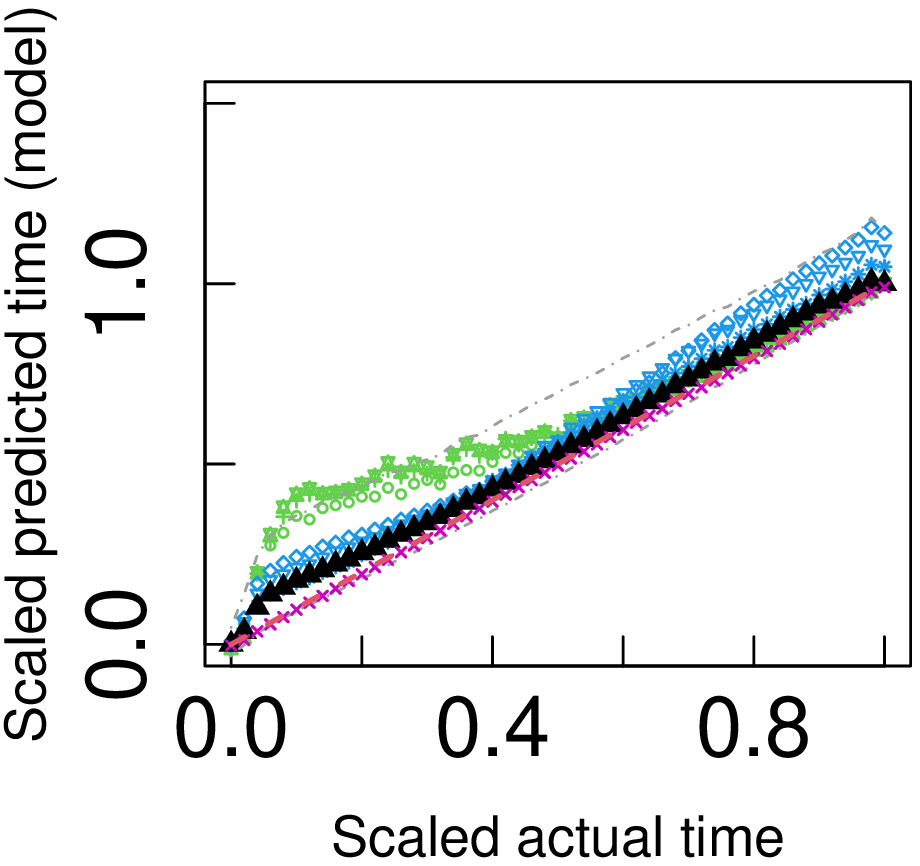}
            %\put(22,55){(a)}
            \put(23,72){(b)}
        \end{overpic}
    \end{minipage}
    \begin{minipage}{0.19\hsize}
    \begin{overpic}[width=3.8cm]{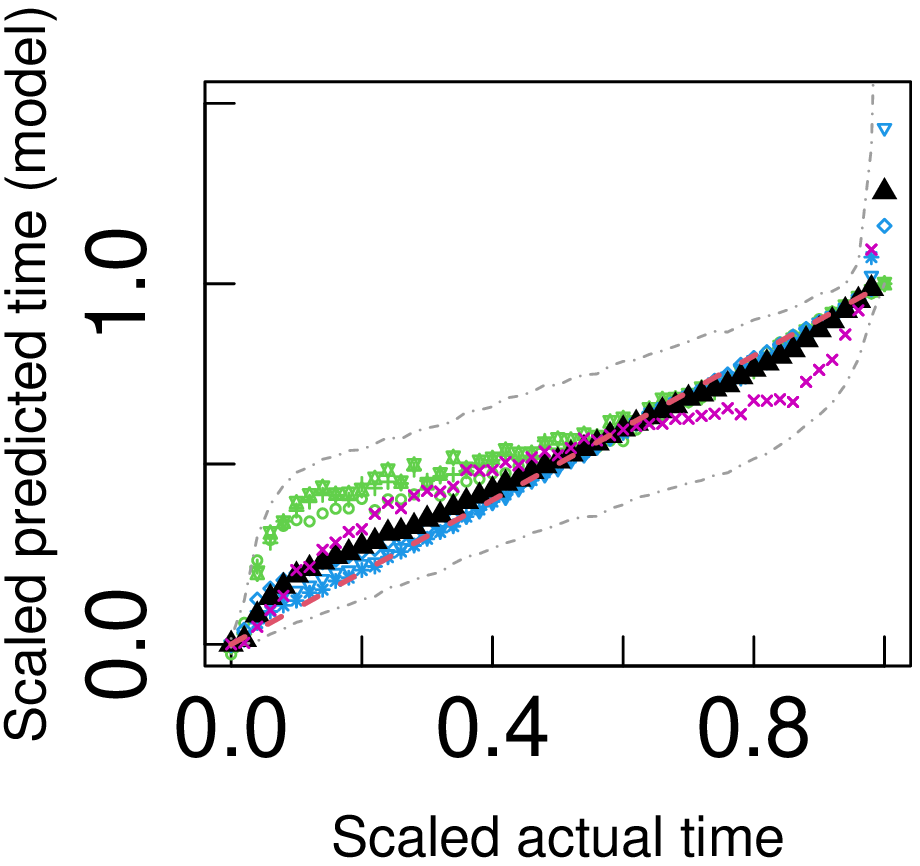}
        %\put(22,55){(a)}
        \put(23,72){(c)}
    \end{overpic}
    \end{minipage}
    \begin{minipage}{0.19\hsize}
    \begin{overpic}[width=3.8cm]{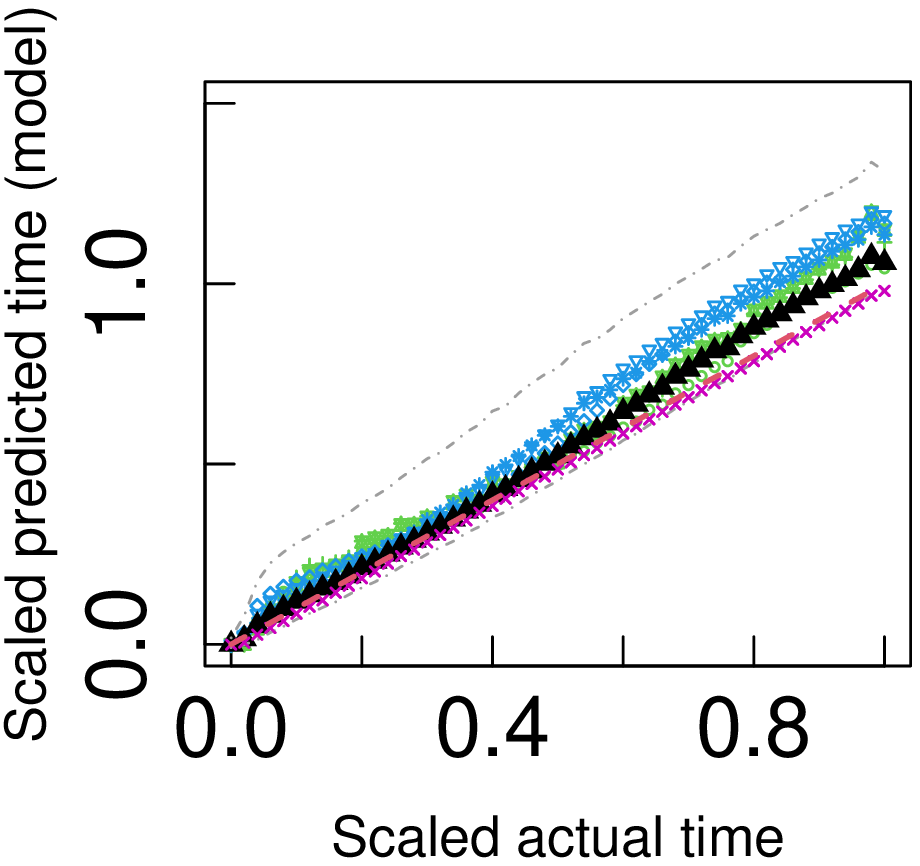}
        %\put(22,55){(a)}
        \put(23,72){(d)}
    \end{overpic}
    \end{minipage}
    \begin{minipage}{0.19\hsize}
    \begin{overpic}[width=3.8cm]{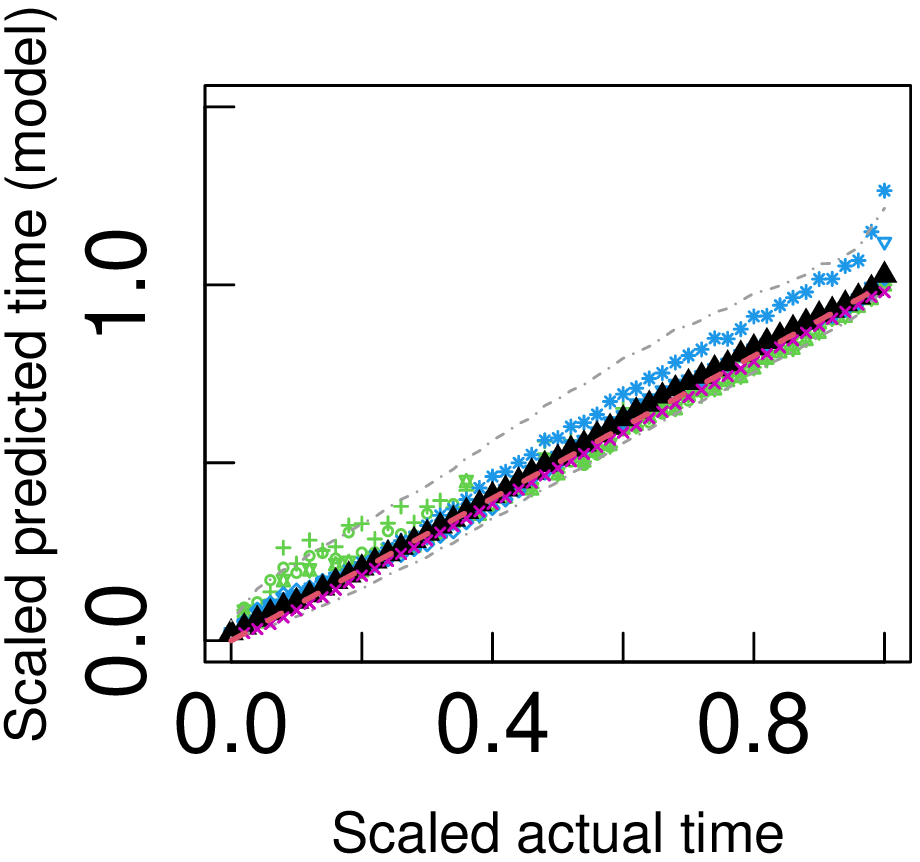}
        %\put(22,55){(a)}
        \put(23,72){(e)}
    \end{overpic}
    \end{minipage}
    \begin{minipage}{0.19\hsize}
        \begin{overpic}[width=3.8cm]{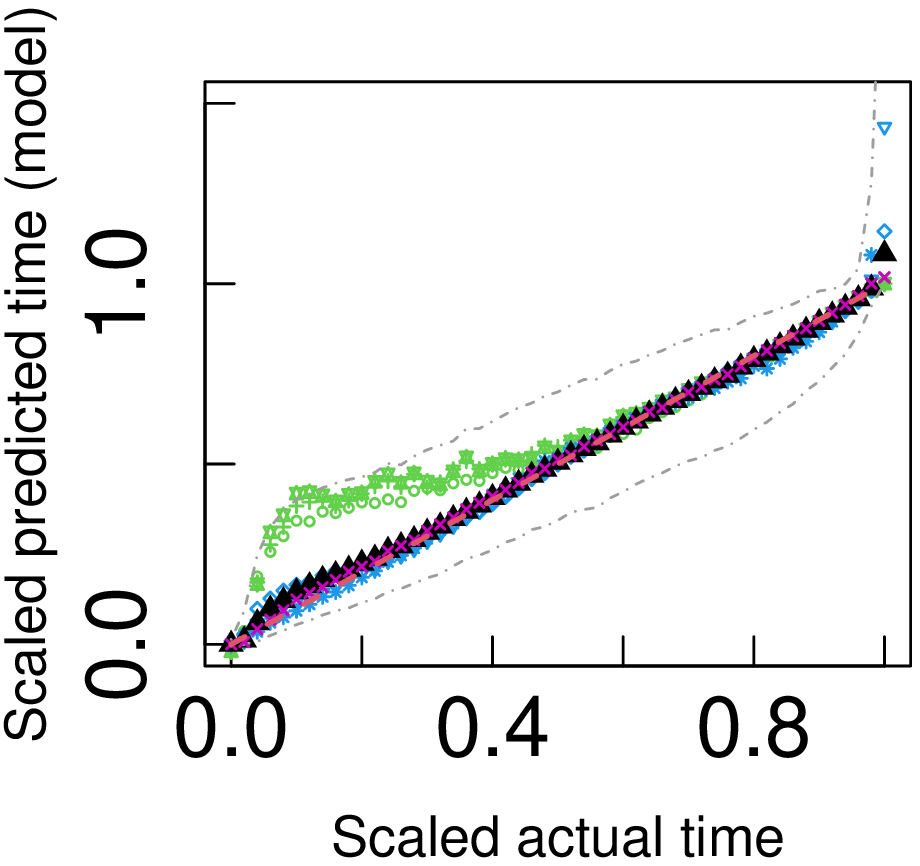}
            %\put(22,55){(a)}
            \put(23,72){(f)}
        \end{overpic}
    \end{minipage}
    \begin{minipage}{0.19\hsize}
        %\includegraphics[width=3.8cm]{transformed_count_ave_t_thb.eps}
        %\begin{overpic}[width=3.8cm]{transformed_count_ave_t_thb.eps}
        \begin{overpic}[width=3.8cm]{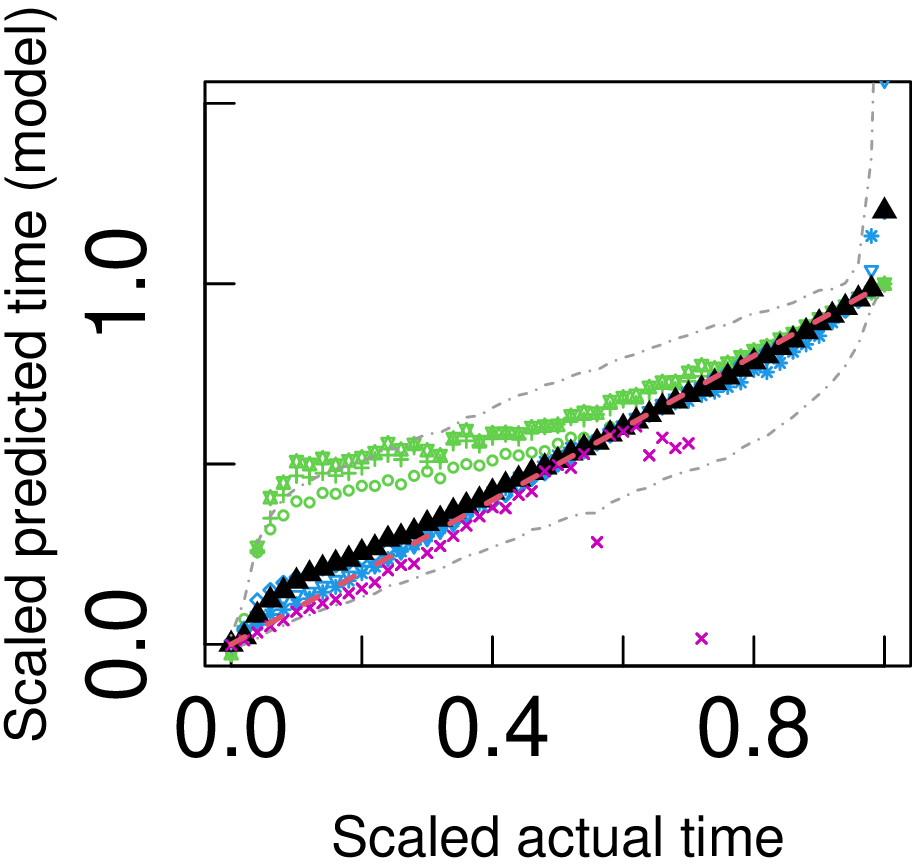}
            %\put(22,55){(a)}
            \put(23,72){(g)}
        \end{overpic}
    \end{minipage}
    \begin{minipage}{0.19\hsize}
        \begin{overpic}[width=3.8cm]{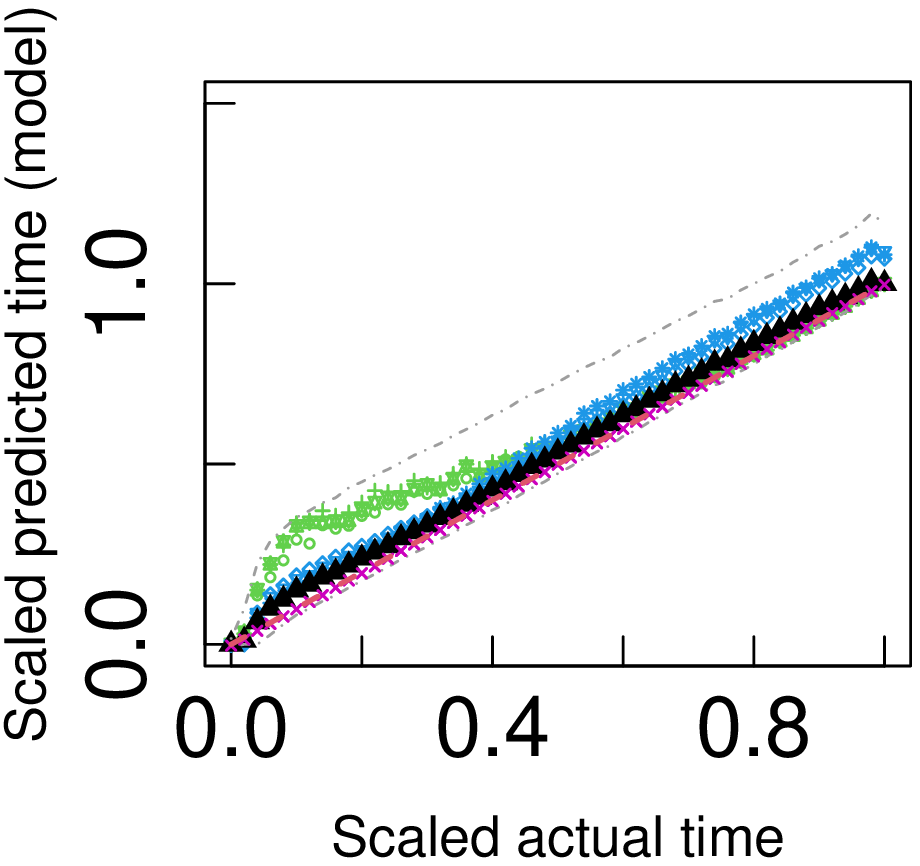}
            %\put(22,55){(a)}
            \put(23,72){(h)}
        \end{overpic}
    \end{minipage}
    \begin{minipage}{0.19\hsize}
        \begin{overpic}[width=3.8cm]{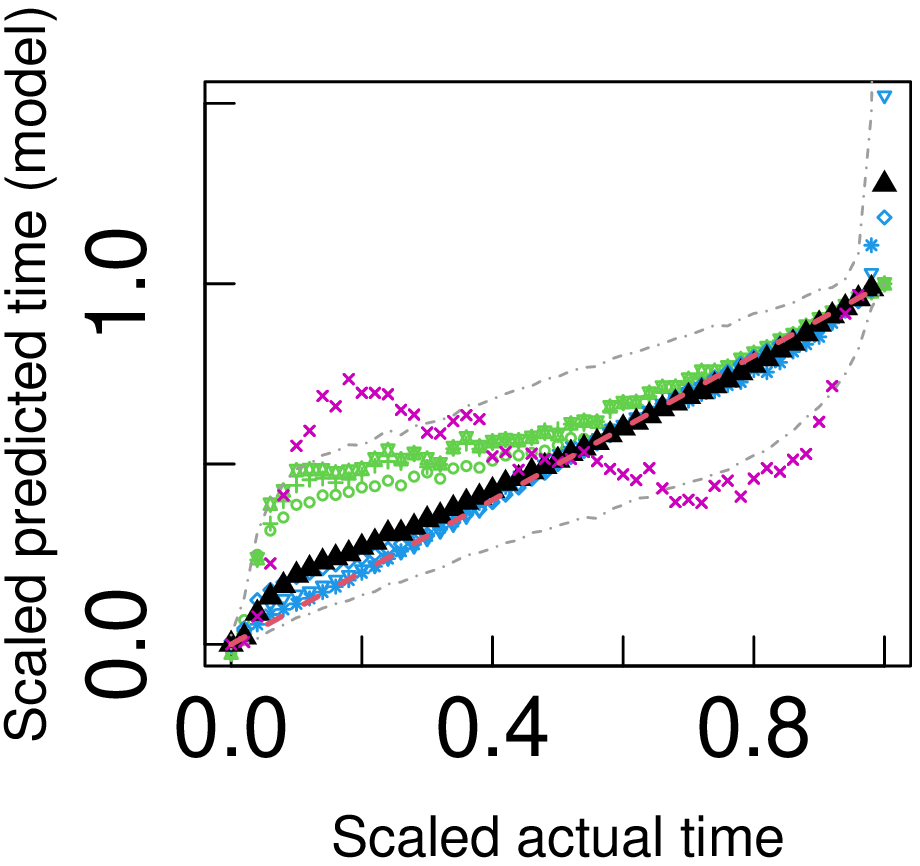}
            %\put(22,55){(a)}
            \put(23,72){(i)}
        \end{overpic}
        \end{minipage}
    \begin{minipage}{0.19\hsize}
    \begin{overpic}[width=3.8cm]{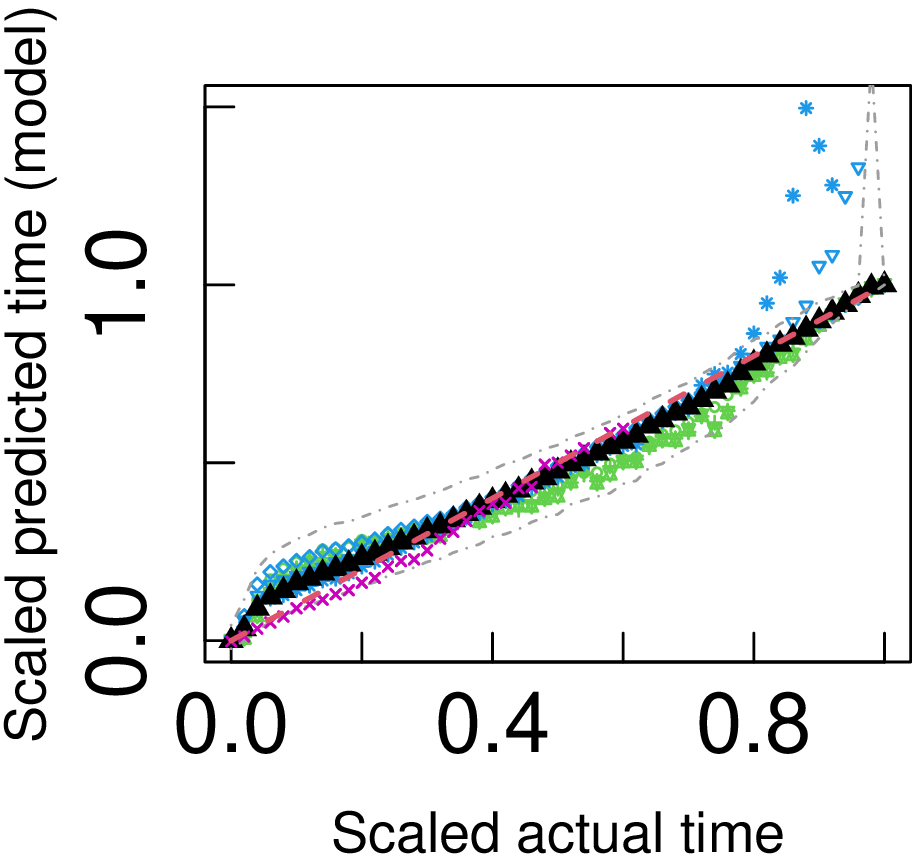}
        %\put(22,55){(a)}
        \put(23,72){(j)}
    \end{overpic}
    \end{minipage}
    \caption{
        Comparison of the descriptiveness of the dynamics of word count time series in various models using the scaled inverse functions of solution of differential equation, $F^{(-1)}(t|\alpha_j,Y_j)/T_j$.
        This figure is corresponding figure of Fig. \ref{app_fig_comp_t} for the time scale normalised to 1 for the entire growth period (see Eq. \ref{app_eq_q_ts}). 
    (a) Logistic equation given by Eq. \ref{app_base_logi},
    (b) $Y$-sign extended logistic equation given by Eq. \ref{app_base_logi2},
    (c) Single factor power-law model given by Eq. \ref{app_base_pow},
    (d) ($Y$-sign extended) Bass model given by Eq. \ref{app_base_bass},
    (e) Constant model given by Eq. \ref{app_base_const},
    (f) First factor power-law model given by Eq. \ref{app_base_first},
    (g) Second factor power-law model (Proposed model) given by Eq. \ref{app_base_second},
    (h) Inside power-law model given by Eq. \ref{app_base_inside},
    (i) $Y$-sign positive-restricted second factor power-law model given by Eq. \ref{app_base_posi} and
    (j) $Y$-sign negative-restricted second factor power-law model given by Eq. \ref{app_base_nega}.
        }
    \label{app_fig_comp_ts}
    \end{figure*}

\begin{figure*}
        \begin{minipage}{0.19\hsize}
            \begin{overpic}[width=3.8cm]{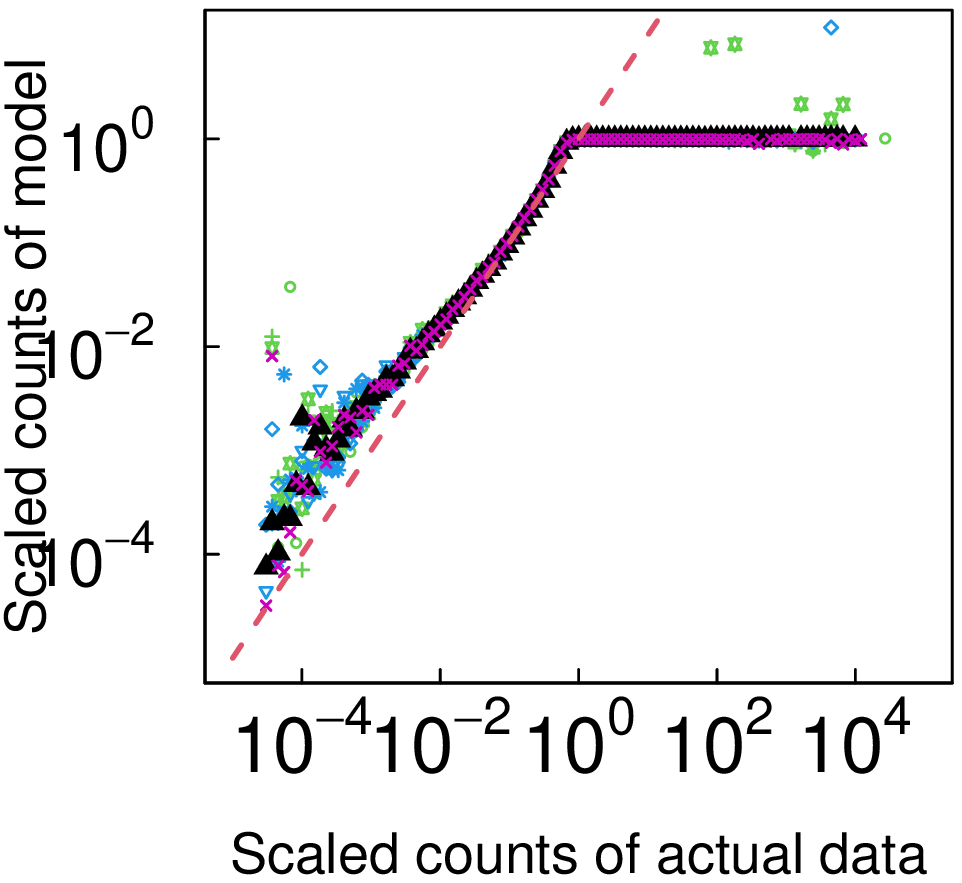}
                %\put(22,55){(a)}
                \put(23,72){(a)}
            \end{overpic}
        \end{minipage}
        \begin{minipage}{0.19\hsize}
            \begin{overpic}[width=3.8cm]{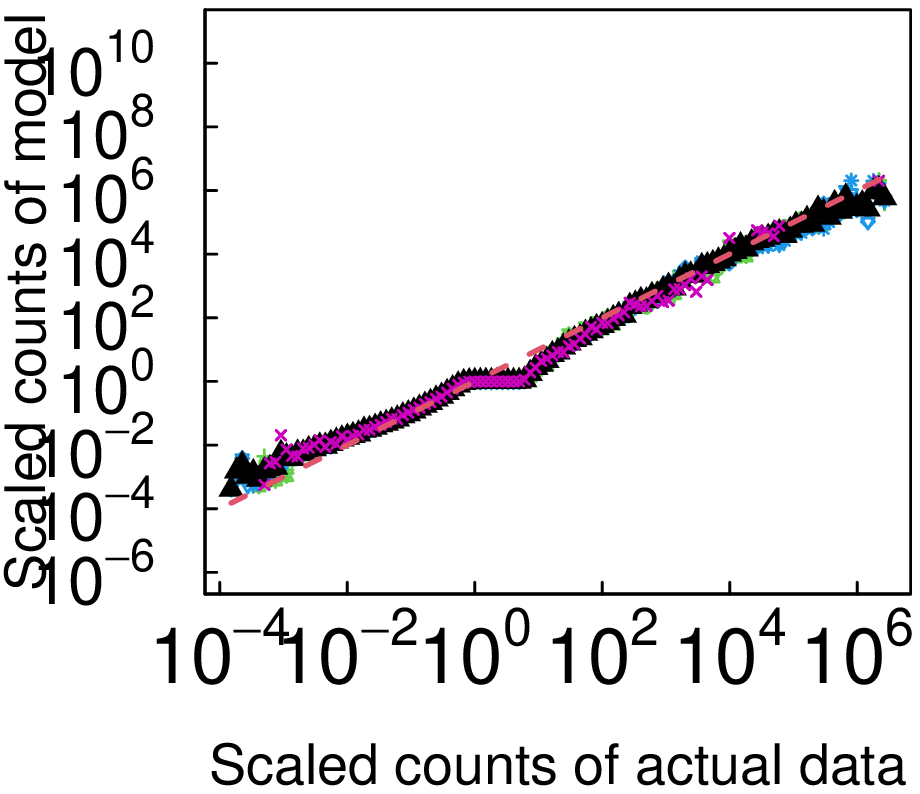}
                %\put(22,55){(a)}
                \put(23,72){(b)}
            \end{overpic}
        \end{minipage}
        \begin{minipage}{0.19\hsize}
        \begin{overpic}[width=3.8cm]{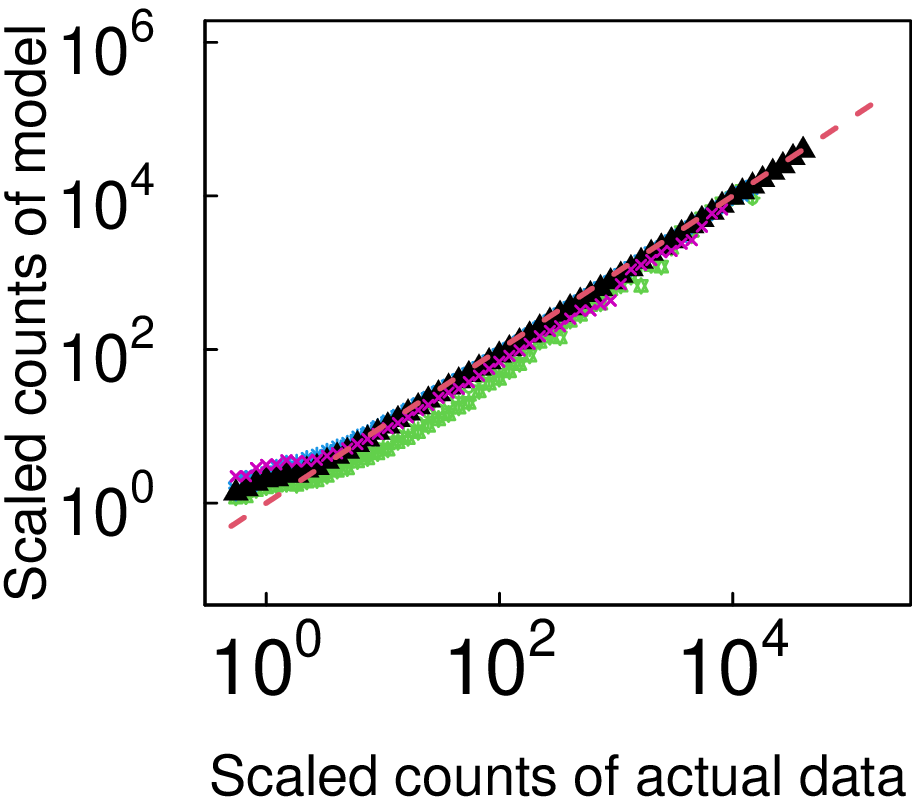}
            %\put(22,55){(a)}
            \put(23,72){(c)}
        \end{overpic}
        \end{minipage}
        \begin{minipage}{0.19\hsize}
        \begin{overpic}[width=3.8cm]{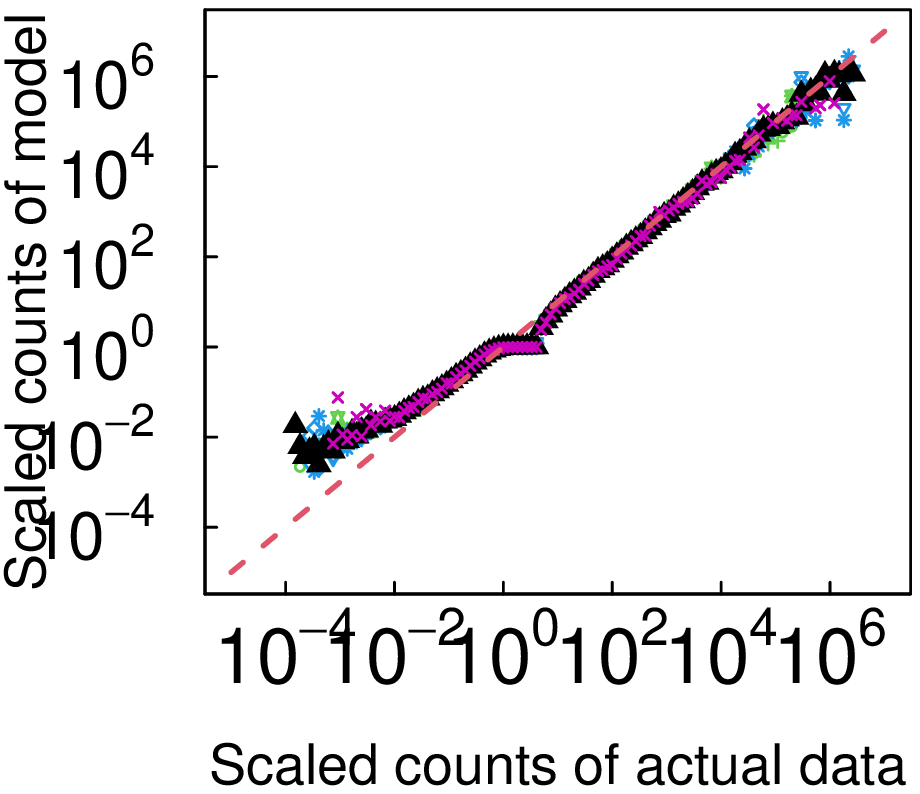}
            %\put(22,55){(a)}
            \put(23,72){(d)}
        \end{overpic}
        \end{minipage}
        \begin{minipage}{0.19\hsize}
        \begin{overpic}[width=3.8cm]{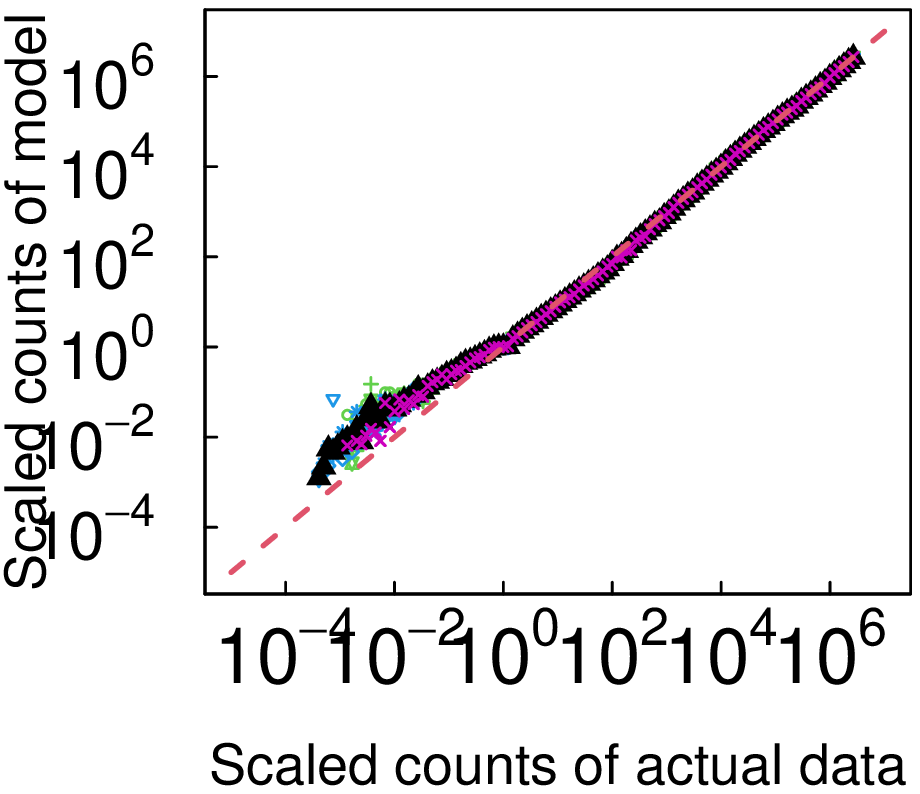}
            %\put(22,55){(a)}
            \put(23,72){(e)}
        \end{overpic}
        \end{minipage}
        \begin{minipage}{0.19\hsize}
            \begin{overpic}[width=3.8cm]{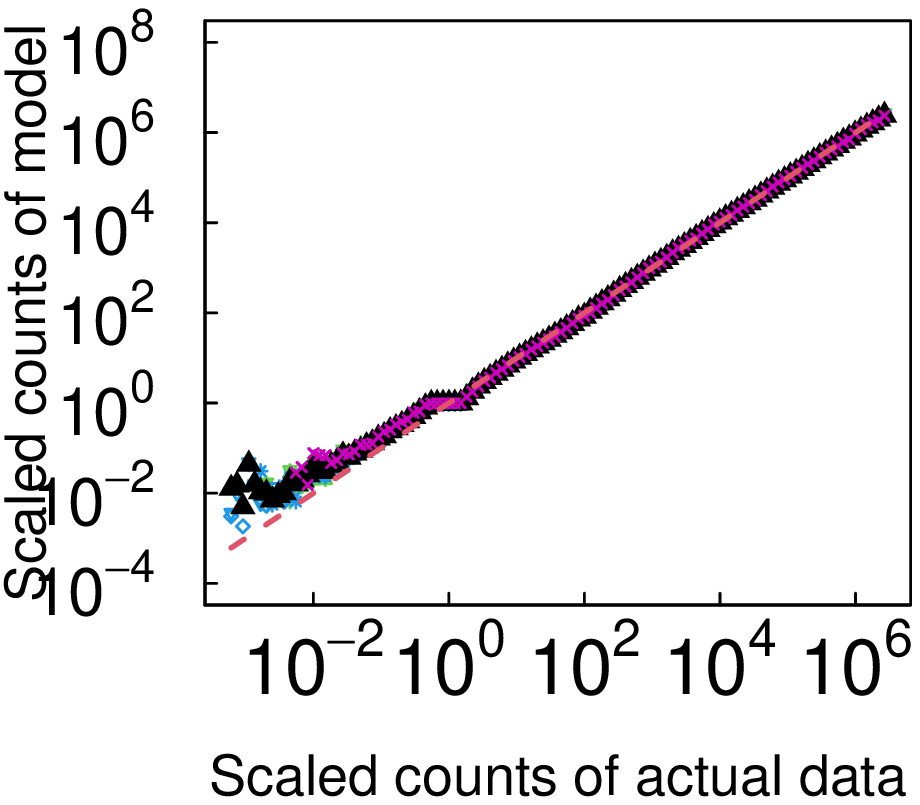}
                %\put(22,55){(a)}
                \put(23,72){(f)}
            \end{overpic}
        \end{minipage}
        \begin{minipage}{0.19\hsize}
            %\includegraphics[width=3.8cm]{transformed_count_ave_y_th.eps}
            %\begin{overpic}[width=3.8cm]{transformed_count_ave_y_th.eps}
            \begin{overpic}[width=3.8cm]{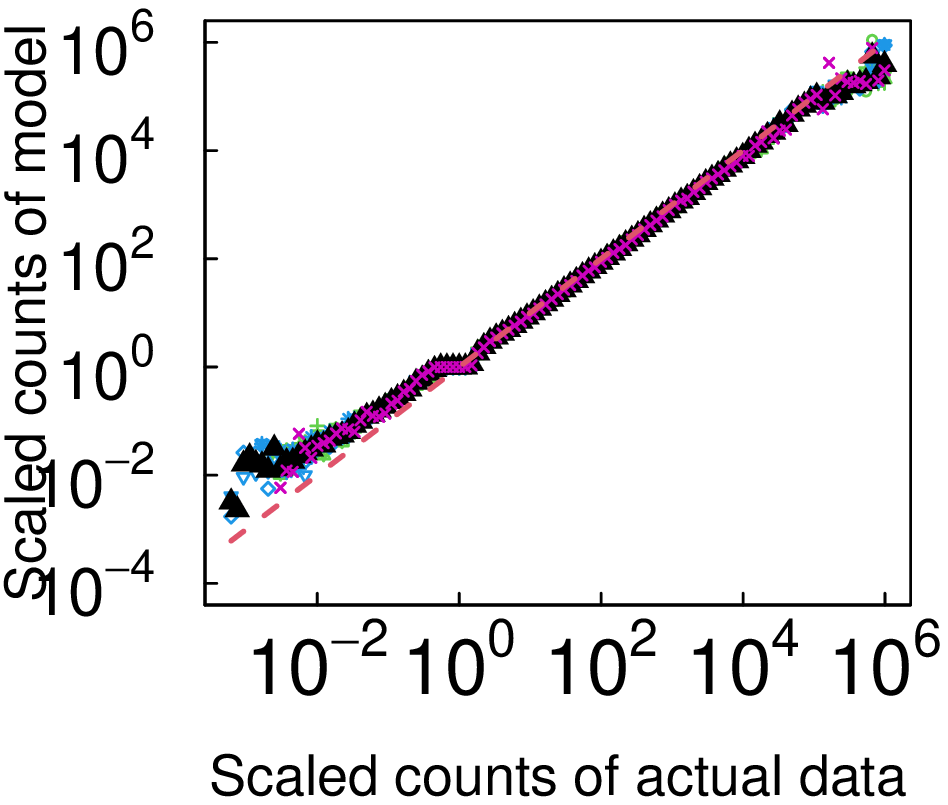}
                %\put(22,55){(a)}
                \put(23,72){(g)}
            \end{overpic}
        \end{minipage}
        \begin{minipage}{0.19\hsize}
            \begin{overpic}[width=3.8cm]{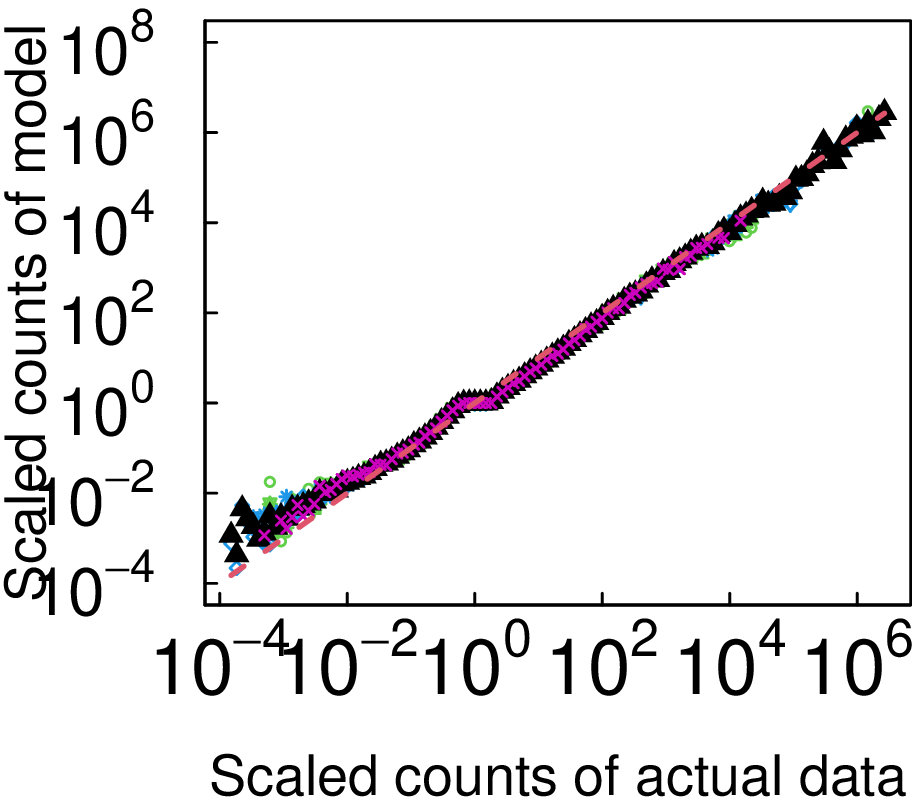}
                %\put(22,55){(a)}
                \put(23,72){(h)}
            \end{overpic}
            \end{minipage}
        \begin{minipage}{0.19\hsize}
        \begin{overpic}[width=3.8cm]{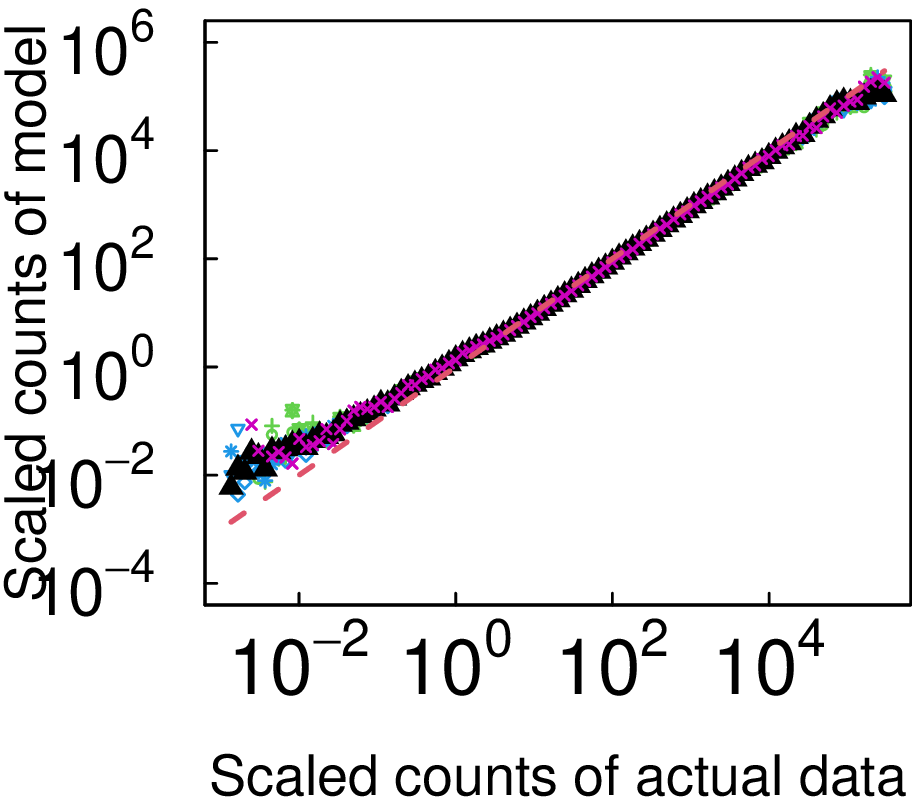}
            %\put(22,55){(a)}
            \put(23,72){(i)}
        \end{overpic}
        \end{minipage}
        \begin{minipage}{0.19\hsize}
            \begin{overpic}[width=3.8cm]{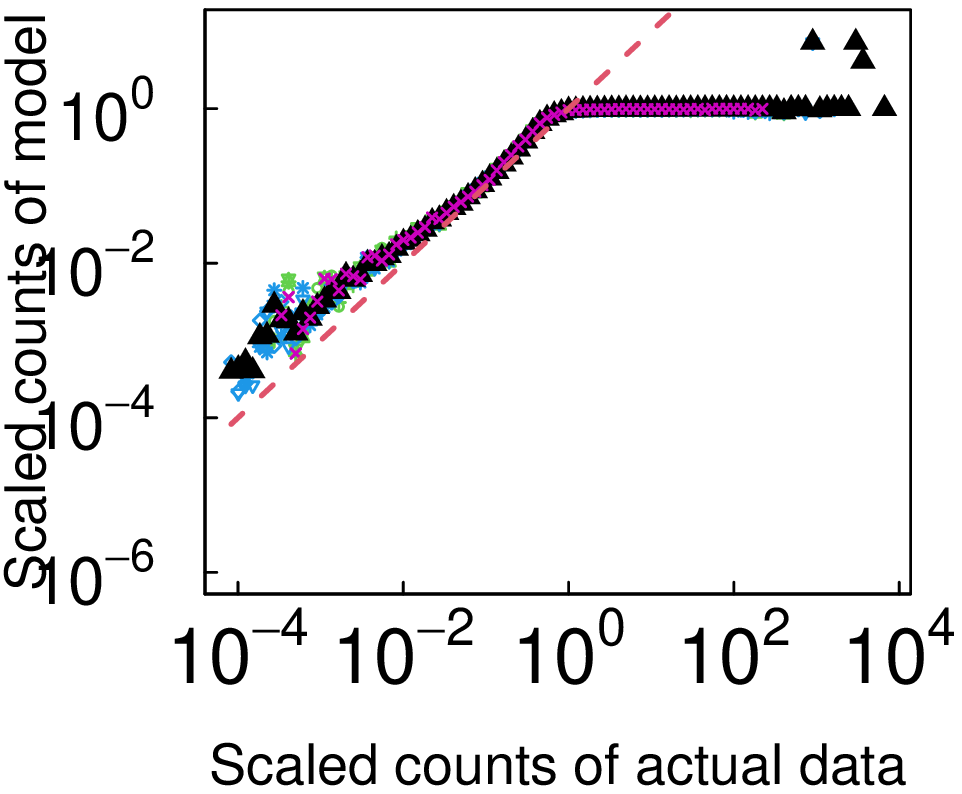}
                %\put(22,55){(a)}
                \put(23,72){(j)}
            \end{overpic}
        \end{minipage}
        \caption{
            Comparison of the descriptiveness of the dynamics of word count time series in various models using the solution of (scaled) differential equation $F(t|\alpha_j,Y_j)/|Y_j|$. If the model explains the data well, the graph will be close to the $y=x$ line shown by the red dashed line. 
            The y-axis corresponds to the left-hand side of $F(t|\alpha_j,Y_j)/|Y_j|$ (Theoretical value from $t$) and the x-axis to the right-hand side of $y_j(t)/|Y_j|$ (observed value), where the differential equation is written as $\frac{dy(t)}{dt}=f(y(t)|\alpha_j,Y_j)$ and the its solution as $y=F(t|\alpha_j,Y_j)$.
            Specifically, we plot the median quantity with respect to words given by Eq. \ref{app_eq_medi} using Eq. \ref{app_eq_p_y}, which is scaled by the model's scale parameter $Y$.
            %Specifically, we plot the median quantity scaled by the model's scale parameter $Y_j$ with respect to words given by Eq. \ref{xxx}.
            %Specifically, we plot the median quantity with respect to words given by Eq. \ref{app_eq_medi} (See also the subsection \ref{app_sec_medi} in sec. \ref{app_sec_model}). 
            The ensemble median over all data is plotted in black triangles, and also the ensemble median grouped by the proposed model's  (Eq. \ref{app_base_eq}) parameters $\alpha_j^{(0)}$ and $Y_j^{(0)}$ is plotted.
            In case of $Y_j^{(0)}>0$, we plot in green circles for $\alpha_j^{(0)}=0.5$, green pluses for $\alpha_j^{(0)}=1.0$, for green triangles up and down $\alpha_j^{(0)}=1.5$, blue diamonds for $\alpha_j^{(0)}=-0.5$, blue triangles pointing down for $\alpha_j^{(0)}=-1.0$ and in the case of $Y_j^{(0)}<0$ magenta crosses. 
        (a) Logistic equation given by Eq. \ref{app_base_logi},
        (b) $Y$-sign extended logistic equation given by Eq. \ref{app_base_logi2},
        (c) Single factor power-law model given by Eq. \ref{app_base_pow},
        (d)($Y$-sign extended) Bass model given by Eq. \ref{app_base_bass},
        (e) Constant model given by Eq. \ref{app_base_const},
        (f) First factor power-law model given by Eq. \ref{app_base_first},
        (g) Second factor power-law model (Proposed model) given by Eq. \ref{app_base_second},
        (h)Inside power-law model given by Eq. \ref{app_base_inside},
        (i)$Y$-sign positive-restricted second factor power-law model given by Eq. \ref{app_base_posi} and
        (j)$Y$-sign negative-restricted second factor power-law model given by Eq. \ref{app_base_nega}.
        }
        \label{app_fig_comp_y}
    \end{figure*}

%Up to the previous section, the explanatory power of the model was compared with the prediction of the velocity $dy(t)/dt$.
%Here, we compare the models using theoretical relationships rather than the differential equation relationship used in Fig. \ref{app_fig_comp_v}. 
\textcolor{black}{We discuss theoretical relationships that are not differential equations as discussed in the previous section (Fig .\ref{app_fig_comp_v}). }
In particular, we use the relation of the raw time $t_j=F^{(-1)}(y_j|\alpha_j,r_j,Y_j)$ in Fig. \ref{app_fig_comp_t},
scaled time $t_j/T_j=F^{(-1)}(y_j|\alpha_j,r_j,Y_j)/T_j$ in Fig. \ref{app_fig_comp_ts}, and the scaled word counts $y_j/Y_j=F(t_j|\alpha_j,r_j,Y_j)/Y_j$ in Fig. \ref{app_fig_comp_y}, where the solution of $\frac{dy(t)}{dt}=f(t|\alpha,r,Y)$ is denoted as $y(t)=F(t|\alpha,r,Y)$, its inverse function is denoted as $t=F^{(-1)}(y(t)|\alpha,r,Y)$ and $T_j$ is the length of the growth period of j-th word. In these figures, we plotted the ensemble median values in the same way as in Eq. \ref{app_eq_medi} \par
%  the explanatory power of the model for time $t$ and word count $y(t)$ is compared graphically. 
%前節までは，速度の予測で比較したが，時間の予測と$y$の予測でも精度をグラフィカルに比較することができる。
%これらより，$dy/td \propto y(t)^\alpha $に漸近するべき乗系のモデル（べき乗，前べき、後ろべき，後ろべき正限定）以外は
From these figures, it can be seen that both the first factor power-law model given in Eq. \ref{app_base_first} shown in the panel (f) and the second factor power-law model (the proposed model) given by Eq. \ref{app_base_second} shown in the panel (g) can explain the real data consistently. \par
Note that in Fig. \ref{app_fig_comp_ts} (g),  for the small $p$, even the proposed model cannot explain well the growth curve of the finite-time divergence (i.e., $\alpha^{(0)}>0, Y^{(0)}>0$) shown in green points. The reason is expected to be that the curve of finite-time divergence grows very slowly at the beginning; therefore, it is difficult to distinguish the growth dynamics from noise for a small $p$. Thus, the subsequent stagnation for small $p$ in Fig. \ref{app_fig_comp_ts} (g)  is considered stagnation until the growth dynamics of the model exceed the noise level. \par
Although in Fig. \ref{app_fig_comp_v} we could not distinguish between the proposed model shown in panel (g) and the positive model shown in panel (i), we can distinguished these models in Fig. \ref{app_fig_comp_ts}. In particular, as shown in Figs. \ref{app_fig_comp_ts} (i) the positive model given by Eq .\ref{app_base_posi} can not explain the growth curve of the S-curve ($Y^{(0)}<0$) shown in magenta crosses. \par  
% This result corresponds well with the velocity analysis. \par 
%なお,  Figs. \ref{xxx}, \ref{xxx} and  \ref{xxx} では，式 \ref{xxx}の$p_j(t)$と$q_j(t)$を以下の量に変えてプロットしている。

%Figs. \ref{}, \ref{xxx} and \ref{xxx} we have replaced $p_j(t)$ and $q_j(t)$ in Eq. \ref{xxx} with the following quantities.
The specific calculations used in Figs \ref{app_fig_comp_t}, \ref{app_fig_comp_ts} and \ref{app_fig_comp_y}, respectively, are as follows: We replace $p_j(t)$ as given in Eq. \ref{app_eq_p} and $q_j(t)$ given in Eq. \ref{app_eq_q} with the following quantities, 
%In the following, the solution of $\frac{dy(t)}{dt}=f(t)$ is denoted as $y(t)=F(t)$ and its inverse function is denoted as $t=F^{(-1)}(y)$ \par
%$t=F^{(-1)}(y(t))$.
%なお、$\frac{dy(t){dt}=f(t)}$の解を$y(t)=F(t)$とし，その逆関数を
%$t=F^{(-1)}(y(t))$としている。
in Fig. \ref{app_fig_comp_t}, we use the time $t$,  
%図\ref{xxx}は，時刻$t$の予測を上昇期間を１としてスケールしたものである。
\begin{equation}
    p_j(t)= F^{(-1)}(y_j(t)|Y_j,r_j,\alpha_j),  \label{app_eq_p_t} 
\end{equation}
\begin{equation}
    q_j(t)= t, \label{app_eq_q_ts} 
\end{equation}
in Fig. \ref{app_fig_comp_ts} we use the scaled time $t$, 
\begin{equation}
    p_j(t)= F^{(-1)}(y_j(t)|Y_j,r_j,\alpha_j)/T_j,   \label{app_eq_p_ts}
\end{equation}
\begin{equation}
    q_j(t)= t/T_j, \label{app_eq_q_ts} 
\end{equation}
and in Fig. \ref{app_fig_comp_y}, we use scaled word count $y$, 
%図\ref{xxx}は，時刻$y$の予測をスケールしたものである。
\begin{equation}
    p_j(t)= F(y_j(t)|Y_j,r_j,\alpha_j)/|Y_j|, \label{app_eq_p_y} 
\end{equation}
    \begin{equation}
    q_j(t)= y_j(t)/|Y_j|. \label{app_eq_q_y} 
\end{equation}
%\par
%これらの統計量を網羅的にうまく説明できてないことがわかる。
\subsection{Comparison of forecasting ability between models}
We compared the forecasting ability of the models introduced in section \ref{app_sec_model}. 
Specifically, we estimated the model parameters using the first 70 percent of the time series from the beginning of the growth period and predicted the remaining 30 percent of the time series.
%具体的には，７５パーセントの時系列を使いパラメーターを推定し，残り２５パーセントの予測精度を提案モデルとそのほかの時系列モデルと比較した.
%表\ref{table_pred}は，単語のうち，提案モデルのほうが予測が中央値絶対誤差の意味で改善した割合を満たしたものである. 
%中央値誤差は
As the same way as Eq. \ref{eq_delta} in subsection \ref{sec_pred} in section \ref{sec_vali}, the absolute mean error was used as a measure of fit to the data, 
\begin{equation}
\delta_j^{(model)}=Mean_{\{t|0.70T_j \leq t \leq T_j \}}[F^{(model)}(t|\alpha_j,r_j,Y_j)-y_j(t)| ],   
\end{equation}
where $F^{(model)}(t|\alpha_j,r_j,Y_j)$ is the solution of models, such as the logarithmic equation given by Eq. \ref{app_base_logi} and the negative model given by Eq .\ref{app_base_nega}.
%で定義した.
%where $Median_{\{t|t \in S\}}[x(t)]$は，集合$S$をみたす$t$に関するデータの中央値. 
%$Mean_{\{t|t \in S\}}[x(t)]$ is the mean of the data for $t$ satisfying the set $S$. 
%$\hat{y}_j(t)$ is the predicted value from the models such as the proposed model or the logistic equation.
%In the case of the proposed model, $F^{(model)}(t|\alpha_j,r_j,Y_j)$ the solutions of models which are given in Appendix \ref{xxx}, 
%SARIMA model, Prophet model (see \sec ...) and $\alpha$を正則化しない提案モデル, and we note $\delta_j^{(0)}$ as the absolute mean error of the extended logsitic equation. \par
Table \ref{app_table_pred} shows the winning ratio of the proposed model (i.e., the number ratio of the word-count time series for which the proposed model has a higher prediction accuracy than the comparison model).
%This ratio corresponds to the winning ratio of the proposed model and was defined as 
As the same way as Eq. \ref{eq_win}, the winning ratio against the proposed model is defined as 
\begin{eqnarray}
&&R^{(model)}= \\ \nonumber
&&\frac{\sum_{\{j| \delta_j^{(model)} > \delta_j^{(0)}, i \in W_s \}}1}{\sum_{\{i|\delta_j^{(model)} > \delta_j^{(0)}, j \in W_s \}}1+\sum_{\{i|\delta_j^{(model)} < \delta_j^{(0)}, j \in W_s \}}1}, 
\end{eqnarray}
where we estimated the parameters of each word using data from $t=1$ to $t=0.7 \times T_j$.
%Note that 
\par
%where $S$ is the sets of focused words, which corresponds to the limitation by the number of training data. 
%For this analysis, we used limited word count time series that have more than 12 training points.　\par
In this table, we also show the winning ratio for training errors using all data. 
\begin{eqnarray}
&&S^{(model)}= \\ \nonumber 
&&\frac{\sum_{\{j| \Delta_j^{(model)} > \Delta_j^{(0)}, j \in W_s \}}1}{\sum_{\{j|\Delta_j^{(model)} > \Delta_j^{(0)}, i \in W_s \}}1+\sum_{\{j|\Delta_j^{(model)} < \Delta_j^{(0)}, j \in W_s \}}1},
\end{eqnarray}
where the training error is defined as: 
\begin{equation}
    \Delta_j=Mean_{\{t|1 \leq t \leq T_j \}}[|F^{(model)}(t|\alpha_j,r_j,Y_j)-y_j(t)| ],  
\end{equation}
where we estimated the parameters of each word using data from $t=1$ to $t=T_j$.
\par
%をプロットしている。
\textcolor{black}{In addition to the models (a)-(g) mentioned in section \ref{app_sec_model}, we added the SARIMA model and Prophet model to the table \ref{app_table_pred}.} \par 
% In contrast to other models that use Eqs. \ref{app_eq_esitmate}, the parameters of No $\alpha$-reg model are estimated by minimizing the error without regularization of $\alpha$ given by Eq. \ref{app_eq_esitmate_no_reg}.
%\begin{equation}
 %   Median_{\{t=1,2S,\cdots,T_i\}}[y_i(t)-G^{(-1)}(t|\alpha,r,Y,y_0))]+\lambda \cdot (|r|+|Y|). 
%A\label{app_eq_esitmate2} 
%\end{equation}
%\par
\textcolor{black}From the table, we can see that the proposed model is relatively more accurate than the compared models in terms of prediction error $R$ except for the single power-law model (c) and the positive model (i). 
\textcolor{black}{The reason for these exceptions is probably that the two models (c) and (i) remove the $Y<0$ region from the parameter estimation and it prevents overfitting. 
Since the frequency of the words with $Y<0$ is small (see table \ref{table_para}), negative effects of this limitation of $Y>0$ on the winning ratio $R$ can be expected to be relatively small.
}
%これらの理由はおそらくこの２つのモデルが$Y<0$が負の領域を予測から除去してることが原因である。表\ref{table_para} より$Y<0$は少なく，これをはじめから無視することで過剰学習の可能性を落としていると思われる。　 }
We can also see from the table that the proposed model is as accurate or more accurate than the compared models in terms of the training error $S$ except for some models.
%The Prophet model has a significantly  prediction error $R^{(prophet)}<1$ than the training error win rate $S>1$. 
%This implies that
Here, we discuss the reasons for the exceptions:  
%For the No $\alpha$-reg model $S^{(no-\alpha \quad reg)}=0.44 <0.5$, this is an obvious consequence because of the property of regularization. 
%It is the obvious consequence that the training winning ratio of the proposed model against the No $\alpha$-reg model is low,. 
(i) For the Prophet model $S^{(Prophet)}=0.34 <0.5$, the result that the predictive ability is not as good as that of the proposed model, $R^{(Prophet)}=0.59 >0.5$ suggests overfitting, 
% Implies that the overfitting because the   $R^{(Prophet)}=0.64 >0.5$. 
%In cases where $S$ is clearly less than $0.5$,
%In 
and (ii) for the inside power-law model, taking $S^{(Inside)}=0.39 < 0.5$, 
it is unclear why the model has more explanatory power for training than that of the proposed model. \par
%Which remains in cases where $S$ is clearly less than 0.5, 
%it is unclear why the model has more explanatory power for training than the proposed model. \par
Note that because the proposed model includes the logistic model as a special case,  $S^{(logistic)}$ should take 1 theoretically, but in reality, $S^{(logistic)} \approx 0.85$. 
The reason for this is the limitation of the numerical optimization performed in this study, which does not yield a theoretically perfect minimum solution.
%With those exceptions, we can also see from the table that the proposed model is as accurate or more accurate than the compared models in terms of prediction error and training error $S$.
%Both in terms of prediction error and training error. 
%With those exceptions, 
%ノモデルは説明精度は高いが予測精度が低いため、少ないデータではオーバーフィットしている可能性もある。

    %\begin{figure}
        %\includegraphics[width=4.5cm]{"hist_powalpha_rY03.eps"}
        %\caption{Probability density function for $\alpha$ of the single factor power-low %model given by Eq. \ref{app_base_pow}. From the figure, we can see that the most %typical values are  $\alpha \sim -0.5$ shown in a red vertical dashed line.}
    %    \label{app_fig_onpow_hist}    
    %\end{figure}    

\begin{table*}
\centering
%\caption{Comparison of the fitted potential energy surfaces and ab initio benchmark electronic energy calculations}
    \begin{tabular}{lcc}
    Model & Prediction error R & Training Error S  \\
        \hline
    (a) Logistic & 0.64 [0.63,0.65] & 0.73 [0.73,0.74]   \\ 
    (b) $Y$-sign extended Logistic & 0.69 [0.68,0.70] & 0.65 [0.64,0.66]  \\ 
    (c) Single power-law & 0.45 [0.44,0.46] & 0.63 [0.63,0.64] \\
    (d) $Y$-sign extended Bass model & 0.75 [0.74,0.76] & 0.73[0.72-0.74]  \\
    Bass model ($Y<0$) & 0.67 [0.66,0.68] & 0.72 [0.71,0.73] \\
    (e) Constant factor & 0.65 [0.64,0.66] & 0.53 [0.52,0.54]   \\     
    (f) First power-law & 0.61 [0.60,0.62] & 0.49 [0.48,0.50]  \\
    (g) Second power-law (Proposed model) & - & - \\  
    (h) Inside power-law  & 0.63[0.62,0.64] & 0.39 [0.38,0.40] \\
    (i) Positive model & 0.46 [0.45,0.47] & 0.54 [0.53-0.55]  \\
    (j) Negative model & 0.64 [0.63,0.65] & 0.69 [0.68-0.70] \\
    SARIMA  & 0.52 [0.51,0.53] & 0.61 [0.60-0.62]  \\
    Prophet & 0.59 [0.58,0.60] & 0.34 [0.33-0.34]   \\
    \hline
    \multicolumn{3}{l}{95\% confidence intervals in brackets.}
    \end{tabular}
    \caption{Winning ratio of prediction and training errors of the proposed model compared to other models}
    \label{app_table_pred}
\end{table*}

\section{Examples of words in the four categories}
\label{app_example}
\textcolor{black}{In the section \ref{sec_model}, we presented the following four categorizations of words:
\begin{enumerate}
\item Convergence to a constant (S-curve) ($Y<0$, $\alpha>0$). 
\item Finite-time divergence (Deadline effects) ($Y>0$, $\alpha>0$)
\item Divergence after infinite time (Asymptotic power-law function) ($Y>0$, $\alpha<0$) 
\item Finite-time divergence of first-order derivatives ($Y<0$, $\alpha<0$) 
\end{enumerate}
In this section we give examples of words in each of those four categories for blog data. Specifically, (i) $\alpha>0$ and $Y<0$ are shown in table \ref{tab:alphapym}, (ii) $\alpha>0$ and $Y>0$ in table \ref{tab:alphapyp}, (iii) $\alpha<0$ and $Y>0$ in table \ref{tab:alphamyp}, and (iv) $\alpha<0$ and $Y<0$ in table \ref{tab:alphamym}.}
\begin{table*}[h]
\centering
\begin{tabular}{lp{4cm}p{4cm}p{4cm}}
%\hline
\textbf{Word} & \textbf{English Translation} & \textbf{Romanization} & \textbf{Note} \\
\hline
\Ja{タブレット端末}  & Tablet terminal & TaburretoTanmatsu & Information equipment \\
\Ja{リツイート}  & Retweet & RiTsuiito & SNS term \\
\Ja{HKT48}  & HKT48 & HKT48 &  Japanese idol girl group \\
\Ja{スカイツリー}  & Sky Tree & SukaiTsurii & Architectural monument \\
\Ja{LINEスタンプ}  & LINE stamp & LINEsutanpu & LINE is the most used mobile messenger application in Japan \\
\Ja{東京スカイツリー}  & Tokyo Sky Tree & ToukyouSukaiTsurii & Architectural monument \\
\Ja{イクメン}  & Ikumen & Ikumen & Catchphrase: men raising children \\
\Ja{黒子のバスケ}  & Kuroko's basketball & KurokonoBasuke & Cartoon title \\
\Ja{アンジュルム}  & Anjurum & Anjyurumu & Dance group \\
\Ja{ウェアラブル端末}  & Wearable device & UearaburuTanmatsu & Technology term \\
\Ja{国家戦略特区}  & National Strategic Special Zone & KokkaSenryakuTokku & Areas with special legal treatment \\
\Ja{USTREAM}  & USTREAM & USTREAM & Video distribution service \\
\Ja{小島瑠璃子}  & Ruriko Kojima & KojimaRuriko & Name of TV Personality \\
\Ja{坂口杏里}  & Anri Sakaguchi & SakaguchiAnri & Name of TV Personality \\
\Ja{ウサイン・ボルト}  & Usain Bolt & UsainBoruto & Track and field player \\
\Ja{アスキー・メディアワークス}  & ASCII Media Works & AsukiiMediaWaakusu & Publisher name \\
\Ja{神ってる}  & Be godly & Kamitteru & Baseball buzzword \\
\Ja{株式会社KADOKAWA}  & KADOKAWA Co., Ltd. & KabushikiGaishiya KADOKAWA & Media company name \\
\Ja{日笠陽子}  & Yoko Hikasa & HikasaYouko & Voice actor name \\
\Ja{オワコン}  & Owakon & Owakon & Slang for out of fashion \\
\Ja{人工多能性幹細胞}  & Artificial pluripotent stem cells & JinkouTanouseiKanSaibou & Biological term \\
\Ja{カイロ・レン}  & Cairo Len & KairoRen & Person name in the movie \\
\Ja{シシド・カフカ}  & Shishid Kafka & ShishidoKafuka & Actress name \\
\Ja{サッスオーロ}  & Sassuolo & Sassuoro & Name of soccer team \\
\Ja{肉食系女子}  & Carnivorous girls & NikuShyokuKeiJyoshi & Buzzwords for a woman who is active in romance \\
\Ja{水卜麻美}  & Asami Miura & MiuraAsami & The one of most famous female announcer name in Japan \\
\Ja{Astell\&Kern}  & Astell \& Kern & Astell \& Kern & Audio maker name \\
\Ja{井浦新}  & Arata Iura & IuraArata & Actor name \\
\Ja{美人時計}  & Beautiful woman clock & BijinDokei & Website that displays a beautiful woman every hour \\
\Ja{南野拓実}  & Takumi Minamino & MinaminoTakumi & Soccer player \\
\Ja{八戸学院光星}  & Hachinohe Gakuin Kousei & HachinoheGakuinKousei & High school famous for a baseball \\
\Ja{クリス・ヘムズワース}  & Chris Hemsworth & KurisuHemuzuwaasu & Actor name \\
\Ja{ソン・ジュンギ}  & Song Junggi & SonJyungi & Actor name \\
\Ja{ウェアラブルカメラ}  & Wearable camera & UearaburuKamera & Technology term \\
\Ja{稲嶺進}  & Susumu Inamine & InemineSusumu & Governor of okinawa prefecture name \\
\Ja{標的型メール}  & Target email & HyoutekiGataMeiru & Malicious Targeted Email \\
\Ja{スカパーJSAT}  & SKY PerfecTV JSAT & SukapaaJSAT & Satellite TV company name \\
\Ja{大塚角満}  & Kadoman Otsuka & OotsukakaKadoman & Game writer name \\
\Ja{ミライース}  & Daihatsu Mira e:S & Miraiisu & Car name \\
\Ja{ヤンマースタジアム長居}  & Yanmar Stadium Nagai & YanmaaSutajiamuNagai & Soccer stadium name \\
\Ja{涙活}  & Tear-jerking activities & RuiKatsu & Activities to relieve stress through conscious crying \\
\hline
\end{tabular}
\caption{Example of words for $\alpha>0 $ and $Y<0$ (Convergence to a constant or S-curve) for blog data. Specifically, the top 40 words in frequency that satisfy the following conditions: (i) a growth period of at least 48 timestamps (about 4 years) (ii) $\alpha> 0.1$ (iii) $Y<0$.  }
\label{tab:alphapym}
\end{table*}

\begin{table*}[h]
\centering
\begin{tabular}{lp{4cm}p{4cm}p{4cm}}
%\hline
\textbf{Word} & \textbf{English Translation} & \textbf{Romanization} & \textbf{Note} \\ 
\hline
\Ja{ビットコイン}  & Bitcoin & BittoKoin & Cryptocurrency name \\ 
\Ja{リオ五輪}  & Rio Olympics & RioGorin & Sport event \\ 
\Ja{リオデジャネイロ五輪}  & Rio de Janeiro Olympics & Riodejyaneirogorin & Sport event \\ 
\Ja{ソチ五輪}  & Sochi Olympics & SochiGorin & Sport event \\ 
\Ja{清武弘嗣}  & Hiroshi Kiyotake & KiyotakeHiroshi & Soccer player name \\ 
\Ja{ピコ太郎}  & Pico Taro & PikoTarou & Comedian whose videos attracted worldwide attention \\ 
\Ja{iPhone7}  & iPhone7 & iPhone7 & Smartphone model \\ 
\Ja{森脇稔}  & Minoru Moriwaki & MoriwaKiminoru & Baseball director name \\ 
\Ja{EU離脱}  & Withdrawal from the EU & EURidatsu & Political terms \\ 
\Ja{IFTTT}  & IFTTT & IFTTT & Web service name \\ 
\Ja{シャドバ}  & Shadowverse & Shyadoba & Game name \\ 
\Ja{ソードアート・オンライン}  & Sword Art Online & SoodoAatoOnrain & Game name \\ 
\Ja{コウノドリ}  & Kounodori: Dr. Stork & Kounodori & TV drama name \\ 
\Ja{山本美月}  & Mizuki Yamamoto & YamamotoMizuki &Young actoress name \\ 
\Ja{川内原発}  & Kawauchi nuclear power plant & KawauchiGenpatsu & Nuclear power plant name \\ 
\Ja{S660}  & S660 & S660 & Car name \\ 
\Ja{マクロン}  & Macron & Makuron & President French name \\ 
\Ja{ユニットコム}  & Unitcom & Yunnitokomu & PC shop name \\ 
\Ja{カズレーザー}  & Kaz laser & Kazureezaa & Comedian name \\ 
\Ja{オバマケア}  & Obamcare & ObamaKea & Political terms \\ 
\Ja{シェールガス}  & Shale gas & SheiruGasu & Economic term \\ 
\Ja{広州恒大}  & Guangzhou Kenzu & KoushyuuKoudai & Soccer team name \\ 
\Ja{Jアラート}  & J alert & JAraato & Japan emergency alert system \\ 
\Ja{Cortana}  & Cortana & Cortana & Digital assistant system name \\ 
\Ja{リオデジャネイロオリンピック}  & Rio de Janeiro Olympics & RiodejyaneiroOrinppiku & Sport event \\ 
\Ja{貴ノ岩}  & Takanoiwa & Takanoiwa & Sumo player name \\ 
\Ja{ソチオリンピック}  & Sochi Olympics & SochiOrinppiku & Sport event \\ 
\Ja{南シナ海問題}  & South China Sea problem & Minamishinakaimondai & Political terms \\ 
\Ja{オクタコア}  & Octa-core processor & OkutaKoa & Computer parts \\  
\Ja{海街diary}  & Sea Town DIARY & UmigaiDaiarii & Movie name \\ 
\Ja{社会保障と税の一体改革}  & Integrated reform of social security and tax & ShyakaiHoshiyouTo ZeinoIittaiKaikaku & Political term \\ 
\Ja{山田哲人}  & Tetsuto Yamada & YamadaTetsuto & Baseball player name \\ 
\Ja{清宮幸太郎}  & Kotaro Kiyomiya & SeimiyaKoutarou & Baseball player name \\ 
\Ja{ジュラシック・ワールド}  & Jurassic World & JyurasshikuWaarudo & Movie name \\ 
\Ja{マイナス金利政策}  & Negative interest rate policy & MainasuKinriSeisaku & Economic term \\ 
\Ja{永野芽郁}  & Meiku Nagano & NaganoMeiku &Young actoress name \\ 
\Ja{日本エレキテル連合}  & Japan Electric Generator Union & NipponErekiteruRengou & Comedian who won a buzzword award \\ 
\Ja{KCON}  & KCON & KCON & Events on Korean popular culture \\ 
\Ja{噴火警戒レベル}  & Eruption alert level & FunkaKeikaiReberu & Disaster term \\ 
\Ja{グノシー}  & Gnossy & Gunoshii & Information curation services on smartphones \\ 
\hline
\end{tabular}
\caption{Example of words for $\alpha>0 $ and $Y>0$ (Finite-time divergence or Deadline effects)  for blog data. Specifically, the top 40 words in frequency that satisfy the following conditions: (i) a growth period of at least 48 timestamps (about 4 years) (ii) $\alpha> 0.1$ (iii) $Y>0$. }
\label{tab:alphapyp}
\end{table*}

\begin{table*}[h]
\centering
\begin{tabular}{lp{4cm}p{4cm}p{4cm}}
%\hline
\textbf{Word} & \textbf{English Translation} & \textbf{Romanization} & \textbf{Note} \\
\hline
\Ja{スマホ}  & Smartphone & Sumaho & Smartphone \\
\Ja{モデルプレス}  & Model press & ModeruPuresu & Net news site \\
\Ja{インスタグラム}  & Instagram & Insutaguramu & SNS term \\
\Ja{Xperia}  & Xperia & Xperia & Smartphone model \\
\Ja{妊活}  & Pregnancy preparation activity & Ninkatsu & Buzz word \\
\Ja{SKE48}  & SKE48 & SKE48 & Japanese idol girl group \\
\Ja{タブレット端末}  & Tablet terminal & TaburretoTanmatsu & Information equipment \\
\Ja{bit.ly}  & bit.ly & bit.ly & Short URL service \\
\Ja{ニコニコ生放送}  & Nico Nico Live Broadcast & NikonikoNamaHousou & Internet live broadcasting Service \\
\Ja{ビッグデータ}  & Big data & Biggudeeta & Information technical term \\
\Ja{公式Twitter}  & Twitter verification mark & KoushikiTwitter & SNS term \\
\Ja{メルカリ}  & Mercari & Merukari & Online flea market application \\
\Ja{クラウドファンディング}  & Crowdfunding & KuraudoFuandingu & Economic term \\
\Ja{習近平}  & Xi Jinping & ShyuuKinpei & Chinese leader \\
\Ja{goo.gl}  & goo.gl & goo.gl & Short URL service \\
\Ja{熊本地震}  & Kumamoto earthquake & KumamotoJishin & Earthquake name \\
\Ja{ももクロ}  & Momokuro & Momokuro &  Japanese idol girl group \\
\Ja{ふるさと納税}  & Hometown tax payment & FurusatoNouzei & Tax deduction system \\
\Ja{TechinsightJapan}  & TechinsightJapan & TechinsightJapan & Net news site \\
\Ja{ニコ生}  & Nico Nico Live Broadcast & Nikonama & Internet live broadcasting Service \\
\Ja{スマートフォンアプリ}  & Smartphone app & SumaatofonApuri & Smartphone term \\
\Ja{ガラケー}  & Flip phone & Garakee & Feature phone before smartphone \\
\Ja{日馬富士}  & Hima Fuji & KusamaFujisaki & Sumo wrestler name \\
\Ja{朝日新聞デジタル}  & Asahi Shimbun Digital & AsahiShinbunDejitaru & News site \\
\Ja{Androidアプリ}  & Android app & Androidapuri & Smartphone term \\
\Ja{nanapi}  & nanapi & nanapi & Online know-how sharing service \\
\Ja{指原莉乃}  & Rino Sashihara & SasuharaRino & Young singer and actoress name \\
\Ja{スマホアプリ}  & Smartphone app & SumahonApuri & Smartphone term \\
\Ja{格安スマホ}  & Low-Cost smartphone & KakuyasuSumaho & Smartphone term \\
\Ja{みんなの党}  & Everyone's party & MinnanoTou & Political party name \\
\Ja{松山英樹}  & Hideki Matsuyama & MatsuyamaHideki & Golf player name \\
\Ja{スマートウォッチ}  & Smart watch & SumaatoUochi & Smartphone term \\
\Ja{pixiv}  & pixiv & pixiv &  Online community service for artists \\
\Ja{格安SIM}  &  Low-Cost Sim Card & KakuyasuSIM &  Smartphone term   \\
\Ja{スマホゲーム}  & Smartphone game & SumahoGemu & Smartphone term \\
\Ja{レコチョク}  & RecoChoku & Rekochiyoku & Music distribution site \\
\Ja{SHINee}  & Shinee & SHINee & Korean idol boy group \\
\Ja{AppStore}  & AppStore & AppStore & Smartphone term \\
\Ja{Twitterアカウント}  & Twitter account & TwitterAkaunto & SNS term \\
\Ja{広瀬すず}  & Hirose Suzu & HiroseSuzu & Young Actoress name \\
\hline
\end{tabular}
\caption{Examples of words for $\alpha<0 $ and $Y>0$ (Divergence after infinite time  or Asymptotic power-law function)  for blog data. Specifically, the top 40 words in frequency that satisfy the following conditions: (i) a growth period of at least 48 timestamps (about 4 years) (ii) $\alpha< -0.1$ (iii) $Y>0$.  }
\label{tab:alphamyp}
\end{table*}

\begin{table*}[h]
\centering
\begin{tabular}{lp{4cm}p{4cm}p{4cm}}
%\hline
\textbf{Word} & \textbf{English Translation} & \textbf{Romanization} & \textbf{Note} \\
\hline
\Ja{Android携帯}  & Android mobile & Androidkeitai & Smartphone term \\
\Ja{CULEN}  & CULEN & CULEN & Entertainment office name \\
\Ja{メッセニア}  & Messenia & Messenia & Place name \\
\hline
\end{tabular}
\caption{Example of words for $\alpha <0 $ and $Y<0$ (Finite-time divergence of first-order derivatives)  for blog data.  Specifically, these are words that satisfy the following conditions: (i) a growth period greater than 48 timestamps (about 4 years) (ii) $\alpha> 0.1$ (iii) $Y<0$.  Words in this category have a large discrepancy between the theoretical line and the real data, i.e., the model does not work well.}
\label{tab:alphamym}
\end{table*}
\end{document}